\documentclass[12pt]{article}
\usepackage{color}
\usepackage{amsmath} %Never write a paper without using amsmath for its many new commands 
\usepackage{amssymb} %Some extra symbols 
\usepackage[english]{babel}
\usepackage{graphics}
\usepackage[pdftex]{graphicx}
\usepackage{xspace}
\usepackage[gen]{eurosym}
\usepackage[font=footnotesize,format=plain,labelfont=bf,up,textfont=it,up]{caption}
\usepackage{graphics}
\usepackage{subfig}
\usepackage{latexsym}
\usepackage[mathscr]{eucal} 
\usepackage{latexsym}
\usepackage[left]{lineno}
\newcommand{\nue}{\ensuremath{\nu_{e}}\xspace}

\newcommand{\numu}{\ensuremath{\nu_{\mu}}\xspace}

\newcommand{\nubarmu}{\ensuremath{\overline{\nu}_{\mu}}\xspace}

\newcommand{\numunue}{\ensuremath{\nu_\mu \rightarrow \nu_e}\xspace}
\newcommand{\numunust}{\ensuremath{\nu_\mu \rightarrow \nu_s}\xspace}

\newcommand{\boss}[2]{\ensuremath{\rlap{\kern-2.5pt\ensuremath{\overset{\scriptscriptstyle(-)}{\phantom{#1}}}}{\ensuremath{{#1}_{#2}}}}}

\newcommand{\BARII}                   {1}
\newcommand{\BARIU}                 {2}
\newcommand{\BOLOGNAI}         {3}
\newcommand{\BOLOGNAU}       {4}
\newcommand{\CERN}                  {5}
\newcommand{\FRASCATI}          {6}
\newcommand{\LECCEI}             {7}
\newcommand{\LECCEU}           {8}
\newcommand{\LECCEING}       {9}
\newcommand{\LEBEDEV}          {10}
\newcommand{\MSU}                 {11}
\newcommand{\PADOVAI}          {12}
\newcommand{\PADOVAU}        {13}
\newcommand{\ROMAU}            {14}
\newcommand{\ZAGREB}             {15}
\newcommand{\NessieInstitutes}{
\BARII        . INFN, Sezione di Bari, 70126 Bari, Italy \\
\BARIU        . Dipartimento di Fisica dell'Universit\`a  di Bari, 70126 Bari, Italy \\
\BOLOGNAI     . INFN, Sezione di Bologna, 40127 Bologna, Italy \\
\BOLOGNAU     . Dipartimento di Fisica dell'Universit\`a  di Bologna, 40127 Bologna, Italy \\
\CERN             .  European Organization for Nuclear Research (CERN), Geneva, Switzerland \\
\FRASCATI    . Laboratori Nazionali di Frascati dell'INFN, 00044 Frascati (Roma), Italy \\
\LECCEI        . INFN, Sezione di Lecce, 73100 Lecce, Italy \\
\LECCEU        . Dipartimento di Matematica e Fisica dell'Universit\`a  del Salento, 73100 Lecce, Italy \\
\LECCEING        . Dipartimento di Ingegneria dell'Innovazione dell'Universit\`a  del Salento, 73100 Lecce, Italy \\
\LEBEDEV     . Lebedev Physical Institute of Russian Academy of Science, Leninskie pr., 53, 119333 Moscow, Russia.\\
\MSU       . Lomonosov Moscow State University (MSU SINP), 1(2) Leninskie gory, GSP-1, 119991 Moscow, Russia\\
\PADOVAI      . INFN, Sezione di Padova, 35131 Padova, Italy \\
\PADOVAU      . Dipartimento di Fisica e Astronomia dell'Universit\`a  di Padova, 35131 Padova, Italy \\
\ROMAU        . Dipartimento di Fisica dell'Universit\`a  di Roma ``La Sapienza" and INFN, 00185 Roma, Italy \\
\ZAGREB             . Rudjer Boskovic Institute, Bijenicka 54, 10002 Zagreb, Croatia\\
\ddag~Also at Centre de Recherche en Astronomie Astrophysique et Geophysique, Alger, Algeria\\
}
\newcommand{\NessieAuthorList}{
\noindent 
A.~Anokhina$^{\MSU}$,
A.~Bagulya$^{\LEBEDEV}$,
M.~Benettoni$^{\PADOVAI}$,
P.~Bernardini$^{\LECCEU, \LECCEI}$,
R.~Brugnera$^{\PADOVAU, \PADOVAI}$,
M.~Calabrese$^{\LECCEI}$,
A.~Cecchetti$^{\FRASCATI}$,
S.~Cecchini$^{\BOLOGNAI}$,
M.~Chernyavskiy$^{\LEBEDEV}$,
P.~Creti$^{\LECCEI}$,
F.~Dal~Corso$^{\PADOVAI}$,
O.~Dalkarov$^{\LEBEDEV}$,
A.~Del~Prete$^{\LECCEING}$,
G.~De~Robertis$^{\BARII}$,
M.~De~Serio$^{\BARIU, \BARII}$,
L.~Degli~Esposti$^{\BOLOGNAI}$,
D.~Di~Ferdinando$^{\BOLOGNAI}$,
S.~Dusini$^{\PADOVAI}$,
T.~Dzhatdoev$^{\MSU}$,
C.~Fanin$^{\PADOVAI}$,
R.~A.~Fini$^{\BARII}$,
G.~Fiore$^{\LECCEI}$,
A.~Garfagnini$^{\PADOVAU, \PADOVAI}$,
S.~Golovanov$^{\LEBEDEV}$,
M.~Guerzoni$^{\BOLOGNAI}$,
B.~Klicek$^{\ZAGREB}$,
U.~Kose$^{\CERN}$,
K.~Jakovcic$^{\ZAGREB}$,
G.~Laurenti$^{\BOLOGNAI}$,
I.~Lippi$^{\PADOVAI}$,
F.~Loddo$^{\BARII}$,
A.~Longhin$^{\FRASCATI}$,
M.~Malenica$^{\ZAGREB}$,
G.~Mancarella$^{\LECCEU, \LECCEI}$,
G.~Mandrioli$^{\BOLOGNAI}$,
A.~Margiotta$^{\BOLOGNAU, \BOLOGNAI}$,
G.~Marsella$^{\LECCEU, \LECCEI}$,
N.~Mauri$^{\FRASCATI}$,
E.~Medinaceli$^{\PADOVAU, \PADOVAI}$,
A.~Mengucci$^{\FRASCATI}$,
R.~Mingazheva$^{\LEBEDEV}$,
O.~Morgunova$^{\MSU}$,
M.~T.~Muciaccia$^{\BARIU, \BARII}$,
M.~Nessi$^{\CERN}$,
D.~Orecchini$^{\FRASCATI}$,
A.~Paoloni$^{\FRASCATI}$,
G.~Papadia$^{\LECCEING}$,
L.~Paparella$^{\BARIU, \BARII}$,
L.~Pasqualini$^{\BOLOGNAU,\BOLOGNAI}$,
A.~Pastore$^{\BARII}$,
L.~Patrizii$^{\BOLOGNAI}$,
N.~Polukhina$^{\LEBEDEV}$,
M.~Pozzato$^{\BOLOGNAU, \BOLOGNAI}$,
M.~Roda$^{\PADOVAU, \PADOVAI}$,
T.~Roganova$^{\MSU}$,
G.~Rosa$^{\ROMAU}$,
Z.~Sahnoun$^{\BOLOGNAI \ddag}$,
S.~Simone$^{\BARIU, \BARII}$,
C.~Sirignano$^{\PADOVAU, \PADOVAI}$,
G.~Sirri$^{\BOLOGNAI}$,
M.~Spurio$^{\BOLOGNAU, \BOLOGNAI}$,
L.~Stanco$^{\PADOVAI, a}$,
N.~Starkov$^{\LEBEDEV}$,
M.~Stipcevic$^{\ZAGREB}$,
A.~Surdo$^{\LECCEI}$,
M.~Tenti$^{\BOLOGNAU, \BOLOGNAI}$,
V.~Togo$^{\BOLOGNAI}$,
M.~Ventura$^{\FRASCATI}$ and
M.~Vladymyrov$^{\LEBEDEV}$.\\
{\em (a)} Spokesperson
}
  
\begin{document}

%\linenumbers

\renewcommand{\thefootnote}{\alph{footnote}}
  \thispagestyle{empty}
\title{\bf Prospects for the measurement of $\numu$~disappearance at the FNAL--Booster}
\date{}
\maketitle

%\centerline{\em September 2011 v1.32. DRAFT+3 (11 October)}
\vspace{-2 cm}

\centerline{\em The NESSiE Collaboration}

\vspace{1 cm}

\abstract{

Neutrino physics is nowadays receiving more and more attention as a possible source of information for the long--standing problem
of new physics beyond the Standard Model. The recent measurement of the mixing angle $\theta_{13}$ in the standard 
mixing oscillation scenario encourages us to pursue the still missing results on leptonic CP violation and absolute neutrino
masses. However, puzzling measurements exist that deserve an exhaustive evaluation.

The NESSiE Collaboration has been setup to undertake conclusive experiments to clarify the  {\em muon--neutrino disappearance} measurements at small $L/E$,
which will be able to put severe constraints to models with more than the three-standard neutrinos, or even to robustly measure 
the presence of a new kind of neutrino oscillation for the first time.

To this aim the use of the current FNAL--Booster neutrino beam for a Short--Baseline experiment
has been carefully evaluated. This proposal  refers to the use of magnetic spectrometers at two different sites, Near and Far.
Their positions have been extensively studied, together with the possible performances of two OPERA--like spectrometers.
The proposal is constrained by availability of existing hardware and a time--schedule compatible with the CERN 
project for a new more performant neutrino beam, which will nicely extend the physics results achievable at the Booster.

The possible FNAL experiment will allow to clarify the current $\numu$ disappearance tension with $\nue$ appearance
and disappearance at the eV mass scale. Instead, a new CERN neutrino beam would allow a further span in the parameter space together 
with a refined control of systematics and, more relevant, the measurement of the antineutrino sector, by upgrading the spectrometer with detectors currently under R\&D study.
}

\clearpage

\author{\noindent \\ \NessieAuthorList }

\begin{flushleft}
\footnotesize{\NessieInstitutes }
\end{flushleft}

\clearpage
\tableofcontents
  
\clearpage

\section{Introduction and Physics overview}\label{sec:intro}

The unfolding of the physics of the neutrino is a long and exciting story spanning the last 80 years. Over this time the interchange of 
theoretical hypotheses and experimental facts has been one of the most fruitful demonstrations of the progress of knowledge in physics.
The work of the last decade and a half finally brought a coherent picture within the Standard Model (SM) (or some small extensions of it),
namely the mixing of three neutrino flavour states with three  $\nu_1$, $\nu_2$ and $\nu_3$ mass eigenstates. 
The last unknown mixing angle, $\theta_{13}$, was recently measured~\cite{theta13-DB, theta13-RE, theta13-DC, theta13-T2} but still
many questions remain unanswered to completely settle the scenario: the absolute masses, 
the Majorana/Dirac nature and the existence and magnitude of leptonic CP violation.
Answers to these questions will beautifully complete the (standard) three--neutrino model but they will hardly provide an insight into new physics 
Beyond the Standard Model (BSM). 
Many relevant questions will stay open: the reason for neutrinos, the relation between the
leptonic and hadronic sectors of the SM, the origin of Dark Matter and, overall, where and how to look for BSM physics.
Neutrinos may be an excellent source of BSM physics and their story is supporting that at length.

There are actually several experimental hints for deviations from the ``coherent'' picture described above.
Many unexpected results, not statistically significant on a single basis, appeared also in the last decade and a half,
bringing attention to the hypothesis of the existence of {\em sterile neutrinos}~\cite{pontecorvo}. We refer to a White Paper~\cite{whitepaper},
which contains a comprehensive review of these issues. We also refer to the very recent discovery of B-modes in the polarization pattern of the 
Cosmic Microwave Background by BICEP--2~\cite{bicep2} and its possible
implication for sterile neutrinos (see e.g. ~\cite{ster-bicep2} and references therein).
In particular we would like to focus about tensions in many phenomenological models that grew up with experimental results on 
neutrino/antineutrino oscillations at Short--Baseline (SBL) and with the more recent, carefully recomputed, antineutrino fluxes from 
nuclear reactors. The main source of tension corresponds to the lack so far of any \numu disappearance signal~\cite{kopp-tension}.

This scenario promoted several proposals for new, exhaustive evaluations of the neutrino behaviour
at SBL. Since end 2012 CERN is undergoing a study to setup a Neutrino Platform, with a new
infrastructure at the North Area that may eventually include a new neutrino beam~\cite{edms}. Meanwhile FNAL is
welcoming proposal of experiments to exploit the physics potentialities of their two existing neutrino beams. 
Two very recent proposals have been submitted for experiments of SBL at the Booster beam, to complement
the soon starting of MicroBooNE~\cite{microboone}. The two proposals 
from the LAr1--ND~\cite{LAr1-ND} and the ICARUS~\cite{ICARUSFNAL} Collaborations
are both based on the Liquid Argon technology and aim to measure the \nue appearance at SBL, with some possibilities to 
study also the \numu disappearance.

\begin{figure}[htbp]
\centering
\includegraphics[width=12cm]{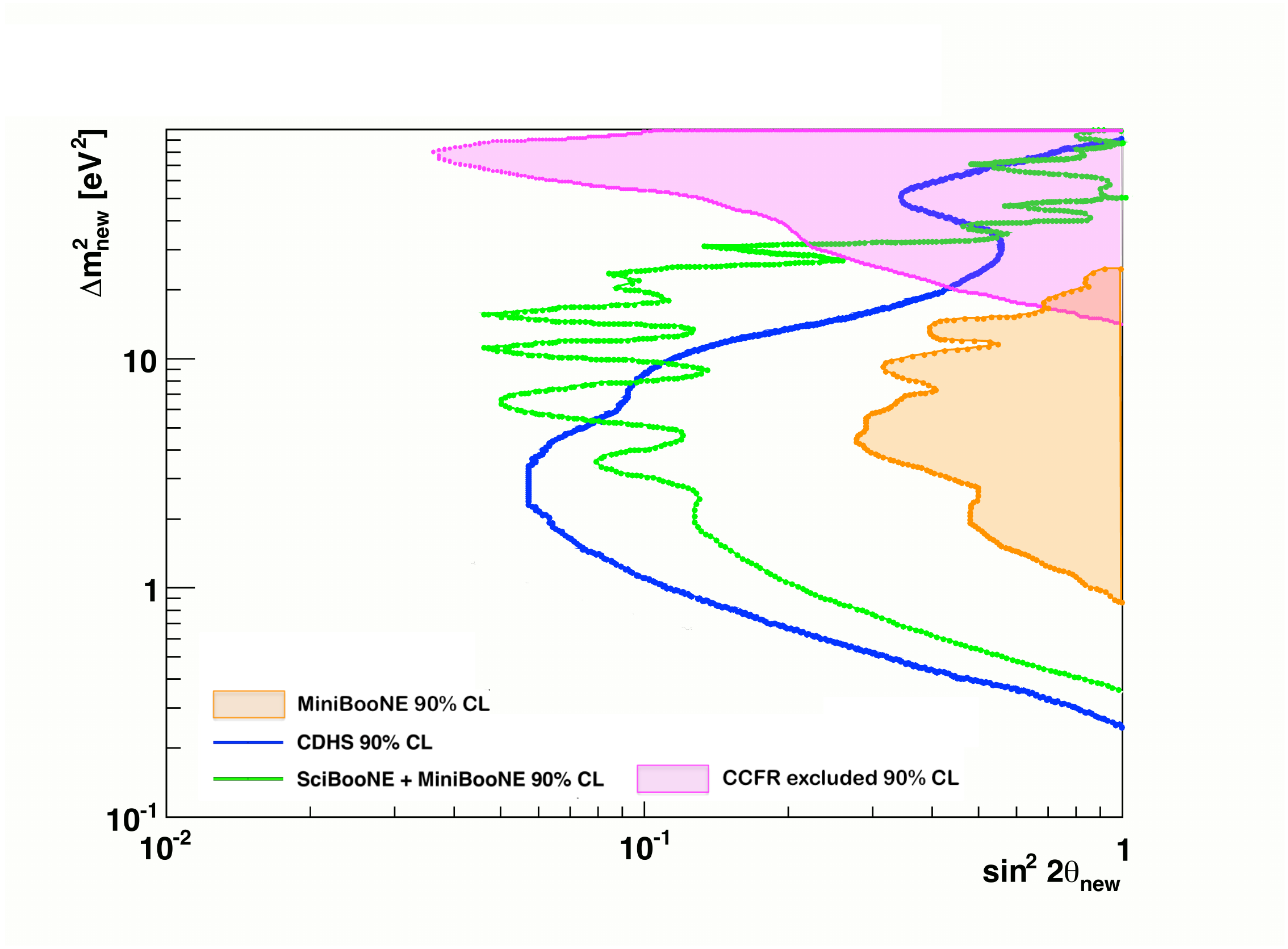}
\caption{The current exclusion limits on the \numu disappearance searches at the eV$^2$ scale.
Blue (green) line: old (recent) exclusion limits on \numu from previous CDHS~\cite{CDHS} and recent MiniBooNE/SciBooNE~\cite{mini-sci-mu1, mini-sci-mu2} measurements.
The two filled areas correspond to the exclusion limits on the \nubarmu from CCFR~\cite{ccfr} and MiniBooNE--alone~\cite{mini-mu} experiments (at 90\% C.L.).}
\label{fig:old-res}
\end{figure}

The present proposal is based on the following considerations:
\begin{itemize}
\item the measurement of \numu behaviour is mandatory for a correct interpretation of the \nue data, even in case of 
a null result for the latter;
\item a decoupled measurement of \nue and \numu interactions will allow to keep at low level systematics due to the different cross--sections;
\item very massive detectors are mandatory to collect a large number of events and therefore improve the disentangling
of systematic effects.
\item the current experimental knowledge of the \numu disappearance at SBL is limited by the dated CDHS experiment~\cite{CDHS} 
and the more recent results from MiniBooNE~\cite{mini-mu}, a joint MiniBooNE/SciBooNE analysis~\cite{mini-sci-mu1, mini-sci-mu2} 
and MINOS~\cite{minos}. The latter results slightly extend the $\nu_\mu$ disappearance exclusion region by CDHS.
Fig.~\ref{fig:old-res} shows the excluded regions in the space parameters for the \numunust oscillation, obtained through
\numu disappearance experiments. The mixing angle is denoted as $\theta_{new}$ and the squared mass difference as
$\Delta m^2_{new}$. The region with $\sin^2(2\theta_{new})<0.1$ is largely still unconstrained.
%however still leaving out the mixing angle region below 0.1 in $\sin^2(2\theta)$ (see Fig.~\ref{fig:old-res}).
\end{itemize}

Motivated by this scenario a detailed study of the physics case for the FNAL--Booster beam was performed. 
The study follows the similar analysis performed for the CERN--PS and CERN--SPS cases~\cite{nessie,larnessie} and the study in~\cite{stancoetal}.
However, we tackled many specific detector configurations investigating experimental aspects not fully covered by the LAr detection. This includes 
the measurements of the lepton charge on event--by--event basis and its energy over a wide range.
Indeed, the muons from Charged Current (CC) neutrino interactions play an important role in disentangling 
different phenomenological scenarios provided their charge state is determined. Also,
the study of muon appearance/disappearance can benefit of the large statistics of CC muon events from the primary 
neutrino beam. In the FNAL--Booster beam the antineutrino contribution is rather small and it then becomes a systematics
effects to be taken into account.

Results of our study are reported in detail in this proposal. We aim to design, construct and install two spectrometers at each site,
``Near'' (110 m in line with the beam) and ``Far'' (710 m on surface), of the SBL FNAL--Booster, fully compatible with the already 
proposed LAr detectors. 
Profiting of the large mass of the two spectrometer--systems their stand--alone performances have been exploited for the \numu 
disappearance study. Besides, complementarity measurements with LAr can be undertaken to increase their control of
systematic errors.

Some important practical constraints were assumed in order to draft the proposal on a
conservative, manageable basis, and maintain it sustainable in terms of time--scale and cost.
Well known technologies were considered as well as re--using large parts of existing detectors. 

The momentum and charge state measurements of muons in a wide energy range, from few hundreds MeV/c to several 
GeV/c, over a $> \ 50\ {\rm m}^2$ surface, is an extremely challenging task if constrained by an order of a million~\euro\ budget
for construction and installation.  Running costs have to be kept at low level, too.

%The detection of muons (momentum and charge state measurements) in a wide energy range (from few hundreds MeV to several 
%GeV) over a very large surface ($> \ 50\ m^2$) is made extremely challenging by a 10 (and not 100) millions~\euro budget constrain
%or construction and installation.  Running costs have to be kept at low level, too.

%The experiment is identified throughout the proposal with the acronym NESSiE (Neutrino Experiment with SpectrometerS in Europe).

We believe to have succeeded to develop a substantial proposal that, by keeping the systematic error at the level of $1\div 2$\% for the measurements of the 
\numu interactions, will allow to:
\begin{itemize}
	\item
measure \numu disappearance in the almost entire available momentum range ($p_{\mu}\ge 500$~MeV/c). This is a key information in 
rejecting/observing the anomalies over the whole expected parameter space of sterile neutrino oscillations, since the latter range drives the
$\Delta m^2_{new}$ interval;
\item collect a very large statistical sample to allow to span the oscillation mixing parameter up to till un--explored regions ($\sin^2(2\theta_{new})\gtrsim 0.01$);
	\item
measure the neutrino flux at the Near detector, in the relevant muon momentum range, to keep the systematic errors at the lowest possible values;
\item measure the sign of the muon charge to
separate \numu from \nubarmu to control systematics.
\end{itemize}

In the next Section a detailed and exhaustive study using different simulations of the FNAL--Booster
is reported, to work out a realistic choice for the detector configuration, and to correctly identify some sources of the systematic effects.
In Sect.~\ref{sec:spect1} the choice and the outlook of the spectrometers from OPERA~\cite{bopera} are discussed. 
After a brief discussion about the possibility to add a small target section (Sect.~\ref{sec:scinti}) to allow an on--site
separation of CC and neutral current (NC) events, the full details of the Monte Carlo simulation and reconstruction for neutrino events (Sect.~\ref{sec:MC}) 
are illustrated.
Sections~\ref{sec:mech-struc} and ~\ref{sec:mag-SC}  deal with the technical definition of the mechanical structure and
electrical setting--up for the magnets. Sect.~\ref{sec:rpc} illustrates the use of the already available detectors from OPERA. 
Sect.~\ref{sec:bck} debates about background levels to be taken into account
for the data taking, which is described in Sect.~\ref{sec:daq}. 
The following Section reports about FNAL setting--up, schedule and costs. 
The physics performances are extensively described in
Sect.~\ref{sec:spect2}. To this regard several different approaches of the statistical methods to get the achievable sensitivities
have been taken into account. Comprehensive discussions on the results have been outlined.
Finally conclusions are recapped.

%Start-ANDREA%%%%%%%%%%%%%%%%%%%%%%%%%%%%%%%%%%%%%%%%%%%%%%%%%%%%%%%%

\section{Beam evaluation and constraints}\label{sec:beam}

In this section we walked--through the exhaustive study on the characteristics of the FNAL--Booster $\nu_{\mu}$ beam we underwent. Beam convolution with the muon detection systems described later was also carefully taken into account.

\subsection{The Booster Neutrino Beam (BNB)}

The neutrino beam~\cite{G4BNBflux} is produced using protons with a
kinetic energy of 8 GeV extracted from the Booster and directed to a
Beryllium cylindrical target with a length of 71~cm and a diameter of
1~cm.  The target is surrounded by a magnetic focusing horn pulsed
with a 170~kA current at a rate of 5~Hz. Secondary mesons are
projected into a 50~m long decay pipe where they are allowed to decay
in flight before being absorbed by an absorber and the ground
material.  An additional absorber can be placed in the decay pipe at
about 25~m from the target\footnote{\noindent This configuration,
  which is not currently in use, could eventually alter the beam
  properties (i.e. providing a more point--like source for the Near
  site) thus allowing for extra experimental constraints on the
  systematic errors.}.  Neutrinos travel about horizontally at a depth
of about 7~m underground.

Proton batches typically contain about $4.5\times 10^{12}$
protons. They have a duration of 1.6 $\mu$s and are subdivided
into 84 bunches. Bunches are about 4~ns wide and separated by about
19~ns. The rate of batch extraction is limited by the horn pulsing at
5~Hz. This timing structure provides a very powerful constraint to the
background from cosmic rays.

%In the BNB 8 GeV protons are extracted from the Booster and steered to strike a 71~cm long, 1~cm
%diameter beryllium target. This target sits at the upstream end of a magnetic focusing horn that is pulsed with 
%170~kA to focus the mesons produced by the p-Be interactions. Following the horn is a 50~m long decay pipe
%that gives the pions a chance to decay and produce neutrinos, before the mesons encounter an absorber and then
%dirt which serve to remove all but the neutrinos from the beam.

%In its current mode of operation, the horn focuses $\pi^+$ and
%defocuses $\pi^-$ thus producing a $\nu_\mu$ beam.  By reversing the
%polarity of the horn current, $\pi^-$ are focused and a predominantly
%$\bar{\nu}_\mu$ beam is created.  

\subsection{The Far--to--Near ratio (FNR)}

\begin{figure}[htbp]
\centering
\includegraphics[scale=0.5,type=png,ext=.png,read=.png]{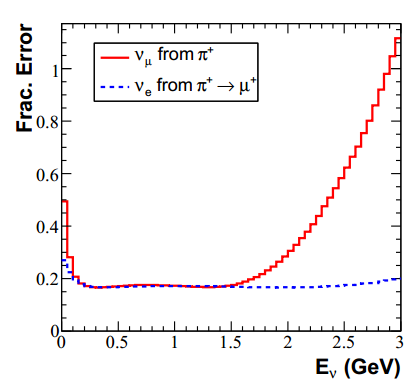}%
\includegraphics[scale=0.3,type=png,ext=.png,read=.png]{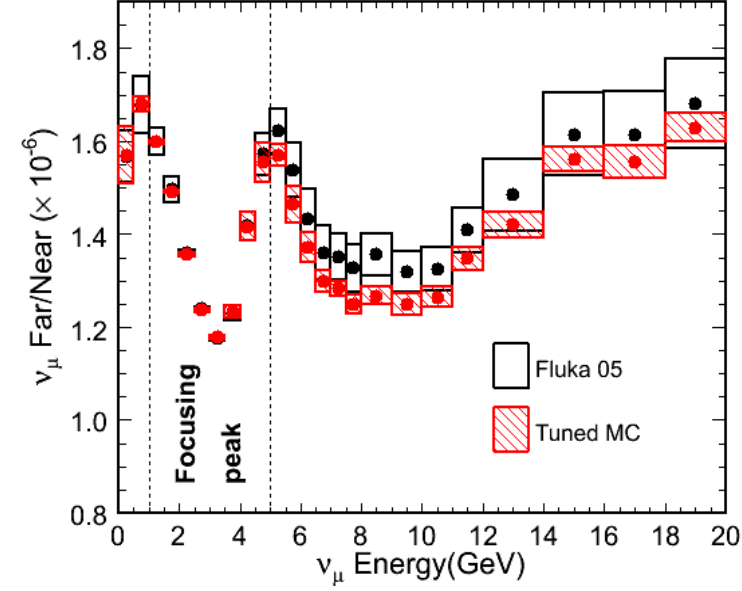}
\caption{Left: uncertainties on the absolute flux of $\nu_\mu$ at MiniBooNE;
right: FNR at NuMI and the associated uncertainties given by box sizes; taken from~\cite{G4BNBflux} and~\cite{numikopp}, respectively.}
\label{fig:hpmb}
\end{figure}

The uncertainty on the absolute $\nu_\mu$ flux at MiniBooNE is shown
in Fig.~\ref{fig:hpmb} (left) (from~\cite{G4BNBflux}). It stays below
20\% for energies below 1.5~GeV while it increases drastically above that energy.
The uncertainty is dominated by the knowledge of hadronic interactions
of protons on the Be target, which modifies the angular and momentum
spectra of neutrino parents emerging from the target. The result of
Fig.~\ref{fig:hpmb} is based on experimental data obtained by the HARP
and E910 collaborations.

This large uncertainty makes the use of two or more identical
detectors at different baselines mandatory for the search of small
disappearance phenomena. The ratio of the event rates at the Far and
Near detectors as function of neutrino energy (FNR) is a convenient
variable since it benefits at first order from cancellation of
systematics due to the common effects of proton--target and neutrino
cross--sections and the effects of reconstruction efficiencies.

Thanks to these cancellations the uncertainty on the FNR or,
equivalently, on the Far spectrum extrapolated from the Near spectrum
is usually at the percent level. As an example the uncertainty on FNR for the NuMI
beam is shown in bins of neutrino
energy in Fig.~\ref{fig:hpmb} (right) (from~\cite{numikopp}). The
uncertainty ranges in the interval 0.5--5.0\%.

It can be noted that, even in the absence of oscillations, the energy
spectra in the two detectors are different, thus leading to a non--flat
FNR. This is especially true if the distance of the Near detector is
comparable to the length of the decay pipe. It is therefore essential
to master the knowledge of the FNR for physics searches.

Compared to the Far site the solid angle subtended by the Near detector. Moreover
neutrinos coming from meson decays at the end of the decay pipe
have a higher probability of being detected.  In the Far detector, on
the contrary, only neutrinos produced in a narrow forward cone
will be visible.

\begin{figure}[htbp]
\centering
\includegraphics[scale=0.55,type=pdf,ext=.pdf,read=.pdf]{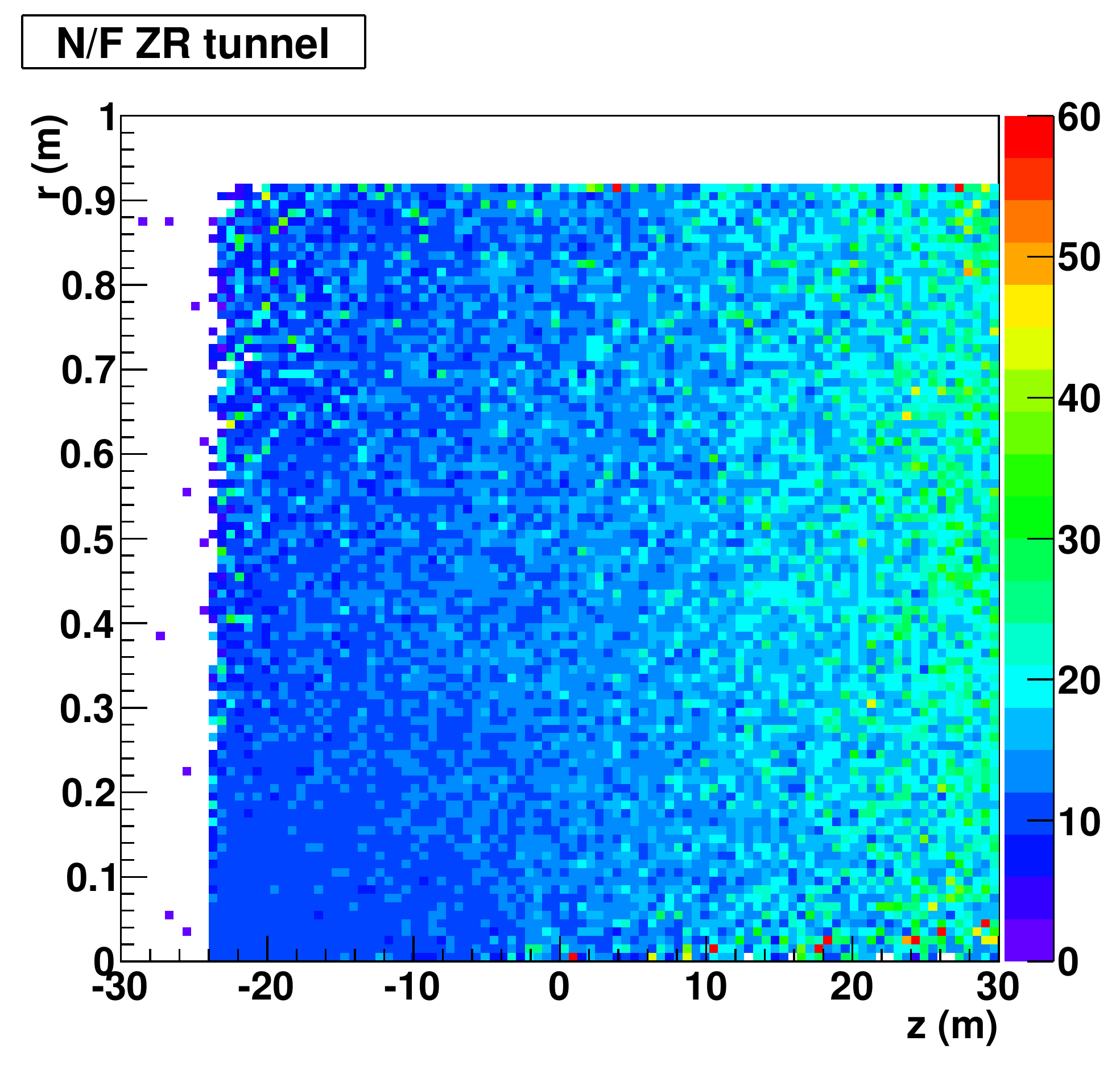}
\caption{Ratio between the $Z$--$R$ distributions of neutrino
  production points for neutrinos observed in a Near detector over
  neutrinos observed in a Far detector.  Two effects are most relevant: there is no apparent dependence on the
  radial $R$ distribution; and, as expected, the Near detector has a
  higher acceptance for neutrinos produced in the most downstream part
  of the decay pipe, i.e. at high $Z$.}
\label{fig:teff}
\end{figure}

The effect of the increased acceptance of the Near detector for
neutrinos from late decays
%different acceptance at the two sites in terms of the production point in the decay pipe 
is illustrated in Fig.~\ref{fig:teff}. It is shown the ratio of the
distributions of the neutrino production points (radius $R$ vs
longitudinal coordinate $Z$) for a sample crossing a Near and a Far
detector placed at 110 and 710 m from the target, respectively.
Neutrinos produced at large $Z$ can be detected in the Near detector even if they are
produced at relatively large angles.  This tends to enhance the low energy
part of the spectrum.  On the other hand neutrinos coming from meson decays  late in
the decay pipe are coming from the fast pion component which is more 
forward--boosted. The former effect is the leading one so the net effect is a
softer energy spectrum at the Near site.

\begin{figure}
\centering
\includegraphics[scale=0.7,type=pdf,ext=.pdf,read=.pdf]{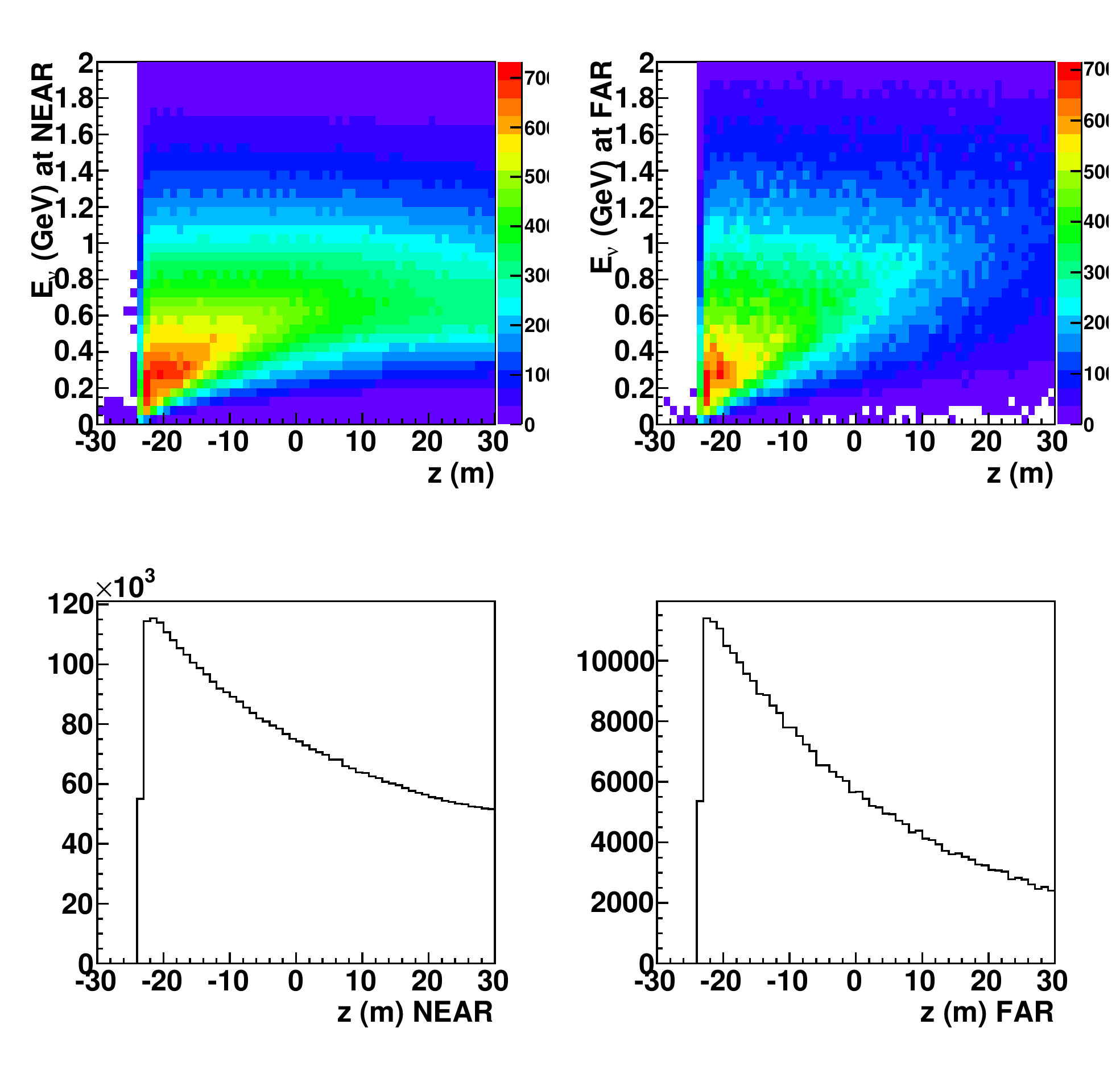}
\caption{Distribution of $E_\nu$ vs $Z$ for neutrinos seen in the Near (top left)
and Far (top right) detectors. Distribution of $Z$ for neutrinos seen in the Near 
(bottom left) and Far (bottom right) detectors. The points at low energy and very $Z$ are due to decays of pions emitted backward.}
\label{fig:teff1}
\end{figure}

The top plots of Fig.~\ref{fig:teff1} show the distribution of $E_\nu$ vs
$Z$ for neutrinos crossing the Near (top left) and the Far site (top
right). As anticipated, the energy spectrum at the Near site is softer, the additional
contribution at low energy being particularly important for neutrinos
coming from meson decays late in the decay pipe.  The distribution of $Z$ is also
shown in Fig.~\ref{fig:teff1} for neutrinos crossing the Near (bottom
left) and Far site (bottom right).

From these qualitative considerations it becomes clear that the
prediction of the FNR is a delicate task requiring a full simulation
of the neutrino beamline and of the detector acceptance. We will now
consider the sources of systematic uncertainties on the FNR.

\subsection{Systematic Uncertainties on the FNR}

All the contributions to the systematic uncertainties have been
studied in detail by the MiniBooNE collaboration in~\cite{G4BNBflux}. The results are summarized in
Tab.~\ref{tab:uncertainties}.  The dominant contribution comes from
the knowledge of the hadroproduction double differential ($p$,
$\theta$) cross--sections in 8 GeV $p$--Be interactions.
At first order these contributions factorize out using a double site. Due
to its importance we have here focused on the largest contribution.

\begin{table}
\centering
\begin{tabular}{|c|c|}
\hline
source & \% error \\
\hline
$p$--Be $\pi^+$ production &13.8\%\\
\hline
2$^{ry}$ nucleons interactions &6.2\\
\hline
$p$--delivery & 2\%\\
\hline
2$^{ry}$ pions interactions &1.5\\
\hline
magnetic field & 1.5\%\\
\hline
beamline geometry & 1\%\\
\hline
\end{tabular}
\caption{Systematic uncertainties on the $\nu_\mu$ BNB flux prediction. Taken from~\cite{G4BNBflux}.}
\label{tab:uncertainties}
\end{table}

\subsection{Monte Carlo beam simulation}

In order to understand the hadroproduction uncertainty impact on the
knowledge of the FNR for the specific case of our experiment we have
developed a new beamline simulation.
The angular and momentum distribution of pions exiting the Be target have been simulated using:
\begin{itemize}
\item FLUKA 2011.2b~\cite{fluka},
\item GEANT4 (v4.9.4 p02, QGSP 3.4 physics list),
\item a Sanford--Wang parametrization determined from a fit of the HARP and E910 data sets in~\cite{G4BNBflux},
\end{itemize}
\begin{equation}
\frac{d^2\sigma}{dpd\Omega} = c_1 p^{c_2}\left(1-\frac{p}{p_B-1}\right)\exp\left(-\frac{p^{c_3}}{p_B^{c_4}}-c_5\theta(p-c_6p_B\cos^{c_7}\theta)\right)
\label{eq:SW}
\end{equation}
$p_B$ being the proton beam momentum in GeV/c.

For the propagation and decays of secondary mesons a simulation using GEANT4 libraries has been developed. A simplified version of
the geometry, which was derived from the literature, was adopted.  Despite the
approximations a fair agreement with the official simulation
of the MiniBooNE Collaboration~\cite{G4BNBflux} has been obtained. This tool is
sufficient for the purpose of the site optimization that will be
described in the following. In order to fully take into account
finite--distance effects, fluxes and spectra are derived after
extrapolating neutrinos up to the detector volumes without using
weighing techniques. A total number of $7 \times 10^8$ protons on
target (p.o.t.), $2.1\times 10^8$ p.o.t. and $1 \times 10^9$ pions have been
simulated with FLUKA, GEANT4 and Sanford--Wang parametrization,
respectively.

\begin{figure}
\centering
\includegraphics[scale=0.7,type=pdf,ext=.pdf,read=.pdf]{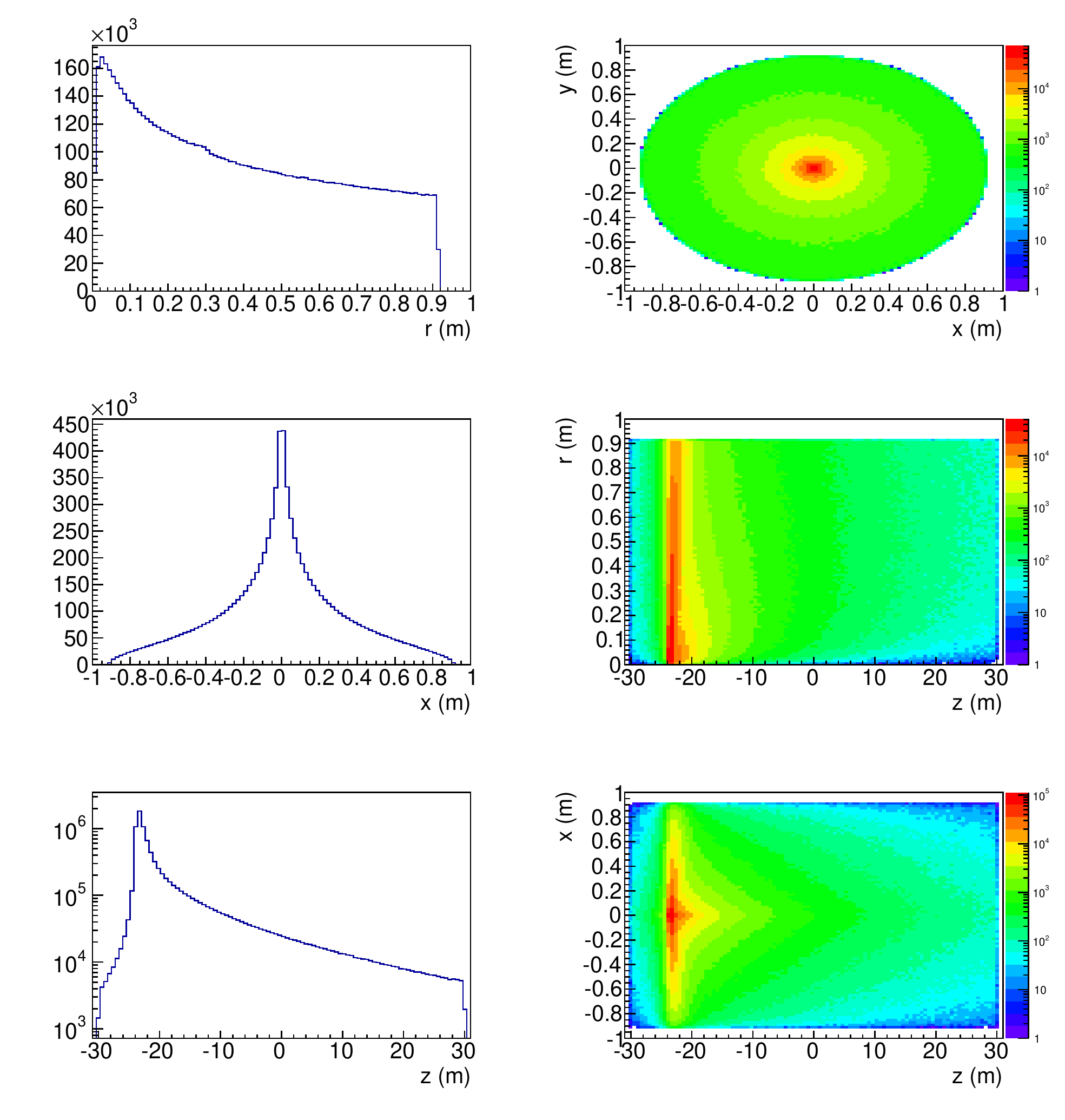}
\caption{
Distributions of the coordinates of neutrino production points in the decay pipe. 
Top left: radius. 
Top right: $X$ vs $Y$.
Middle left: $X$. 
Middle right: $R$ vs $Z$.
Bottom left: $Z$.
Bottom right: $X$ vs $Z$.
}
\label{fig:fig_0}
\end{figure}

Fig.~\ref{fig:fig_0} shows the spatial distributions of neutrinos at their production point in the
decay pipe. From top left in clockwise order are shown: the distance from the axis ($R$), 
$X$ vs $Y$ coordinates\footnote{The reference frame used here is such that $Y$ points upward and $Z$
is along the proton beam direction.}, $X$, $R$ vs $Z$, $Z$ and $X$ vs $Z$.

\begin{figure}
\centering
\includegraphics[scale=0.7,type=pdf,ext=.pdf,read=.pdf]{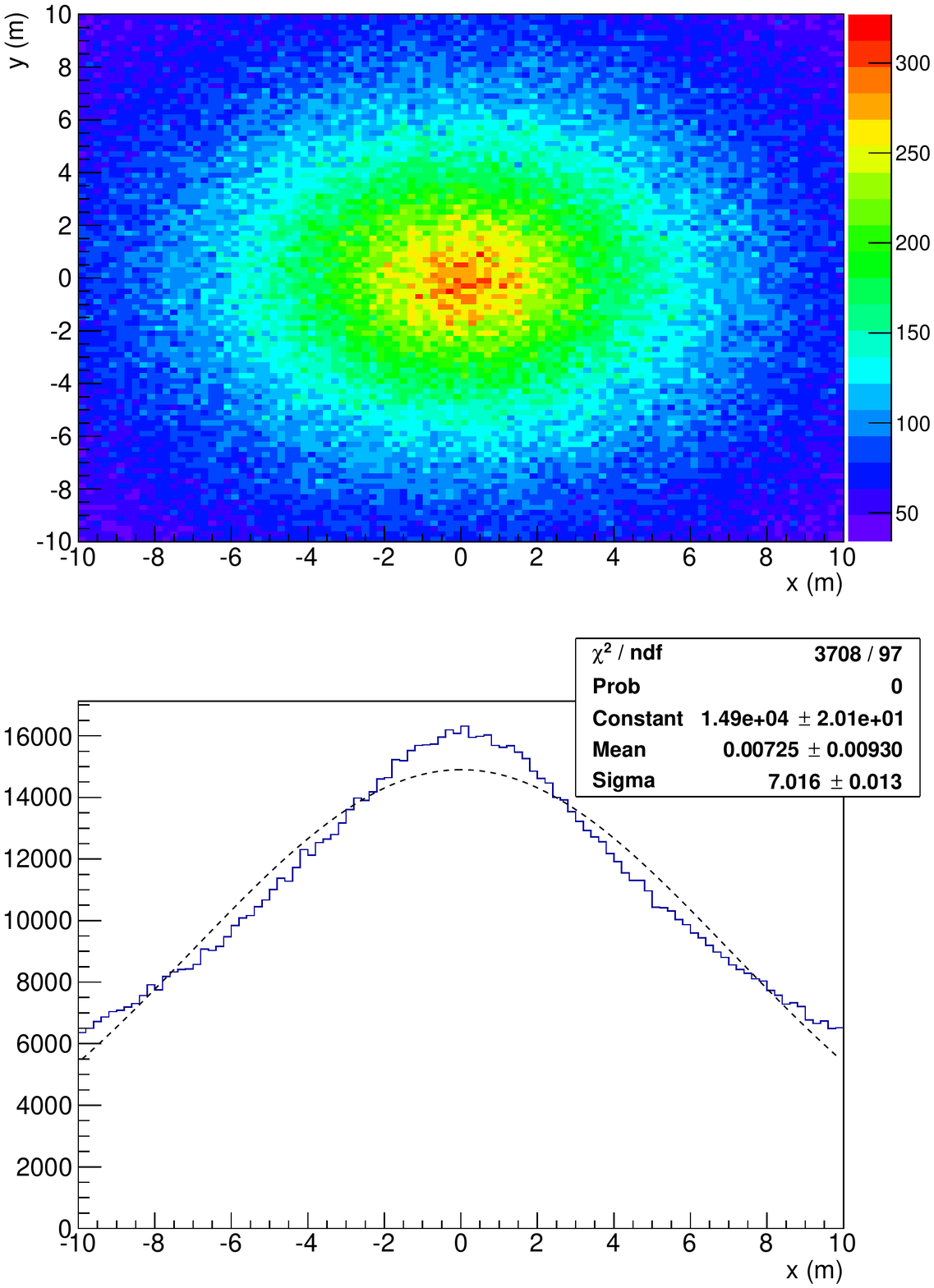}
\caption{Beam profile at 110~m from the target.}
\label{fig:fig_1}
\end{figure}

Fig.~\ref{fig:fig_1} shows the spatial distributions of neutrinos at a distance of 110~m from the
target. The r.m.s. of the distribution is about 5~m. The projected coordinate is shown in the bottom
plot of Fig.~\ref{fig:fig_1} with a Gaussian fit superimposed. This plot indicates that placing
the Near detector on surface would severely limit the statistics (furthermore this would make the
angular acceptance of the Far and Near detectors too different).

\subsection{Choice of the experimental sites}

By supposing to have already decided the geometry and mass of the detectors, several
considerations influence the choice of the experimental 
sites location. 

The ultimate figure of merit is the power of exclusion (or discovery) 
for effects induced by sterile neutrinos in a range of parameters as 
wide as possible in a given running time. 

To achieve good performances an essential point is the reliability of 
the simulation of the spectra of neutrinos at the Near and Far sites. 
As soon as the detectors are further away from the target they ``see'' 
more similar spectra since the production region can be better approximated as 
a point--like source. This helps in reducing the systematic uncertainty. 
This point is further addressed in Sect.~\ref{subsectsysconf}.

On the other hand increasing the distances introduces the loss in the 
collectable event sample and reduces the lever--arm for oscillation studies.

Coming to practical constraints we note that increasing the depth of
detectors impacts considerably the civil engineering
costs. Furthermore the space along the BNB beamline is already
partially occupied by existing or proposed experimental facilities
(SciBooNE/LAr1--ND, T150--Icarus, MiniBooNE, MicroBooNE, LAr1 and ICARUS).

%\begin{itemize}
%\item achievable statistics
%\item similarity of near and far spectra
%\end{itemize}

\subsubsection{Dependence on the detector position of $\nu_\mu$ CC rates and energy spectra}

\begin{figure}
\centering
\includegraphics[scale=0.7,type=pdf,ext=.pdf,read=.pdf]{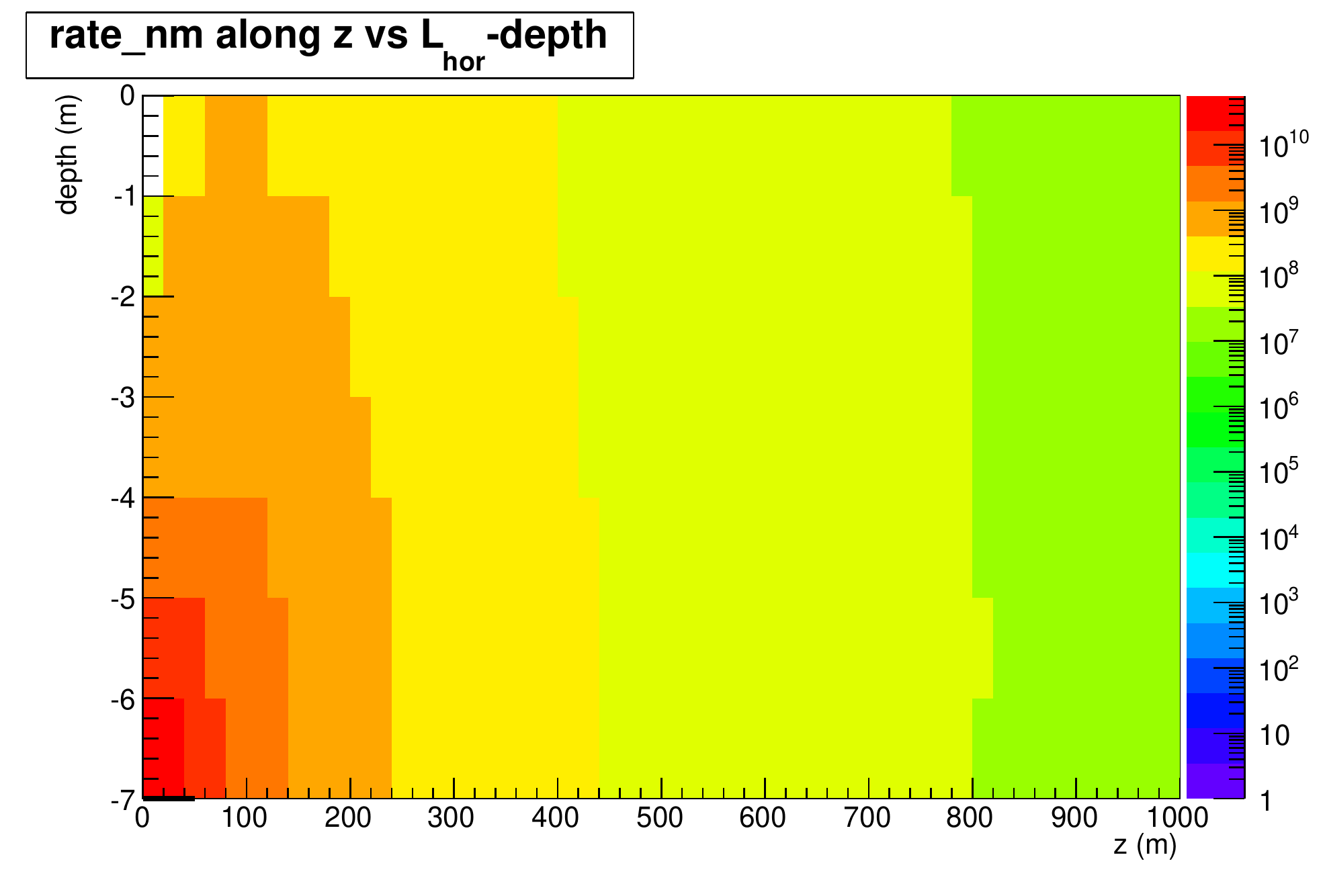}\\
\includegraphics[scale=0.7,type=pdf,ext=.pdf,read=.pdf]{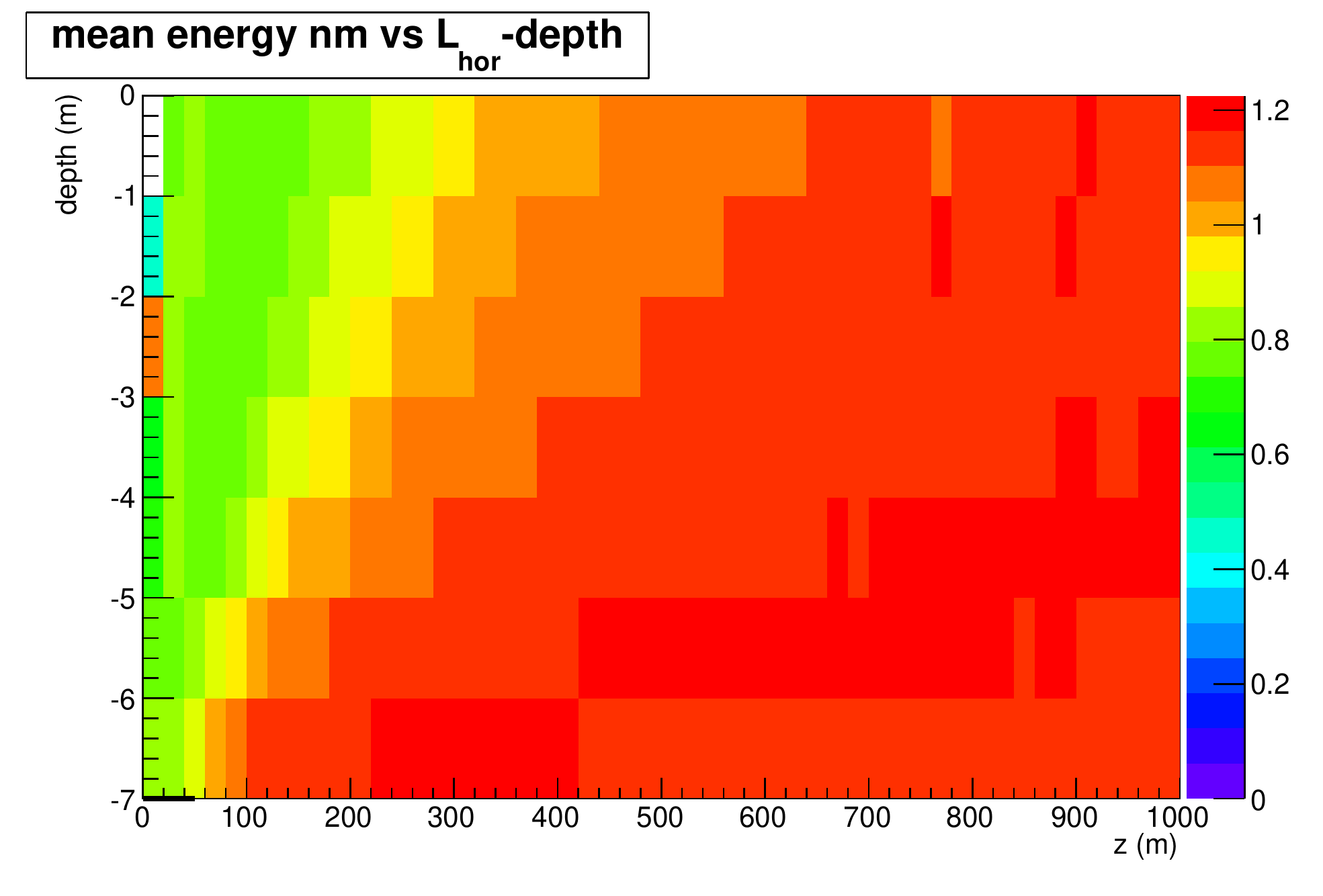}
\caption{Top: $\nu_\mu$ CC rate (a.u.) vs $z$ and depth. Bottom: mean energy of $\nu_\mu$ CC interactions (GeV) vs $Z$ and depth.
The upstream side of the target is the origin of the reference system. The black line close to the origin represents the longitudinal 
extension of the decay pipe.}
\label{fig:figscan}
\end{figure}

Fig.~\ref{fig:figscan} shows how the rates of $\nu_\mu$ CC
interactions (top) and their mean energy (bottom) depend on the
position with respect to the proton target. The horizontal axis shows
the distance from the target in the horizontal direction ($Z$) while
the vertical axis contains the depth from the ground surface.

It can be seen that at about 700 m distance the rate and mean energy
is barely affected when moving from an on--axis to an off--axis
position. This observation supports the idea of putting the Far
detector at surface thus reducing the experiment cost.

\subsubsection{Systematics in the Far--to--Near ratio for a set of detector configurations.}
\label{subsectsysconf}

\begin{figure}
\centering
\includegraphics[scale=0.42,type=pdf,ext=.pdf,read=.pdf]{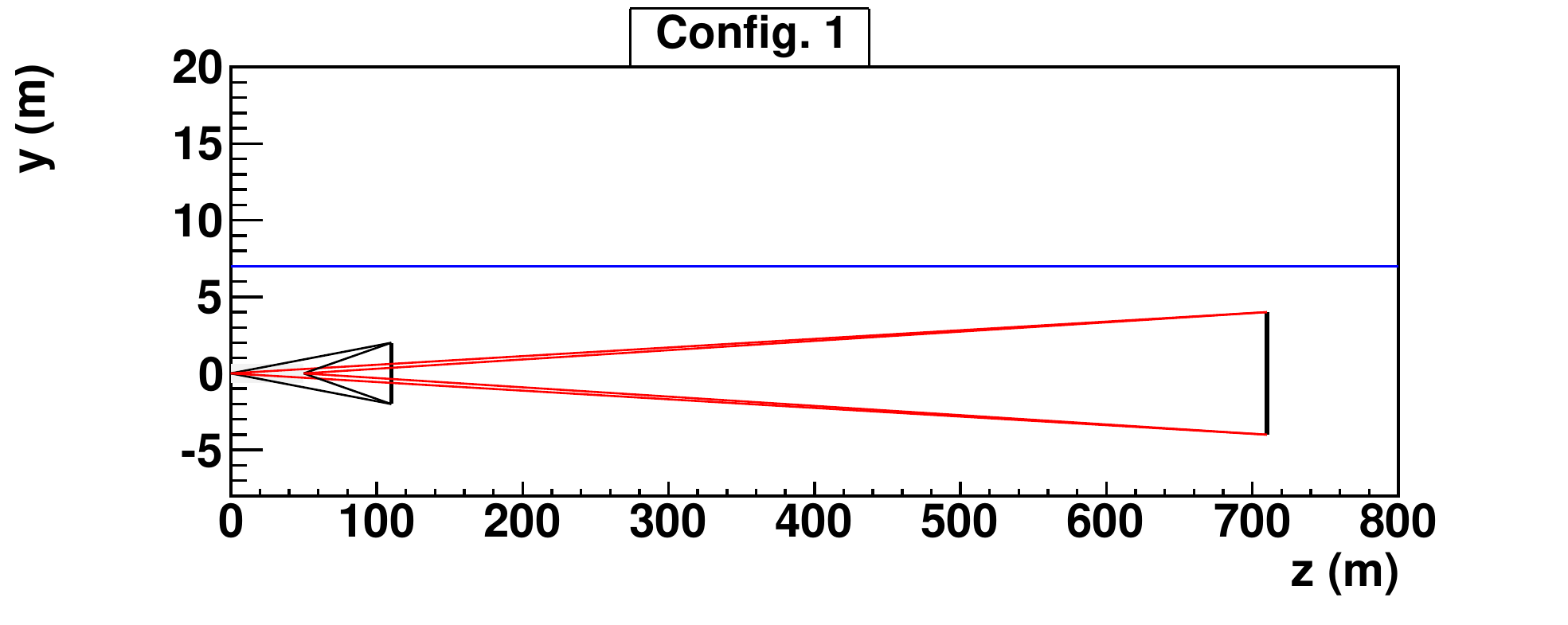}%
\includegraphics[scale=0.42,type=pdf,ext=.pdf,read=.pdf]{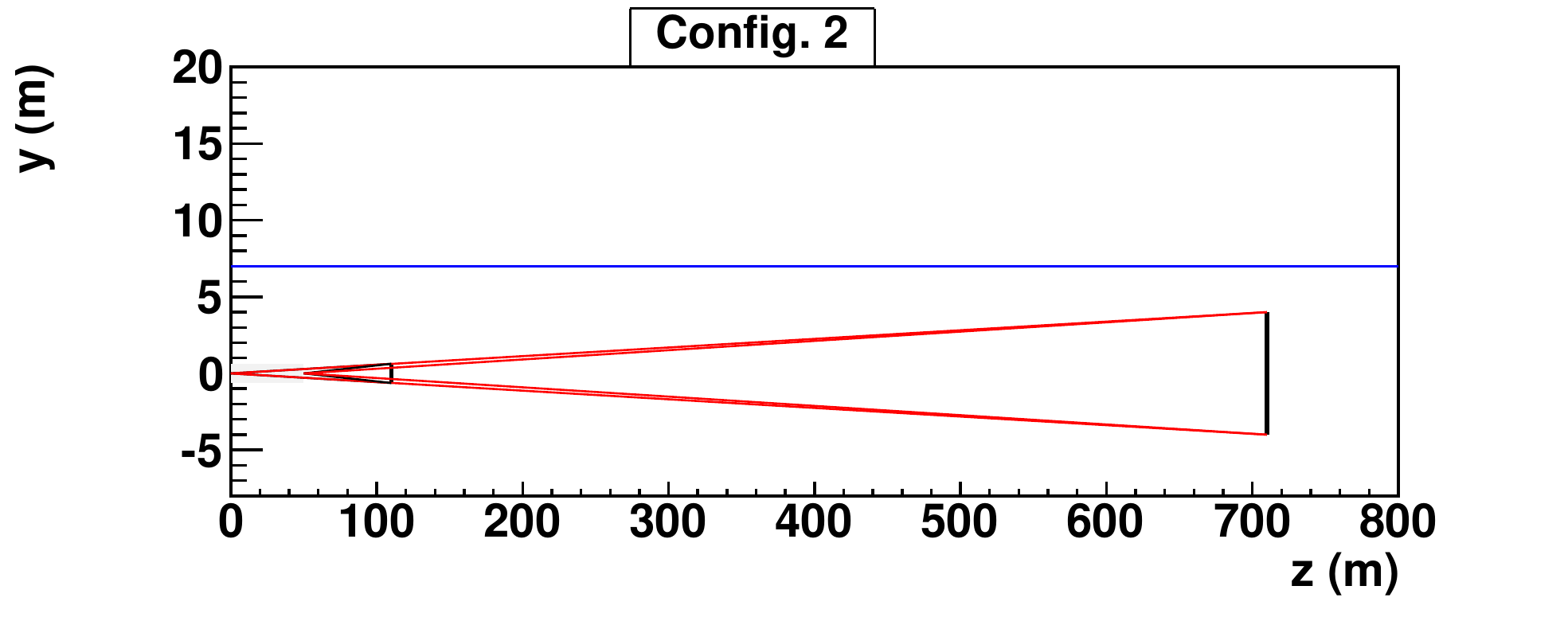}
\includegraphics[scale=0.42,type=pdf,ext=.pdf,read=.pdf]{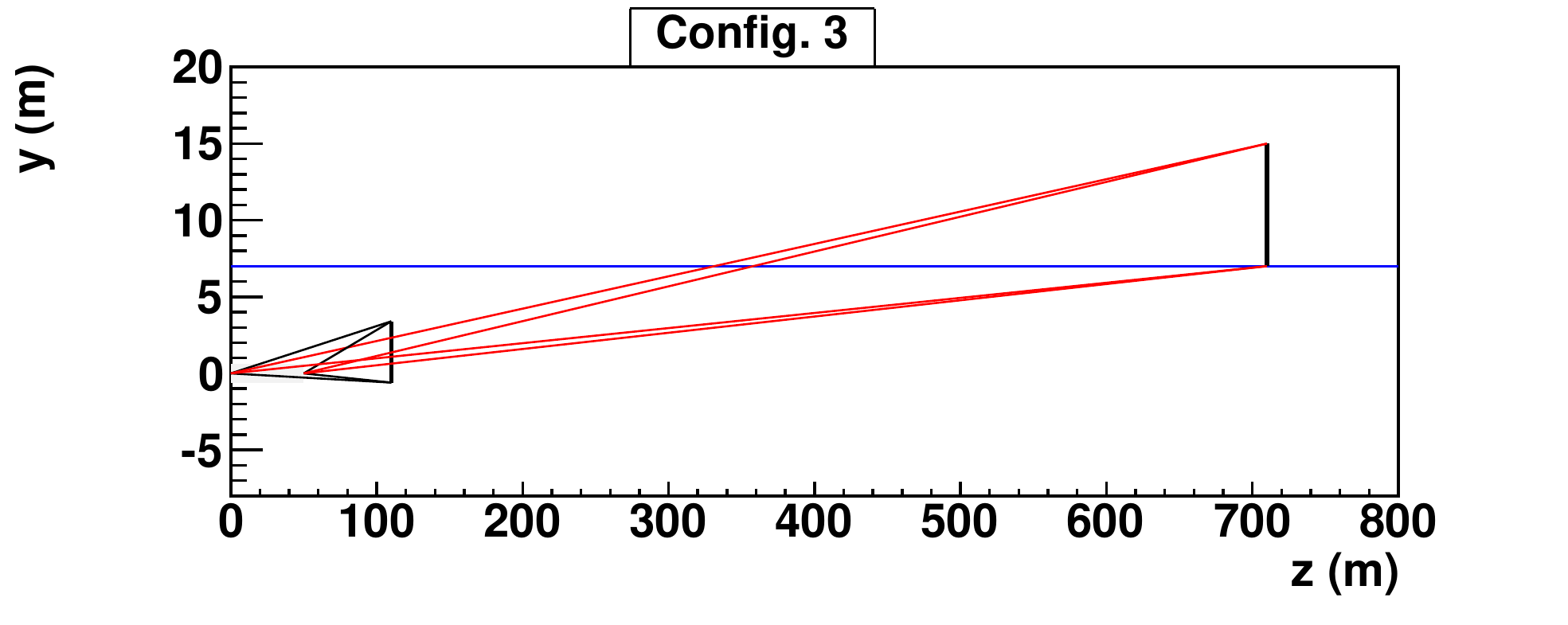}%
\includegraphics[scale=0.42,type=pdf,ext=.pdf,read=.pdf]{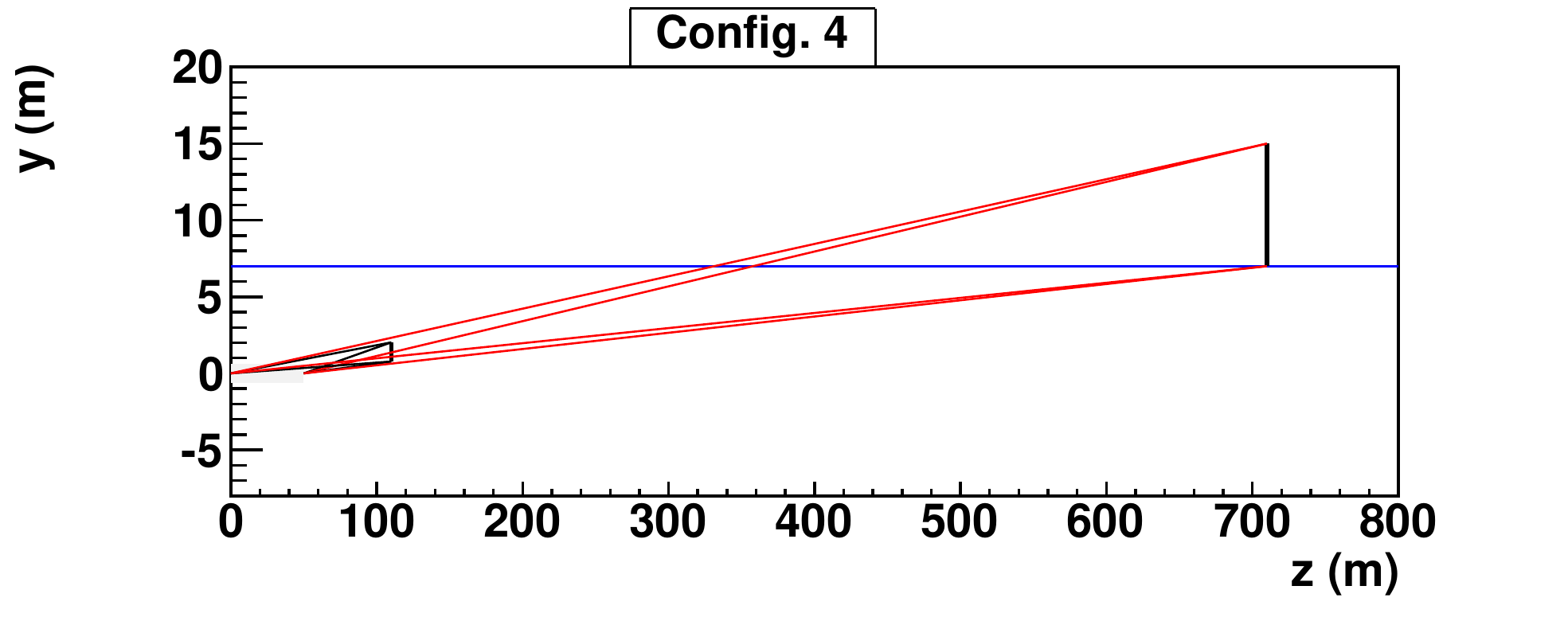}
\includegraphics[scale=0.42,type=pdf,ext=.pdf,read=.pdf]{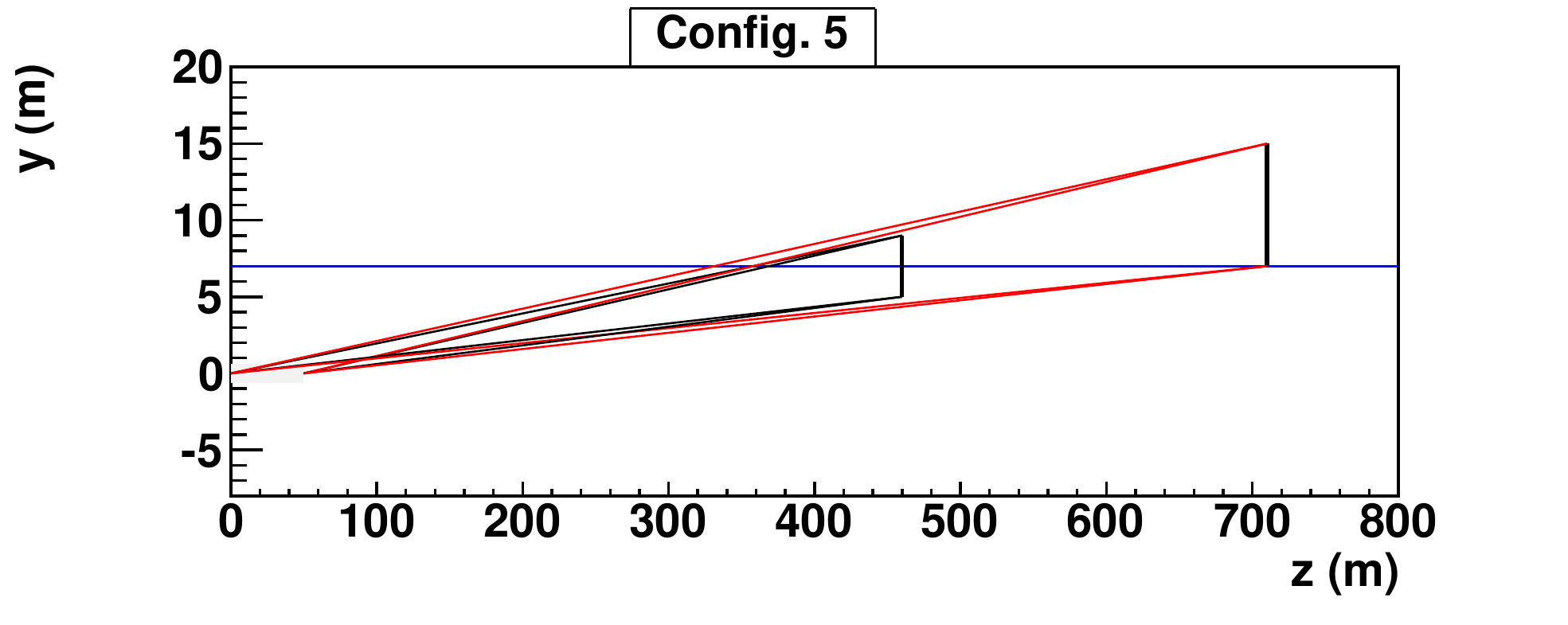}%
\includegraphics[scale=0.42,type=pdf,ext=.pdf,read=.pdf]{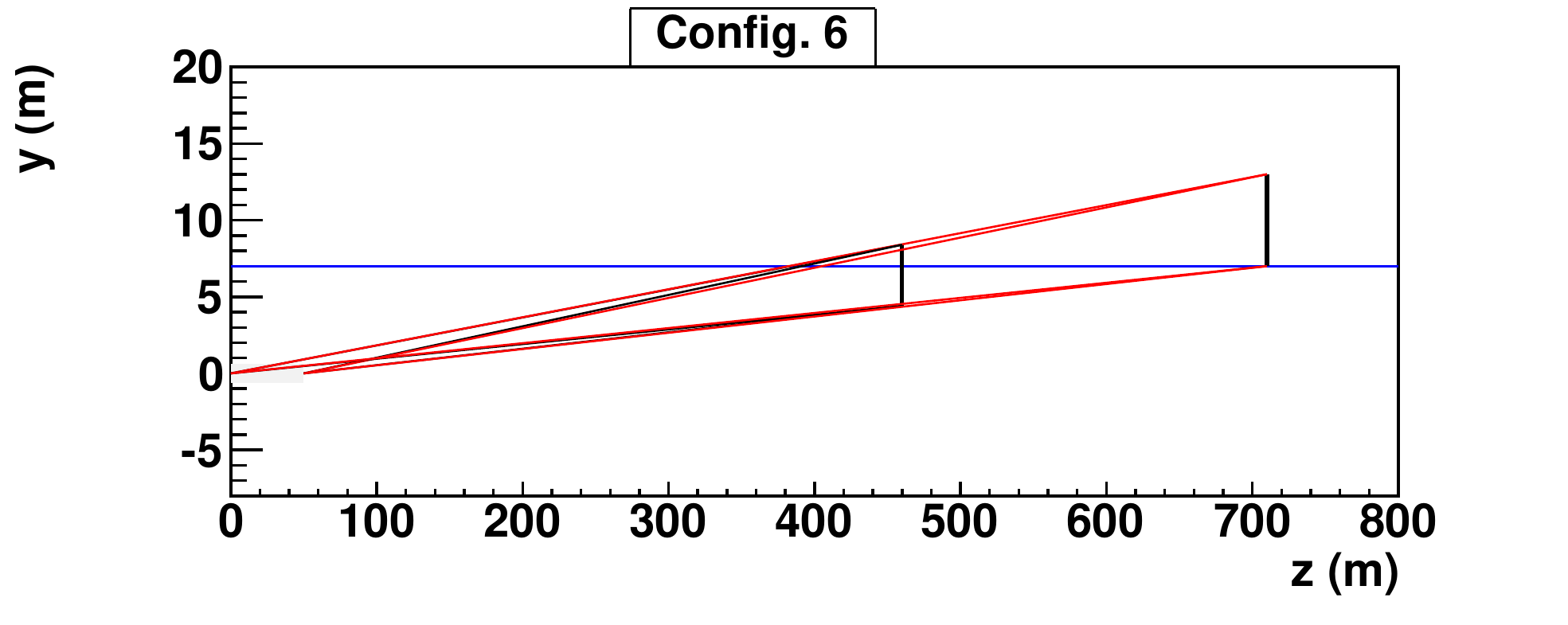}
\caption{Configurations of Far and Near detectors in the $Y$--$Z$ plane (see also Tab.~\ref{tab:confs}).
The blue horizontal line marks the ground level, the vertical black lines mark the detectors and
the red lines show the angle subtended by the detectors at the beginning and the end of the decay pipe.}
\label{fig:confs}
\end{figure}

\begin{figure}
\centering
\includegraphics[scale=0.7,type=pdf,ext=.pdf,read=.pdf]{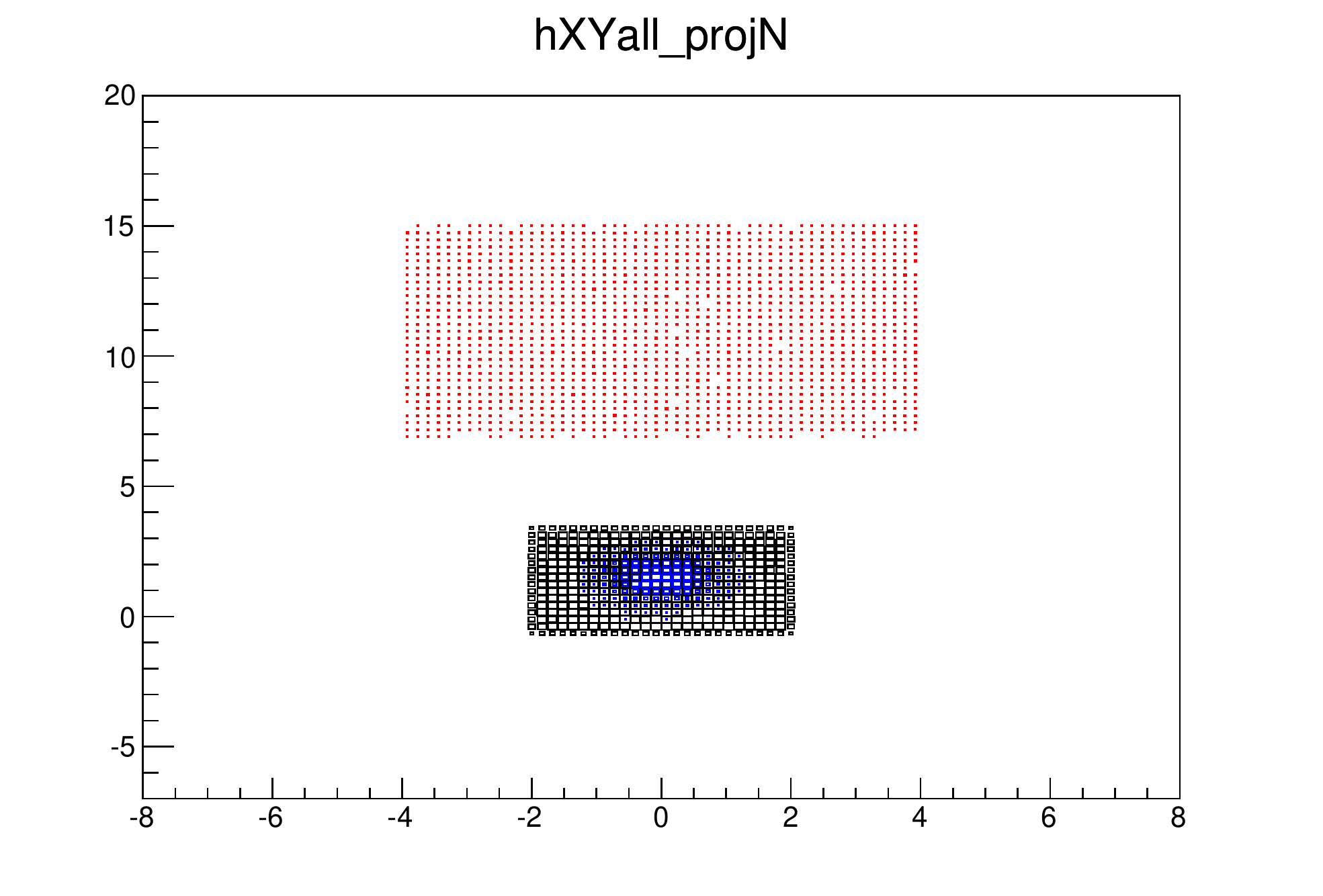}%
\caption{Distribution of the impact points of neutrinos at the Near (black) and Far (red) sites.
Units are in meters. The blue histograms shows the same distribution for the subsample of neutrinos crossing
both the Far and Near detectors. Here we consider configuration 3 (see Tab.~\ref{tab:confs}).}
\label{fig:figconfs}
\end{figure}

A set of six configurations have been studied considering a
combination of distances (110, 460 and 710 m), on--axis or off--axis
configurations and different fiducial sizes of the detectors.  Their
geometrical parameters are given in Tab.~\ref{tab:confs} and
illustrated schematically in Fig.~\ref{fig:confs}.
\begin{itemize}  
\item Configuration 1 considers two on--axis detectors at 110 and 710 m with squared active
areas of $4 \times 4$ m$^2$ (Near) and $8 \times 8~\rm{m}^2$ (Far).
By selecting the subsample of neutrinos crossing the Near detector
which are also crossing the Far detector, a well defined region is
selected in the Near detector. This ``shadow''
(Fig.~\ref{fig:figconfs}) is not sharp due to the fact that the source
is not point--like. 
\item Configuration 2 uses a reduced Near detector area,
limited to $1.25 \times 1.25$ m$^2$, chosen as such to increase the overlap of
neutrinos seen in the two detectors.  
\item Configurations 3 and 4 replicate
the same pattern as for 1 and 2 but having the Far detector at surface and the Near
detector sharing the same off--axis angle (instead of both being on--axis). 
\item Configurations 4 and 5 are similar to 3 and 4 but for a more distant the Near site (460 m).
\end{itemize}

\begin{table}
\centering
\begin{tabular}{|c|c|c|c|c|c|c|}
\hline
configuration &$L_N$ (m)&$L_F$ (m)&$y_N$ (m)& $y_F$ (m)& $s_N$ (m)& $s_F$ (m)\\
\hline
1 &110&710&0& 0 &4  &8 \\
\hline
2 &110&710&0& 0 &1.25  &8 \\
\hline
3 &110&710&1.4& 11 & 4 & 8 \\
\hline
4 &110&710&1.4& 11 &1.25  &8 \\
\hline
5 &460&710&7& 11 & 4 & 8 \\
\hline
6 &460&710&6.5& 10 & 4 & 6 \\
\hline
\end{tabular}
\caption{Near--Far detectors configurations. $L_{N(F)}$ is the distance of the Near (Far) detector
from the target. $y_{N(F)}$ is the vertical coordinate of the center of the fiducial area of the Near (Far) detector with respect to the beam axis
which lies at about -7~m from the ground surface. 
$s_{N(F)}$ is the dimension of the fiducial area of the Near (Far) detector.}
\label{tab:confs}
\end{table}

\begin{figure}
\centering
\includegraphics[scale=0.7,type=pdf,ext=.pdf,read=.pdf]{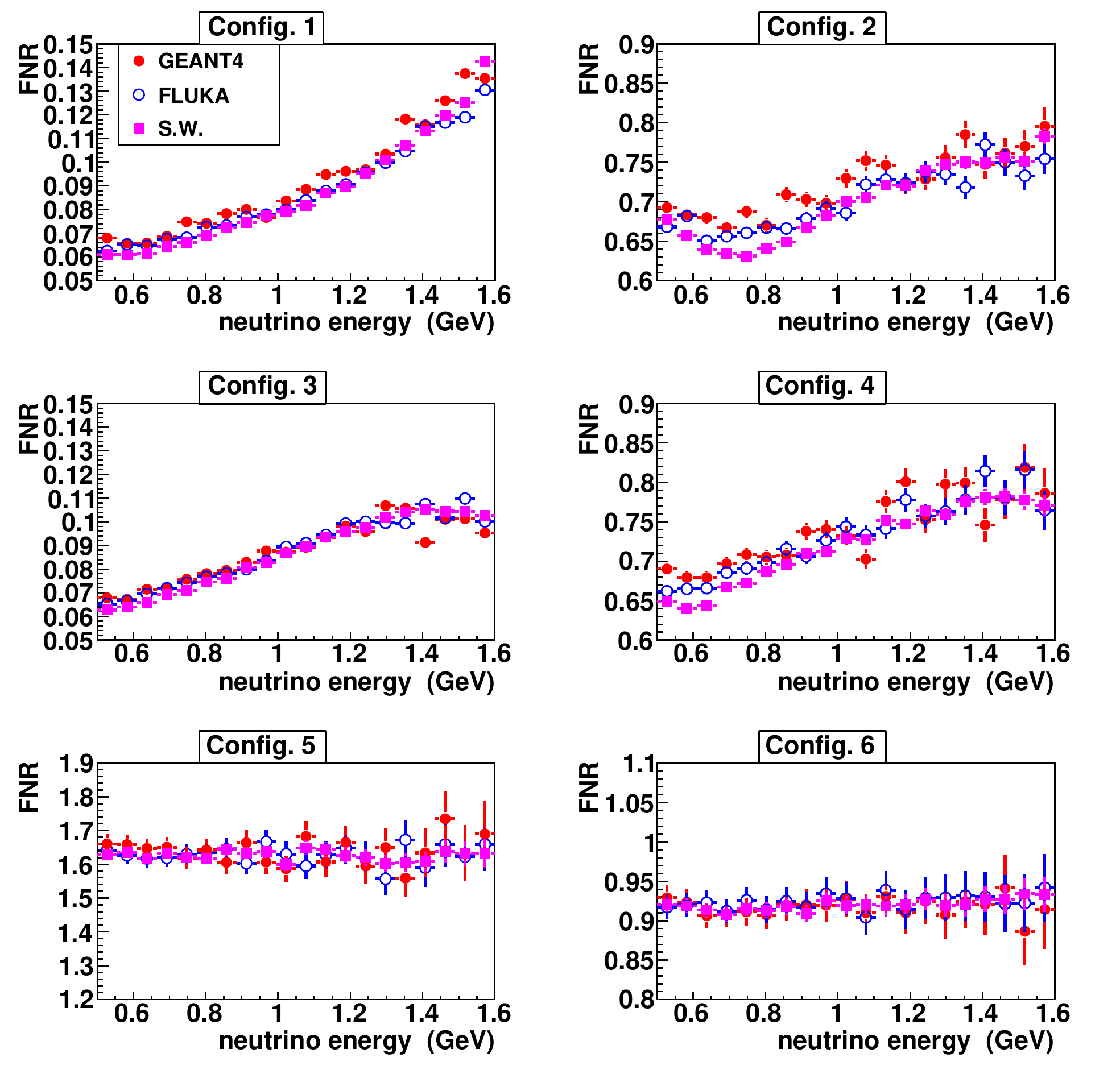}
\caption{Far--to--Near ratios for the six considered configurations. Comparison of FLUKA and GEANT4 for hadroproduction.}
\label{fig:fighadp}
\end{figure}

The FNRs for the six considered configurations using either FLUKA,
GEANT4 or the Sanford--Wang parametrization for the simulation of
$p$--Be interactions are shown in Fig.~\ref{fig:fighadp}. The error
bars here only indicate the uncertainty introduced by the limitations
in Monte Carlo samples.  

Configuration 1 (with on--axis detectors and a large Near detector)
produces a FNR increasing with energy as expected from the
considerations presented above and largely departing from a flat
curve. By restricting the fiducial volume in the Near detector
(configuration 2) the FNR flattens as expected. This behavior is also
confirmed using off--axis detectors (configurations 3 and
4). Configurations with a Near detector at larger baselines (5 and 6)
tend to produce quite flat FNRs, as expected.

\begin{figure}
\centering
\includegraphics[scale=0.7,type=pdf,ext=.pdf,read=.pdf]{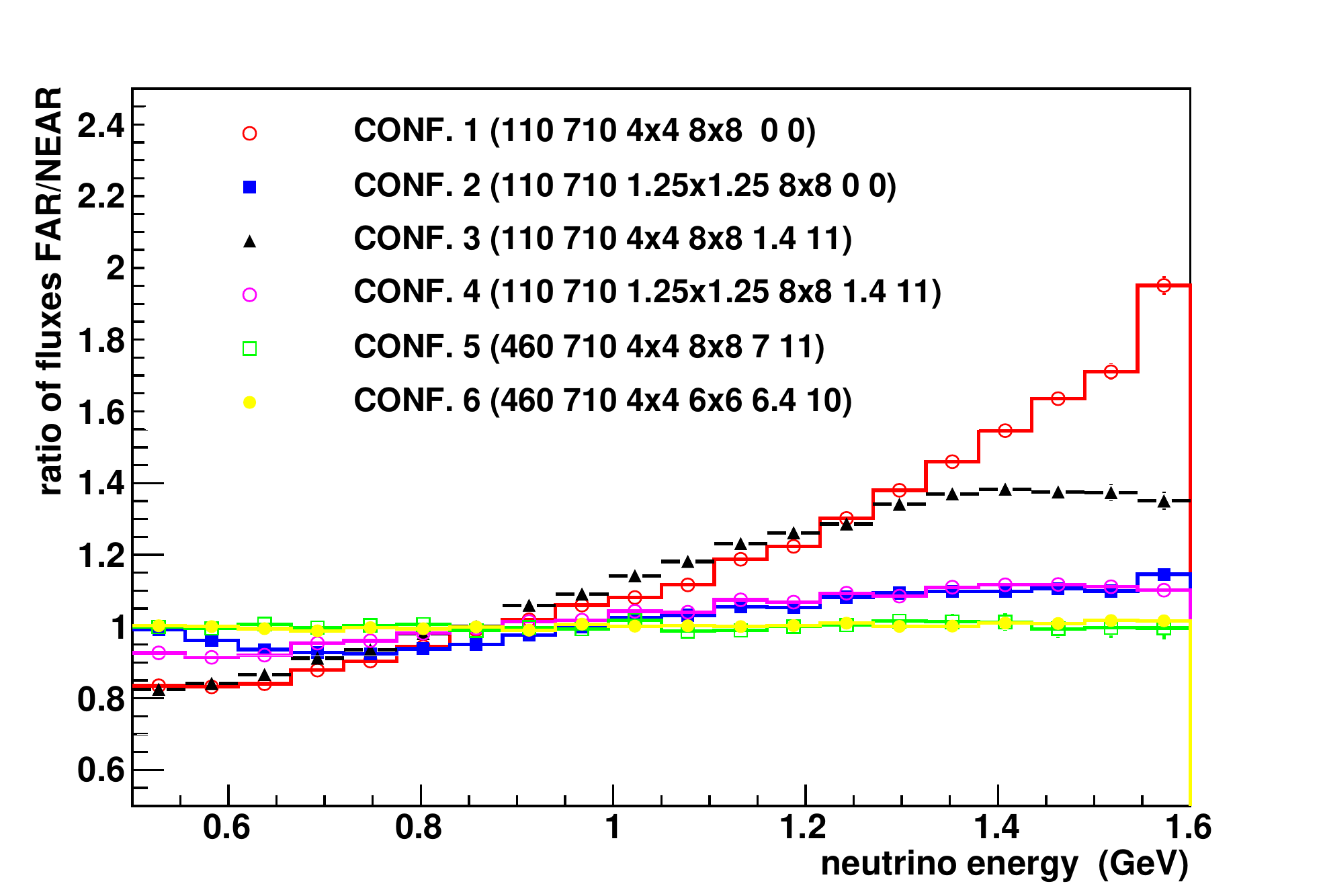}
\caption{Far--to--Near ratios for the six considered configurations using the Sanford--Wang parametrization.}
\label{fig:fig1}
\end{figure}

The different behaviors are more easily visible in Fig.~\ref{fig:fig1} where FNRs, normalized to each other, are compared
 (taking the Sanford--Wang parametrization).

\begin{figure}
\centering
\includegraphics[scale=0.7,type=pdf,ext=.pdf,read=.pdf]{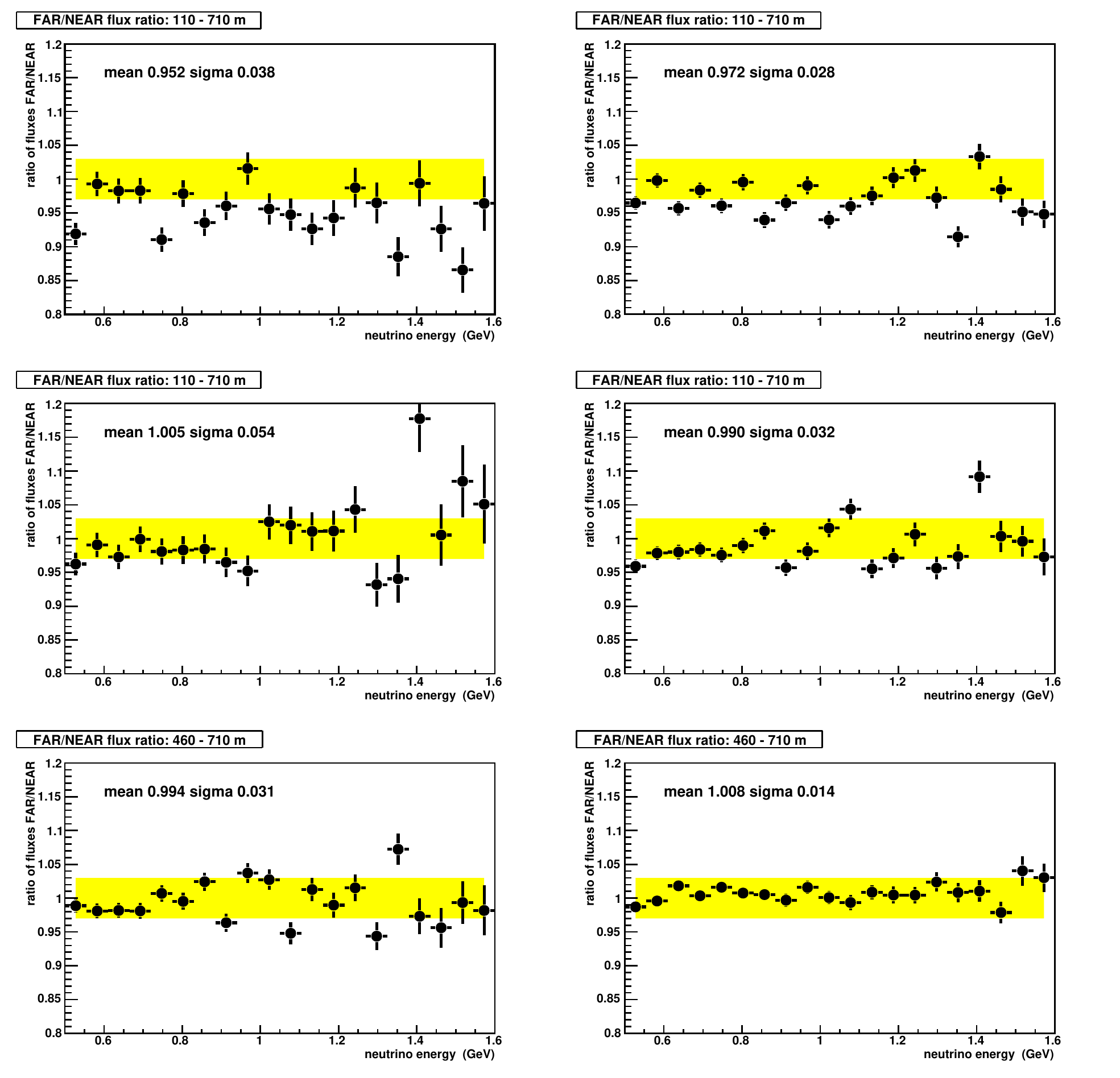}
\caption{Comparison of FLUKA and GEANT4 for hadroproduction.}
\label{fig:fighadp1}
\end{figure}

We have used the differences in the hadronic models implemented in the
FLUKA and GEANT4 generators to estimate the impact of the
hadroproduction uncertainties on the FNR. The results are shown in
Fig.~\ref{fig:fighadp1} where we show the bin--by--bin ratios between
the FNRs predicted by these two Monte Carlos for the six considered
configurations.
The yellow band visualizes the magnitude of a fixed 3\% error on the FNR. It
can be seen that the two simulations provide results which are in general  
in agreement at a level between 1 and 3\% for configurations where the 
common region between the Far and Near detectors are considered. 

%Red bullets represent the result obtained with GEANT4 and the blue hollow
%bullets the results given by FLUKA. The yellow band visualizes the
%magnitude of a fixed 3\% error on the FNR for the GEANT4 points. It
%can be seen that in all energy bins the two simulations provide
%consistent results within xx \%. 
%The presence of structures has a
%physical origin related to the specific geometry of the focusing
%system \cite{Kopp}.  These sharp structures are present in both FLUKA
%and GEANT4.

%\begin{equation}
%\sigma(R)=\sqrt{\frac{F}{N^2}+\frac{F}{N^3}(1-2 N_{common})}
%\end{equation}

\begin{figure}
\centering
\includegraphics[scale=0.5,type=pdf,ext=.pdf,read=.pdf]{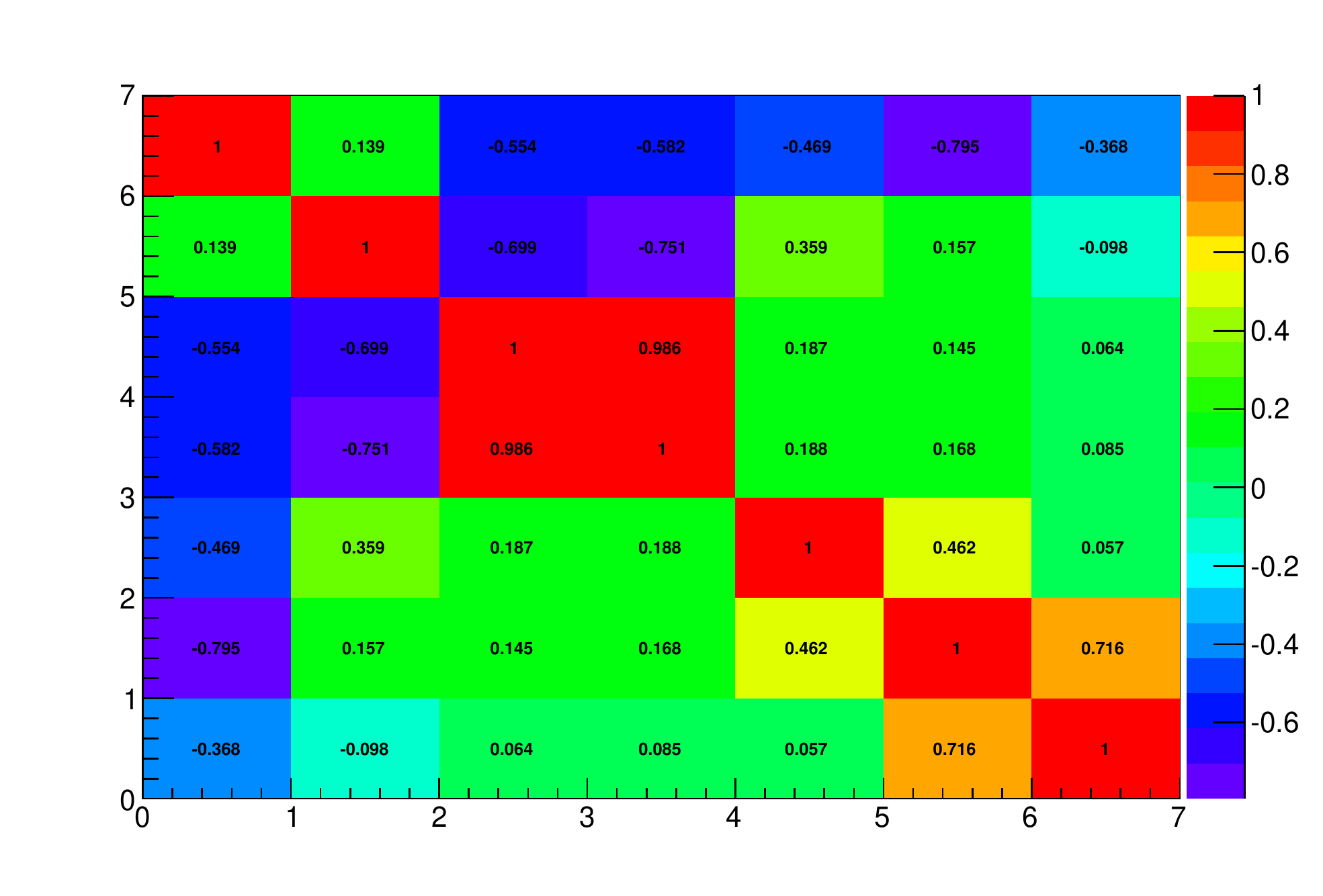}
\caption{Covariance matrix of the $c_i$ coefficients of the Sanford-Wang parametrization
of pion production data from HARP and E910 in Eq.~\ref{eq:SW}.}
\label{fig:covmat}
\end{figure}

\begin{figure}
\centering
\includegraphics[scale=0.7,type=pdf,ext=.pdf,read=.pdf]{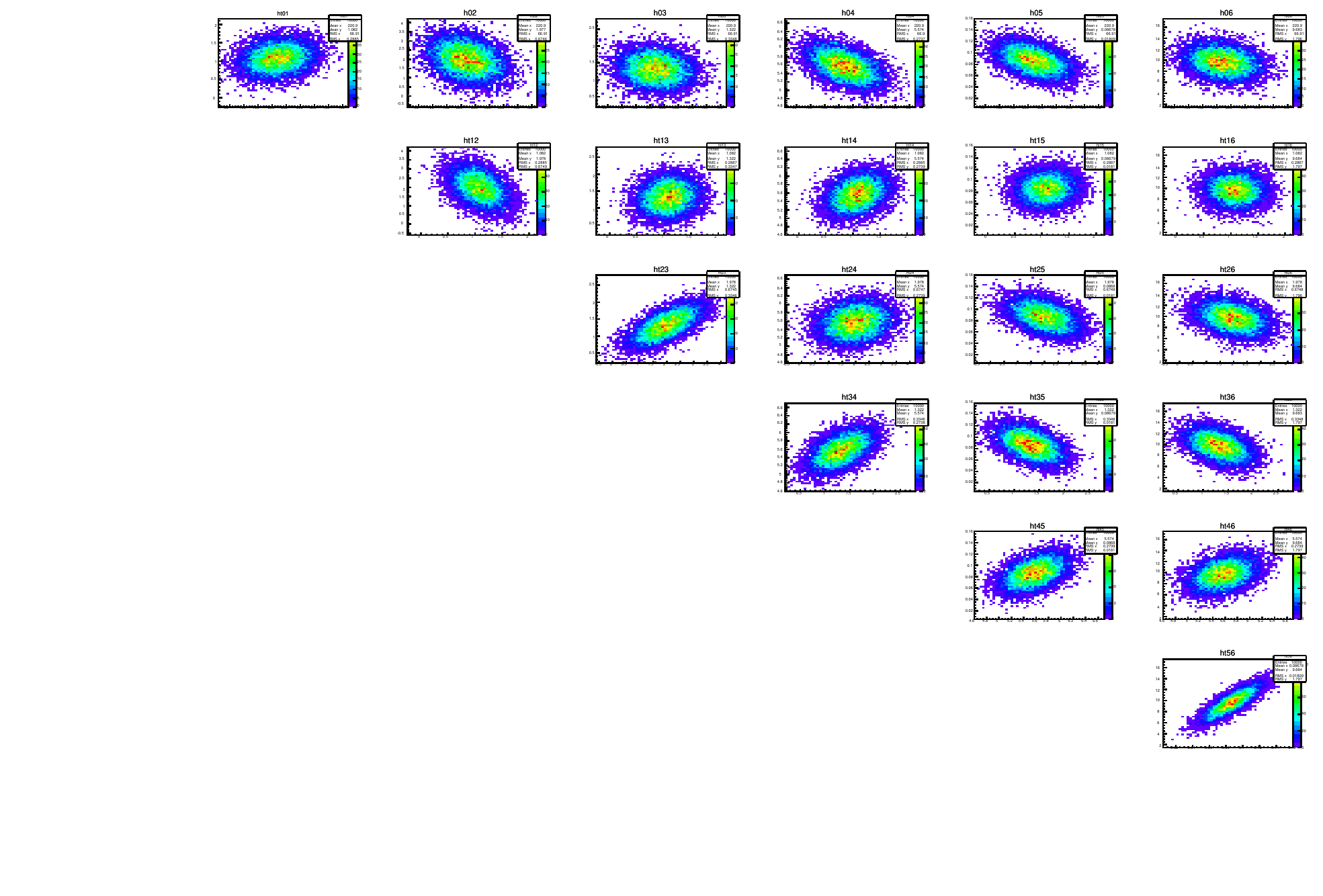}
\caption{Correlation of the $c_i$ parameters in Eq.~\ref{eq:SW}.}
\label{fig:cholesky}
\end{figure}

Since specific hadroproduction measurements for the BNB Be
target replica exist, a more appropriate way of addressing the
hadroproduction related uncertainty on the FNR has also been adopted as
follows. The coefficients $c_i$ of the Sanford--Wang parametrization of
pion production data from HARP and E910 in Eq.~\ref{eq:SW} have been
sampled.  Their correlations have been taken into account using
the covariance matrix published in~\cite{G4BNBflux} and shown here in
Fig.~\ref{fig:covmat}.
The sampling of these correlated variables has been performed using
the Cholesky decomposition of the covariance matrix. The correlation
of the sampled variables reflects the information of the original
covariance matrix as shown in Fig.~\ref{fig:cholesky}.

\begin{figure}
\centering
\includegraphics[scale=0.4,type=pdf,ext=.pdf,read=.pdf]{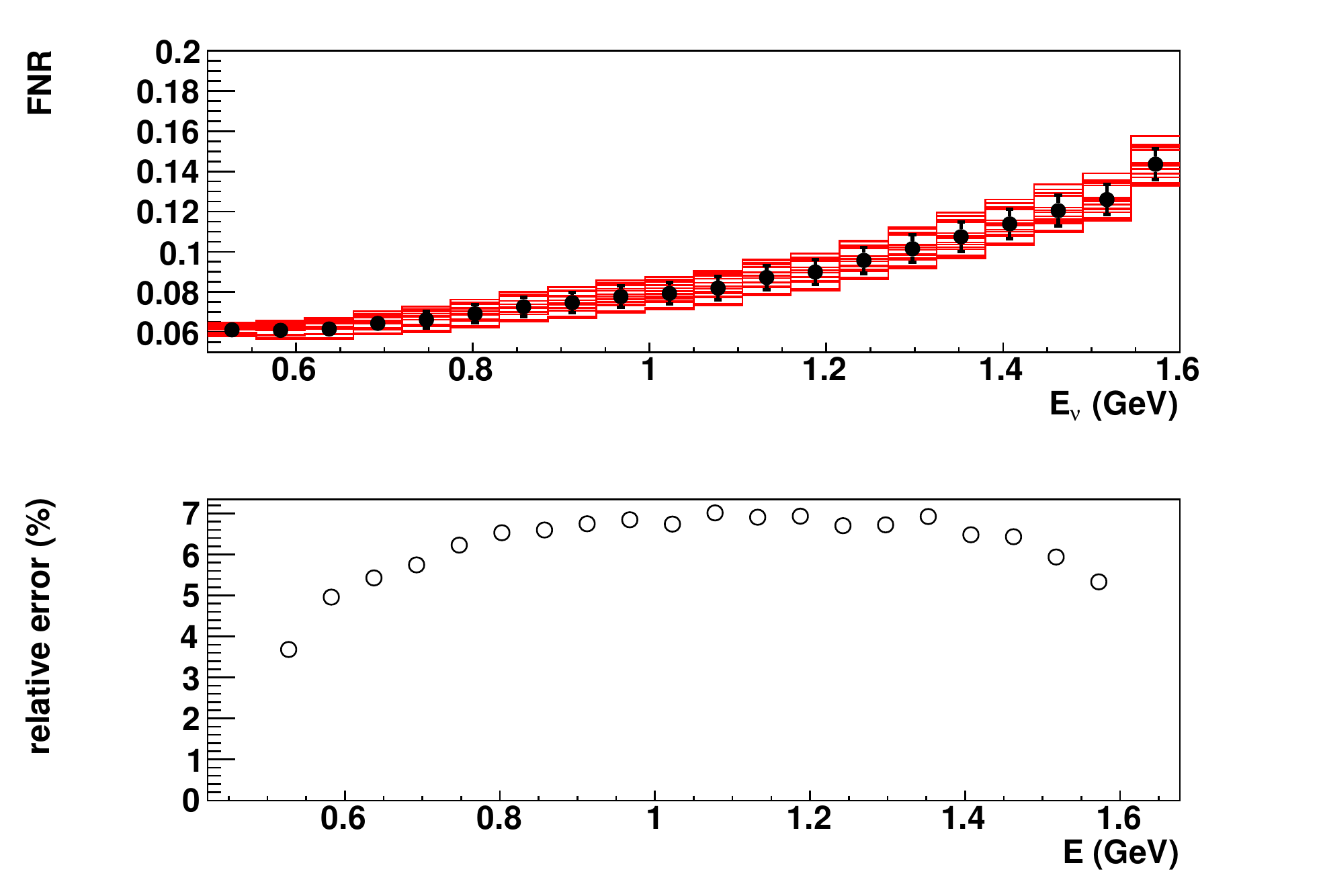}%
\includegraphics[scale=0.4,type=pdf,ext=.pdf,read=.pdf]{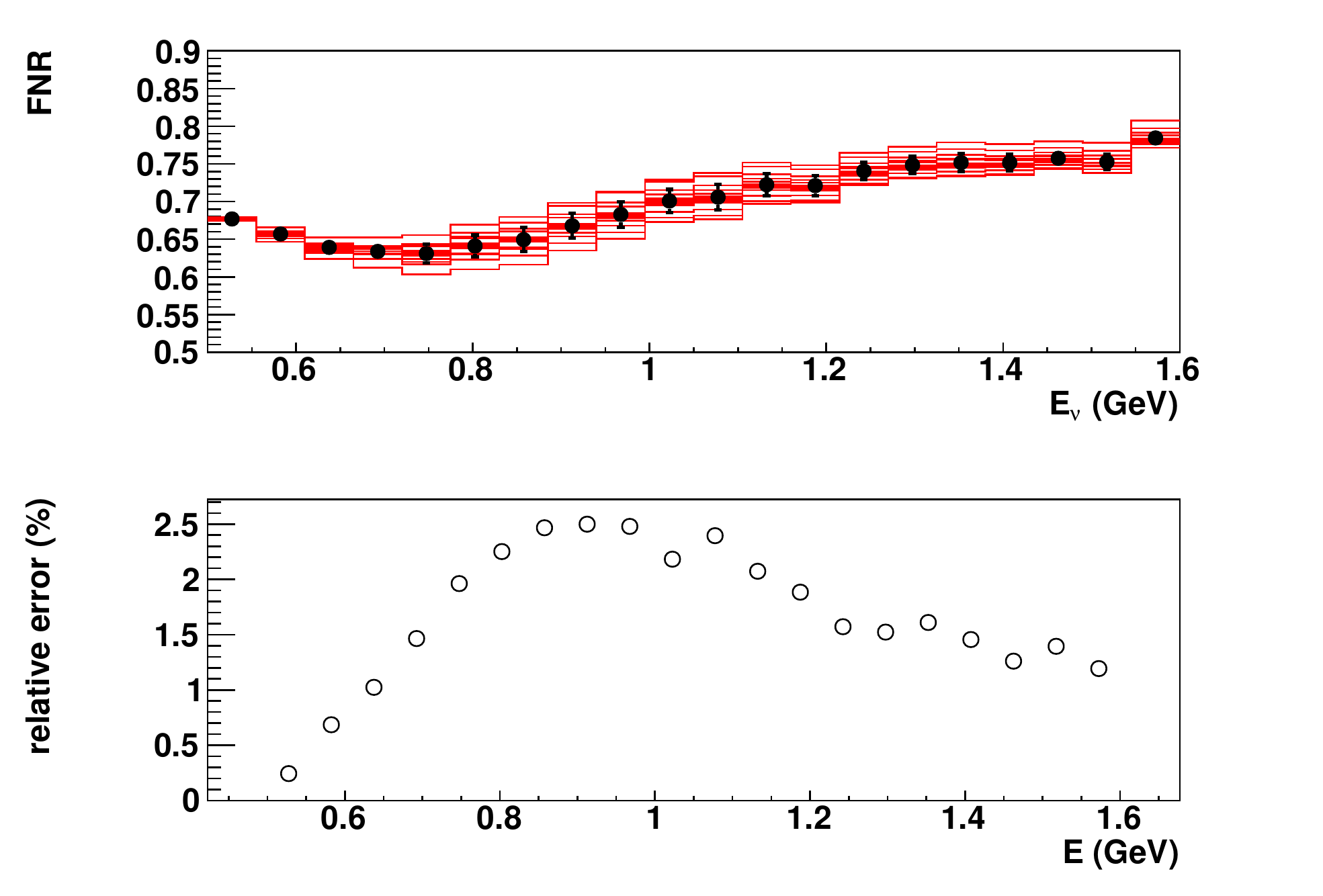}
\includegraphics[scale=0.4,type=pdf,ext=.pdf,read=.pdf]{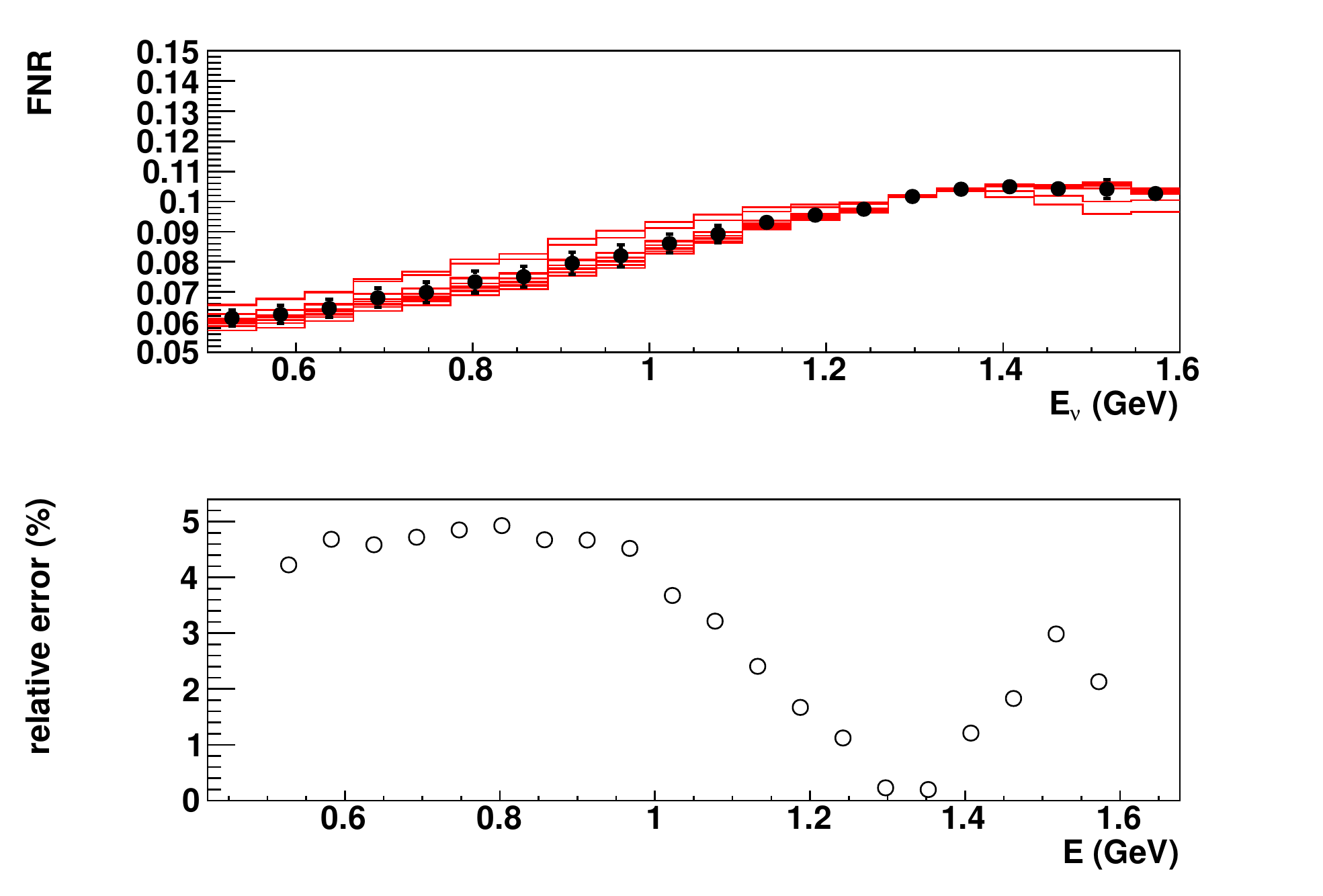}%
\includegraphics[scale=0.4,type=pdf,ext=.pdf,read=.pdf]{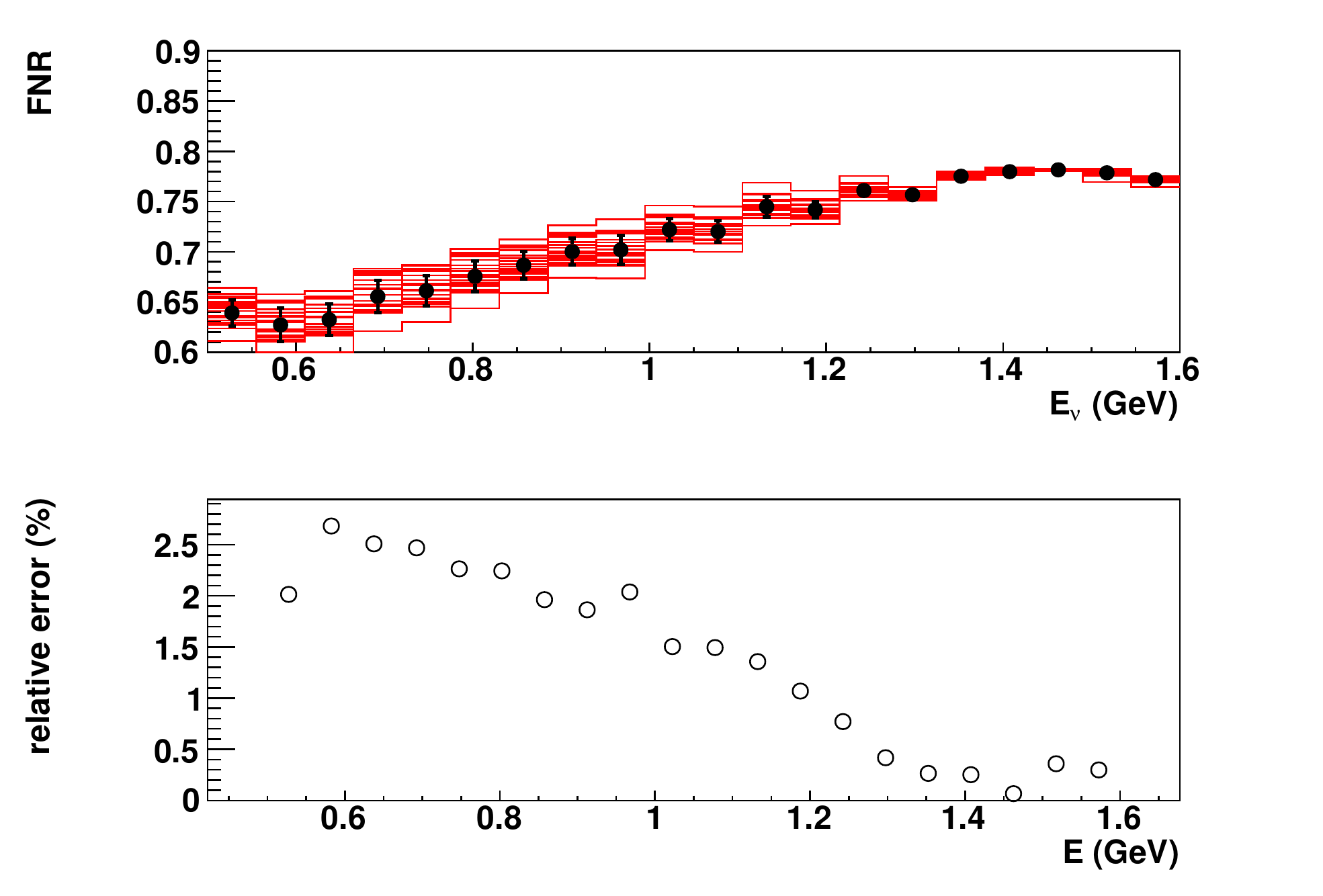}
\includegraphics[scale=0.4,type=pdf,ext=.pdf,read=.pdf]{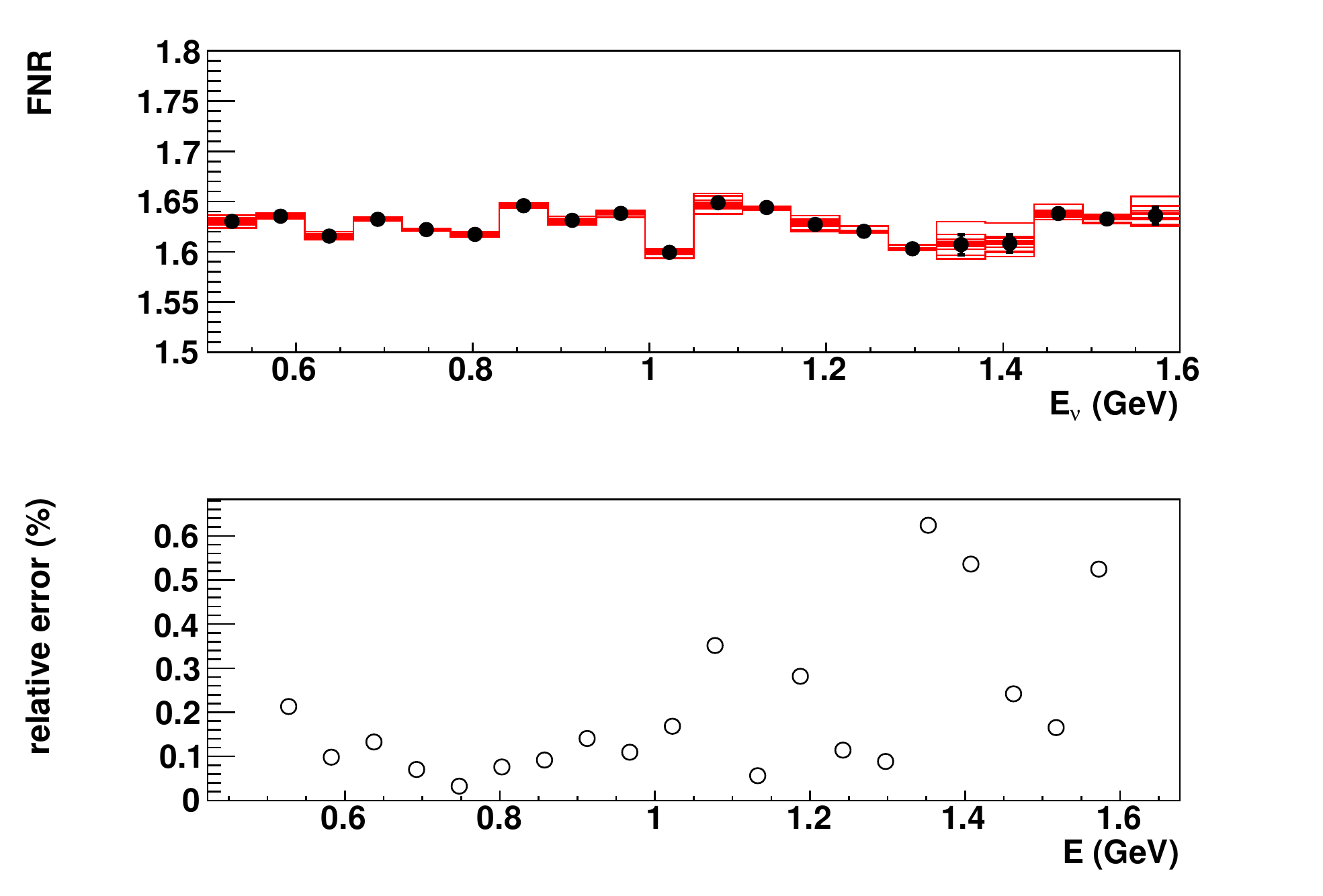}%
\includegraphics[scale=0.4,type=pdf,ext=.pdf,read=.pdf]{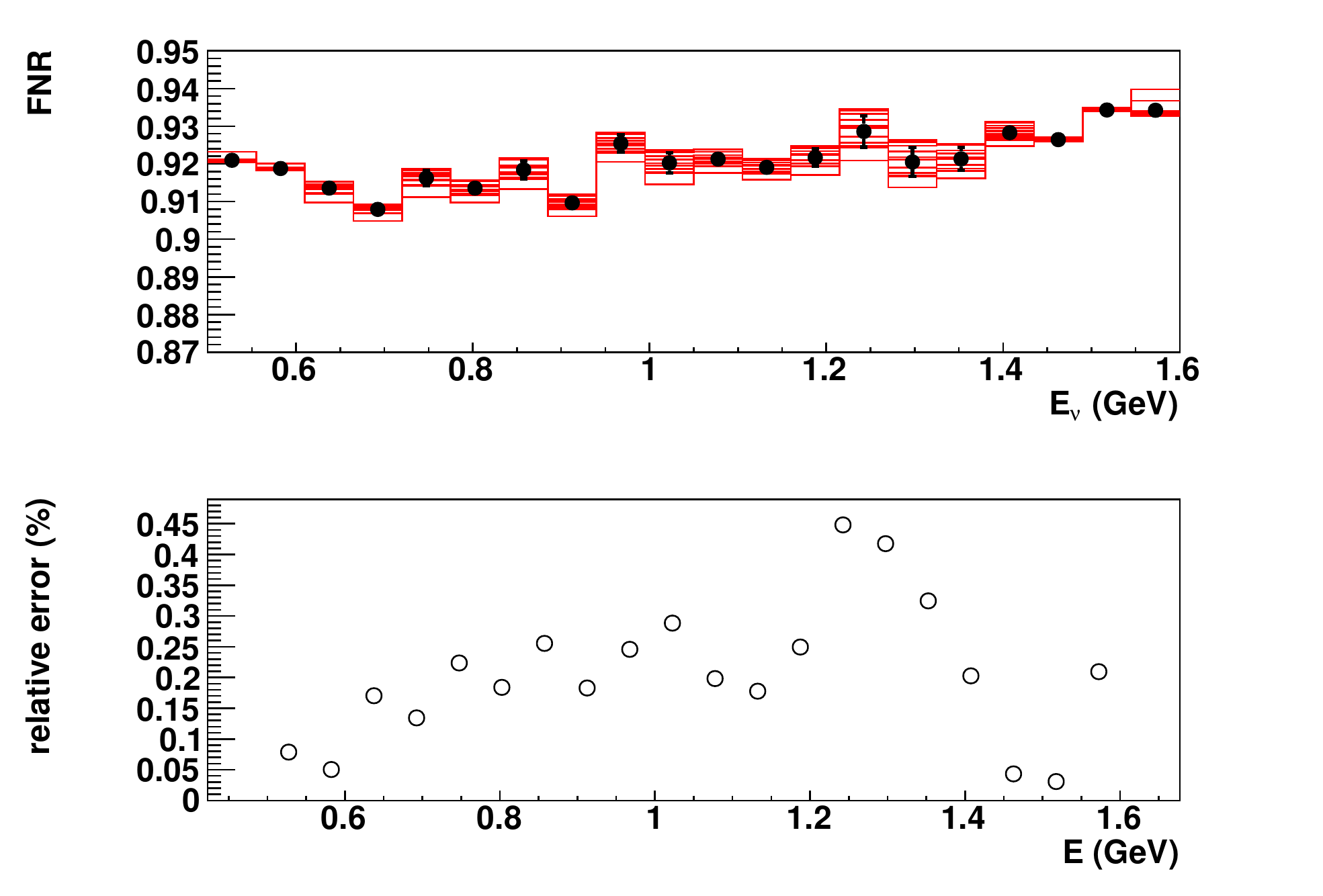}
\caption{Effect of data--driven hadroproduction uncertainties on the FNR calculated using the 
Sanford--Wang parametrization of for the six considered configurations. Red histograms
show different samplings of the $c_i$ parameters of Eq.~\ref{eq:SW}. 
Black bullets show the
average value in each bin while the error bars represent the r.m.s. of the samplings.
Bottom plots (hollow bullets) in each configuration show the ratio of the r.m.s. over the central value
giving an estimate of the fractional systematic error.
}
\label{fig:SWrew}
\end{figure}

For each sampling of the $c_i$ coefficients, neutrinos have been weighted with a factor:
\begin{equation}
w(p_\pi, \theta_\pi)=\frac{\frac{d^2\sigma}{dpd\theta}(c_i)}{\frac{d^2\sigma}{dpd\theta}(c^0_i)}
\end{equation}
depending on the momentum ($p_\pi$) and angle ($\theta_\pi$) of their
parent pion, $c^0_i$ being the best fit values of the fit to the
HARP and E910 data sets. The resulting FNR ratios for a set of
systematic variations of $c_i$ are shown in Fig.~\ref{fig:SWrew} for
the six considered configurations. Black bullets show the
average value in each bin while error bars represent the r.m.s. of the samplings.
Bottom plots (hollow bullets) in each configuration show the ratio of the r.m.s. 
over the central value providing an estimate of the fractional systematic error.
Uncertainties are rather large (5--7\%) when taking the complete area of the Near detector at 110~m
while they decrease significantly by restricting to the central region. In particular, configuration 4,
which is realistic from practical considerations has an uncertainty ranging from 2\% at
low energy decreasing below 1.5\% --0.5\% for neutrino energies above 1 GeV.
The uncertainty is also quite good (generally below 0.5\%) for a Near site at 460~m.

\subsection{Conclusions for the Booster Beam}
Using the constraints from HARP/E910 data sets, we have estimated the uncertainties associated to hadroproduction 
the FNR being of order 1--2\% for a configuration with the Far detector at surface and the Near detector with 
a similar off--axis angle and a fiducial volume tailored to match the acceptance of the Far detector (``configuration 4''). 
Given also the high available statistics and the large lever--arm for oscillation studied we consider
such a layout with baselines of 110~m and 710~m as a viable choice. Of course, ``configuration 4'' is a subset of ``configuration 3'',
which could be that to be used in reality by rearranging the OPERA spectrometers (see next Section). Therefore, given
the possibility for an higher statistics collection and the minor concern about the height of the pit (that has to be anyhow centered at the level
of the beam) the following studies are
driven by  ``configuration 3''.

\clearpage

%End-ANDREA%%%%%%%%%%%%%%%%%%%%%%%%%%%%%%%%%%%%%%%%%%%%%%%%%%%%%%%%%

\section{Spectrometer Design Studies}\label{sec:spect1}

The definition of two sites, Near and Far, constitutes a fundamental issue in the sterile neutrino search. Moreover the two detector systems 
at the two sites have to be as similar as possible. The NESSiE Far spectrometer has to be designed to cope with an aggressive time 
schedule and to largely exploit the acquired experience with the OPERA spectrometers in construction, assembling and 
maintenance~\cite{bopera}. Well known technologies have been considered as well as re--using large parts of existing detectors.
The OPERA spectrometers will begin to be dismantled sometime next year and we foresee to use them totally. The relatively low momentum
range of the muon tracks detected in the charged current events produced by the FNAL--BNB beam suggests to couple together the two OPERA spectrometers,
either for the Far or the Near site. Their modularity will allow to take 4/7 of the acceptance region (in height) for the Far site and 3/7 for the 
Near one. Each iron slab (see Sect.~\ref{sec:mech-struc}) will be cut at 4/7 in height to reproduce exactly the Far and Near targets.
In such a way any inaccuracy either in geometry (the single 5 cm iron slab owns a precision of few mm) or in the material will be 
exactly reproduced in the two detection sites.
The Near NESSiE spectrometer will then be an exact clone of the Far one, with identical thickness along the beam but scaled 
transverse size.

The achievements of 5 cm slabs have been analyzed and compared with a possible more performant thickness.

The current OPERA spectrometer design with 5 cm iron slabs has been studied using a complete and detailed simulation, profiting from the technical knowledge acquired with OPERA. 
The detector simulation is organized in two steps: the first one is the particle propagation inside the apparatus, based on GEANT3.21, with the concurrent creation of track hits; the second one is the digitization, i.e. the detector response to track hits creating detector digits.  
At this level, the Resistive Plate Chamber (RPC) efficiencies are implemented in the Monte Carlo simulation, taking into account %the slight difference due to 
the different widths of the horizontal and vertical sets of read--out strips. 
% with a precise description of the RPC and slab layout. 

The achievements of the current geometry with 5 cm iron slabs are evaluated in terms of NC contamination and momentum resolution, and compared to a possible geometry with 2.5 cm slab thickness. 

\begin{figure}[htbp]
\begin{center}
%\resizebox{0.9\columnwidth}{!}{\includegraphics{CRemucos} }
\resizebox{0.8\columnwidth}{!}{\includegraphics{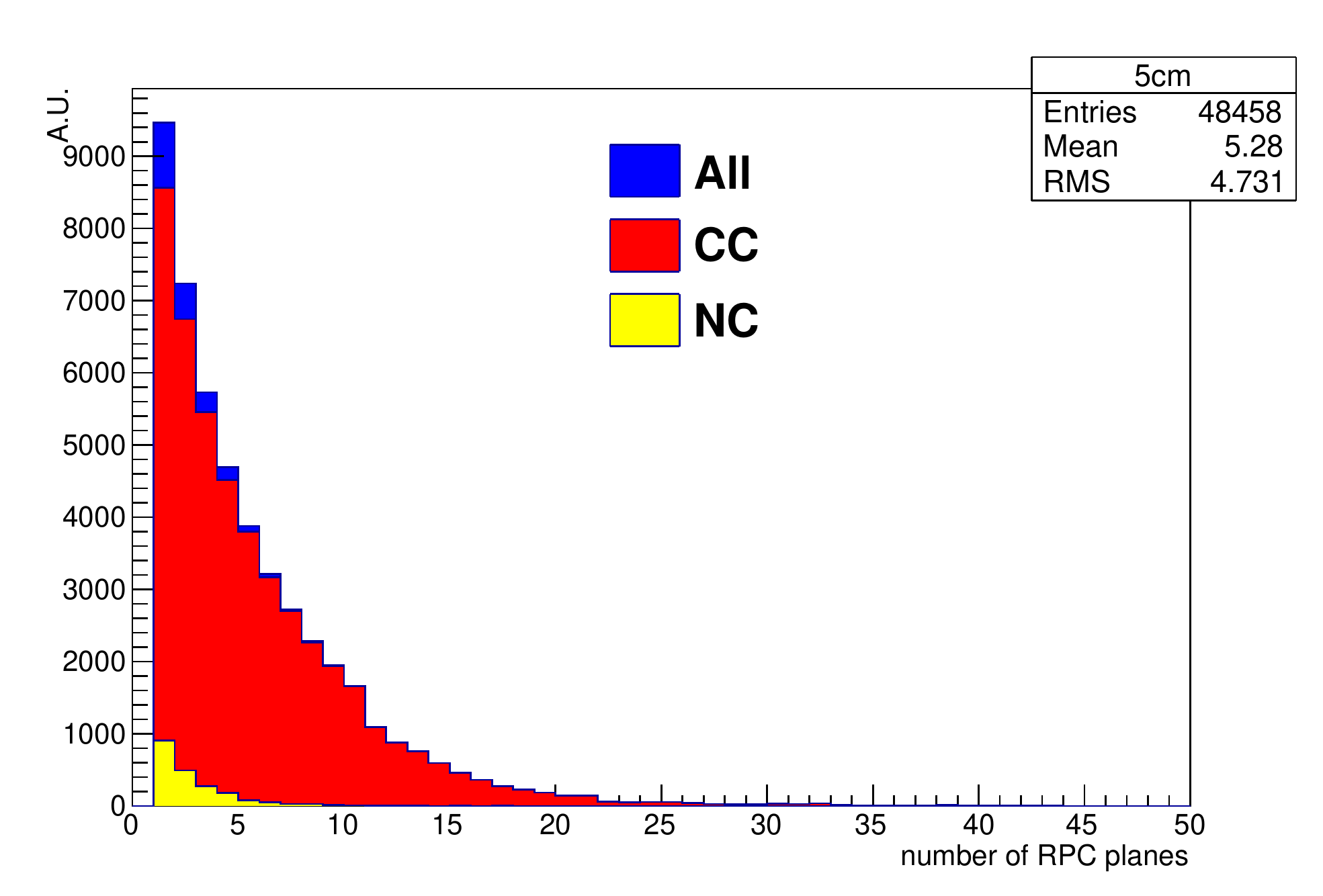} }
\caption{ Number of crossed RPC planes with the 5 cm slab geometry. }
\label{fig:RPCplanes_5cm}
\end{center}
\end{figure}

%\subsection{5 cm iron slabs (OPERA Spectrometer setup)}
Using the current geometry with 5 cm iron slabs, the fraction of neutrino interactions in iron giving a signal in the RPCs ($\varepsilon \equiv \frac{(\geq 1RPC)}{all}$) is 68\%. 
The efficiency %same fraction 
for CC and NC separately is $\varepsilon_{CC} = 86\%$ and $\varepsilon_{NC} = 20\%$. 
This corresponds to a fraction of NC interactions over the total number of interaction $\frac{NC}{all} = 8.1\%$. 
With a minimal cut of 2 crossed RPC planes, the NC contamination is reduced to 4.2\%; requiring 3 RPC planes the NC contamination is 3.0\%. 
%In Fig.~\ref{fig:RPCplanes} 
The distribution of crossed RPC planes is shown in Fig.~\ref{fig:RPCplanes_5cm} for both interaction channels, individually and jointly. 

\begin{figure}[htbp]
\begin{center}
%\resizebox{0.9\columnwidth}{!}{\includegraphics{CRemucos} }
\resizebox{0.8\columnwidth}{!}{\includegraphics{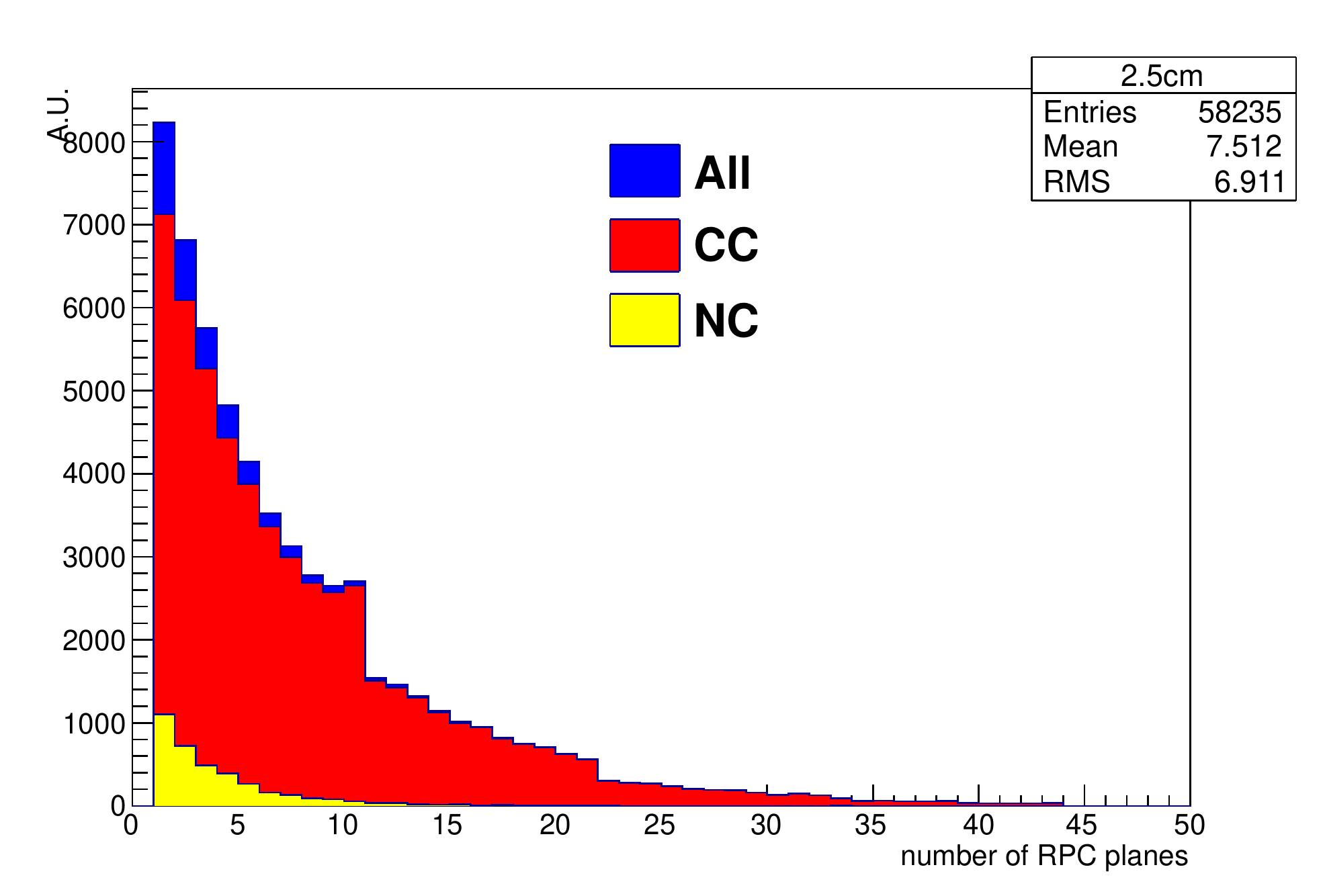} }
\caption{ Number of crossed RPC planes with the 2.5 cm slab geometry. Note that the knee visible at 11 planes is due to the geometry of the OPERA spectrometers, which own a modularity of 11 detector--planes
sandwiched into 12 iron slabs. This effect is present in any configuration but it comes more visible in the 2.5 cm case.}
\label{fig:RPCplanes_2.5cm}
\end{center}
\end{figure}

\begin{figure}[htbp]
\begin{center}
%\resizebox{0.9\columnwidth}{!}{\includegraphics{CRemucos} }
%\resizebox{0.8\columnwidth}{!}{\includegraphics{eff_pur_tot_gridy} }
\resizebox{0.8\columnwidth}{!}{\includegraphics{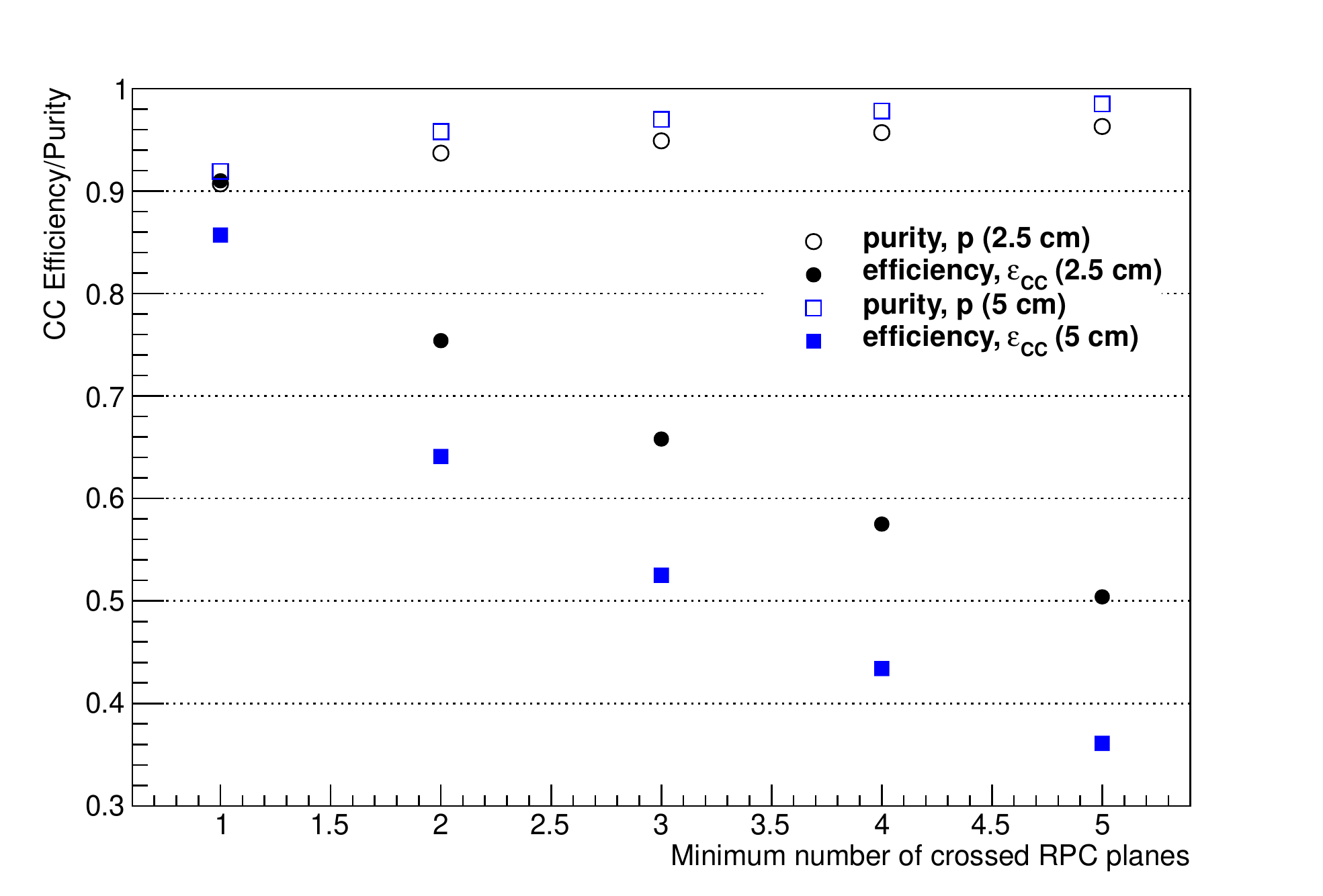} }
\caption{CC efficiency ($\varepsilon_{CC}$, points) and purity ($p$, open circles) as a function of the minimum number of 
RPC planes for the two spectrometer geometries, 5 cm slabs (in blue) and 2.5 cm slabs (in black). 
For a given level of purity $p$ the efficiencies for the two geometries are similar, therefore no advantage in statistics 
is taken requiring the same NC contamination suppression.}
\label{fig:eff_pur}
\end{center}
\end{figure}

Using 2.5 cm thick slabs, the fraction of neutrino interactions giving a signal in the RPCs increases for both NC and CC events. 
The distribution of crossed RPC planes in this configuration is shown in Fig.~\ref{fig:RPCplanes_2.5cm}. %for both interaction types individually and jointly. 
%Thus, 
Both the efficiency $\varepsilon_{CC}$ and the NC contamination are higher with respect to the reference 5 cm geometry. 
In Fig.~\ref{fig:eff_pur} the efficiency $\varepsilon_{CC}$ and the purity $p \equiv \frac{CC}{all} = 1 - \frac{NC}{all}$ are shown 
as a function of the minimum number of crossed RPC planes for both slab thicknesses. 
It can be noted that for a given level of purity $p$ the efficiencies for the two geometries are similar. 
%Therefore 
No advantage in statistics is taken requiring the same NC contamination suppression. 
%The second geometry 

The range in iron is reconstructed in three dimensions, merging the two RPC projection (horizontal and vertical) information. 
The muon momentum is obtained from the track range, using the continuous--slowing--down--approximation (CSDA)~\cite{pdg}. 
%cThe momentum resolution integrated over all the muon spectrum is $\Delta p / p = 13\%$, %calculated as 
%cevaluated from the %HWHM of the 
%cdistributions shown in Figs.~\ref{fig:mom_reso_5cm} and \ref{fig:mom_reso_2.5cm}  
%c(HalfWidth-HalfMax's of the distributions: HWHM(5 cm) = 0.13, HWHM(2.5 cm) = 0.12; $\sigma_{5 cm} = 0.127$, $\sigma_{2.5 cm} = 0.134$ 
%cfrom the gaussian fit).  
%cHalfWidth-HalfMax's of the distributions: HWHM(2.5 cm) = 0.12; $\sigma_{2.5 cm} = 0.134$ from the gaussian fit).  
%cThe momentum resolution is similar for the two considered geometries. 
%cUsing 2.5 cm slabs, the lower bound of the muon momentum measurement is extended, as shown in Fig.~\ref{fig:pt_vs_pr_5cm} and 
%cFig.~\ref{fig:pt_vs_pr_2.5cm}. 

\begin{figure*}[htbp]
%   \centerline{
\begin{center}
%  \subfloat[5 cm iron slab]{\includegraphics[width=0.6\columnwidth]{p_true_vs_p_reco_5cm} \label{fig:pt_vs_pr_5cm}} 
%  \hfil
%  \subfloat[2.5 cm iron slab]{\includegraphics[width=0.6\columnwidth]{p_true_vs_p_reco_25cm} \label{fig:pt_vs_pr_2.5cm}}
\subfloat[5 cm iron slab]{\includegraphics[scale=0.6]{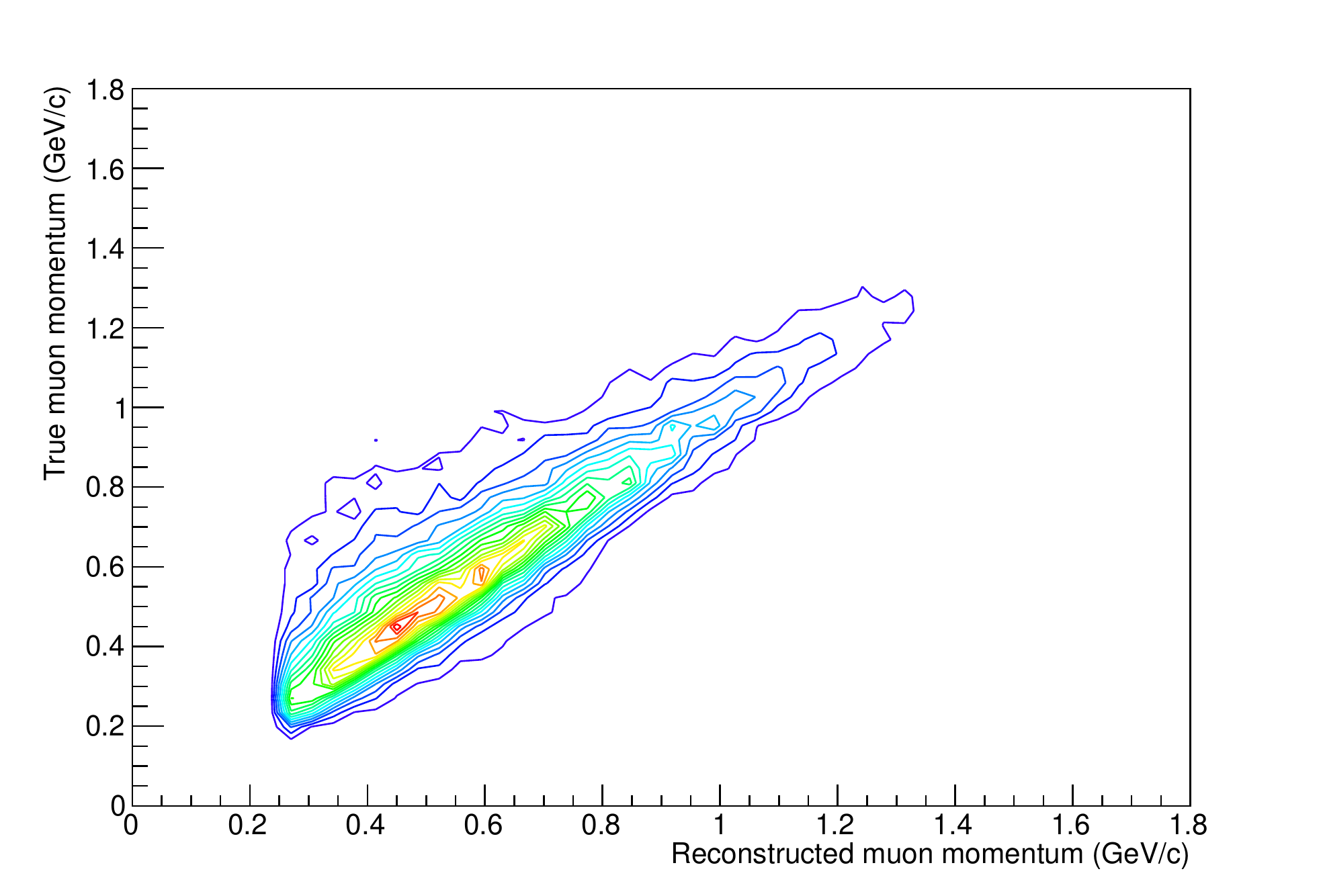}\label{fig:pt_vs_pr_5cm}} 
\end{center}
\begin{center}
\subfloat[2.5 cm iron slab]{\includegraphics[scale=0.6]{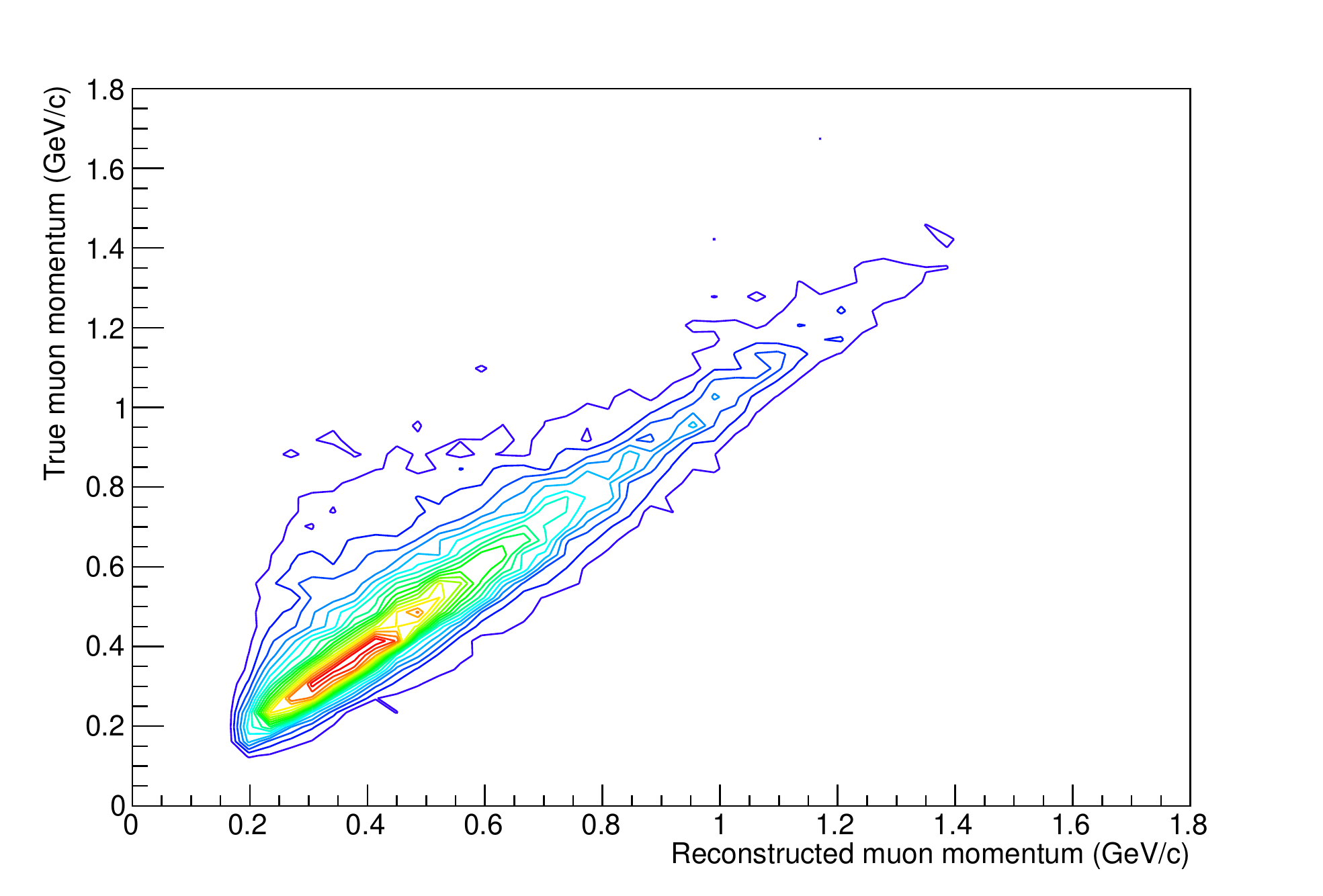}\label{fig:pt_vs_pr_2.5cm}}
 %  \subfloat[5 cm iron slab]{\includegraphics[width=0.6\columnwidth]{p_true_vs_p_reco_5cm} \label{fig:pt_vs_pr_5cm}} 
%   \hfil
%   \subfloat[2.5 cm iron slab]{\includegraphics[width=0.6\columnwidth]{p_true_vs_p_reco_25cm} \label{fig:pt_vs_pr_2.5cm}}
%             }
\end{center}
   \caption{ True muon momentum as a function of the reconstructed muon momentum for the two geometries, 5 cm slabs (a) and 2.5 cm slabs (b). The second geometry would allow to extend the lower bound of the momentum measurement to 200 MeV at the expense of 
   the level of CC purity. The 5 cm geometry will be able to reach anyhow the limit around 400 MeV keeping an high level of purity.}
   \label{fig:pt}
\end{figure*}

The muon momentum estimated in that way is compared to the true momentum in Fig.~\ref{fig:pt_vs_pr_5cm} and Fig.~\ref{fig:pt_vs_pr_2.5cm}
for the two options, 5 cm and 2.5 cm, respectively.
The resolution in either case is similar for the two considered geometries, even if the smaller thickness would allow a good
evaluation at smaller momenta. The 5 cm option is however sufficient to allow estimation from the minimum momentum of 400 MeV/c
or even lower (see also Sec.~\ref{sec:MC}).

%c\begin{figure*}[htbp]
%c   \centerline{\subfloat[5 cm iron slab]{\includegraphics[width=0.6\columnwidth]{mom_reso_5cm} \label{fig:mom_reso_5cm}} 
%c              \hfil
%c             \subfloat[2.5 cm iron slab]{\includegraphics[width=0.6\columnwidth]{mom_reso_25cm} \label{fig:mom_reso_2.5cm}}
%c            }
%c   \caption{ Momentum resolution obtained with the 5 cm slab (a) and the 2.5 cm slab (b) geometries. }
%c   \label{fig:reso}
%c\end{figure*}

%\begin{figure}[h!]
%\begin{center}
%%\resizebox{0.9\columnwidth}{!}{\includegraphics{CRemucos} }
%\resizebox{0.9\columnwidth}{!}{\includegraphics{mom_reso_5cm} }
%\caption{ Momentum resolution obtained  with the 5 cm slab geometry. }
%\label{fig:mom_reso_5cm}
%\end{center}
%\end{figure}
%
%\begin{figure}[h!]
%\begin{center}
%%\resizebox{0.9\columnwidth}{!}{\includegraphics{CRemucos} }
%\resizebox{0.9\columnwidth}{!}{\includegraphics{mom_reso_25cm} }
%\caption{ Momentum resolution obtained  with the 2.5 cm slab geometry. }
%\label{fig:mom_reso_2.5cm}
%\end{center}
%\end{figure}

\clearpage

%Start-STEFANO%%%%%%%%%%%%%%%%%%%%%%%%%%%%%%%%%%%%%%%%%%%%%%%%%%%%%%%

\section{Scintillator target at the Near site}\label{sec:scinti}

The Charged Current (CC) neutrino quasi--elastic (QE) scattering is the dominant process for neutrino interaction in the $\sim 1~{\rm GeV}$ region. In CCQE neutrino scattering the momentum of the produced lepton is strongly correlated with the energy of the neutrino. This implies that at these energies neutrino oscillations produce a distortion in the muon momentum spectrum similar to the one induced in the neutrino energy spectrum.  
Oscillation parameters (see Section~\ref{subsec:cls}) can also be extracted by fitting the muon momentum spectrum $\frac{dN}{dp_{\mu}}$ via the double ratio:
\begin{equation}
R = \frac{	
\left (	
\frac{\left (\frac{dN}{dp_{\mu}}\right )_{\rm Far}}{\left (\frac{dN}{dp_{\mu}}\right )_{\rm Near}}
\right )_{\rm Data}
}{	
\left (	
\frac{\left (\frac{dN}{dp_{\mu}}\right )_{\rm Far}}{\left (\frac{dN}{dp_{\mu}}\right )_{\rm Near}}
\right )_{\rm MC}
}
\end{equation}
on the assumed oscillation model. For the double ratio it is important to have an accurate measurement of the neutrino CC interaction rate to constrain the MC predictions. 

Moreover, as discussed in the previous Section, the contribution of  neutral current (NC) events is limited at the low--momentum range 
and therefore it constitutes 
 a systematic effect to deal with. Any measurement improving the disentanglement of NC and CC events will help in keeping that systematics
under control. Other sub--leading effects may also exist, such as the contamination
of muons coming from neutrino interactions in the rock and infrastructures surrounding the detectors, or even
the possible appearance of the \numunue ``contamination''. 

The study of these systematic effect can not relay only on MC simulations. Indeed neutrino interactions in the $\sim 1~{\rm GeV}$ region are mostly based on measurements of neutrino interactions on deuterium targets 
in bubble chambers~\cite{deuterium}. In this energy region the nuclear effects of the neutrino target material (from Fermi motion and the nuclear potential) are
significant, therefore cross--sections on deuterium target are not directly applicable to  heavier nuclear target materials, like iron.

By an appropriate choice of an active target  design, namely its mass and segmentation, a separation of the quasi--elastic part in the CC sample can also be attained.
As a whole, that target would act as an independent detector to be used for inter--calibration and normalization
of the different data samples.

%MC prediction of neutrino interactions in the $\sim 1{\rm GeV}$ region are mostly based on measurements of neutrino interactions on deuterium targets 
%in bubble chambers ~\cite{deuterium}. In this energy region the nuclear effects of the neutrino target material (from Fermi motion and the nuclear potential) are
%significant, therefore cross sections on deuterium target are not directly applicable to  heavier nuclear target materials, like iron. To better constrain  MC predictions at the Far detector an {\em in situ} measurement of neutrino cross sections are required.

\begin{figure}[htbp]
\begin{center}
\includegraphics[height=1.5in]{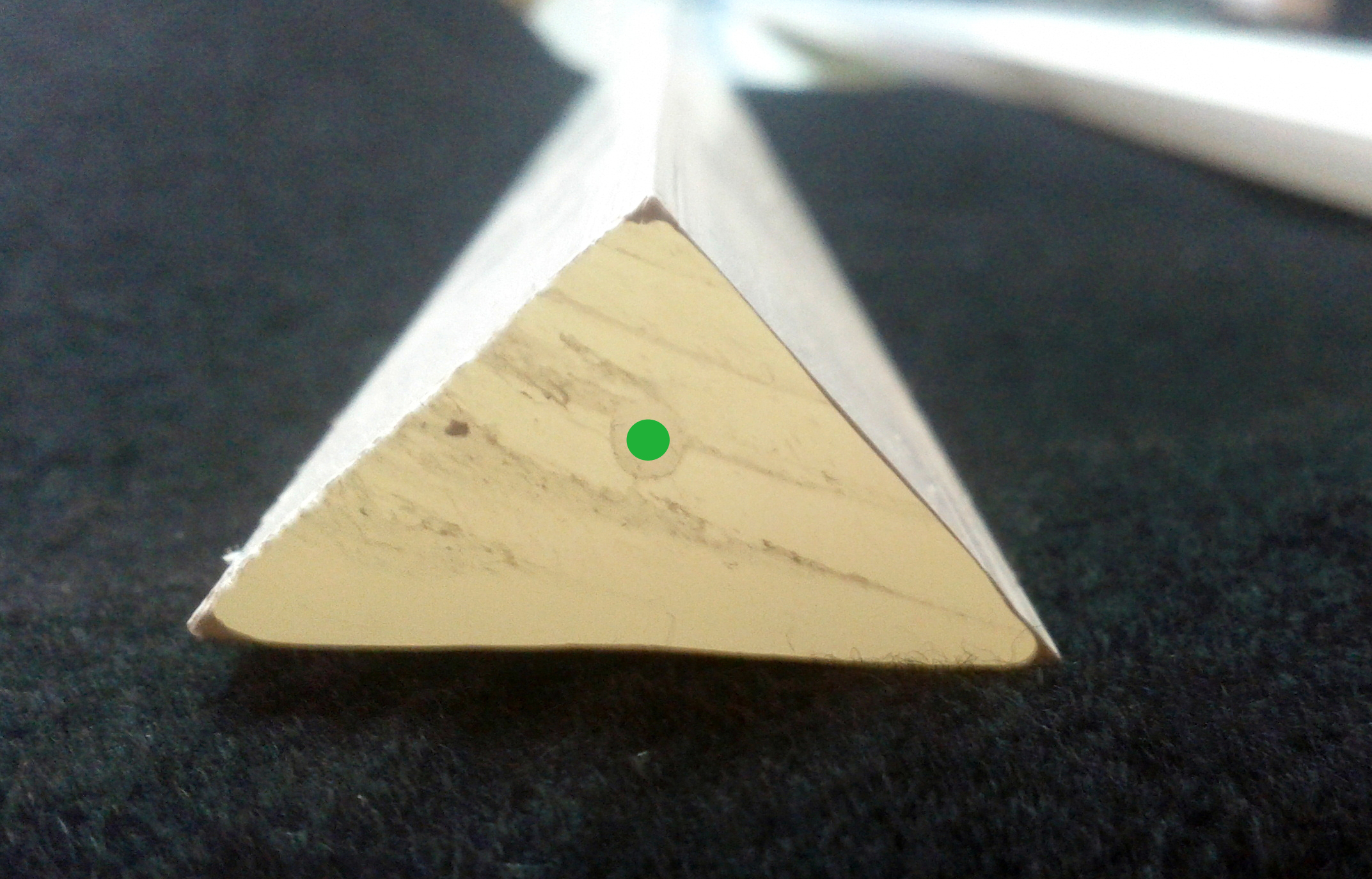}
\includegraphics[height=1.5in]{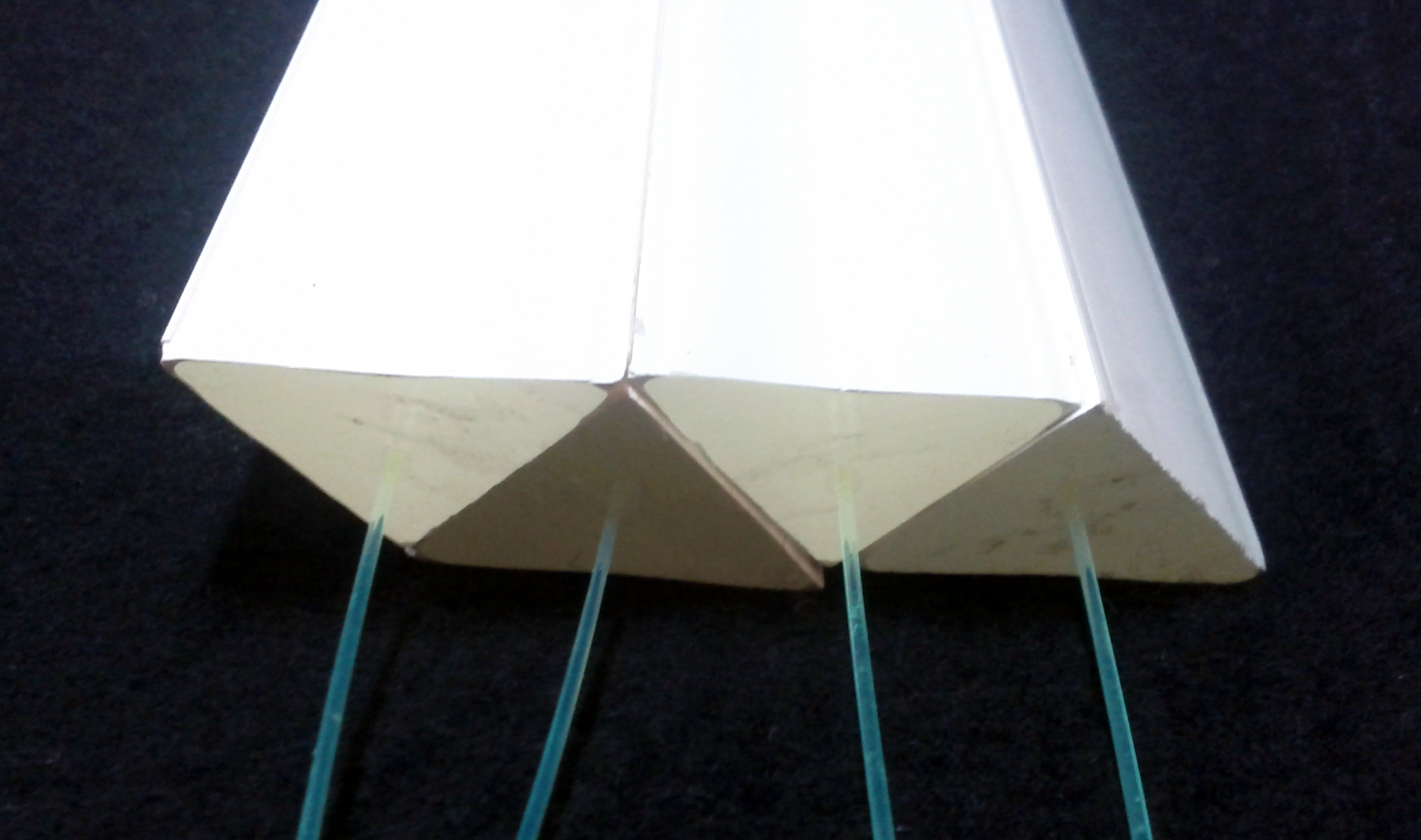}
\caption{Left: triangular scintillator bar and WLS fibers. Right: sketch of a 2 plane scintillator bar tracker.}  
\label{fig:sci-bar}
\end{center}
\end{figure}

To this end  a plastic Scintillator Tracking Detector (STD) to be located at the Near site, upstream of the spectrometer, is under study.
%, with sufficient granularity to study in detail the neutrino interactions.    
%To perform this measurement we plan to equip the near detector with a plastic scintillator tracking detector (calorimeter?) placed in front of the iron spectrometer capable to reconstruct and study the neutrino interactions.   
In its preliminary design it is formed by 20 Target Tracker Modules (TTMs), each  composed of $1~{\rm cm}$ thick iron slab followed by a Tracking Module (TM) section,
%The detector has a modular structure composed of $1~{\rm cm}$ thick iron slabs followed by a total active Tracking Module (TM). The iron slab plus the tracking module form the so called Target Tracker Module (TTM).  
%This detector is composed of  tracking modules interleaved with $1~{\rm cm}$ iron slabs which act as target for the neutrinos. The total number  
based on the MINER${\nu}$A  triangular scintillator bar technology~\cite{bib:minerva}. An TM module consists of two planes of 64 scintillator bars, each 
$211~{\rm cm}$ long with  triangular cross--section ($1.7~{\rm cm}$ high and $3.3~{\rm cm}$ side), and a central $2.6~{\rm mm}$ diameter 
hole to lodge the wavelength shifting (WLS) fiber (Fig. \ref{fig:sci-bar}). Each bar is read by Silicon Photomultipliers (SiPM) with the hit position determined by  analog pulse 
readout and analysis.

%glued together and covered with light tight Lexan sheet. 
%Each planes consist of 64 strips glued together and covered with light tight Lexan sheet. 
Scintillator bars  have vertical or horizontal orientation to provide  $X$ and $Y$ coordinates. Eight consecutive scintillator bi-planes with alternating orientation form a Tracking Module.\\ 
%Each tracking module is composed of two planes of triangular scintillator strips. The module can be oriented with vertical or horizontal strips in order to measure the $x$ and $y$ coordinate.  
%The iron slabs are separated by 4 tracking modules. 

\begin{figure}[htbp]
\begin{center}
\includegraphics[width=0.5\textwidth]{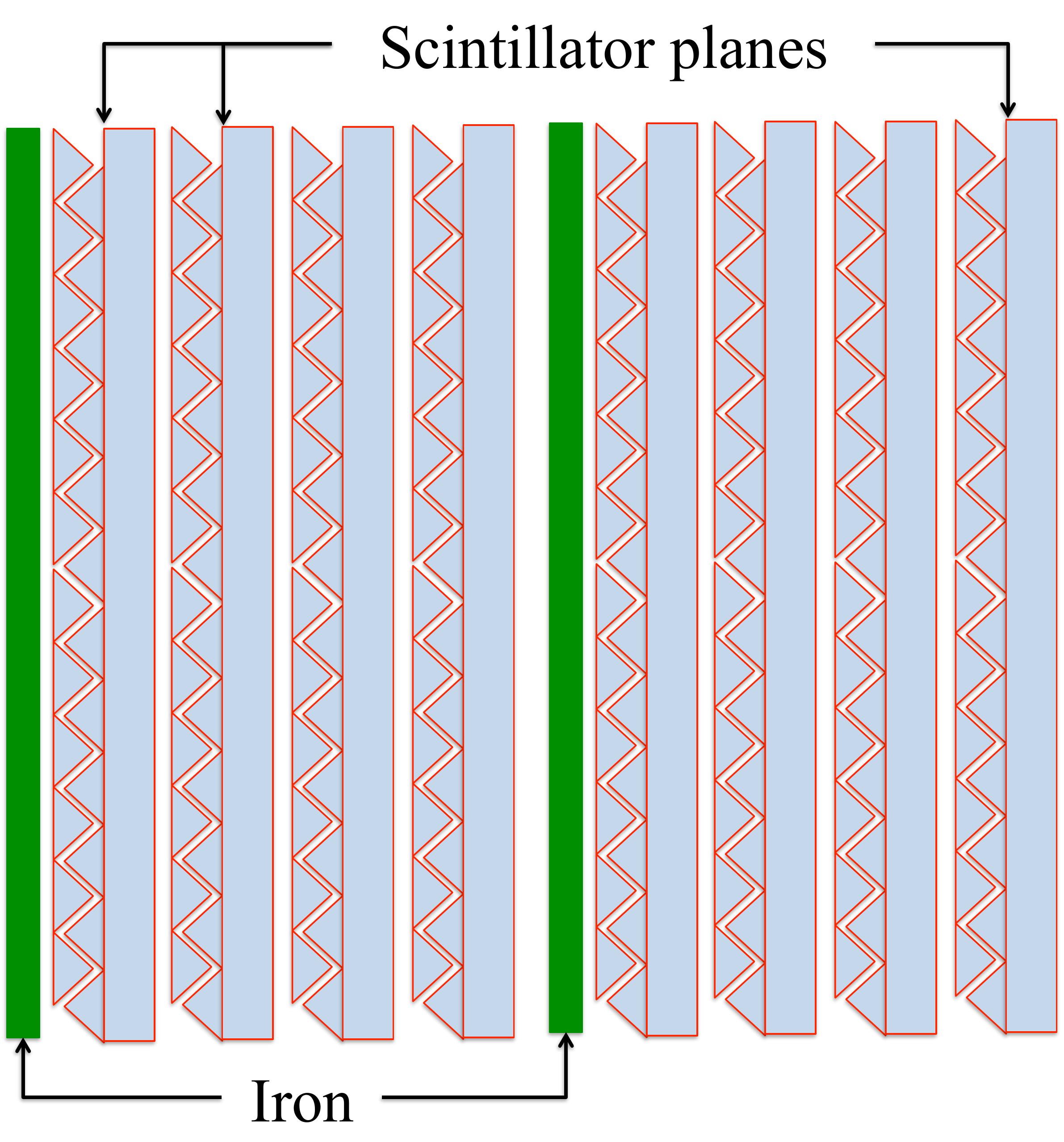}
\caption{Side view of two Target Tracker Modules.}
\label{fig:default}
\end{center}
\end{figure}

The STD allows an accurate  tracking of particles emerging from the interaction down to low energy with a sufficient level arm to reconstruct the neutrino interaction vertex in the iron. Based on the track fine sampling   the particle identification through ${dE}/{dx}$ near the end point on stopping tracks can be obtained.     
%The total-active tracking region between two iron walls  
%In order to track and identify low energy particles produced in the neutrino interaction the iron slabs are separated by 4 tracking modules. This total-active tracking region composed purely of scintillator strips allow a precise tracking of the particle produced in the interaction down to low energy and an accurate reconstruction of the neutrino interaction vertex. The fine sampling of the tracks allow the particle identification thought ${dE}/{dx}$ near the end point on the stopping tracks.\\       
A sketch of the possible detector configuration is shown in Fig. \ref{fig:default}.
Two scintillator planes with horizontal and vertical orientation with a larger cross--section with respect to the TTM is placed upstream would act as veto for neutrino interaction occurring outside the detector.

With a total iron target mass of  $\sim$ 7 tons  the expected CC interaction rate is $\sim 2\times 10^4~\nu_{\mu}~{\rm CC}/10^{20}~{\rm p.o.t.}$. 
According to references~\cite{Nakajima:2010fp} and~\cite{Kurimoto:2009wq}, with an integrated intensity of $2.2\times 10^{20}~{\rm p.o.t.}$
corresponding to 1 year of data taking, the $\nu_{\mu}~{\rm CC}$ rate can be determined with a $3\%$ accuracy and the $NC/CC$ ratio with 
a precision better than $10\%$. 
A detailed Monte Carlo simulation is under development to precisely evaluate the detector performances. In the present document no further reference is made to this subdetector that, at the present stage, was only introduced as a possibility to be considered\footnote{In 
Tab.~\ref{tab:costs} where costs are given, a first estimation for this detector is anyhow provided.}. 

\clearpage

%End-STEFANO%%%%%%%%%%%%%%%%%%%%%%%%%%%%%%%%%%%%%%%%%%%%%%%%%%%%%%%%%%%%

%Start-GABRIELE%%%%%%%%%%%%%%%%%%%%%%%%%%%%%%%%%%%%%%%%%%%%%%%%%%%%%%%%%%%

\newcommand{\includerootgraphics}[1]{\includegraphics[width=7.5cm, trim=5mm 2mm 1mm 10mm,clip=true]{#1}}

\section{Monte Carlo Detector Simulation and Reconstruction}\label{sec:MC}

The present proposal has been extensively developed  using full--detail programming 
and up--to--date software packages to obtain precise
understanding of acceptances, resolutions and physics output. Although not all  possible options 
have been studied, a rather exhaustive list of different magnet configurations and detector designs 
has been adopted as benchmark for further studies once the 
detector structure will be finalized.

\subsection{Simulation}

The aim of the simulation of the apparatus is to help the design studies reported here and to 
understand the main 
features of the proposed experiment. The simulated detector consists of a Near and
and Far magnetic spectrometer (ND and FD, respectively).
%
% BEAM 
%
The muon neutrino and antineutrino Booster fluxes for positive and negative beam 
polarity were considered, with 
a beam of 0 $mrad$ tilted with respect to the horizontal. 
%No angular dependence of the neutrino fluxes has been considered. 
%TODO: Ref: simulazione flusso

%NUINT

\begin{figure}[htp]
\centering
%\includerootgraphics{simulation/pictures-arc/beam-Ev}
%\includerootgraphics{simulation/pictures-arc/beam-inttype}
\includerootgraphics{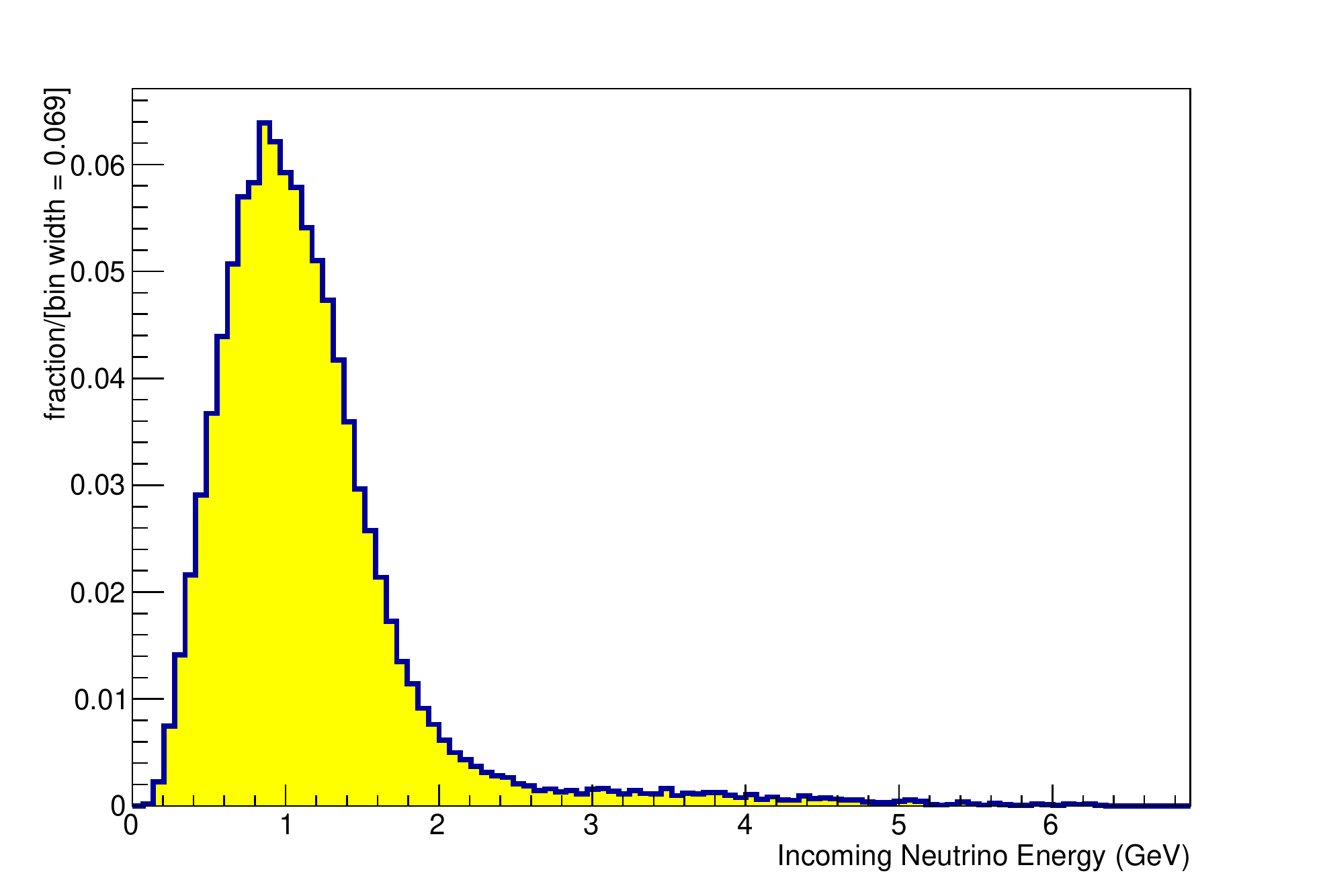}
\includerootgraphics{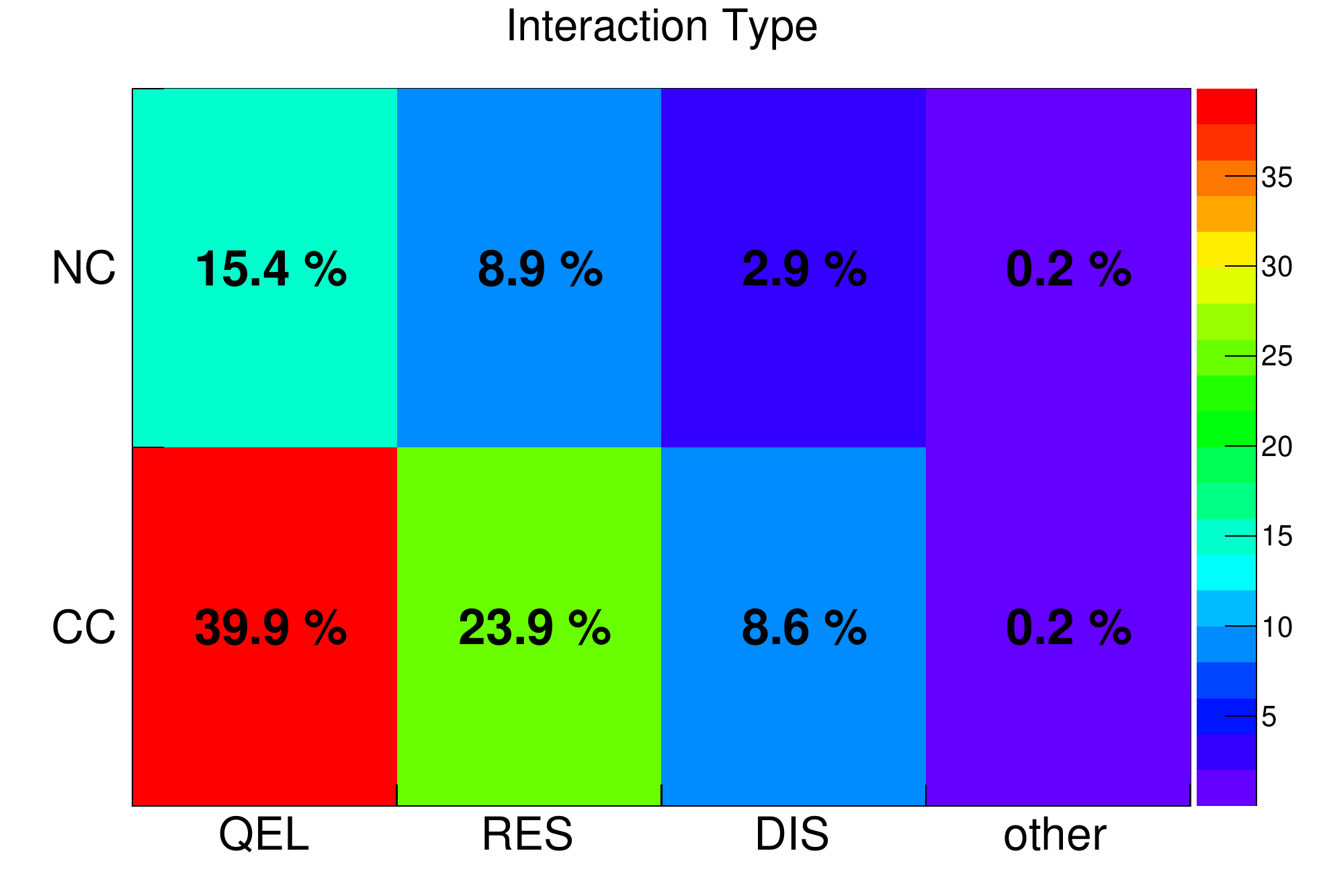}
\caption{ \label{fig:GenieNuInt1} Booster $\nu_\mu$ interactions in iron as generated by GENIE. The distribution of incoming neutrino neutrino energy is shown in the left plot, while the rates of the single interaction processes are reported in the table at right.}
\end{figure}

\begin{figure}[htp]
\centering
%\includerootgraphics{simulation/pictures-arc/beam-hadrons-all}
%\includerootgraphics{simulation/pictures-arc/beam-P-hadrons-all}
\includerootgraphics{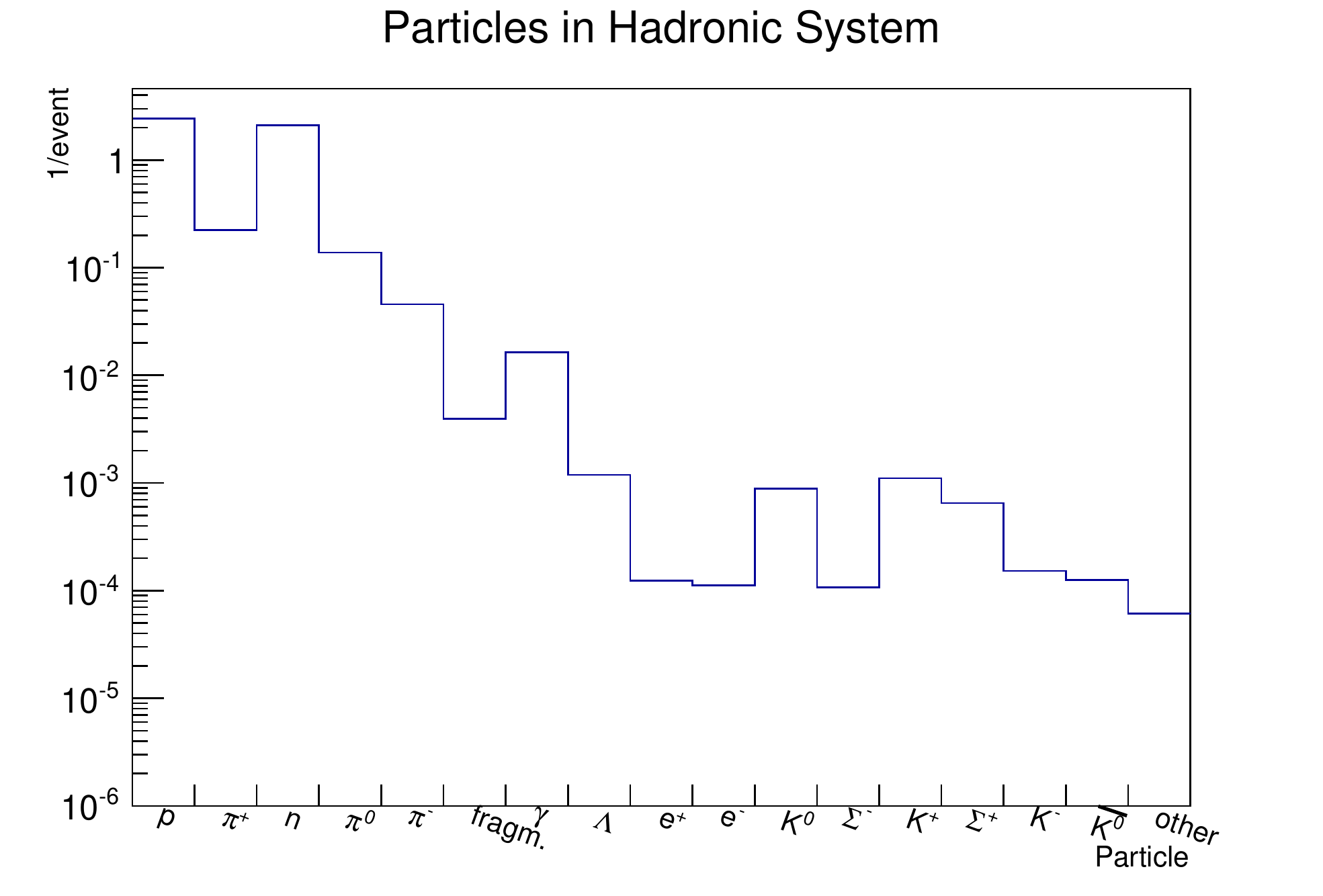}
\includerootgraphics{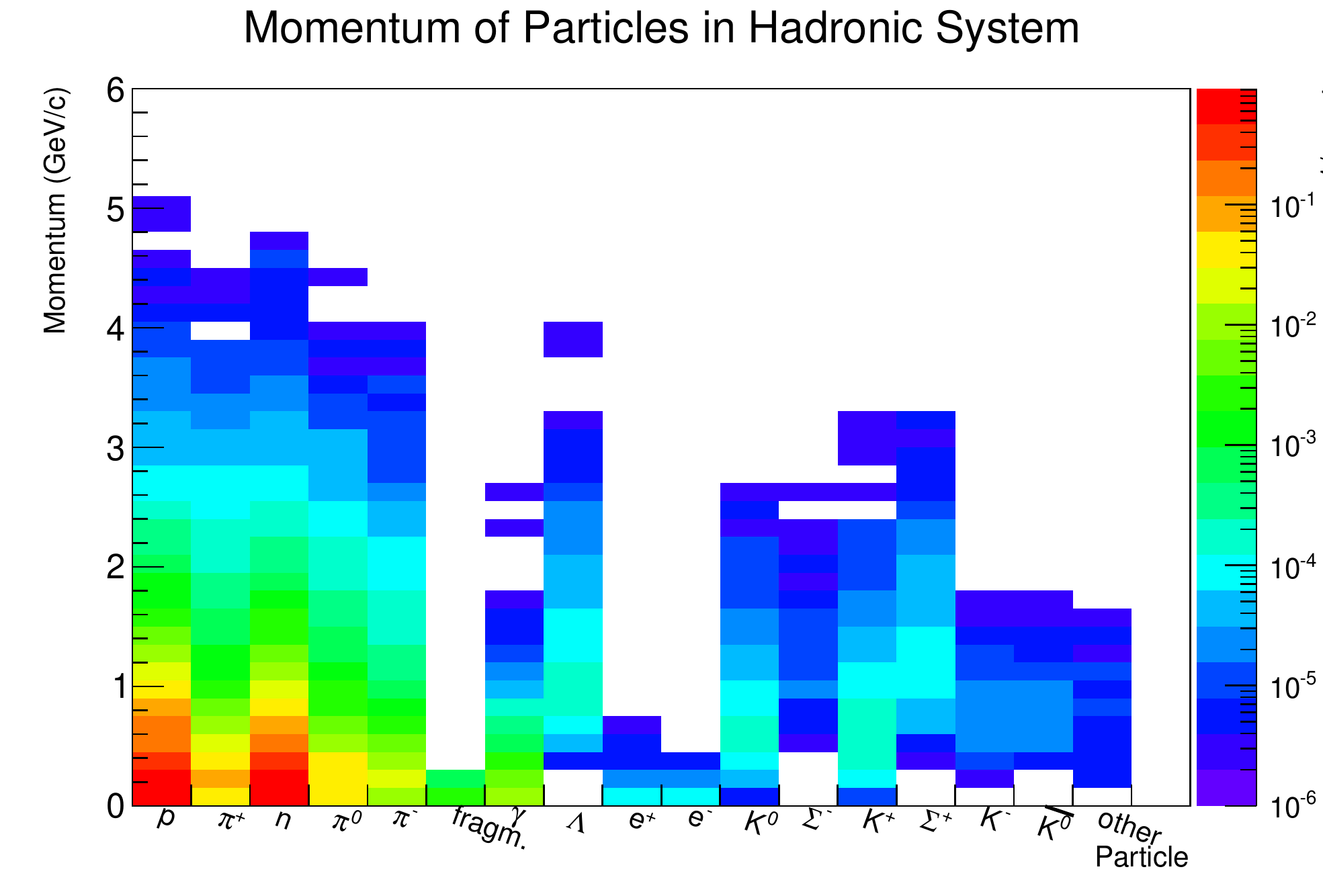}
\caption{ \label{fig:GenieNuInt2} Booster \numu  interactions in iron:  particles in hadronic system (left) and their momentum (right). These distribution are normalized to the total number of events. }
\end{figure}

\begin{figure}[htp]
\centering
%\includerootgraphics{simulation/pictures-arc/beam-pmu}
%\includerootgraphics{simulation/pictures-arc/beam-muang}
\includerootgraphics{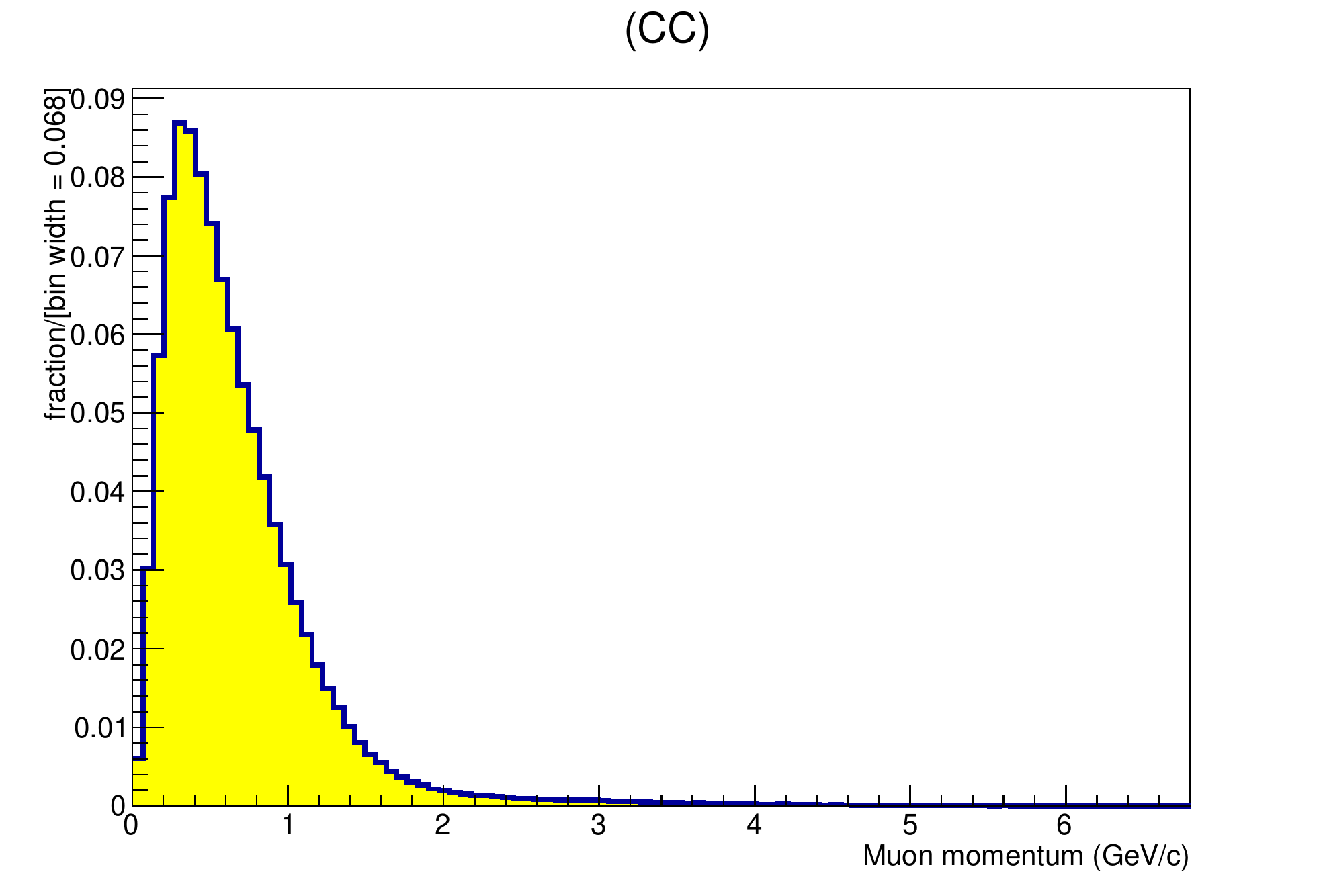}
\includerootgraphics{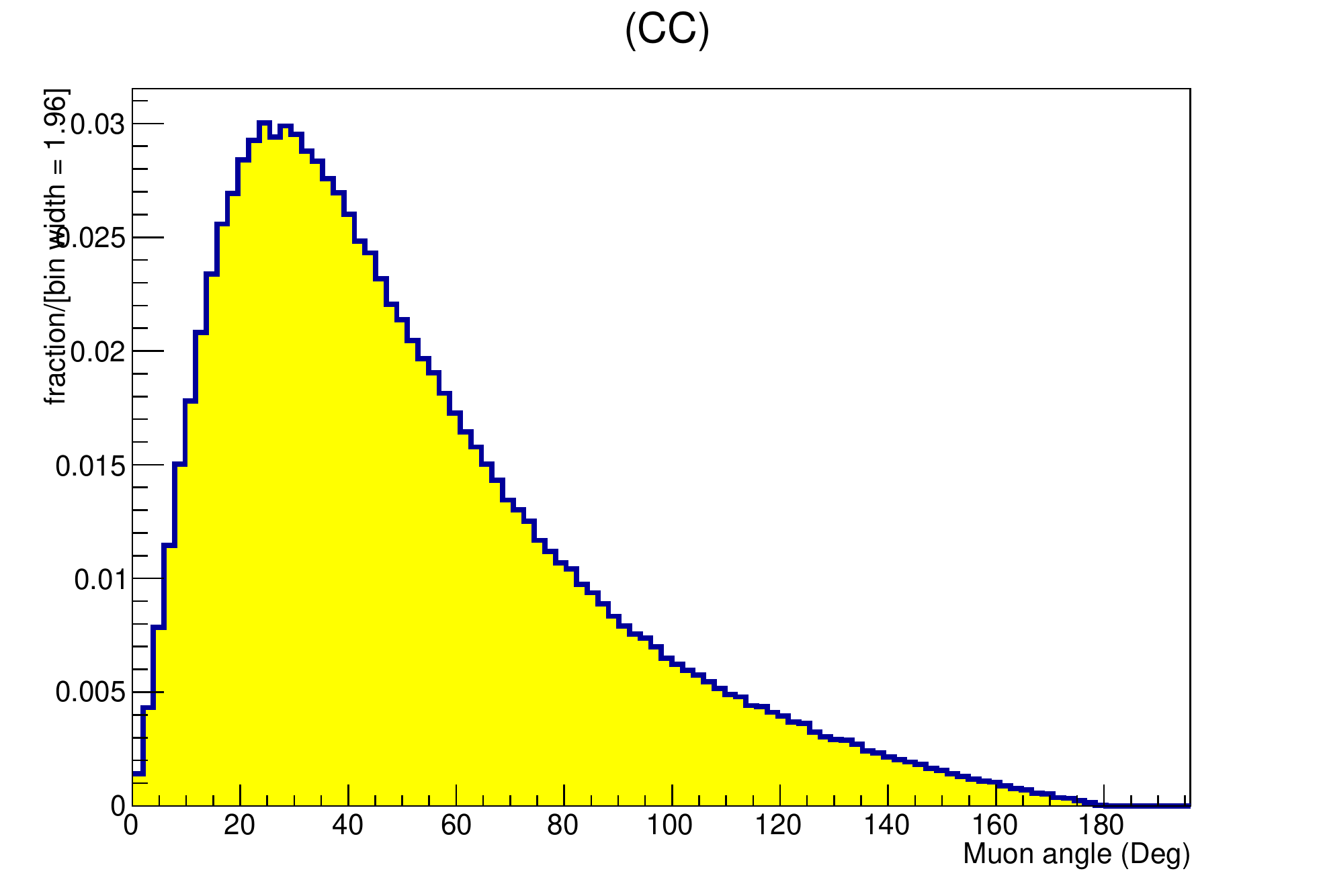}
\caption{ \label{fig:GenieNuInt3} Booster \numu CC interactions in iron: muon momentum (left) and angle (right).  }
\end{figure}

\begin{figure}[htp]
\centering
%\includerootgraphics{simulation/pictures-arc/beam-pmu-Ev-col}
%\includerootgraphics{simulation/pictures-arc/beam-muang-Ev-col}
%\includerootgraphics{beam-pmu-Ev-col}
%\includerootgraphics{beam-muang-Ev-col}
\includegraphics[width=7.5cm]{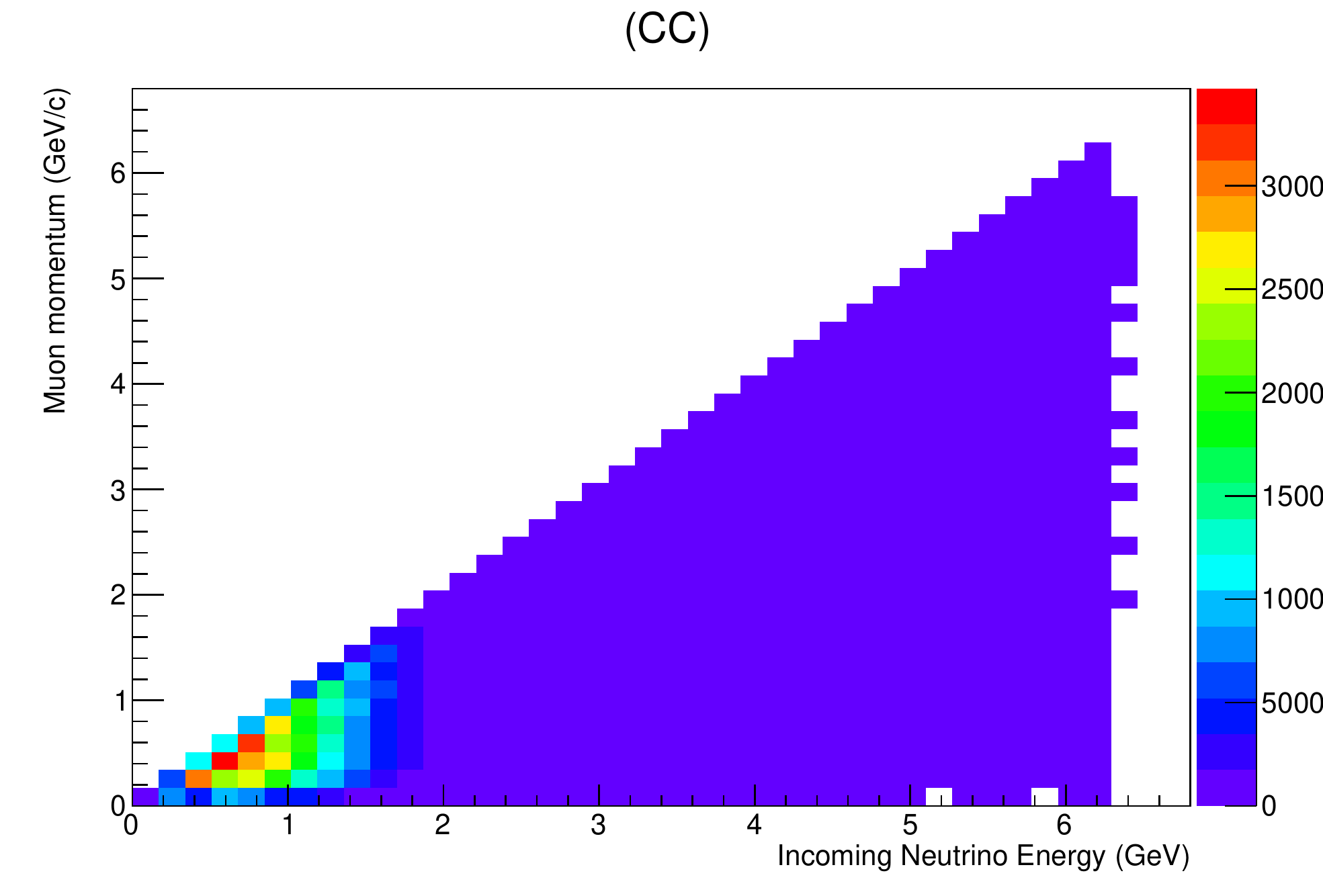}
\includegraphics[width=7.5cm]{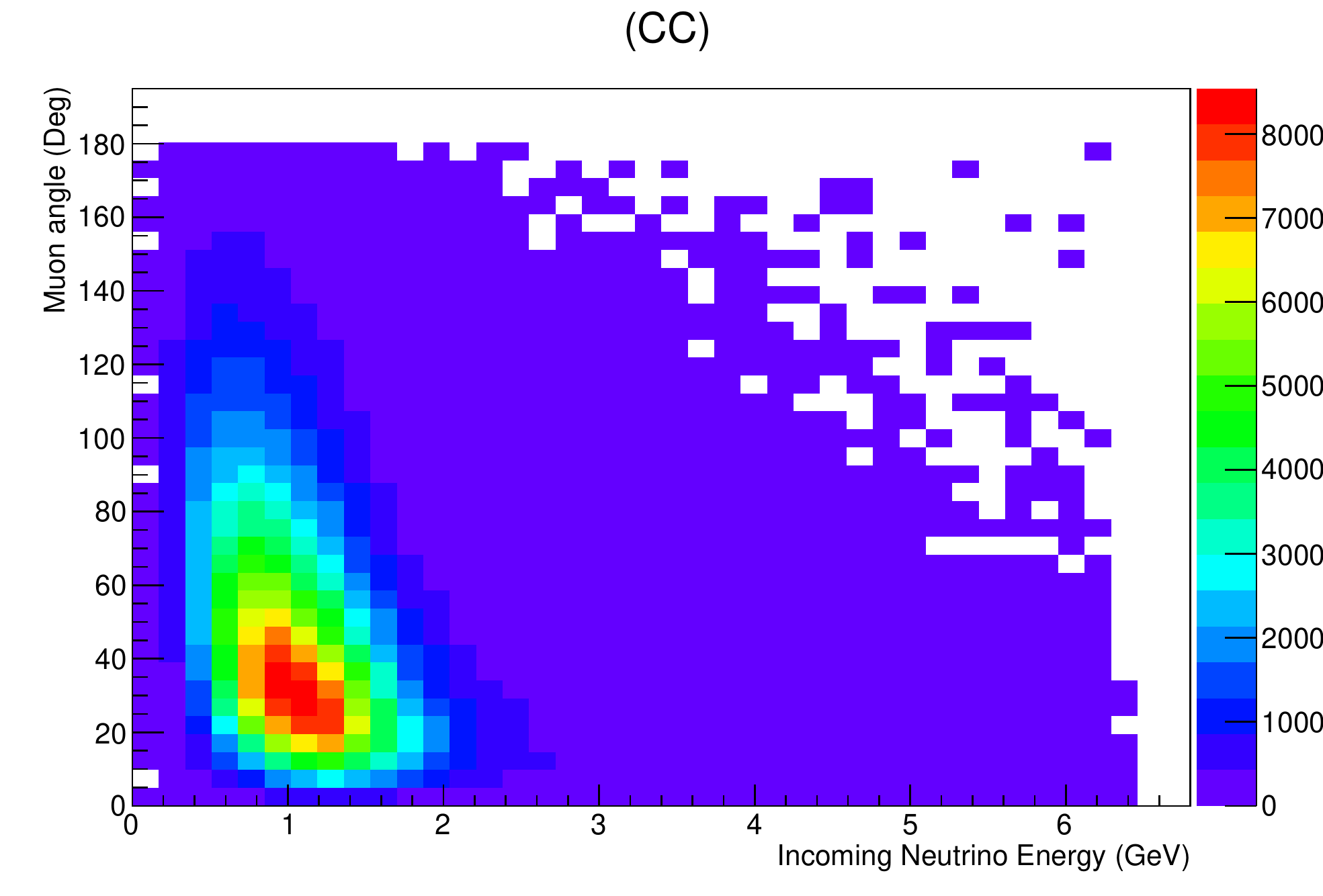}
\caption{ \label{fig:GenieNuInt4} Booster \numu CC interactions in iron: dependency of muon momentum (left) and angle (right) on the incoming neutrino energy.  }
\end{figure}

Neutrino interactions are generated in the Iron target using GENIE~\cite{ref_GEN} 
with standard parameters and including all interaction processes (QE, RES, DIS, NC), as shown in  Fig.~\ref{fig:GenieNuInt1}.
Distribution of particles in the hadronic systems are shown in  Fig.~\ref{fig:GenieNuInt2}; muon momentum and angle in  Fig.~\ref{fig:GenieNuInt3} and~\ref{fig:GenieNuInt4}.

% GEOMETRY

The propagation in the detector is implemented with either GEANT3.21 or FLUKA. The geometry of the detectors is described 
using FLUKA and ROOT geometry packages. The main features of the geometry implemented in the simulation are briefly described below. 
The  spectrometer consists of 2 instrumented dipolar magnets. Each magnet is made of two magnetized 
iron walls producing a field of 1.5 T intensity in the tracking region; 
field lines are vertical and have opposite directions in the two adjacent walls whereas
track bending occurs on the horizontal plane.
The thickness of the iron planes is at present envisaged to be 5 cm.
Planes of bakelite RPC's are interleaved with the iron slabs of each wall
to measure the range of stopping particles and to track penetrating muons. 
Each magnet is equipped with 22 planes of 4 rows, each consisting of 3 
RPC's. The ND spectrometer is assumed to be 
similar to the FD (22 planes each with 3 rows and 2 columns of RPC's).

For the RPC's we assume digital read--out using 2.6~cm strip width and a position resolution of about 1~cm.

\begin{figure}[htp]
\centering
\includegraphics*[width=16cm, trim=1cm 4cm 1cm 2.5cm,clip=true]{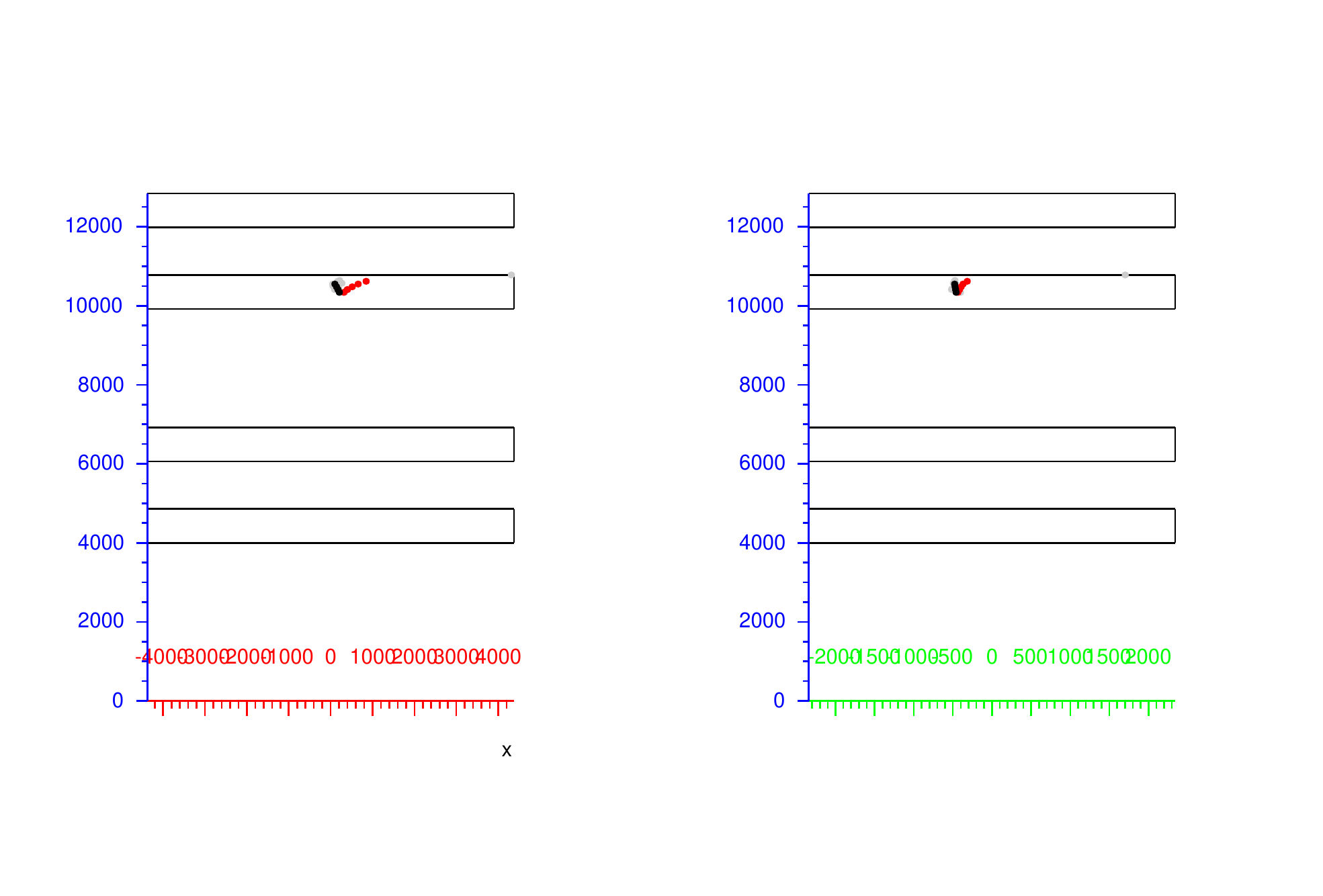}
\caption{ \label{fig:FlukaEventDisplay} A event display of one Booster Neutrino CC Interaction in the spectrometer. The red dots are the hits of a muon track; the blacks of hadrons. The neutrino is coming from the bottom. }
%TODO rifare
\end{figure}

An event display of one booster neutrino \numu CC interaction in the spectrometer is shown in Fig.~\ref{fig:FlukaEventDisplay}.

\subsection{Reconstruction}

% RECONSTRUCTION

A framework based on standard tools (ROOT, C++)  has been developed for the reconstruction in both the Near and Far detectors. 
The reference frame is defined to have the $Z$--axis along the beam direction, $Y$ perpendicular to the floor pointing 
upwards and $X$ completing a right-handed frame. Event reconstruction is performed separately in the two projected views $XZ$ (bending plane) 
and $YZ$ (vertical plane). 

A simple track model is adopted to describe the shape of the trajectory of tracks traveling through the detector. 
The model is based on the standard choice of parameters used in {\em forward} geometry (i.e.  intercepts, slopes, particle momentum, 
particle charge etc.). The reconstruction strategy is optimized to follow a single long track (the muon escaping from the neutrino--interaction 
region) along the $Z$--axis. 

The reconstruction is performed in the usual two steps: Pattern Recognition (Track finding) and Track Fitting. 
The task of the Pattern Recognition is to group hits into tracks. 
% Assuming a read-out capable of avoiding event overlap, 
% we postponed the development of a dedicated algorithm
%{\colorbox{yellow}{For the time being all the hits of the events are associated to a single track}.
The Track Fitting has to compute the best possible estimate of the track parameters according to the track model. 
A parabolic fit is performed in the $XZ$ plane (bending) whereas a linear fit is used for $YZ$ plane (vertical). Particle charge and momentum 
are determined from the track sagitta measured in the bending plane; the track fit is corrected for material interactions (Multiple Scattering 
and energy loss). Each spectrometer arm provides an independent measurement of charge/momentum.  The implementation of a track fitting algorithm based on a {\em Kalman} filter is eventually foreseen.

A better estimation of the momentum is obtained by range for muons stopping inside the spectrometer. 

% ANALYSIS

\begin{figure}[htp]
\centering
%\includerootgraphics{simulation/pictures-arc/nplanes}
%\includerootgraphics{simulation/pictures-arc/range}
\includerootgraphics{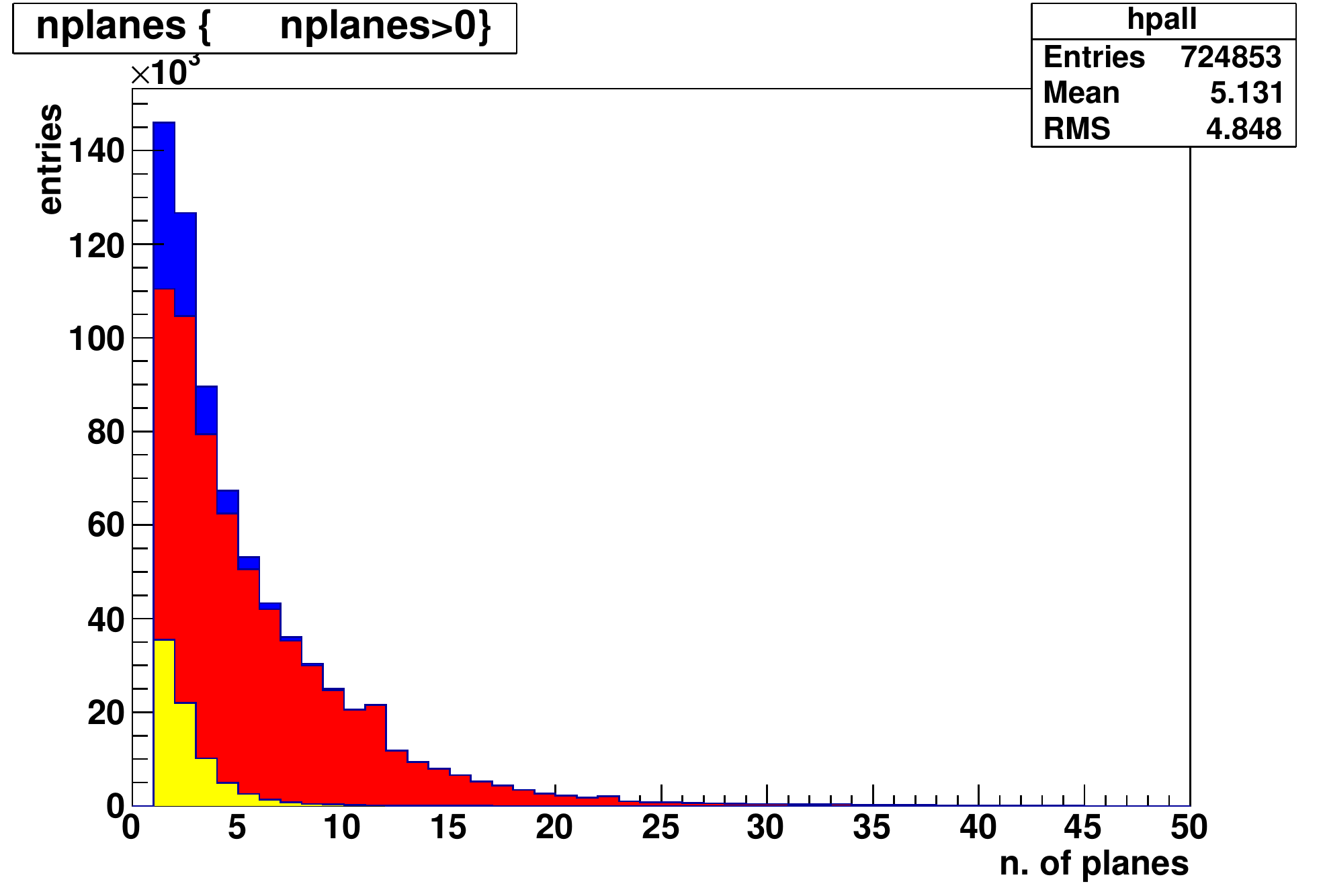}
\includerootgraphics{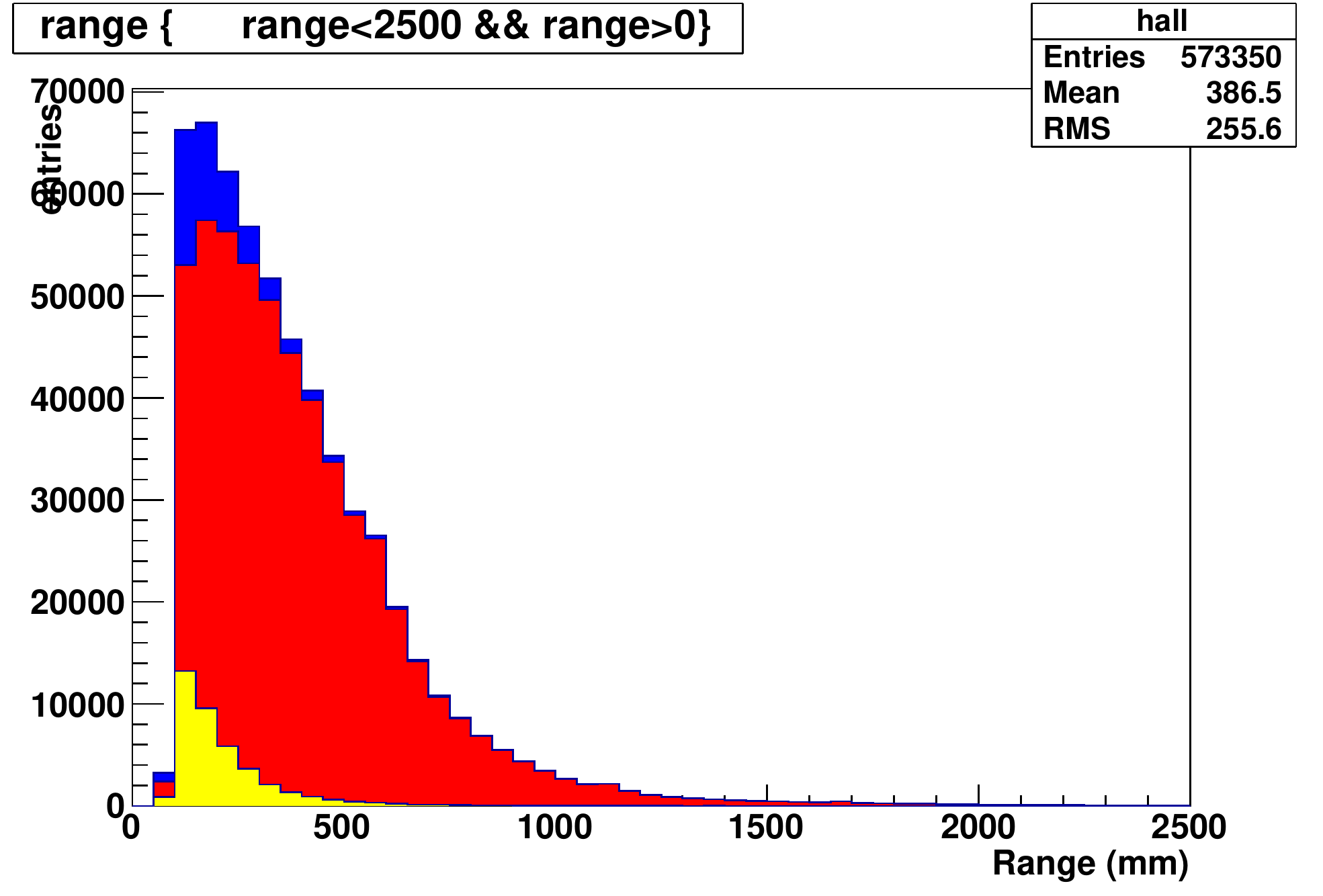}
\caption{ \label{fig:FlukaObservables} Distribution of Number of planes and Range for reconstructed events (FLUKA),
for CC+NC (blue), CC (red) and  NC (yellow)}
\end{figure}

Several observables are evaluated for each reconstructed event: i) number of fired RPC planes (NPLANES); ii) range in iron of the longest particle (RANGE). Their distribution are shown in Fig.~\ref{fig:FlukaObservables}.
%--------------------------------------------------------------------------------------
 
\begin{figure}[htp]
\centering
%\includerootgraphics{simulation/pictures-arc/nplanes-pmu}
%\includerootgraphics{simulation/pictures-arc/range-pmu}
\includerootgraphics{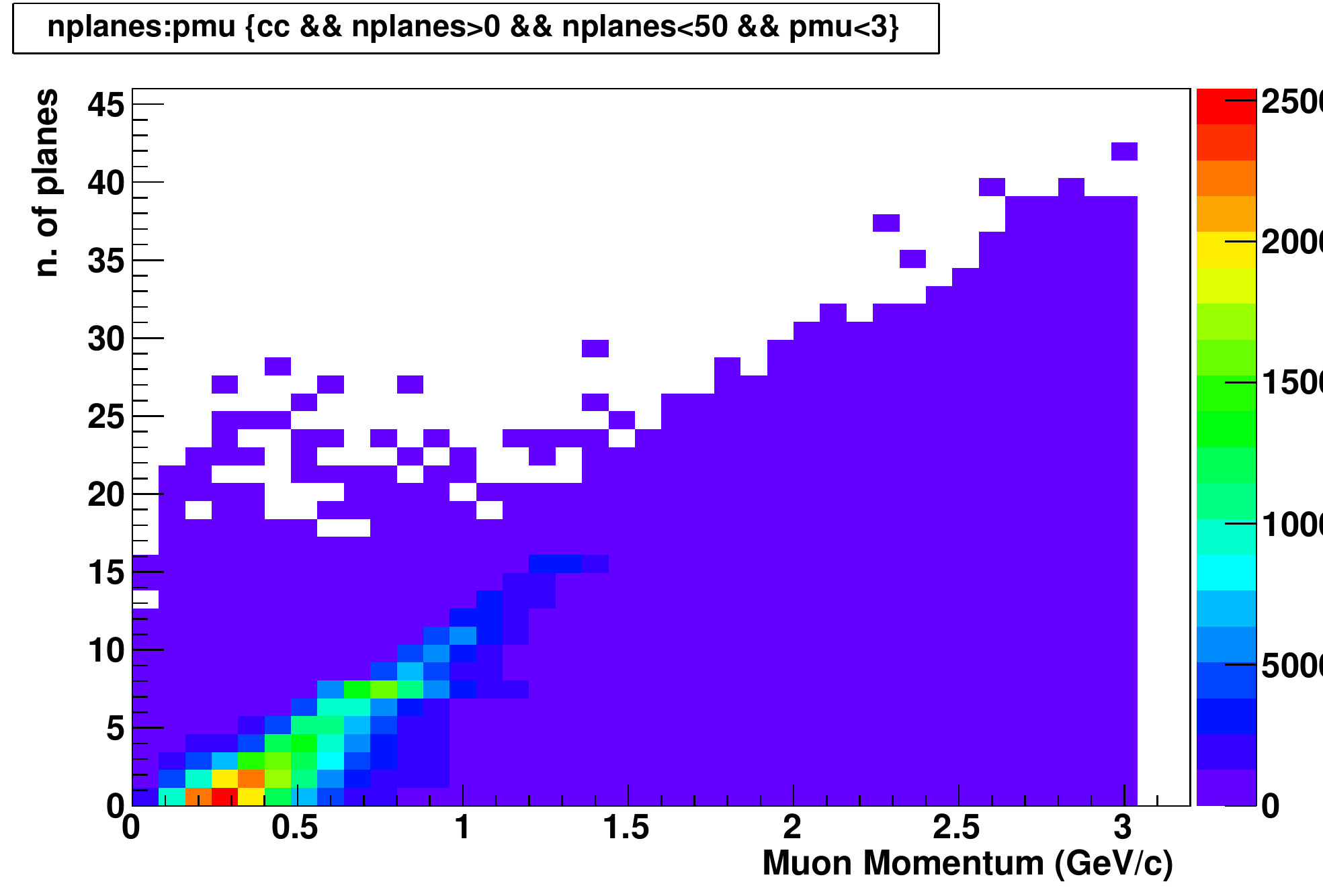}
\includerootgraphics{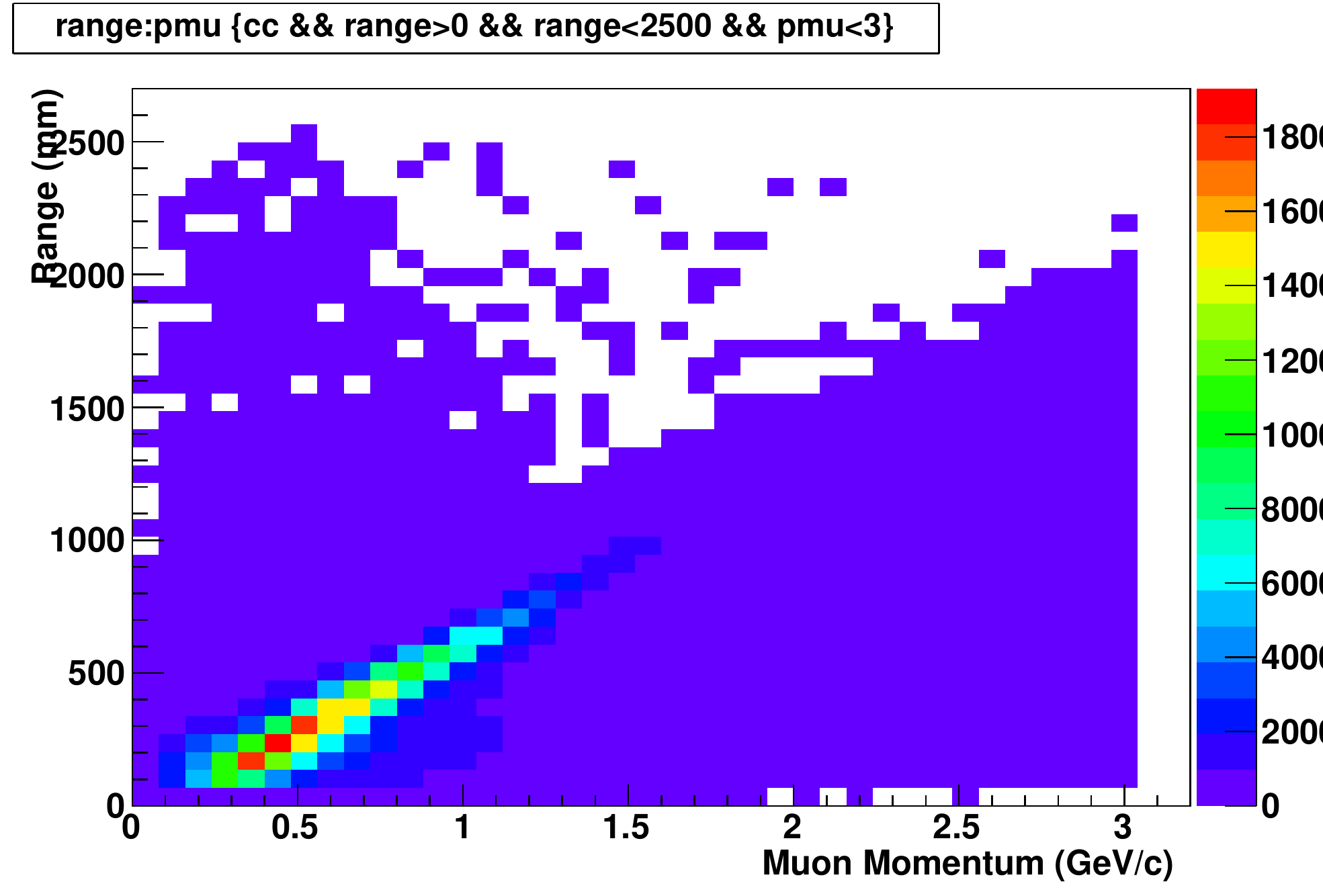}
\caption{ \label{fig:FlukaObservablesCC} Number of planes and Range versus Muon Momentum (CC).}
\end{figure}

\begin{figure}[htp]
\centering
%\includerootgraphics{simulation/pictures-arc/nplanes-pmu-qel}
%\includerootgraphics{simulation/pictures-arc/range-pmu-qel}
\includerootgraphics{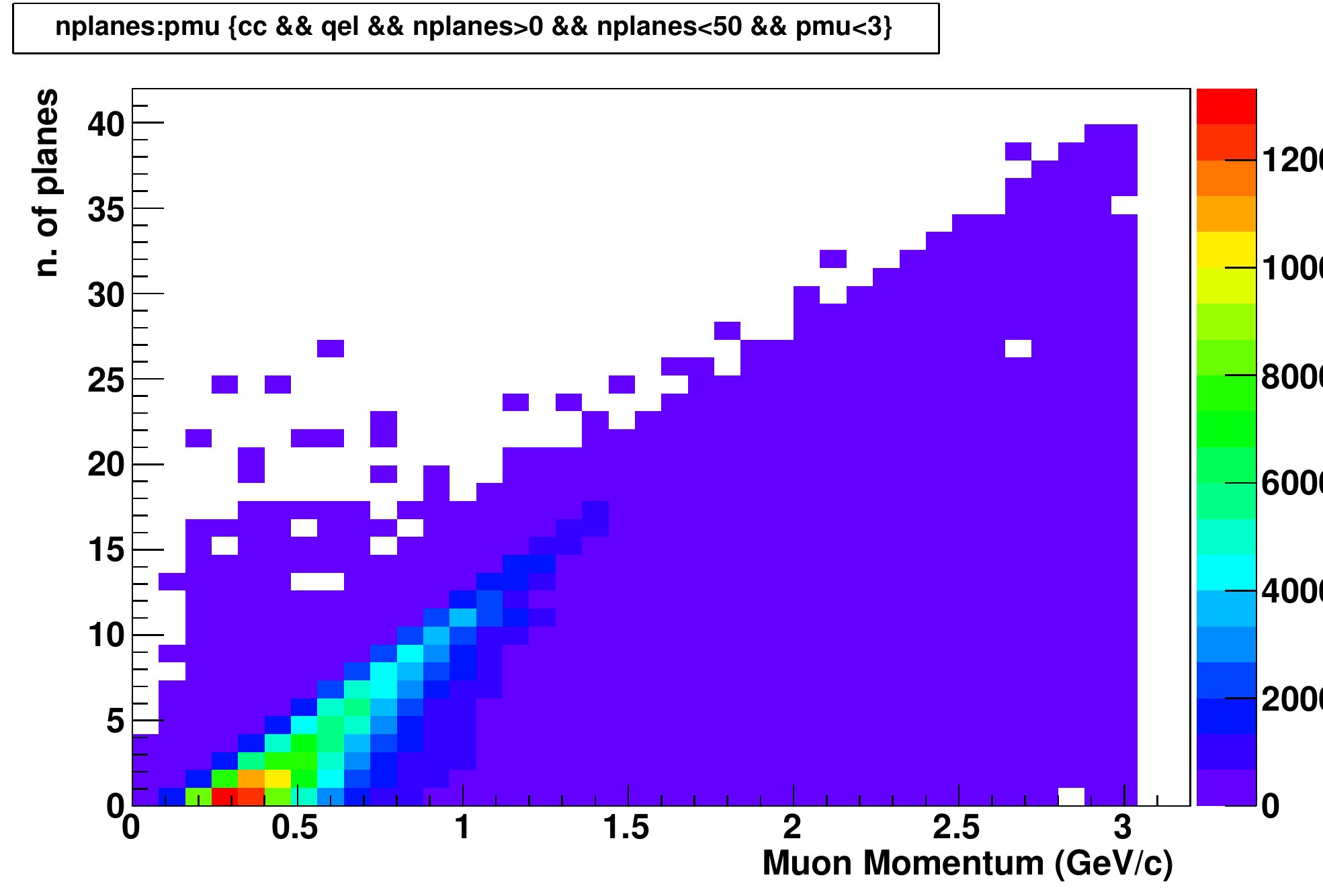}
\includerootgraphics{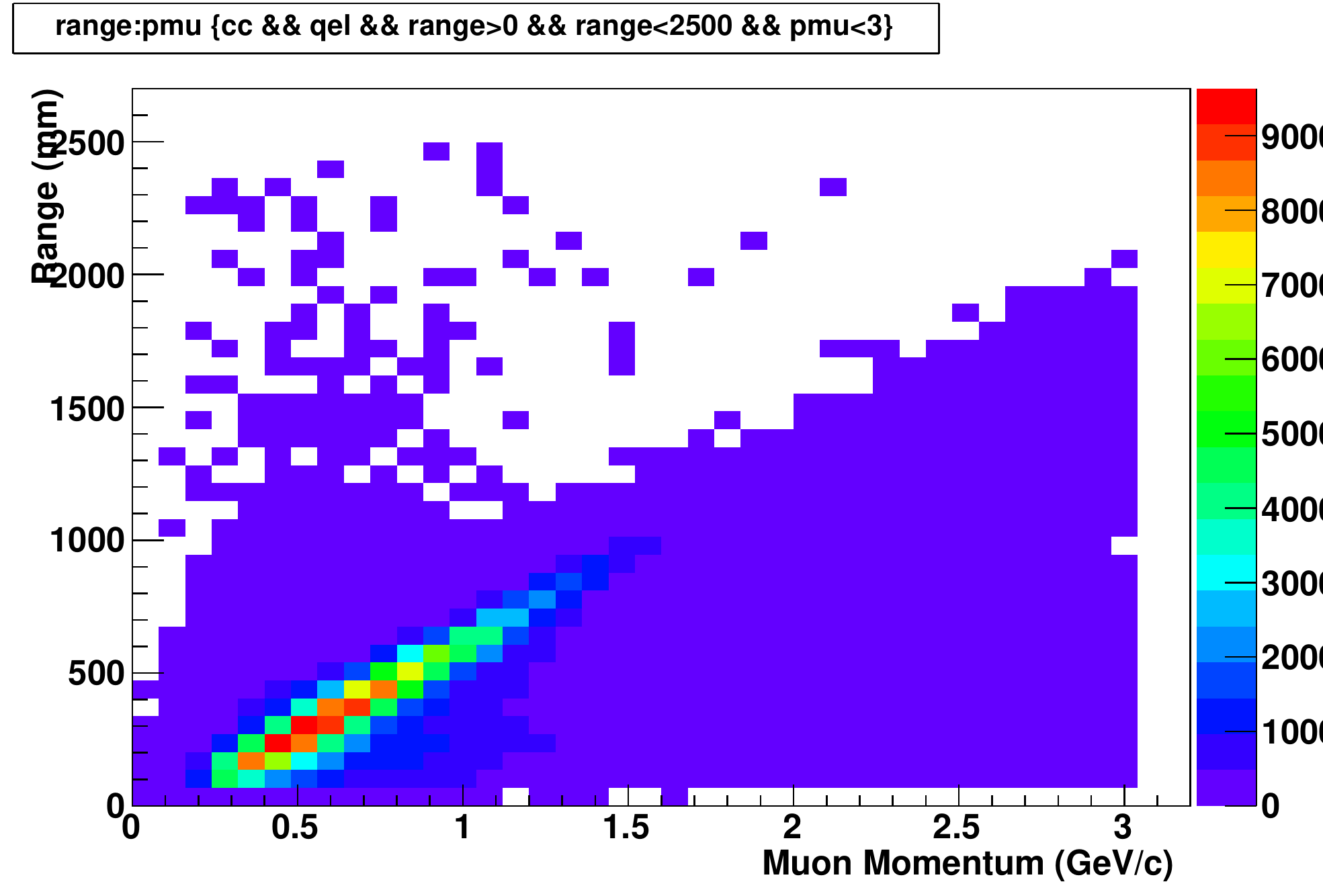}
\caption{ \label{fig:FlukaObservablesCCQE} Number of planes and Range versus Muon Momentum (CCQE).}
\end{figure}

The analysis is performed for three different samples: a) events which have at least 1 hit in the detector (CC+NC); 
b) charged current interactions (CC): c) quasi--elastic charged--current interactions (CCQE). For samples a) and b) we 
assumed perfect identification and background rejection capabilities. The dependency of reconstructed variables on muon 
momentum for sample CC and CCQE are shown respectively in Fig.~\ref{fig:FlukaObservablesCC} and 
Fig.~\ref{fig:FlukaObservablesCCQE}.

%--------------------------------------------------------------------------------------

% SMEARING MATRIX

In order to evaluate the sensitivity of the experiment with the GLoBES tool, the response 
of the detector is expressed in terms of a function $R(E,E')$ ,i.e., a neutrino with a (true)
energy $E$ is reconstructed with an energy between $E'$ and $E' + \delta E'$ with a probability
$R(E,E')\delta E'$. The function $R(E,E')$ is also often called "energy resolution function"; its 
internal representation in the software is a smearing matrix. 

\begin{figure}[htbp]
\centering
\includegraphics[width=7.5cm]{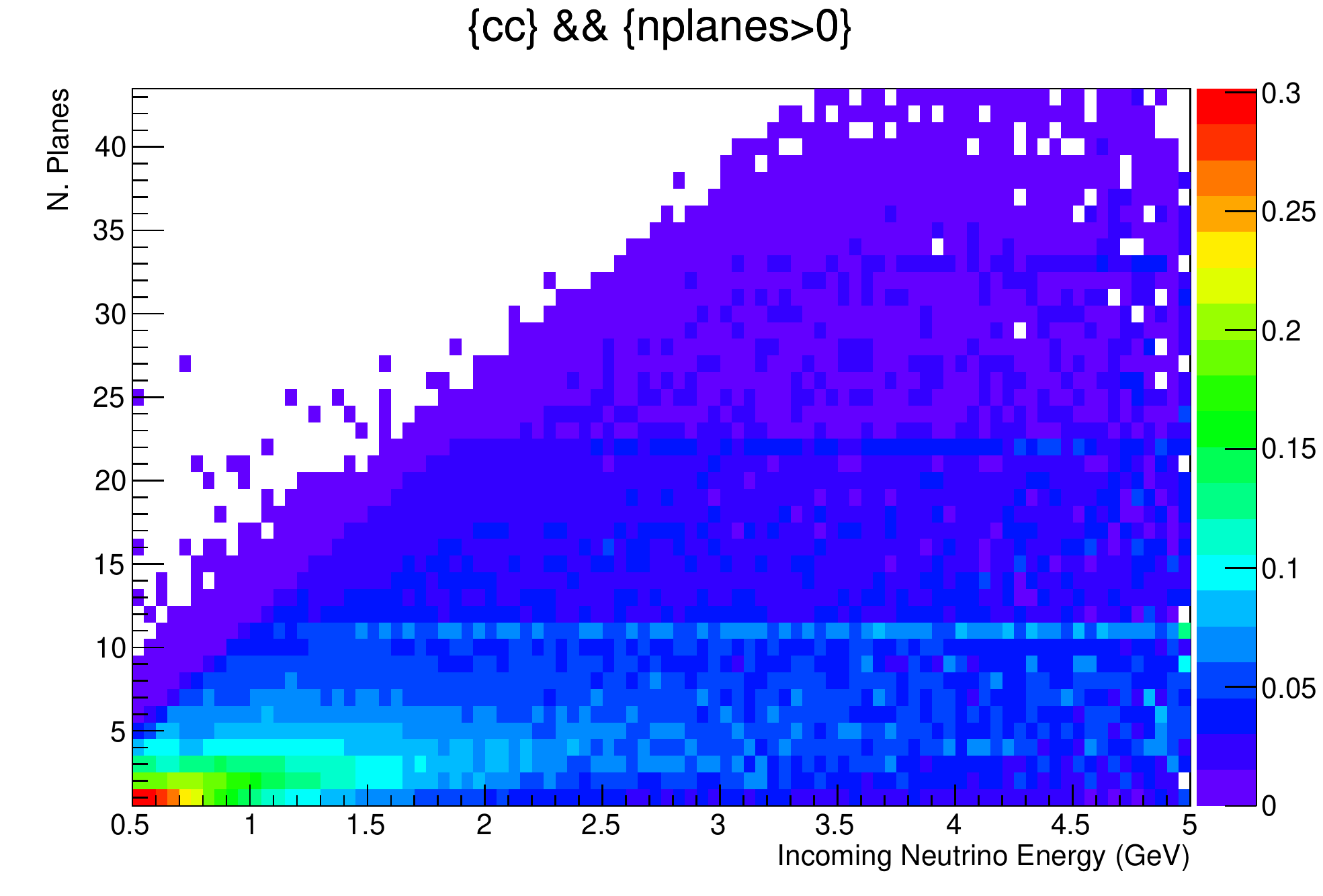}
\includegraphics[width=7.5cm]{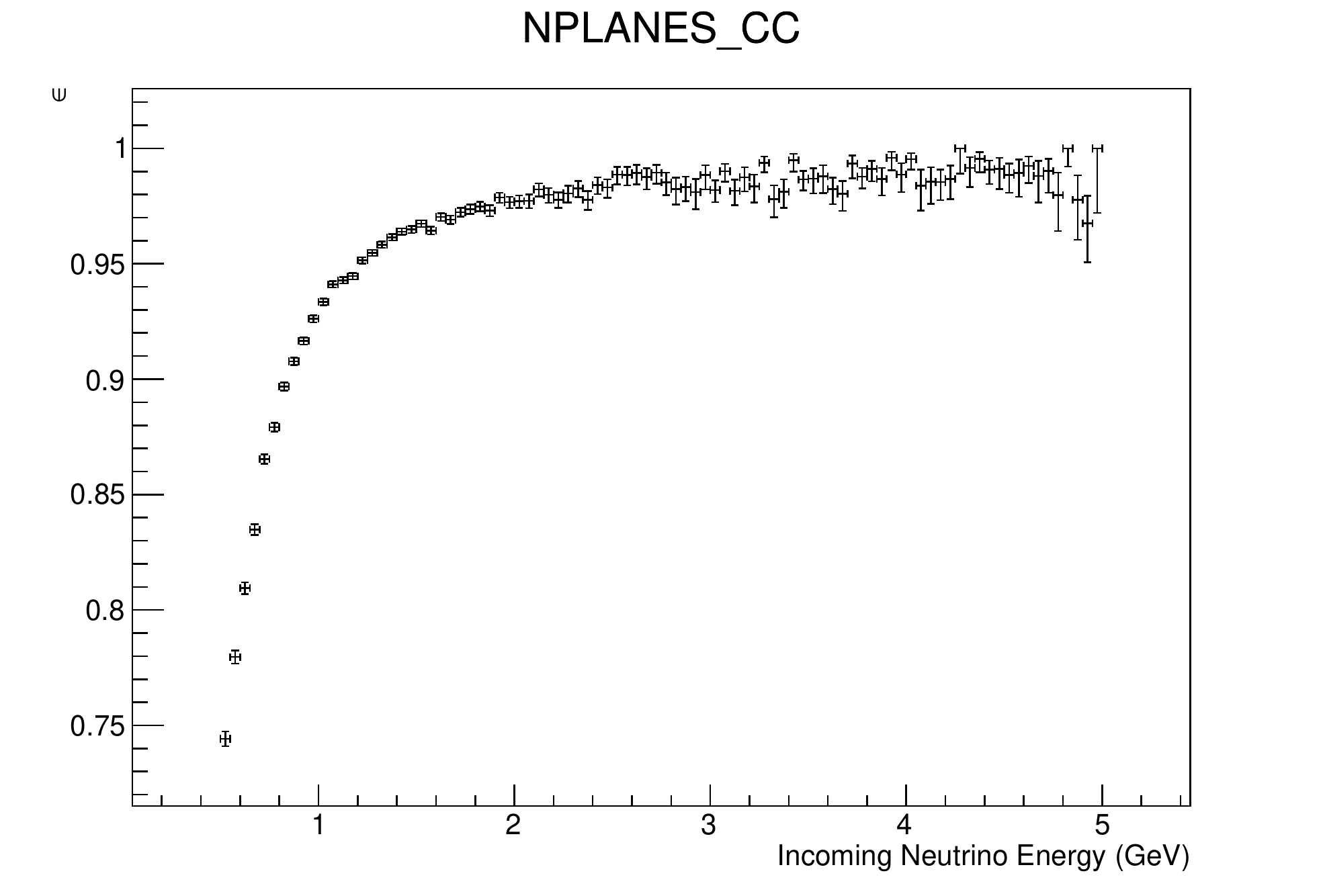}
%\caption{ \label{fig:GenieNuInt} Smearing matrix and efficiency for N.Planes (CC).}
%\end{figure}

\vspace{5mm}

%\begin{figure}[htp]
%\centering
%\includegraphics[width=7.5cm]{simulation/pictures-arc/smear-matrix-NPLANES_CCQEL}
%\includegraphics[width=7.5cm]{simulation/pictures-arc/efficiency-NPLANES_CCQEL}
\includegraphics[width=7.5cm]{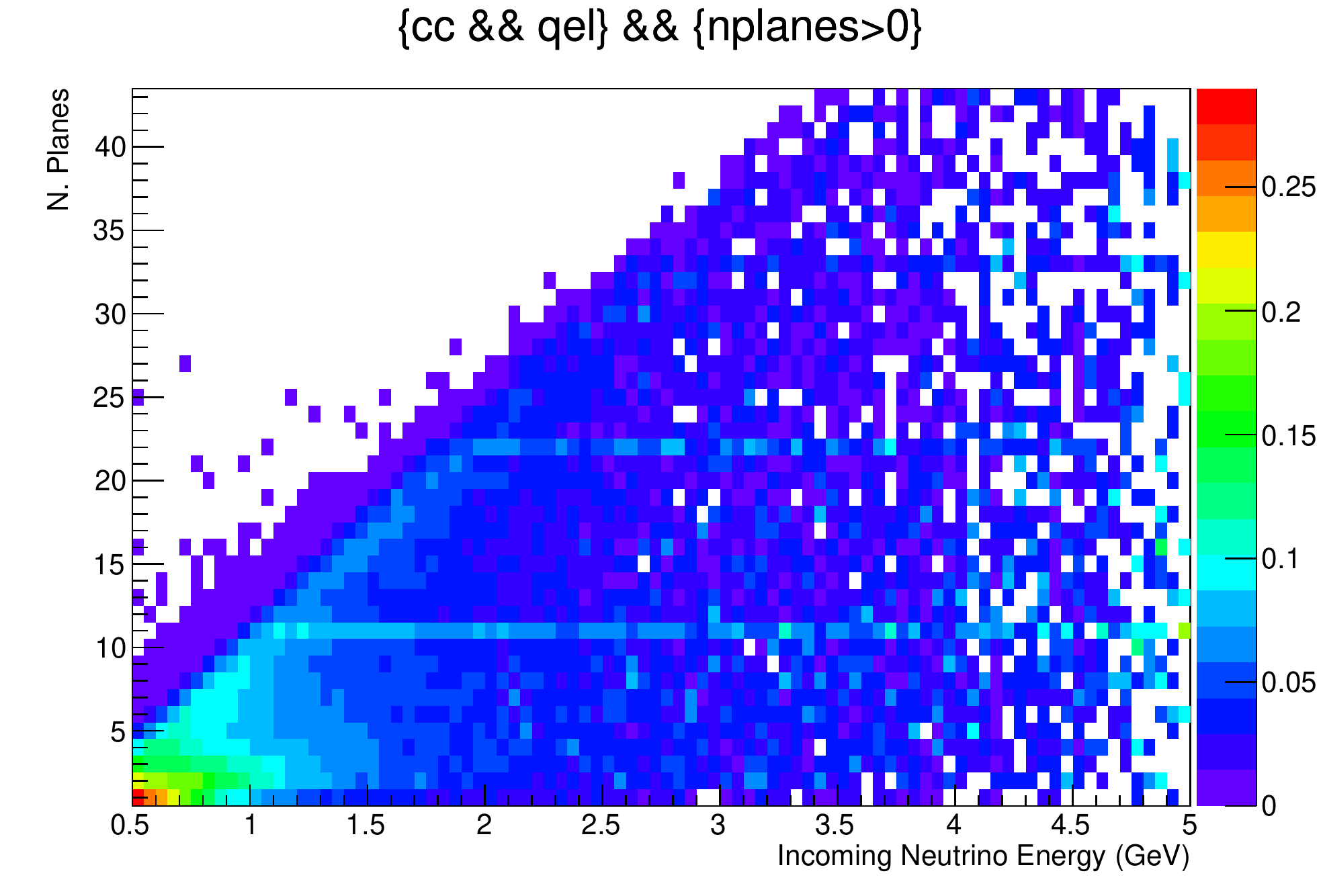}
\includegraphics[width=7.5cm]{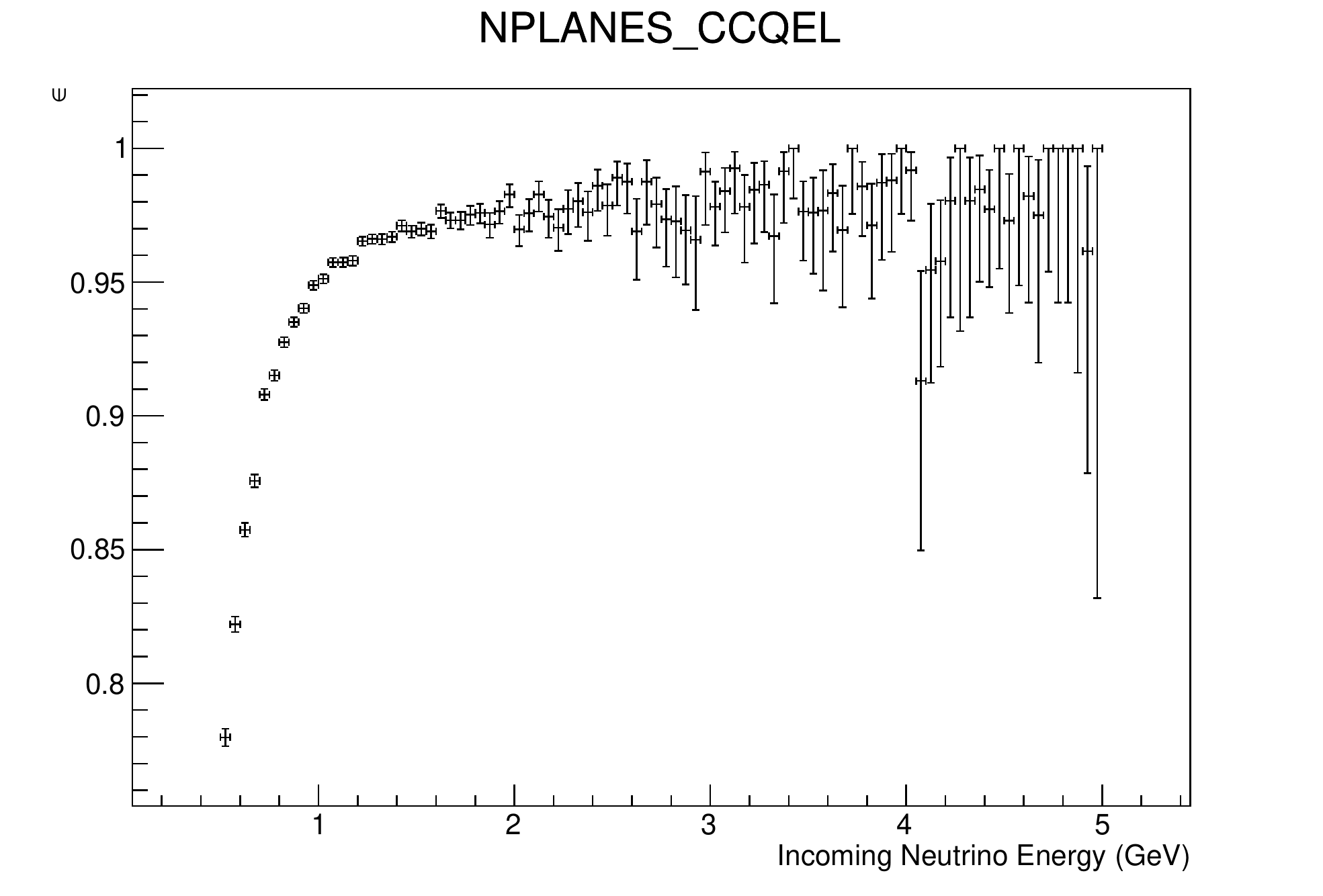}
%\caption{ \label{fig:GenieNuInt} Smearing matrix and efficiency for N.Planes (CC-QE).}
%\end{figure}

\vspace{5mm}

%\begin{figure}[htp]
%\centering

%\includegraphics[width=7.5cm]{simulation/pictures-arc/smear-matrix-NPLANES_NC}
%\includegraphics[width=7.5cm]{simulation/pictures-arc/efficiency-NPLANES_NC}
\includegraphics[width=7.5cm]{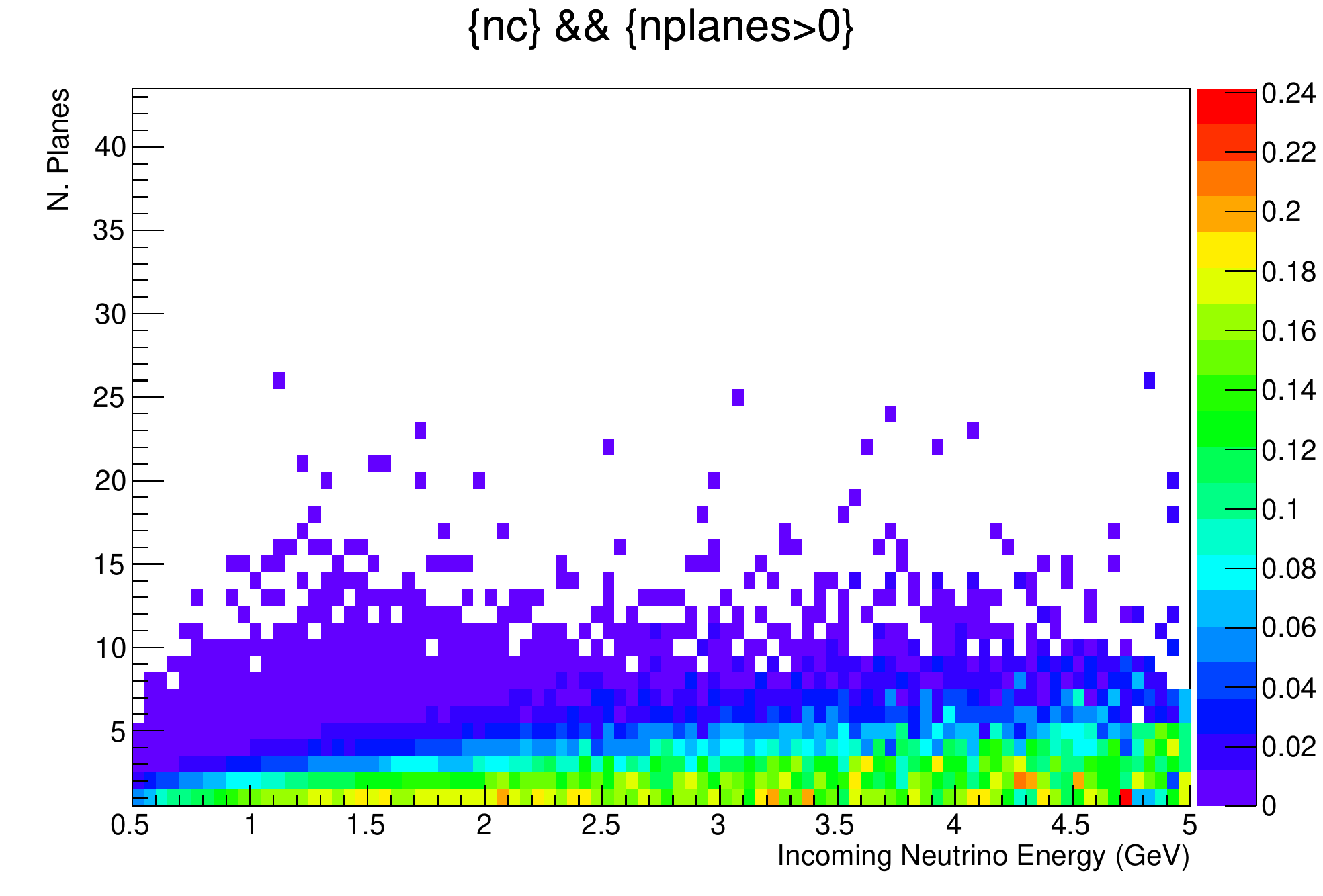}
\includegraphics[width=7.5cm]{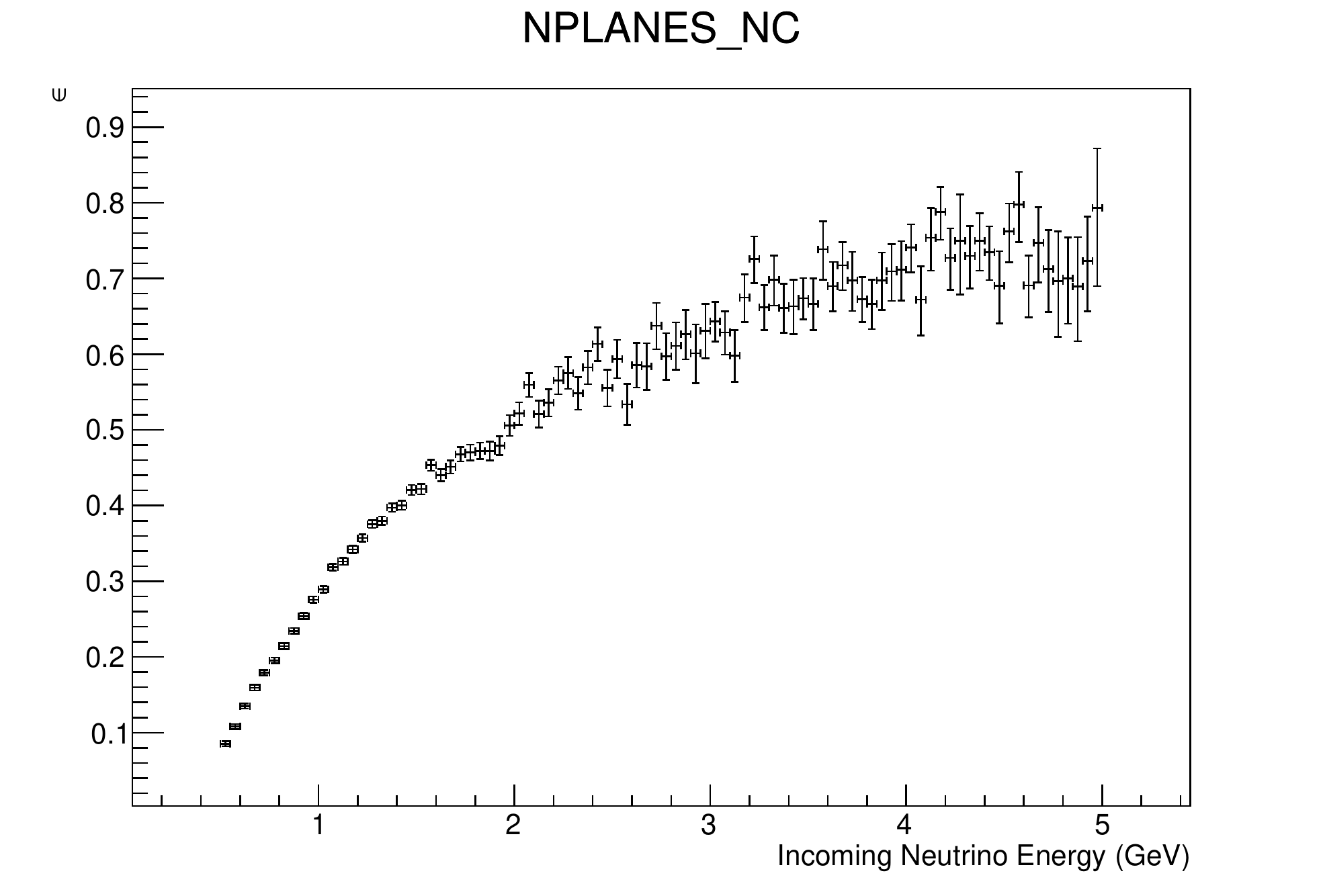}
%\caption{ \label{fig:GlobesMatrixNplanes} Smearing matrix and efficiency for N.Planes (NC).}
\caption{ \label{fig:GlobesMatrixNplanes} Smearing matrix and efficiency for N Planes. In the top figures the whole CC sample is considered.
The middle ones corresponds only to the CCQE sample, while the bottom ones corresponds to the NC channel.}
\end{figure}

\begin{figure}[htbp]
\centering
\includegraphics[width=7.5cm]{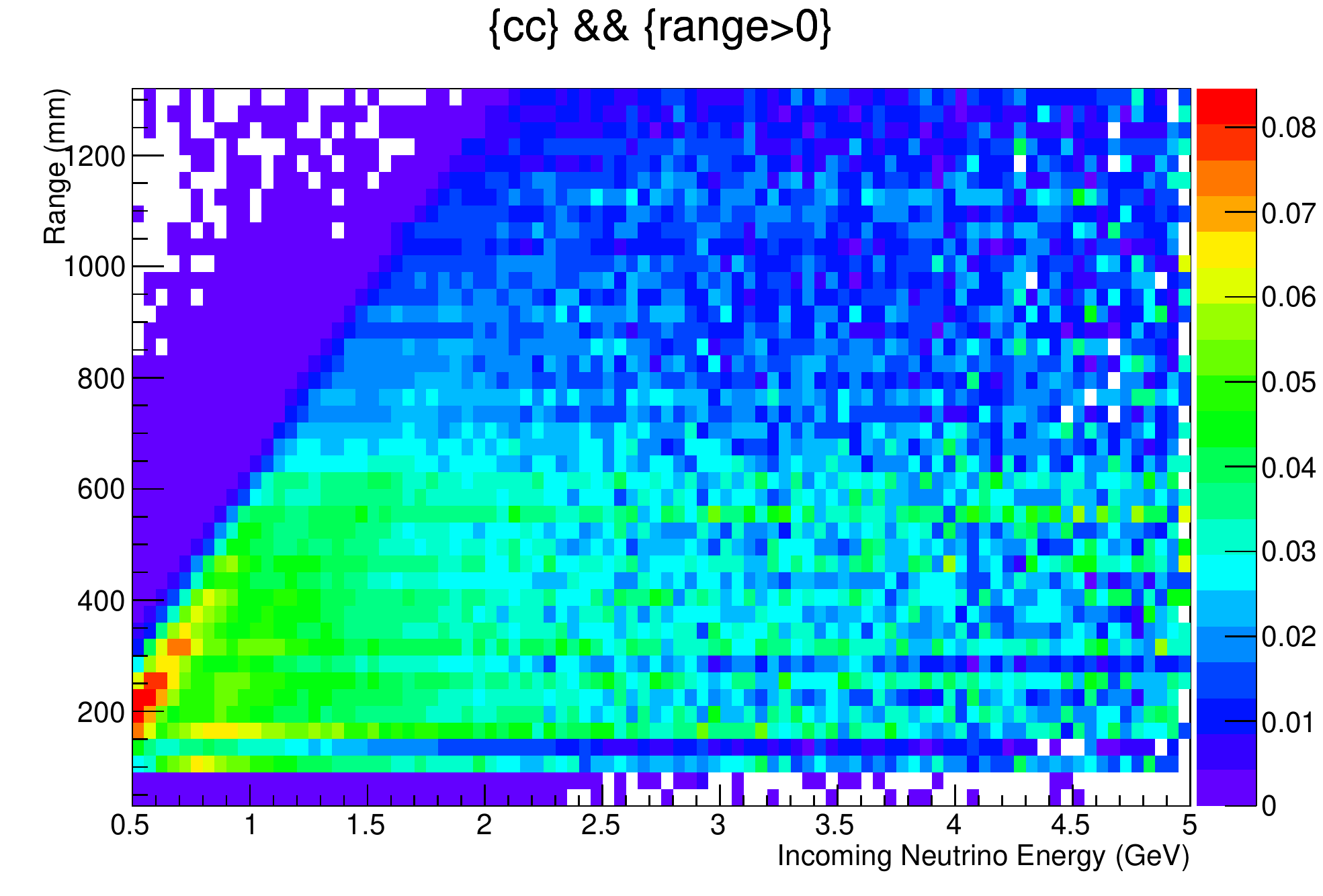}
\includegraphics[width=7.5cm]{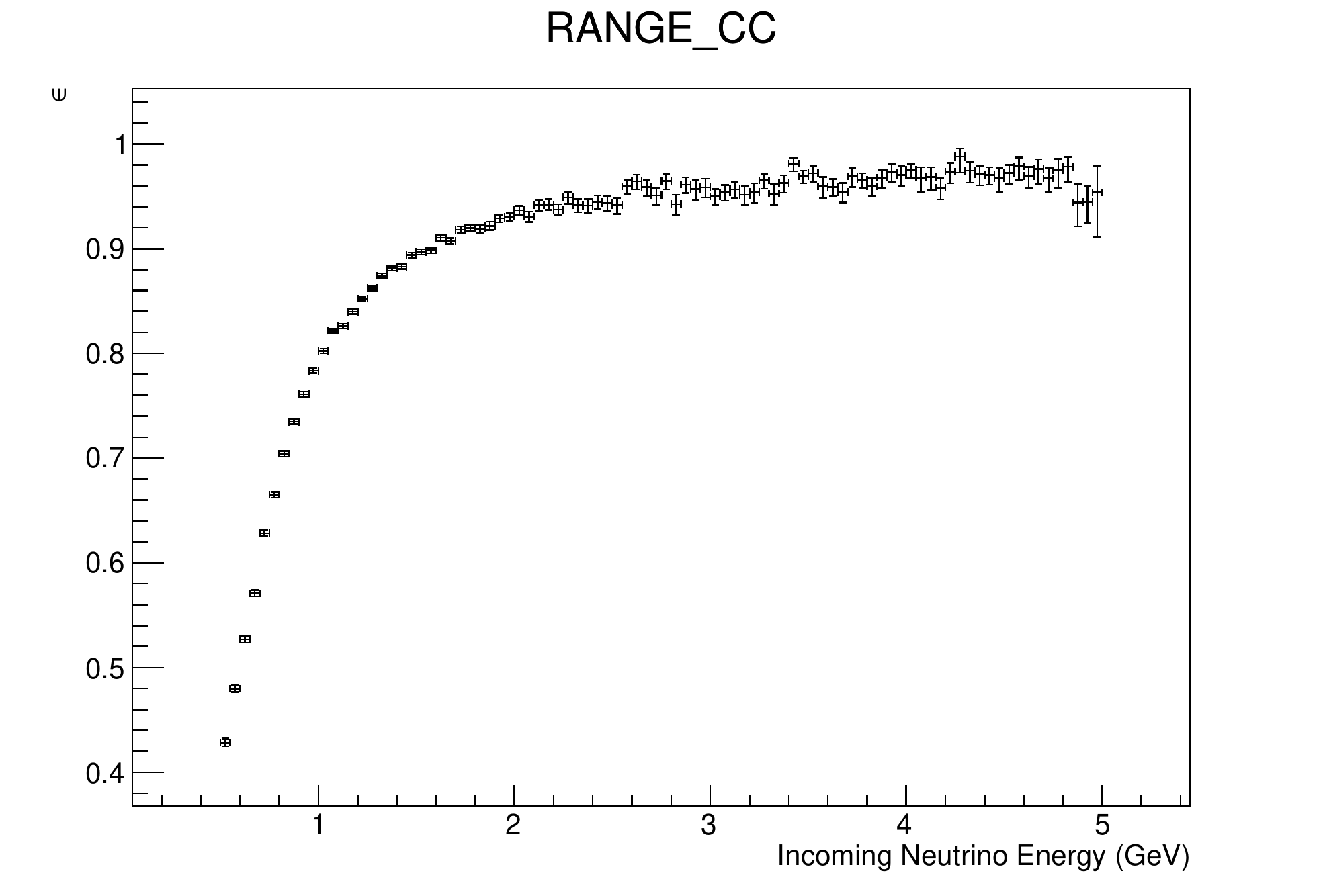}
%\caption{ \label{fig:GenieNuInt} Smearing matrix and efficiency for Range (CC).}
%\end{figure}

\vspace{5mm}

%\begin{figure}[htp]
%\centering
%\includegraphics[width=7.5cm]{simulation/pictures-arc/smear-matrix-RANGE_CCQEL}
%\includegraphics[width=7.5cm]{simulation/pictures-arc/efficiency-RANGE_CCQEL}
\includegraphics[width=7.5cm]{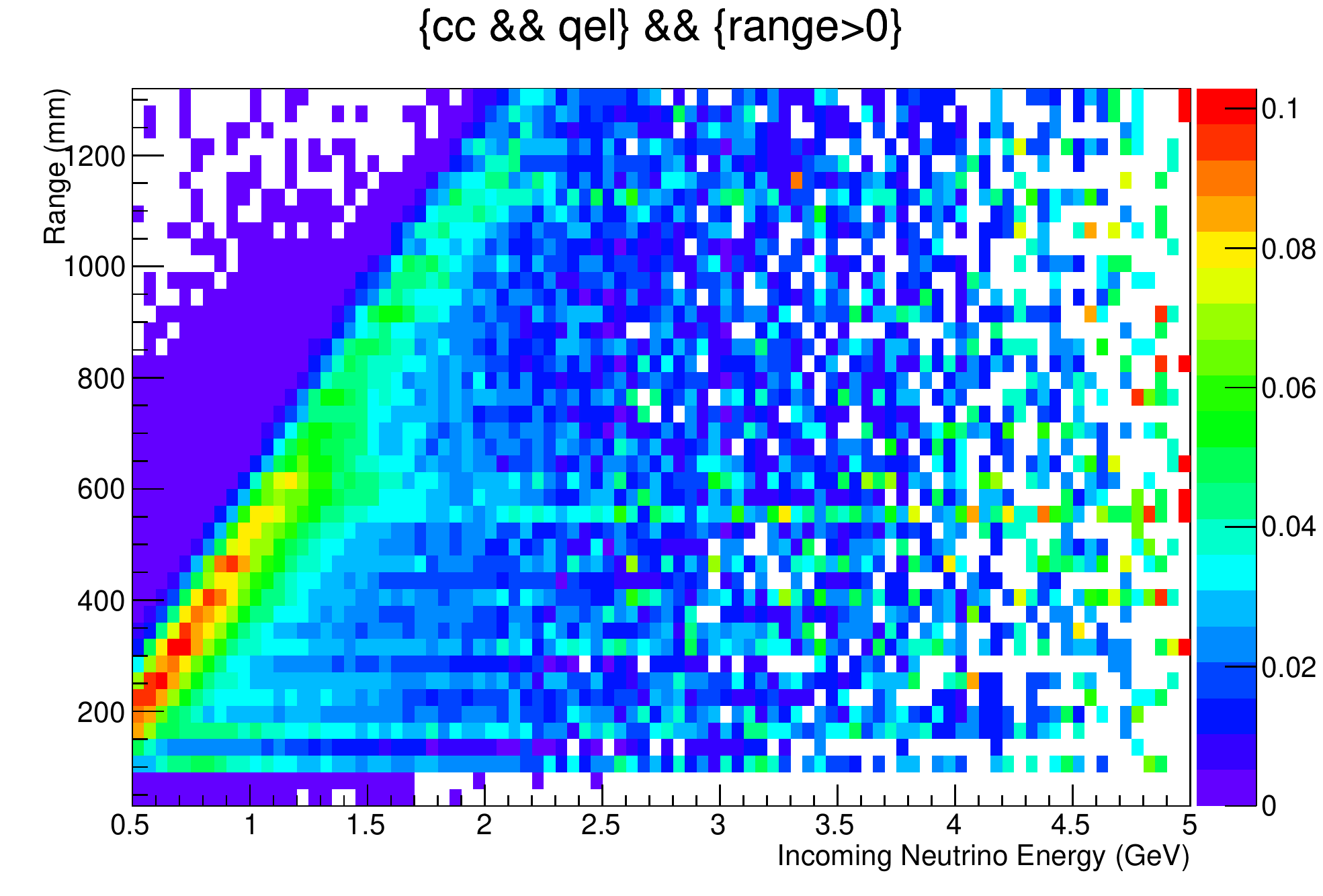}
\includegraphics[width=7.5cm]{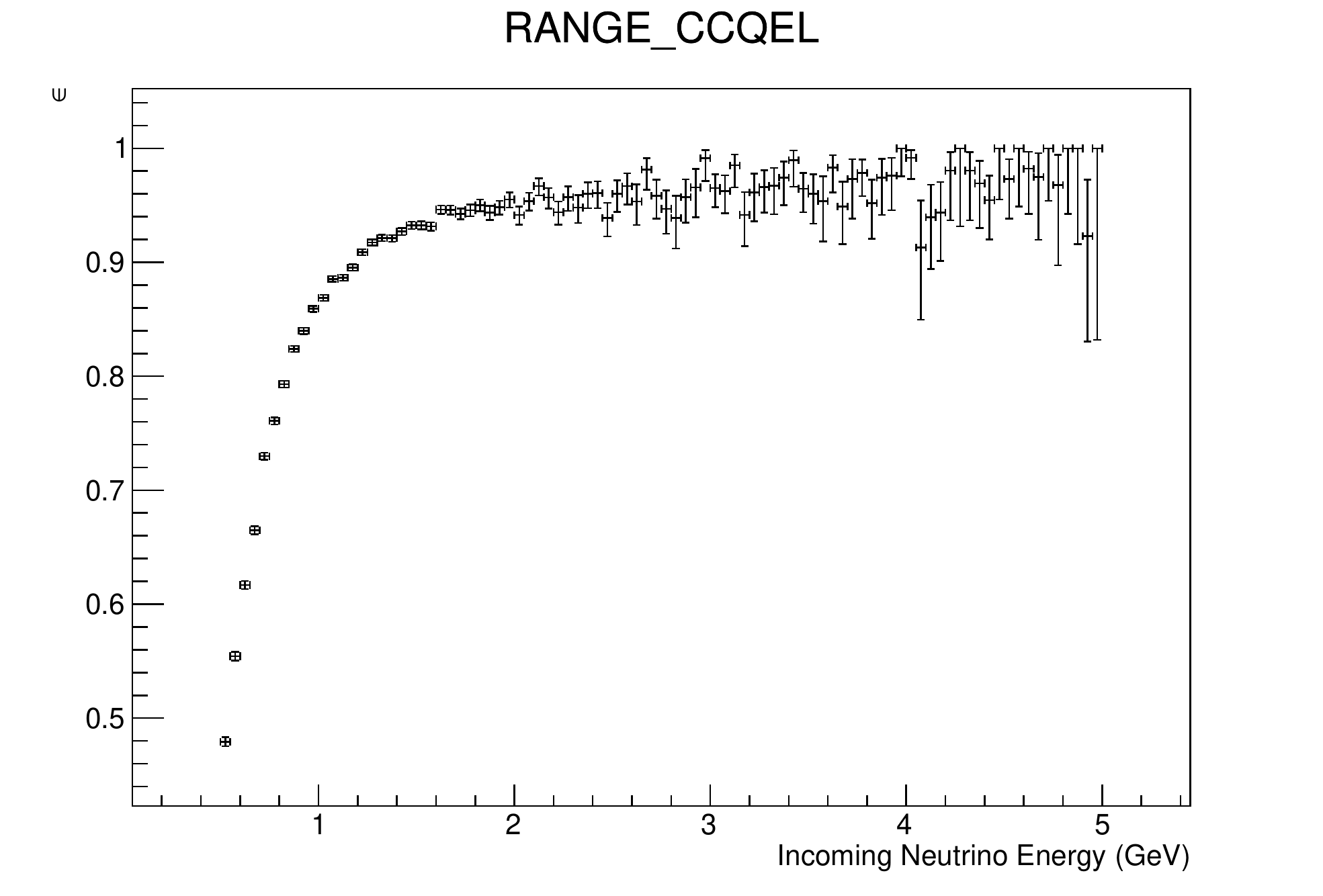}
%\caption{ \label{fig:GenieNuInt} Smearing matrix and efficiency for Range (NC).}
%\end{figure}

\vspace{5mm}

%\begin{figure}[htp]
%\centering
%\includegraphics[width=7.5cm]{simulation/pictures-arc/smear-matrix-RANGE_NC}
%\includegraphics[width=7.5cm]{simulation/pictures-arc/efficiency-RANGE_NC}
\includegraphics[width=7.5cm]{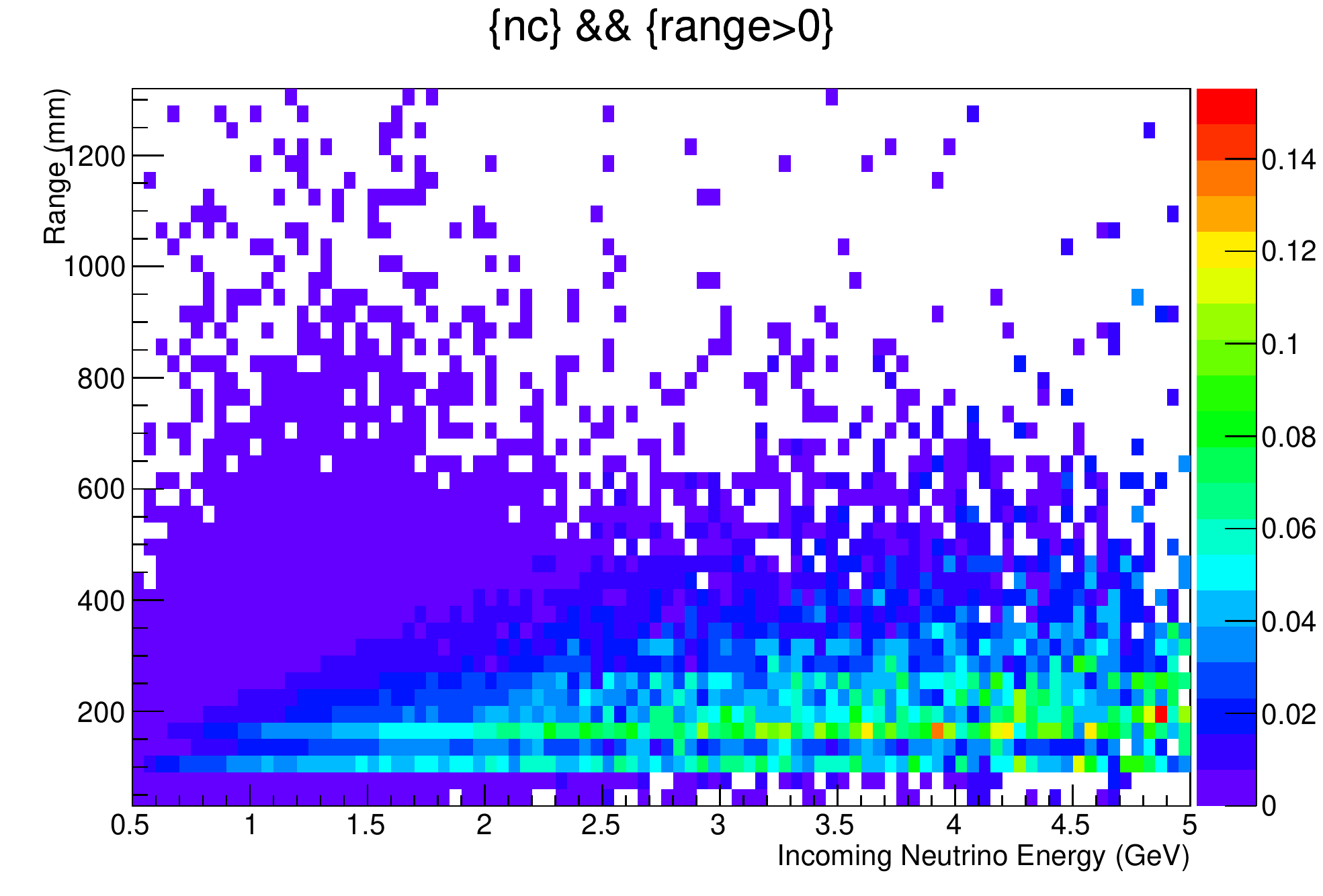}
\includegraphics[width=7.5cm]{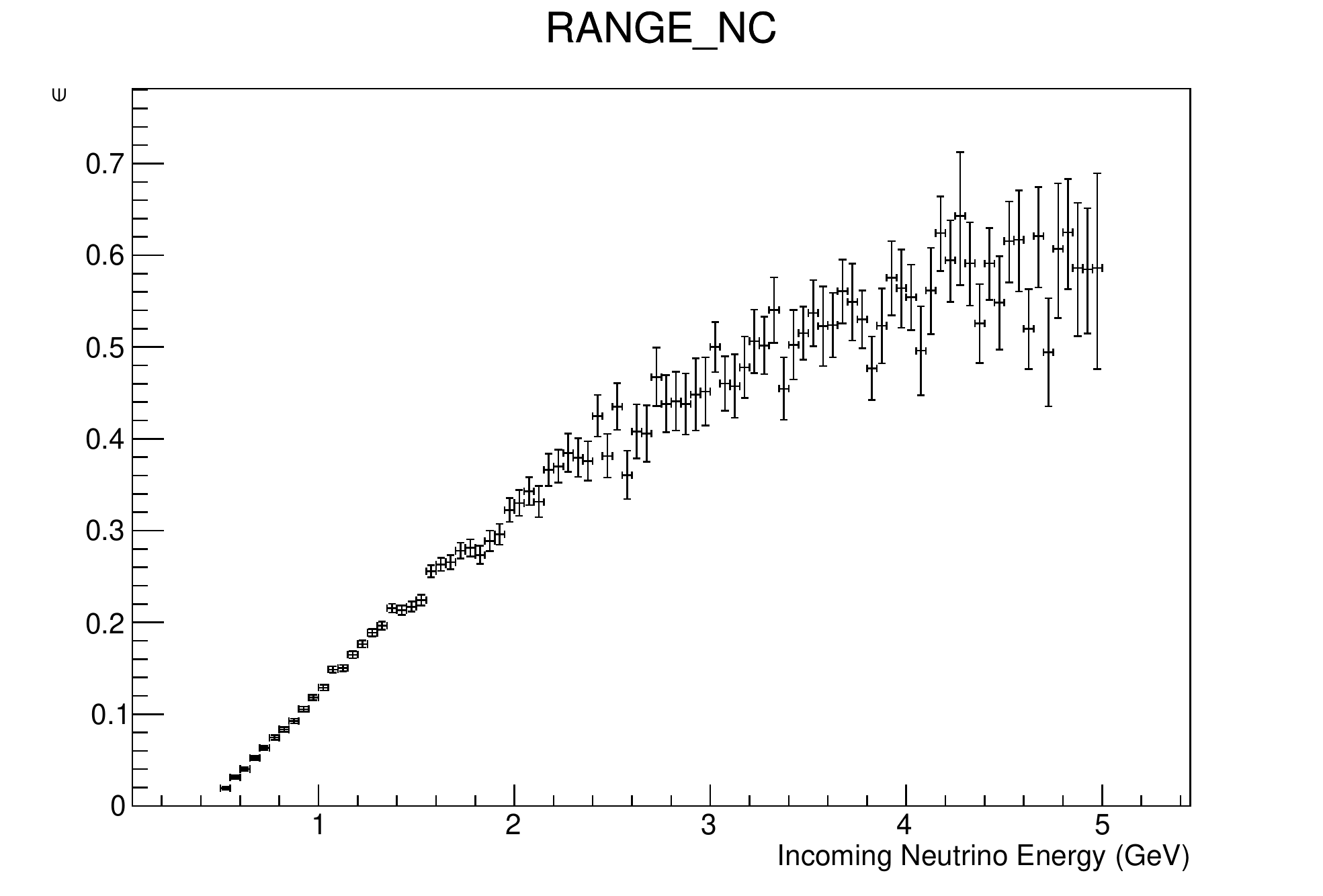}
%\caption{ \label{fig:GlobesMatrixRange} Smearing matrix and efficiency for Range (CC-QE).}
\caption{ \label{fig:GlobesMatrixRange} Smearing matrix and efficiency for Range. In the top figures the whole CC sample is considered.
The middle ones corresponds only to the CCQE sample, while the bottom ones corresponds to the NC channel.}
\end{figure}

In order to exploit the correlation between the number of
planes (or range) and the neutrino energy in the charged current channel, 
we decided to describe directly the response of the detector in terms of a function $R(E,O)$, 
where $O$ is one of such reconstructed variables. Smearing matrices and detector reconstruction efficiencies are shown in Fig.~\ref{fig:GlobesMatrixNplanes} and  Fig.~\ref{fig:GlobesMatrixRange} (RANGE) as a function of NPLANES and RANGE, respectively, for several channels .

%--------------------------------------------------------------------------------------

\clearpage

%end-GABRIELE%%%%%%%%%%%%%%%%%%%%%%%%%%%%%%%%%%%%%%%%%%%%%%%%%%%%%%%%%%%

%Start-LAURA%%%%%%%%%%%%%%%%%%%%%%%%%%%%%%%%%%%%%%%%%%%%%%%%%%%%%%%%%%%

\section{Mechanical Structure}\label{sec:mech-struc}
    
\begin{figure}[htb]
\centering
\includegraphics[width=7cm]{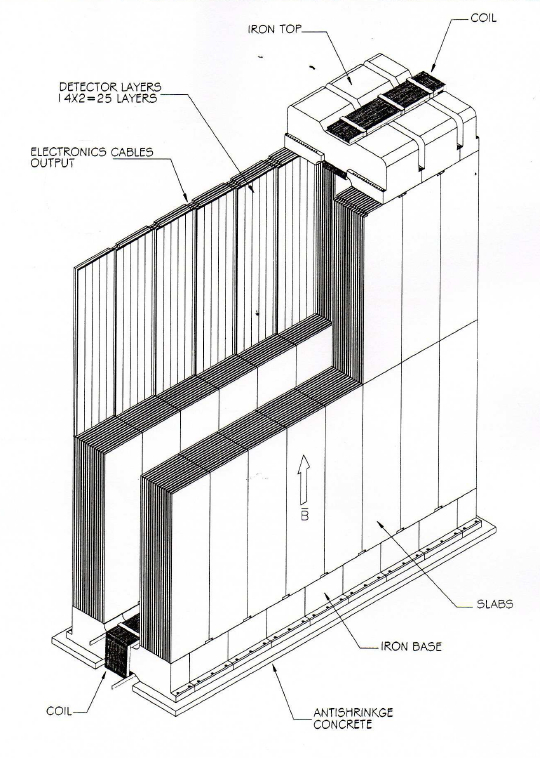}
\caption{OPERA magnet scheme.}
\label{fscheme}
\end{figure}

The design of NESSiE spectrometers follows closely  that of the  OPERA apparatus, where iron dipole magnets are made of two vertical arms with rectangular cross--section and of top and bottom flux return yokes~\cite{bopera} (Fig.~\ref{fscheme}).
Each arm is composed of 12 vertical layers of  iron slabs, $5~{\rm cm}$ thick, interleaved by 11 gaps  $2~{\rm cm}$ wide, hosting RPC detectors.
Each iron layer,  obtained by assembling 7 vertical iron slabs,  covers a surface  of  $(8.75 \times 8.2) \ {\rm m}^2$. 
 The magnet full height is about $10$~m (including the top and bottom return yokes), its length along the beam is $2.85$~m 
and its weight amounts to $\sim 1$~kton.
The slabs, the top and the bottom yokes are held together by means of screws, while more screws (about $1/\mbox{\rm m}^2$) are used to keep slabs straight with
spacers to ensure the thickness uniformity of the gaps hosting the detectors.
The spectrometers are magnetized by coils located at the top and bottom return yokes, as shown in Fig.~\ref{fscheme}.

The installation of  the OPERA magnets was performed  according to the following time sequence:
\begin{itemize}
\item bottom yoke installation;
\item internal support structure construction;
\item iron/RPC layers installation (one plane in each arm at the time in order
to keep the structure balanced);
\item top yoke installation;
\item removal of the internal structure. 
\end{itemize} 
  
\begin{figure}[htb]
\centering
\includegraphics[width=12cm]{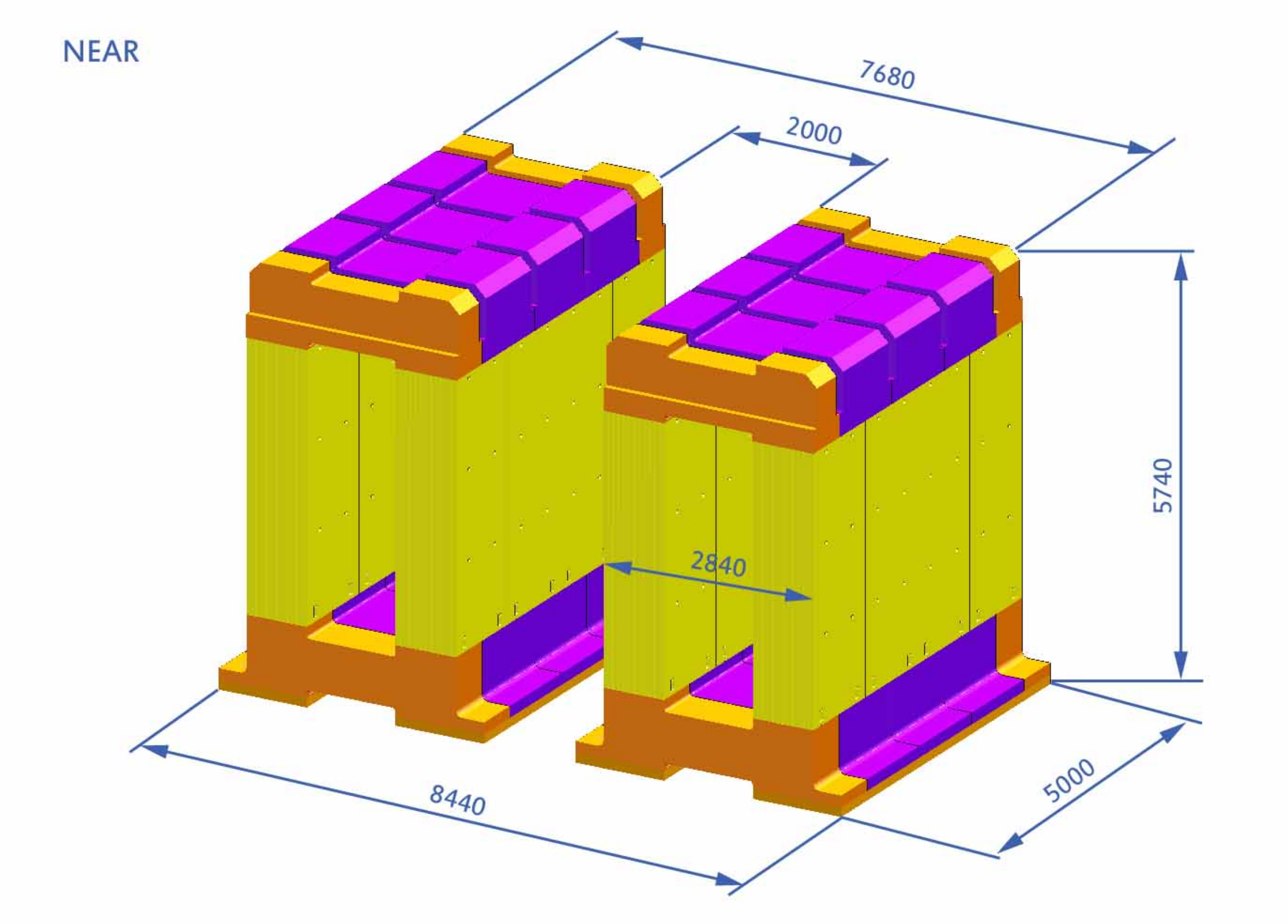}
\vskip 10pt
\includegraphics[width=12cm]{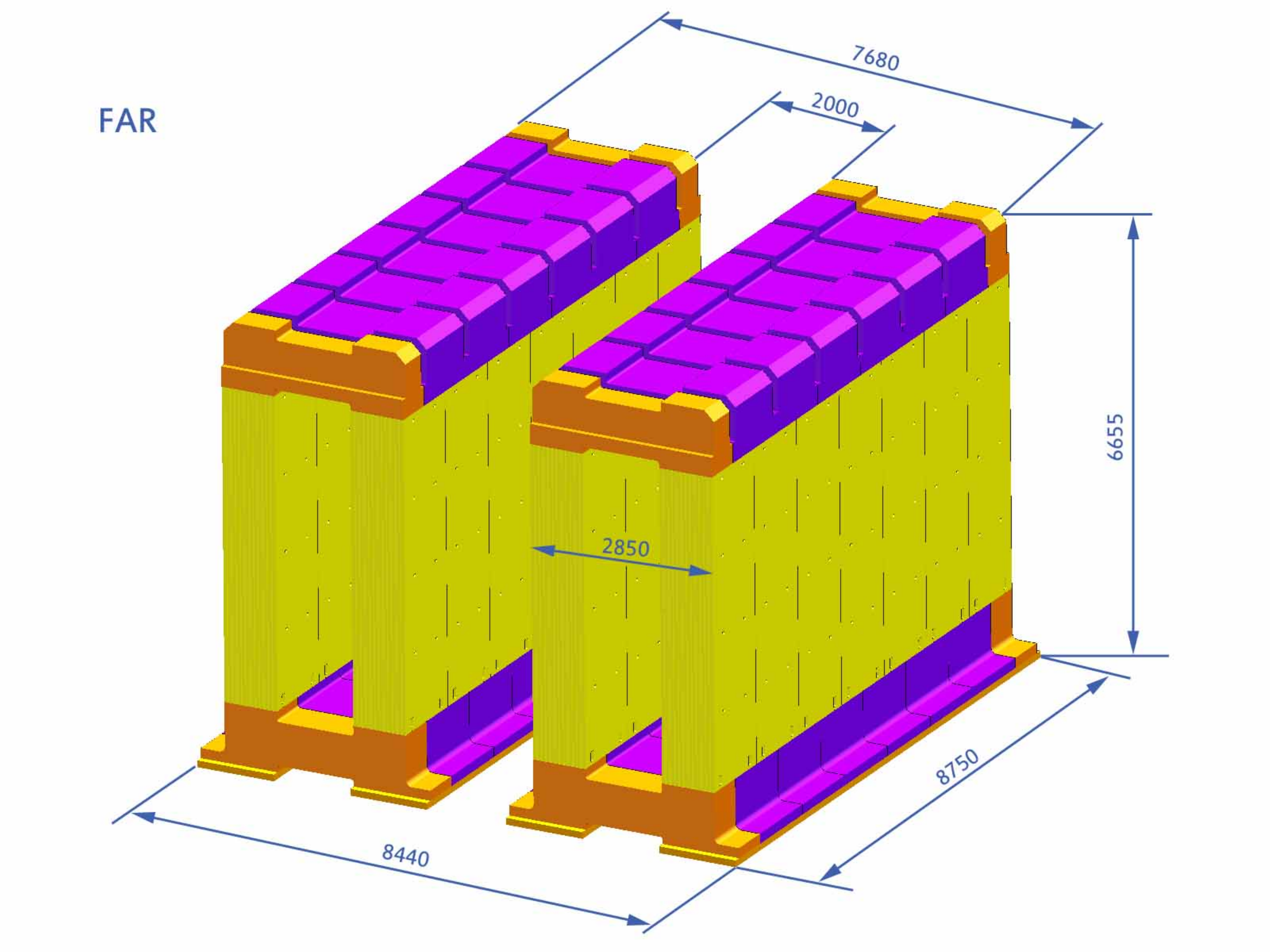}

\caption{NESSiE  Near (top), Far (bottom) spectrometers}
\label{AssemblyFar-Near}
\end{figure}

For the NESSIE experiment a very similar setup should be used in order to profit  of the available detectors, which were designed to provide the maximum acceptance coverage and strip signal configuration. The Far spectrometer is designed  with the same  width as in OPERA (i.e.\ 7 vertical  iron slabs) but smaller  height (4 rows of RPCs). The top and bottom return yokes are the same as the OPERA's ones. The two OPERA spectrometers will be coupled 
in the longitudinal view, by taking the 4/7 in acceptance for the Far site, while the remaining 3/7 will be used for the Near site.
In the Near detector each iron wall, $3.4$~m high, consists of  4  vertical slabs. The  return yokes and the  copper coils must be redesigned accordingly (Fig.~\ref{AssemblyFar-Near}). The size, mass and total instrumented surface of spectrometers are listed in  Tab.~\ref{tab:volmas}.

\begin{table}[ht] \begin{center}
\begin{tabular}{|l|ccccc|} \hline
Single--unit       & Width  & Height & Depth &Total RPC Surface        & Total Iron   \\
Spectrometer & (m)  & (m)  & (m) &  (m$^2$) & Mass (ton) \\ \hline
             &        &        &       &                  &              \\ 
Near         & 5.0    & 3.4    & 2.8   &  380             &  182         \\ 
Far          & 8.8    & 4.5    & 2.8   & 880             &  423         \\ \hline
\end{tabular}
\caption{Dimensions, instrumented surface and iron (slab) mass
        of the single--unit of the Near and Far spectrometers.} \label{tab:volmas}
\end{center} \end{table}

%For the OPERA experiment a total of 336 slabs, (1.25*8.2) $m^2$ wide,
%  were employed.  Such slabs can be cut into the shape needed for the present
%  baseline NESSiE configuration, but a new production will also be needed.

\clearpage

\section{Magnet Power Supply and Slow Control}\label{sec:mag-SC}
The following description is based on the  experience gained with the OPERA iron magnets (see~\cite{bopera} and references therein).
We plan to adopt the same  solution as  for the OPERA spectrometers.

\subsection{Power Supply features}

The magnetomotive force to produce the magnetic field is provided by DC power supplies, located on the top of the magnet. 
They are single-quadrant AC$\rightarrow$DC devices providing a maximum  current of $1700$ A and a maximum voltage of $20$~V. 
As a single--quadrant power supply cannot change continuously the sign of  the voltage, the sign of the current is reversed by ramping down the power supply and inverting the load polarity through a motorized breaker.
The power supplies are connected to the driving coil wound in the return yokes of the magnet by means of short flexible cables. 

\vskip 10pt
\underline{Magnet Ancillary systems}

\begin{itemize}
\item Coils:
They are made of $100 \times 20~{\rm mm}^2$ copper 
(type Cu--ETP UNI 5649--711) bars. The segments are connected through bolts after polishing and gold--plating of the contact surface. Each coil has
20 turns in the upper return yoke connected in series to  20 more turns in the bottom yoke. The two halves are linked by vertical bars running along the arm. Rexilon 
supports provide spacing and insulation of the bars.
\item Magnet  cooling system : 
Water heat exchangers are positioned between these supports and the bars while the vertical sections of the coil are surrounded by protective plates 
to avoid accidental contacts. More than 160 junctions were made for
each coil and the quality of such contacts was tested measuring the overall coil resistance during mounting.
The cooling system provides an operating temperature for  the RPC detectors  lower than 
20 $^0$C.
\end{itemize}

\underline{Status of the OPERA power supply units }

\begin{itemize} 
\item The overall downtime of the two OPERA magnets was about 0.1\%.
The power supply units of the first spectrometer  stopped with a monthly frequency,\footnote{No firm conclusions about  the cause 
of the failures was reached in 6 years of running.}
while the 2$^{nd}$ power supply suffered no failures.
\end{itemize}

The NESSiE Far and Near  magnets will be  powered each by a power supply  unit.

\subsection{Monitored quantities for every magnet}
The  values to be monitored are:
\begin{itemize}
\item Continuous measurement of the magnetic field strength by Hall probes or
pickup coils at magnet ramp down;
\item electrical quantities:
\begin{enumerate}
\item current: check for maximum/minimum range,
\item voltage: check for maximum/minimum range,
\item ground leakage current: check for maximum range (not automatically 
done in OPERA, via webcam or onsite check);
\end{enumerate}

\item temperatures and cooling:

\begin{enumerate}
\item coil temperatures: check for maximum range,
\item cooling water input temperature: check for maximum/minimum range,
\item cooling water output temperature: check for maximum/minimum range,
\item cooling water pressure, cooling water flux: check for maximum/minimum,
values (not automatically done in OPERA, via webcam or onsite check).
\end{enumerate}

\end{itemize}

\subsection{The Slow Control System}

The  slow control system  will monitor the whole hardware related Spectrometers, namely: the magnet power supplies, the active detectors
and  ancillary systems.

As it was done in  the  OPERA experiment~\cite{bib:opera-sc}
the NESSiE slow control system is organized in tasks and data structures developed to acquire
in short time the status of the detector parameters which are important for a
safe and optimal detector running.

The slow control should  provide a set of tools which automatize specific
detector operations (for instance the ramping up of the detector High Voltage
before the start of a physics run) and let people on shift control the
different components during data taking.

Finally, the slow control has to generate alarm messages in the event of a
component failure and react promptly, without human intervention, to preserve
the detector from possible damages. As an example the RPC High Voltage has
to be ramped down at the occurrence of any  gas system failure.

A possible structure of the slow control could be:
\begin{itemize}
\item  a database  is used to store both the slow control data and the detector configurations.
\item the acquisition task is performed by a pool of clients, each serving
a dedicated hardware component. The clients are distributed on various Linux
machines and store the acquired data on the database.
\item the hardware settings are stored in the database and served through a
dynamic web server to all the clients as XML files.
A configuration manager gives the possibility to view and modify the hardware
settings through a Web interface.
\item a supervisor process, the Alarm Manager, retrieves fresh data
from the database, and is able to generate warnings or error messages in case
of detector malfunctioning.
\item the system is integrated by a Web Server for monitoring the
global status of the data taking, the status of the various components, and
to view the latest alarms.
\end{itemize}

\section{Detectors for the Iron Magnets}\label{sec:rpc}

The NESSiE Near and Far spectrometers will be instrumented with large area 
detectors for precision tracking of muon paths
allowing high momentum resolution and charge identification capability. 
A spatial resolution of about $1$~cm is enough.

Suitable active detectors for the ND and FD Iron spectrometers are  RPCs
-- gas detectors widely used in high energy and astroparticle experiments~\cite{bib:origin} -- because: 
\begin{itemize}
\item they can cover large areas;
\item they are relatively simple detectors in terms of construction, flexibility in operation and use;
\item their cost is cheaper than other other large area tracking systems;
\item they have excellent time resolution;
\item and a high counting--rate power (in specific operational modes).
\end{itemize}
Furthermore  by considering the remainders of the OPERA RPC production, about  $1500~{\rm m}^2$ of RPCs are already available.

\subsection{RPC Detectors}\label{subsec:rpc-dect}
We plan to use standard bakelite RPCs:  two electrodes made of $2$~mm plastic laminate kept $2$~mm apart by polycarbonate spacers, of $1$~cm diameter,
in a $10$~cm lattice configuration. The electrodes have high volume resistivity ($10^{11} -- 10^{13}~\Omega\cdot$cm). Double coordinate read--out is obtained by 
copper strip panels. 
The strip pitch can be between 2 and 3.5 cm in order to limit the overall number of read--out channels. An optimization of the strip size and  orientation 
(horizontal, vertical and tilted ones) is required for the best track reconstruction resolution and reduction of ghost hits.
%RPC's are commonly used in streamer mode operation with a digital read-out as described in the following.

\subsection{Detector Ancillary Systems}
The operation of  the RPCs requires:
\begin{itemize}

\item High Voltage system with current monitoring
   performed by dedicated nano--amperometers designed by the Electronics Service of INFN--LNF. 
\item monitoring of several environmental/operational parameters (RPC temperatures, gas pressure and relative humidity).

\item   Gas distribution system.
Since the overall rate (either correlated or uncorrelated) is estimated to be low (see Sect.~\ref{sec:bck}),
standard gas mixtures for RPC  in streamer   operation can be used, like the one used in the OPERA RPCs, namely
 Ar/tetrafluorethane/isobutane and sulfur--hexafluouride in the volume ratios: 75.4/20/4/0.6.

%   Different gas mixtures can be further investigated and an optimization 
%is advisable depending on the adopted read-out system  (digital versus analog read-out) and on the safety regulations at the experimental site.
The OPERA RPCs are flushed with an open flow system at $1500 \ {\rm l/h}$ (5 refills/day). The installation of a recirculating system could also be considered
  if the gas flow has to be increased to prevent detector aging.   
   \end{itemize}

\subsection{The Tracking Detectors for the Near and Far Spectrometers}

The Near spectrometer will be a {\em calorimeter} made of planes of iron interlaced with planes of RPC, each of RPC unit  being $2904\times 1128\ {\rm mm}^2$ in size.
The  RPCs will be arranged in planes of 2 columns and 3 rows,  covering a surface of about 20 m$^2$. 
A total of 44 planes ( 264 RPC units, 800~m$^2$ of detectors) will instrument the spectrometer.

The Far Detector will consist of 3 \ columns  $ \times$ 4\ rows of RPC to form planes of  about 40 m$^2$. 
As for the Near magnet 44 planes of detectors will be interleaved with iron absorbers amounting to 1760 m$^2$ (4528 units) of instrumented surface.

\subsection{RPC Production and Quality Controls}
As it was done for the OPERA experiment, before their installation,  RPC will undergo a full chain of quality control tests, serialized according to the following steps:
\begin{itemize}
\item mechanical tests (gas tightness and spacer adhesion);
\item electrical tests;
\item efficiency measurement with cosmic rays and intrinsic noise determination.
\end{itemize}
The OPERA setup, in part still available at the Gran Sasso INFN Laboratories, was able to validate about 100 m$^2$ of RPCs per week.

\subsection{Costs}
The OPERA RPC are expected to be fully operative again. However, in case the percentage of breakages after dismantling were too high a new production could be foreseen\footnote{About 300 never used RPC detectors are also available from the OPERA contingency, i.e. about 1000~m$^2$.}.
Plastic laminate can be produced in Italy by the Pulicelli company, located near Pavia (the company is currently producing material for the CMS RPC upgrade system).
The cost of the plastic laminate is about 30 \euro/m$^2$.
The RPC chamber assembly can be done in Italy by the renovated General Tecnica company at an estimated cost of about 300 \euro/m$^2$.
The overall cost of an entire new RPC production is estimated of about 1.5 M\euro.

\noindent The cost for 5000 m$^2$ of  read--out strips is about  500 k\euro.

\section{Backgrounds}\label{sec:bck}

The number of expected  events from the Booster neutrino flux (Sect.~\ref{sec:beam}) at the Near site is
0.1 event per spill  (a spill being 1.6 $\mu$s long within a cycle of 0.2~s).
At the Far Detector the event rate is a factor of 20 smaller.

The background rate is  estimated assuming a tracking system of single--gap RPC's.
We distinguish the uncorrelated background due to detector noise and local radioactivity
(dark counting rate) and the correlated background due to cosmic rays.

\subsection{Uncorrelated Background}

The dark counting rate depends on the detector features and ambient radioactivity. At sea level a typical rate for RPC  is 
$300\ {\rm Hz/m}^2$. Therefore the expected rate per plane is
$\lambda_N = 300\ {\rm Hz/m}^2 \times (5 \times 3.4\ {\rm m}^2) \simeq  5.1$~kHz in the Near spectrometer and 
$\lambda_F = 300\ {\rm Hz/m}^2 \times (8.7 \times 4.5\ {\rm m}^2) \simeq 11.7$~kHz on the Far spectrometer. 

With a time coincidence of at least three consecutive RPC planes within a time window of $300~{\rm ns}$ the trigger rate per beam spill due to random coincidence is of the order of $3\times 10^{-6}$ and $4\times 10^{-5}$ for the Near and Far detector, respectively. This rate can be further suppress topological selection of the hits. Dark Noise events can be measured during the inter--spill time and subtracedt statistically.

%Assuming a read-out time window of we expect 0.05 (0.12) hits per RPC plane
%per event in the Near (Far) detector. 
The number of fired strips depends on the strip width:
with $2.6$ and $3.5$~cm strip--wide  the typical cluster size is $\sim 1.5$ strips.

The requirement to be in the beam--spill time--window makes the dark noise contribution to the trigger rate negligible
(see last column in Tab.~\ref{tab:noise}).
\subsection{Cosmic Ray Background}

The contribution of Cosmic Ray  (CR) muons   to the plane--by--plane background is similar  to  the uncorrelated background, but  long muon tracks  can induce a more relevant background.

Assuming a trigger majority of $\ge 3$ fired planes a cosmic 
ray muon can trigger the data--taking if  it can cross at least 2
iron slabs ($2 \times 5$~cm), namely if its momentum exceeds $250$~MeV/c.

The integrated vertical muon flux is $J = 97\ {\rm Hz/m}^2/sr$ at sea level~\cite{bib:cr}.
The integrated vertical flux of the soft component (electrons and positrons), for $E\gtrsim 80$~MeV,
is  $J(> E) = 0.22\ E^{-1.45}~[{\rm Hz/m}^2/{\rm sr}]$, where $E$ is
in GeV. Taking into account this contribution and minor ones due to hadrons,
an  integrated vertical CR flux at sea level of $J = 100~{\rm Hz/m}^2/{\rm sr}$ will be used in the following estimations. Under the conservative hypothesis of an isotropic CR ray flux  equal to the vertical flux,  the total rate $\lambda_{RC}$ on a  detector shaped as a fully efficient box is

$$ \lambda_{RC} = \frac{\pi}{2} S_{tot} J $$

\noindent where $S_{tot}$ is the  detector surface. The expected number of CR 
events in a time window of $1.6\ \mu {\rm s}$ (the beam-spill time) is
reported in Tab.~\ref{tab:noise}. They scale with 
the time window $T$ (by a factor $T/1.6\ \mu {\rm s}$).
The data in Tab.~\ref{tab:noise} conservatively ignore that more detailed
trigger conditions allow a significant reduction of the background.

\begin{table}[ht] \begin{center}
\begin{tabular}{|l|ccc|} \hline
Single       & CR events            & Dark Noise events & Beam events  \\
Spectrometer & in $1.6\ \mu s$     & in $1.6\ \mu s$  & in  $1.6\ \mu s$ \\ \hline
             &                      &           &          \\ 
Near         & $4 \times 10^{-2}$ & $3 \times 10^{-6}$ & $10^{-1}$\\ 
Far          & $7 \times 10^{-2}$ & $4 \times 10^{-5}$ & $5\times 10^{-3}$\\ \hline
\end{tabular}
\caption{Background events (CR muons and dark noise) and beam related events 
        in the Near and Far spectrometers.} \label{tab:noise}
\end{center} \end{table}

\section{Read--out, Trigger and DAQ}\label{sec:daq}

\subsection{DAQ overview}
The aim of the DAQ is to read the signals produced by the electronic detectors and create a database of detected events. 
We recall in Tab.~\ref{tab:prot-beam} the characteristics of the Booster primary proton beam. These are important inputs
to define properly the data acquisition and flow.

\begin{table}[htbp]
\caption{Characteristics of the Booster primary proton beam}
\label{tab:prot-beam}
\begin{center}
\begin{tabular}{|l|c|c|}
\hline
& Booster Proton Beam\\
\hline
Proton beam momentum & 8 GeV\\
\hline
Protons per pulse & $4.5\times 10^{12}$\\	
\hline
Number of bunches & 84\\
\hline
Bunch length & 4\ ns  \\
\hline
Bunch spacing & 19\ ns  \\
\hline
Burst length & 1.6 $\mu$s\\
\hline
Maximum repetition rate &0.2\ s \\
\hline
Beam energy & 5.8\ kJ \\
\hline
Average beam power & 30\ kW \\ 
\hline
\end{tabular}
\end{center}
\label{default-daq}
\end{table}

\vskip 10pt
The foreseen DAQ architecture is in three stages: 
\begin{itemize}
\item the front--end electronics close to the detector (FEB)
\item the read--out interface which together with the trigger board control the FEB read--out 
\item the Event Building which reconstructs events using standard workstations.
\end{itemize}
The whole event reconstruction is based on the time correlation of channels, which depends on the accuracy of the  channel time stamping. 
A time resolution in the range $5\div 10$~ns is sufficient to correctly associate the different hits to the corresponding events. 
A common time 
reference with respect to the Booster extraction time is used to time--stamp the data and to correlate the data of the various detectors. 

\subsection{Data Flow}
The Far detector is designed with 12 RPC/plane $\times$ 11 planes/arm $\times$ 2 arms $\times$ 2 spectrometers
for a total 528 RPC modules, 3~m$^2$ each.
The Near detector is designed with 6 RPC/plane $\times$ 20 planes/arm $\times$ 2 arms $\times$ 2 spectrometers for a 
total of 264 RPC modules.

Assuming a strip size of $2.6$~cm along $Y$ and $3.5$~cm along $X$ each Far detector plane will be equipped with $32\times 4=128$ horizontal strips plus 
$112\times 3 = 336$ vertical strips for a total number of about 500 electronic 
channels per plane. The total number of channels is therefore about 20,000.
In the Near detector the number of electronic channels is about
290 per plane and the total number of channels is about 13,000.

%The expected RPC single rate is of the order of $300~Hz/m^2$ which imply a noise rate per electronic channel of the order of $70~Hz$. 
%Hence the contribution of the random coincidence of the electronic noise with the neutrino burst which last $2.1\mu s$ is negligible. 

The expected background rate due to the RPC single rate and the counts due to cosmic ray within the beam spill are
reported in Sect.~\ref{sec:bck}. With a maximum  beam intensity of 
$4.5\times10^{12}~\mbox{p.o.t.}$ about $0.1$ events are expected in the Near detector. Assuming four hit strips per plane and at 
most $16$~byte of data per hit (channel address, signal and time) the size of the event after zero suppression is expected to be $1$~kbyte.

\subsection{Front--End- Electronics}

The front--end electronics of the RPCs instrumenting the iron magnets was designed to operate with an event rate of the order of several tens of events per spill. A trigger--less logic has been implemented. 

\begin{figure}[h]
\centering
\includegraphics[scale=0.4]{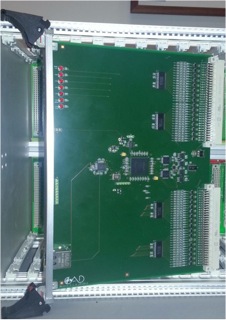}
\vspace{1.5cm}
\includegraphics[scale=0.3]{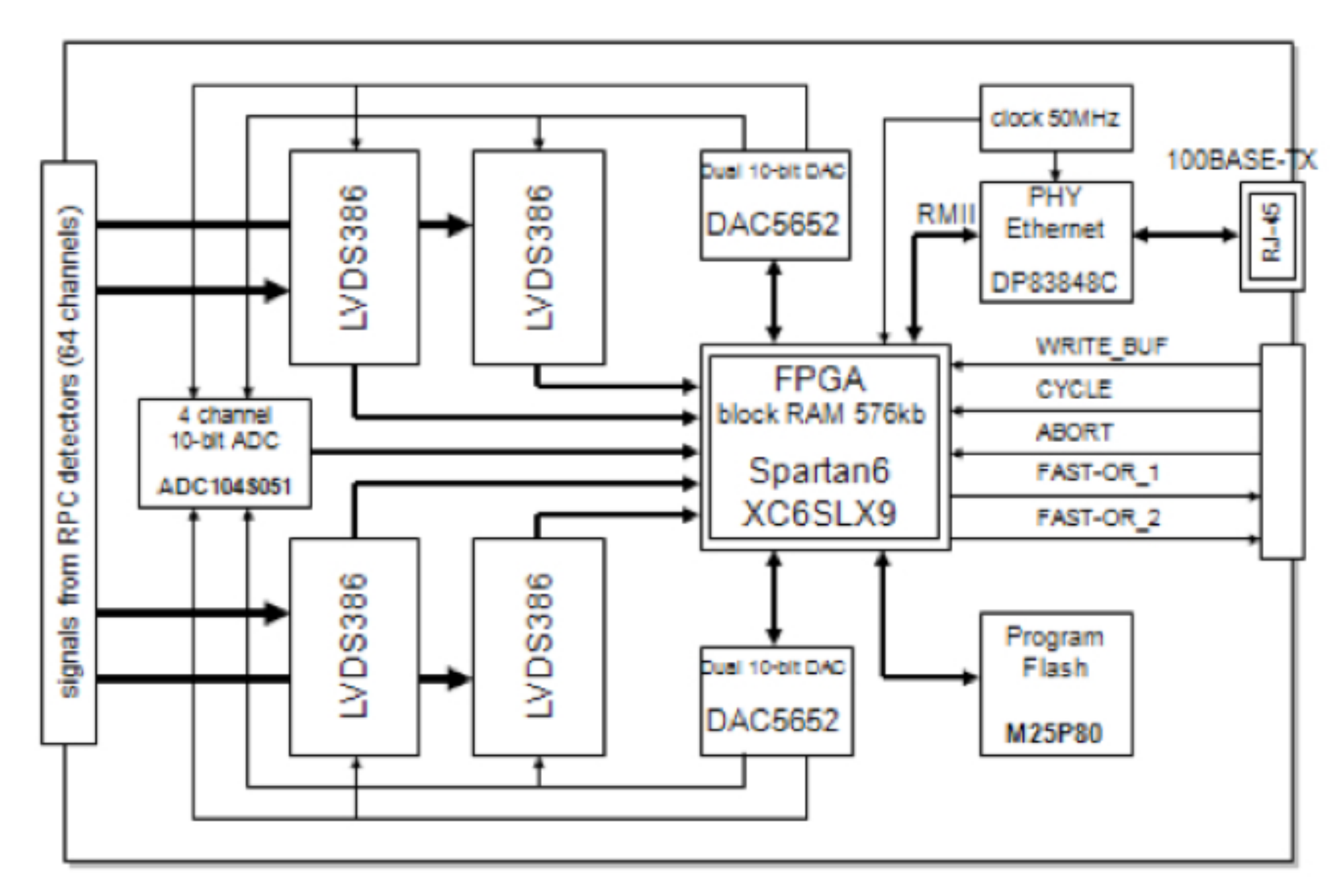}
\vspace{-1.5cm}
\caption{The Front--End board for the RPC readout.}
\label{FE}
\end{figure}

According to the same design scheme adopted for the OPERA experiment~\cite{lvds}, groups of 64 signals coming from the RPCs working in streamer mode are read--out by means of front--end boards (FEB) equipped with 4 16--channel LVDS receivers and an Ethernet configurable Field Programmable Gate Array (FPGA), Fig. \ref{FE}.

The LVDS receivers act as discriminators with programmable thresholds that can be set via Ethernet by 4 integrated 10--bit DACs. The output of each discriminator is sampled with a resolution of $\rm 10 \, ns$ and continuously stored in a 4096--sample circular buffer whenever a write--enable signal is active. A time stamp with a $\rm 10 \, ns$ resolution is provided for each stored signal. Each FEB provides 2 FAST--OR signals implementing the trigger of groups of 32 channels. 

FEBs are housed in crates controlled by a FPGA--based Crate Controller Board (CCB) with several tasks such as power supply management and monitoring, control signal distribution, masking, FAST--OR collection and management. Each CCB is able to manage up to 19 FEBs. The FAST--OR signals coming from the FEBs are stored in a circular buffer in a similar way as described above for the discriminated RPC signals in each FEB. The CCB provides 4 configurable FAST--OR signals as input to a Trigger Supervisor Board (TSB) able to generate a programmable trigger which can be used for the acquisition of cosmic ray muons as well as for monitoring and calibration purposes. Prototypes of FEBs and CCB are currently under test with RPC detectors exposed to cosmic rays .

The total estimated cost of the RPC read--out electronics is about 200 kEuro, taking into account 33000 digital channels.

\subsection{DAQ}
The Data Acquisition system is built like an Ethernet network whose nodes are the FEB equipped with an Ethernet controller. 
The Ethernet network is used to collect the data from the FEB, send them to the event building workstation and dispatch 
the commands to the FEBs for configuration, monitoring and slow control.
This scheme implies the distribution of a global clock signal to synchronize the 
local counters running on the FEBs that are used to time stamp the
data. 
The DAQ clock is synchronized with the FNAL accelerator timing system in order to start the DAQ readout cycle during particle extraction time. 

Given the very short beam spill ($1.6~{\rm \mu sec}$) the beam related events can be acquired in a trigger--less mode. Along the spill duration the FEBs store the status of the discriminators
or the pulse height of the input signals, for digital and analog readout, respectively, 
in a circular buffer driven by an on--board clock. The readout of the buffer by the DAQ is triggered by the {\it end--of--spill} signal sent to each FEB. This signal causes the FEB to disable the writing and time stamp information is generated concurrently on all FEBs. The time stamp is carried out on the trailing edge of the write signal, so the time information can be reconstructed backwards for all data. The duration of the write is also stored.
The time--stamp information and the circular buffer content are then transferred through Ethernet to the event building\footnote{The eventual correlation between spectrometer and LAr data can be achieved using
a common clock signal to time--stamp the events. Data merging can therefore performed offline at the reconstruction level.}.

In the inter--spill time the acquisition of cosmic ray muons and calibration data is triggered by a fake spill gate, possibly validated by a programmable logic (Trigger Board) on the basis of the FAST--OR- signals generated by the FEBs.
The {\it start--of--spill} signal is used to abort all the readout process on the FEBs and to start new data read--out.

Assuming 64 channels per board, a clock of 10 ns and a buffer depth of $160$, needed to acquire $1.6~{\rm \mu sec}$ of data, the time needed to transfer the data to the event building is about $100~{\rm \mu sec}$ on a Fast Ethernet.

The Event Builder should be based on standard commercial workstation. Data spying and monitoring process will
also be implemented at this level.

\section{FNAL Logistics, Schedule and Costs}\label{sec:logistic}

The two experimental halls requested in the ICARUS proposal (\cite{ICARUSFNAL}) to host the Near and Far detectors,
are well suited for the two NESSiE spectrometers, too, even if, following our studies, a surface building is also suitable for our need.
The main constraint is the need for a new pit in the Near site.

\begin{figure}[htbp]
\centering
\includegraphics[scale=0.65]{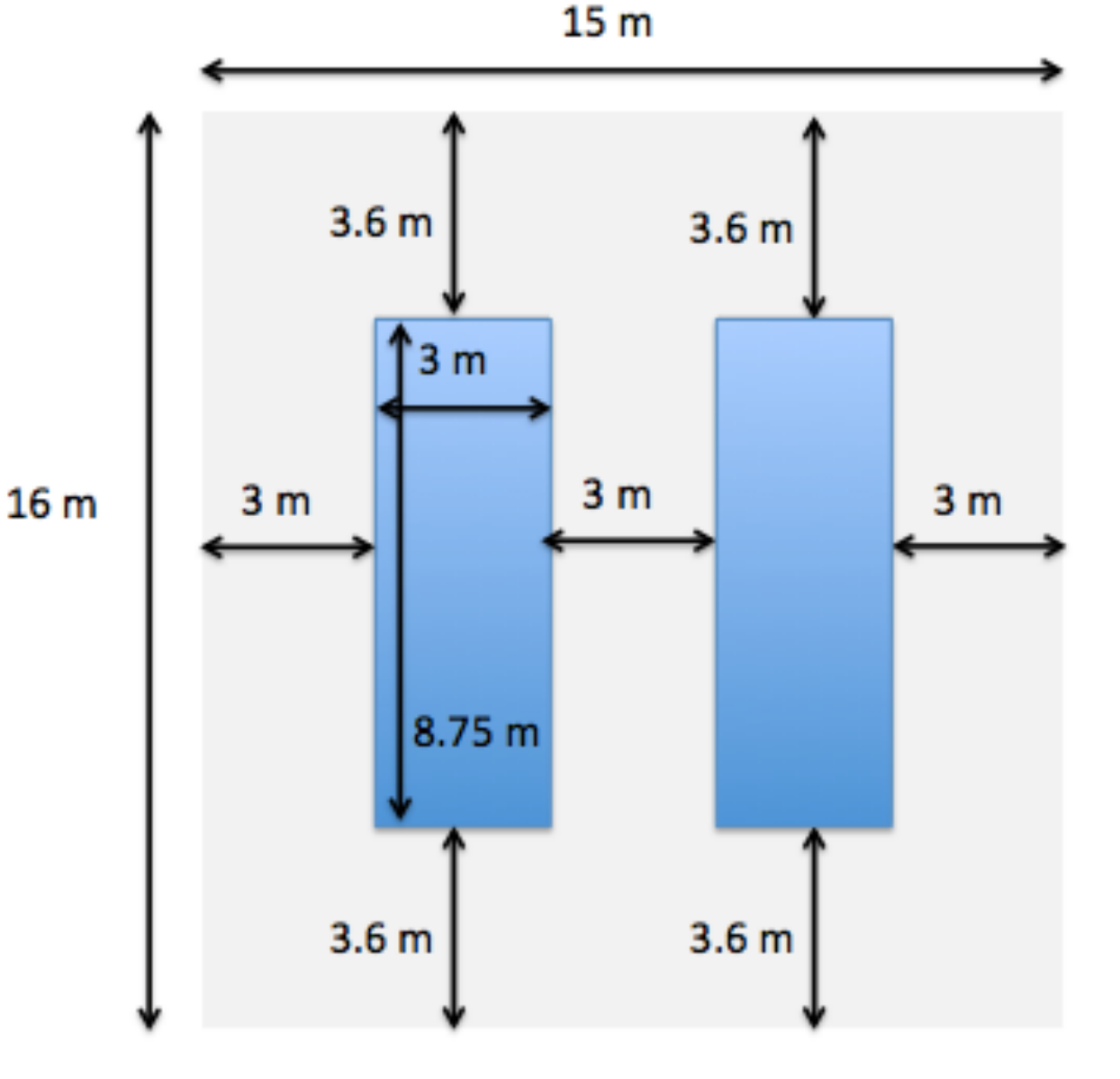}
\caption{Description of the Far area needed for the {\em in--situ} assembling of the two spectrometer. The two 3.63 $m$ lateral
space are needed for the extraction of the internal support structure once the assembling is finished. The 3 $m$ in the other 
direction are requested to ease the installation phase. Note that the minimum distance between the two magnets should be 2 $m$,
as indicated in Fig.~\ref{AssemblyFar-Near}.
Other $50\div 100$~m$^2$ are needed in a {\em close} site to keep the
iron slabs and mechanical structures and to assemble RPCs during installation.}
\label{fig:Far-site}
\end{figure}

The Far detector experimental facility has to be placed 710~m from the production target.In principle this facility will may host two detectors and 
the related infrastructure: ICARUS T600 and NESSiE. NESSiE may be placed downstream the LAr TPC, minimizing the distance from ICARUS. In 
Fig.~\ref{fig:Far-site} the horizontal dimensions requested for the assembling of the spectrometers
are outlined, while a $10\times 10$~m$^2$ area would be finally used. Additional storage area is needed initially for the delivered detector parts and their pre--assembly. For the in--situ assembly of the NESSiE spectrometer a 25 ton crane capacity is required.
The electrical power requirements are of the order of 100 KW for the NESSiE Far site. The experimental hall has to be
equipped with a ventilation system in order to dissipate maximum 20 KW power and keep the room temperature below 27 $^0$C due to
the constraint of the running RPC.

\begin{figure}[htbp]
\centering
\includegraphics[scale=0.65]{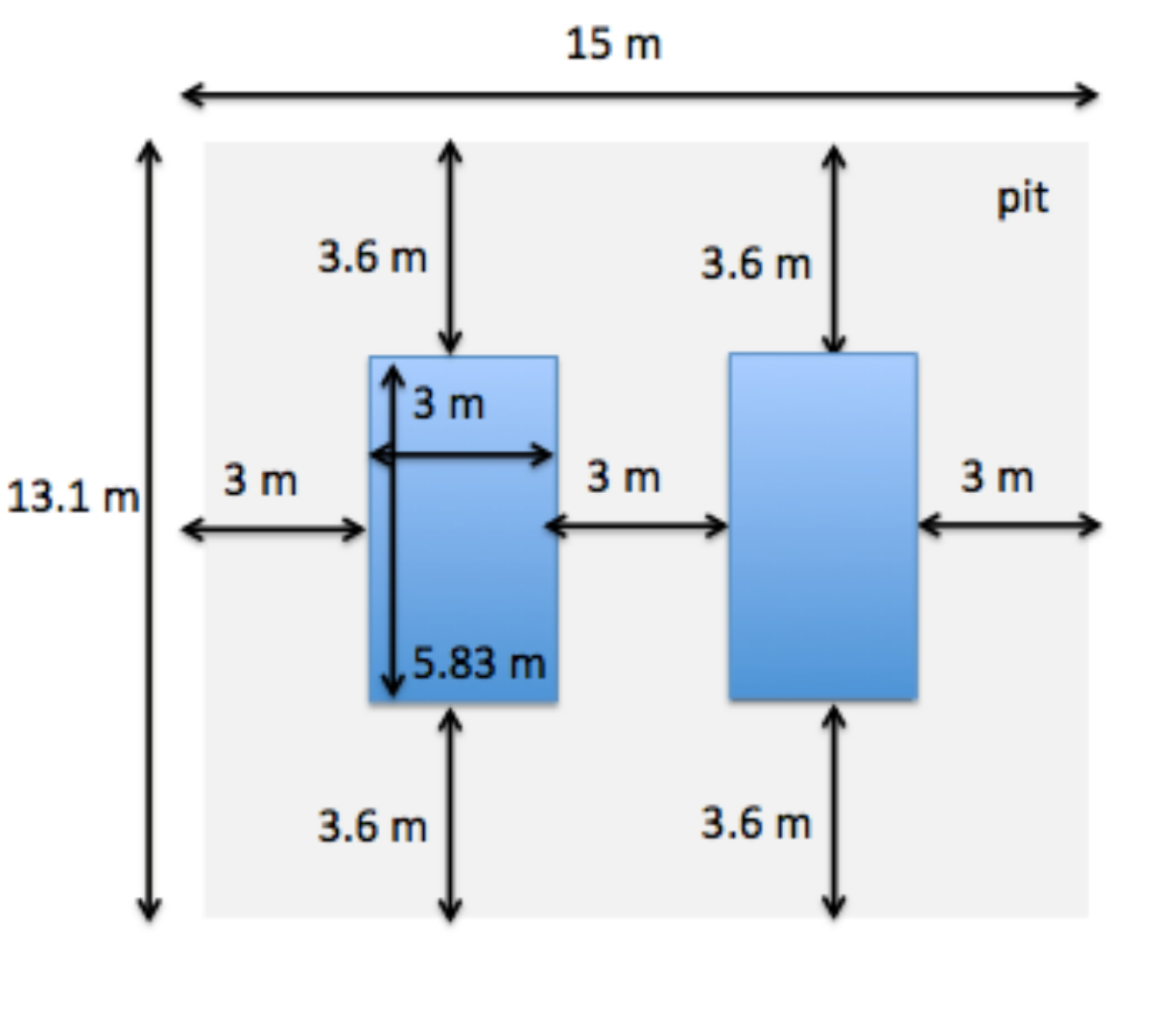}
\caption{Description of the Near area needed for the {\em in-situ} assembling of the two spectrometer. The two 3.63 $m$ lateral
space are needed for the extraction of the internal support structure once the assembling is finished. The 3 $m$ in the other 
direction are requested to ease the installation phase. Note that the minimum distance between the two magnets should be 2 $m$,
as indicated in Fig.~\ref{AssemblyFar-Near}.
The small space needed for the possible addition of a {\em target} is not included.
Other $50\div 100\ m^2$ are needed in a {\em close} site to keep the
iron slabs and mechanical structures and to assemble RPCs during installation.}
\label{fig:Near-site}
\end{figure}

The Near site spectrometer cross--section ($5\times 5.35$~m$^2$) is about half the Far site spectrometer. The requested areas follow
the same constraints of the Far site: they are reduced in $X$ and $Y$ while keeping the same length in $Z$. They are outlined in 
Fig.~\ref{fig:Near-site}.

The spectrometers (the iron magnets and the RPC detectors), will be assembled in their final location, requiring sufficient surface area for the 
pre--assembly of components. The servicing systems are planned to be prepared mainly off line and eventually positioned, cabled and connected 
in the experimental hall.

\subsection{Schedule and Costs}\label{sec:schedule-costs}

The choice of performing a  design study that is {\em reliable} under several aspects led to make conservative, well controlled 
and realistic options. The re--use of systems developed for the OPERA experiment is envisaged. We note that the OPERA Spectrometers 
have been fully funded by INFN, except the Precision Trackers, which is therefore committed to their dismantling and entitled to 
possibly re--use them as well. 
In particular the choice to use detectors like the RPC ones and the development of dipole magnets
which had, at least partly, already been constructed and have been in use, allows us to keep under control both the time schedule and the 
costs.

With respect to the schedule, which is reported in Tab.~\ref{tab:time}, it should be underlined that is based on the  
deep experience acquired with the OPERA
Spectrometers, built up from 2005 to 2006 under critical conditions\footnote{During the period 2005--2007 LNGS laboratory
underwent restoration and safety works enforced by Italian Government, following the temporary seal of May 2005.}.  
The provisional areas and the experimental hall have to be ready for installation by the Fall 2016.
The NESSiE installation will last conservatively 1.5 years, leaving six months commissioning period before the run start (assuming it is done almost simultaneously 
at the Near and Far sites, which is feasible with adequate manpower).
The assumed exposure is $6.6\times 10^{20}$ proton--on--target (p.o.t.) to be delivered by the FNAL--Booster in 3 years of activity.

\begin{table}[htbp]
\begin{center}
\begin{tabular}{|l|l|l|}
\hline
Year(portion)&Action\\
\hline
1$^{rst} $ half 2015&Define tenders/contracts \\
2$^{nd}$ half 2015&Site preparation\\
&    Setting up Detectors Test--stands   \\
\hline
1$^{rst}$ half 2016&Mechanical Structure construction\\
 & Start Magnet installation \\
&                          Start detectors installation \\
\hline
2$^{nd}$ half 2016&End installation \\
\hline
1$^{rst} $ half 2017&Commissioning and Starting Run \\
\hline
2$^{nd} $ half 2019&End Data Taking \\
\hline
\end{tabular}
\end{center}
\caption{ \label{tab:time}
Tentative time schedule for the 2015--2019 years, including detector construction and installation at FNAL and data taking.}
\end{table}

With respect to the cost estimate, the expenses needed for the major items are reported in Tab.~\ref{tab:costs}.

\begin{table}[htbp]
\begin{center}
\begin{tabular}{| l | c c |}
\hline
Item&\multicolumn{2}{|c|}{Cost (in M\euro )} \\
\hline\hline
Far & & \\ \hline
Magnet & & 2.5 (in--kind)\\
RPC detectors & & 0.8 (in--kind)\\
Strips & & 0.3 (in--kind) \\
New Electronics & 0.2 & \\
Data Acquisition & 0.1 & \\ 
\hline\hline
Near & & \\ \hline
Magnet & & 2.0 (in-kind) \\
Top/bottom yokes & 1.0 & \\
Coils, Power Supplies & 0.2 & \\
RPC detectors & & 0.6 (in--kind) \\
New detectors & 0.2 &\\
Strips & & 0.2 (in--kind) \\
New Electronics & 0.1  & \\
Data Acquisition & 0.1 & \\ 
\hline\hline
Transportation & 0.6 & \\
\hline\hline
Total& \multicolumn{2}{|c|}{2.5 $+$ 6.4 (in--kind)}\\
\hline
\end{tabular}
\end{center}
\caption{ \label{tab:costs}
Estimate of the costs of the major items. Note the major costs for the Near site that include the duplicated parts of the
OPERA spectrometers, i.e. the top and bottom closures, the coils and the power supplies. The cost for the possible construction of a
small target with scintillators is indicated as "new detectors" in the Near set.}
\end{table}

\clearpage

\section{Physics Analysis and Performances}\label{sec:spect2}

We developed sophisticated analysis to obtain the sensitivity region that can be achieved with an exposure of 
$6.6\times 10^{20}$ p.o.t., corresponding to 3 years of data collection at FNAL--Booster beam.
Our guidelines have been the maximum extension at small values of the mixing angle parameter,
as well as its dependence on systematic effects.

To this aim, three different analysis have been set up, of different complexity:
\begin{itemize}
\item the usual sensitivity plot based on the Feldman\&Cousins technique (see Section V of~\cite{Fel-Cou}) has been obtained, 
by adding {\em ad hoc} systematic error evaluations;
\item a full correlation matrix based on the full Monte Carlo simulation and reconstructed data;
\item a new approach based on the profile CLs, similar to that used in the Higgs discovery. 
\end{itemize}

Throughout the analysis the following framework has been considered. We assumed two identical muon spectrometers exposed to the Booster Neutrino Beam (BNB)
located at a distance of 110 m (Near) and 710 m (Far) from the target and fiducial mass of 297 tons and 693 tons, respectively, as shown in 
Tab.~\ref{tab:NFD}.

\begin{table}[h]
\centering
\begin{tabular}{|l|c|c|c|c|c|c|}
\hline
 & Fiducial Mass (ton) & Baseline (m)\\
\hline
\hline
Near & 297 & 110\\
\hline
Far & 693 & 710\\
\hline
\end{tabular}
\caption{Fiducial mass and baselines for Near and Far detectors. }
\label{tab:NFD}
\end{table}

\begin{figure}[htbp]
 \centering
  \includegraphics[width=0.80\textwidth]{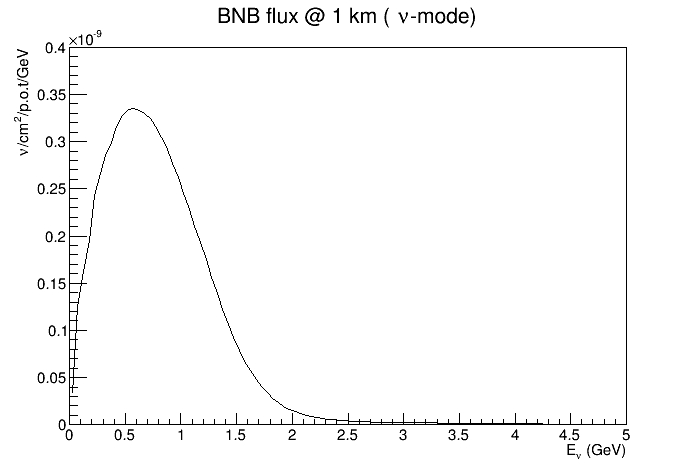}
 \caption{\footnotesize $\nu_{\mu}$ BNB flux at a distance of 1 km from the neutrino source. } 
\label{fig:flux}
\end{figure}

\begin{figure}[htbp]
 \centering
  \includegraphics[width=0.80\textwidth]{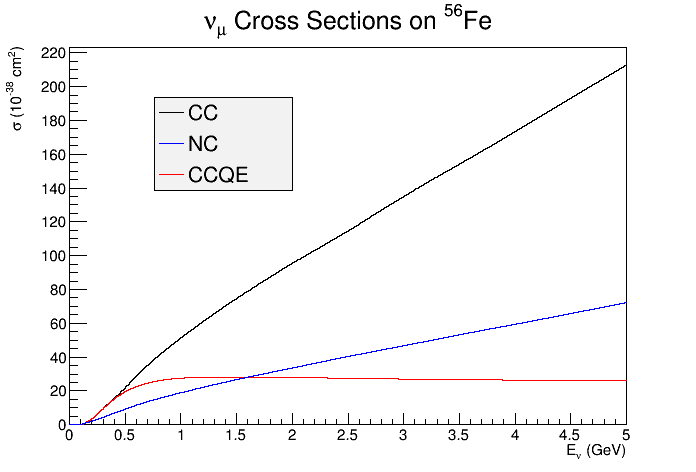}
 \caption{\footnotesize $\nu_{\mu}$ cross--sections on $^{56}Fe$ for all (QE, Res, DIS) charged current (black) and 
 neutral current (blue) interactions and for CCQE interaction (red), as extracted from GENIE~\cite{ref_GEN}.} 
\label{fig:xsec}
\end{figure}

The number of expected events at Near and Far detector constitutes the primary input to compute the achievable sensitivity.
The Booster Neutrino Beam (BNB) flux~\cite{BNB}, as expected at 1 km from the source, is shown in Fig.~\ref{fig:flux},
while the neutrino cross--sections for the different contributions of charged current (CC) and neutral current (NC) interaction 
(quasi--elastic, deep--inelastic--scattering and resonant) compared to the single quasi--elastic charged current (CCQE) 
interactions~\cite{ref_GEN}, 
as function of the incoming neutrino energy, are shown in Fig.~\ref{fig:xsec}.
The convolution of flux and cross--sections implies the relevance of the CCQE component in our analysis that makes use of the 
muon momentum as estimator. To go from $p_{\mu}$ to $E_{\nu}$ either the usual formula  in the CCQE approximation is
applied
\begin{equation}
E_{\nu} = \frac{E_\mu - m^2_{\mu}/(2M)}{1-(E_\mu -p_\mu\cos\theta)/M},
\end{equation}
\noindent ($M$ being the nucleon mass) or it is extracted via Monte Carlo simulation.

For all the analysis the usual two--flavor neutrino mixing scheme is considered. The oscillation probability is therefore given by the formula:
\begin{equation}\label{eq:2flavour}
P = \sin^{2}(2\theta_{new})\sin^{2}(1.27\ \Delta m^{2}_{new}\ L({\rm km})/E({\rm GeV}))
\end{equation}
where $\Delta m^{2}_{new}$ is the mass splitting between a heavy mass state and the heaviest of the three light neutrino mass 
states, and $\theta_{new}$ is the mixing angle between them.
Then the disappearance of flavour $\alpha$ is due to the oscillation of neutrino mass states at the $\Delta m^2$ scale
and at an effective mixing angle $\theta$ that can be simply parametrized as a function of the elements of a $3+1$ extended mixing matrix:
\begin{equation}
P(\nu_\alpha\rightarrow\nu_\alpha)=1-\sin^2(2\theta)\cdot\sin^2(1.267\cdot\Delta m^2\cdot L/E).
\end{equation}
As $L$ is fixed by the experiment location, the oscillation is naturally driven by the neutrino 
energy, with an {\em amplitude} determined by the mixing parameter.

The disappearance of muon neutrinos due to the presence of an additional sterile state
depends only on terms of the extended PMNS~\cite{pmns} mixing matrix ($U_{\alpha i}$ with $\alpha= e,\mu,\tau$ and $i=1$,\ldots,4)
involving the $\nu_{\mu}$ flavor state and the additional fourth mass eigenstate. In a 3+1 model at Short Baseline (SBL) we have:
\begin{equation}
P(\nu_{\mu}\to\nu_{\mu})_{SBL}^{3+1} = 1 - \left[ 4 \vert U_{\mu 4} \vert^2 (1 - \vert U_{\mu 4} \vert^2)\right] \cdot \sin^2 \frac{\Delta m^2_{41} L}{4E},
\end{equation}
where $4 \vert U_{\mu 4} \vert^2 (1 - \vert U_{\mu 4} \vert^2)$ results as an {\em amplitude}.

In contrast, appearance channels (i.e. $\nu_\mu \to \nu_e$) are driven by
terms that mix up the couplings between the initial and final flavour states and
the sterile state yielding a more complex picture:
\begin{equation}
P(\nu_{\mu}\to\nu_e)_{SBL}^{3+1} = 4 \vert U_{\mu 4}\vert^2 \vert U_{e 4} \vert^2  \sin^2 \frac{\Delta m^2_{41} L}{4E}
\end{equation}
This also holds in extended $3 + n$ models.

It is interesting to notice that the appearance channel is suppressed by two more
powers in $\vert U_{\alpha 4}\vert$. Furthermore, since $\nu_e$ or $\nu_\mu$ appearance
requires $\vert U_{e 4}\vert > 0$ and $\vert U_{\mu 4}\vert > 0$, it should be naturally accompanied by
a corresponding $\nu_e$ and $\nu_\mu$ disappearance. In this sense the disappearance
searches are essential for providing severe constraints  on the models of the theory
(a more extensive discussion on this issue can be found e.g. in Sect. 2 of~\cite{winter}).

It must also be noted that the number of $\nu_e$ neutrinos depends on
the $\nu_e\rightarrow\nu_s$ disappearance and $\nu_\mu\rightarrow\nu_e$ appearance, and, naturally,
from the intrinsic $\nu_e$ contamination in the beam. 
On the other hand, the amount of $\nu_\mu$ neutrinos depends only on the
$\nu_\mu\rightarrow\nu_s$ disappearance and $\nu_e\rightarrow\nu_\mu$ appearance but the latter is
much smaller due to the fact that the $\nu_e$ contamination in $\nu_\mu$ beams is
usually at the percent level. 
Therefore in the $\nu_{\mu}$ disappearance channel the oscillation probabilities in both Near and Far detectors can be measured without any interplay 
of different flavours, i.e. by the same probability amplitude.

\begin{figure}[htbp]
\begin{center}
\includegraphics[scale=0.6]{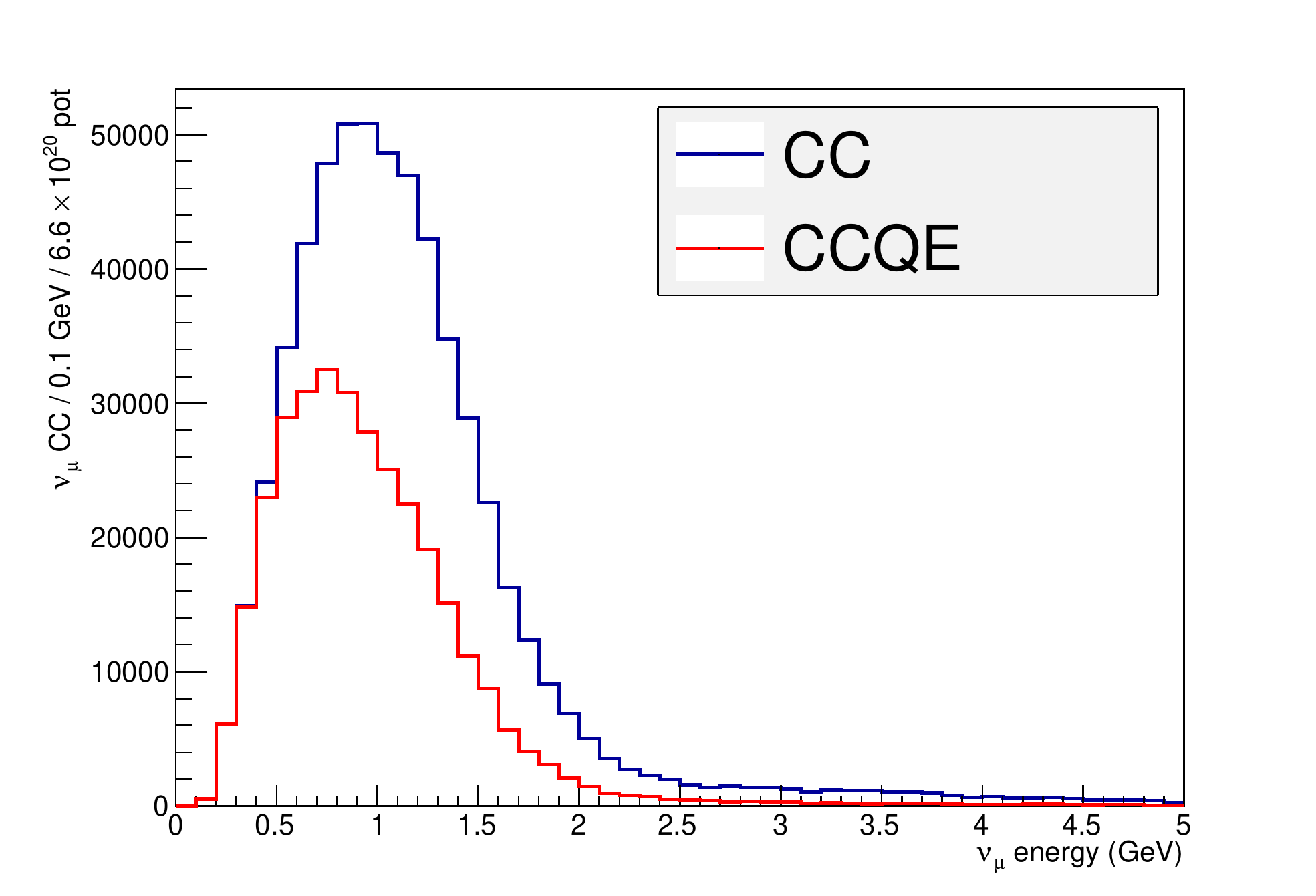}
\includegraphics[scale=0.6]{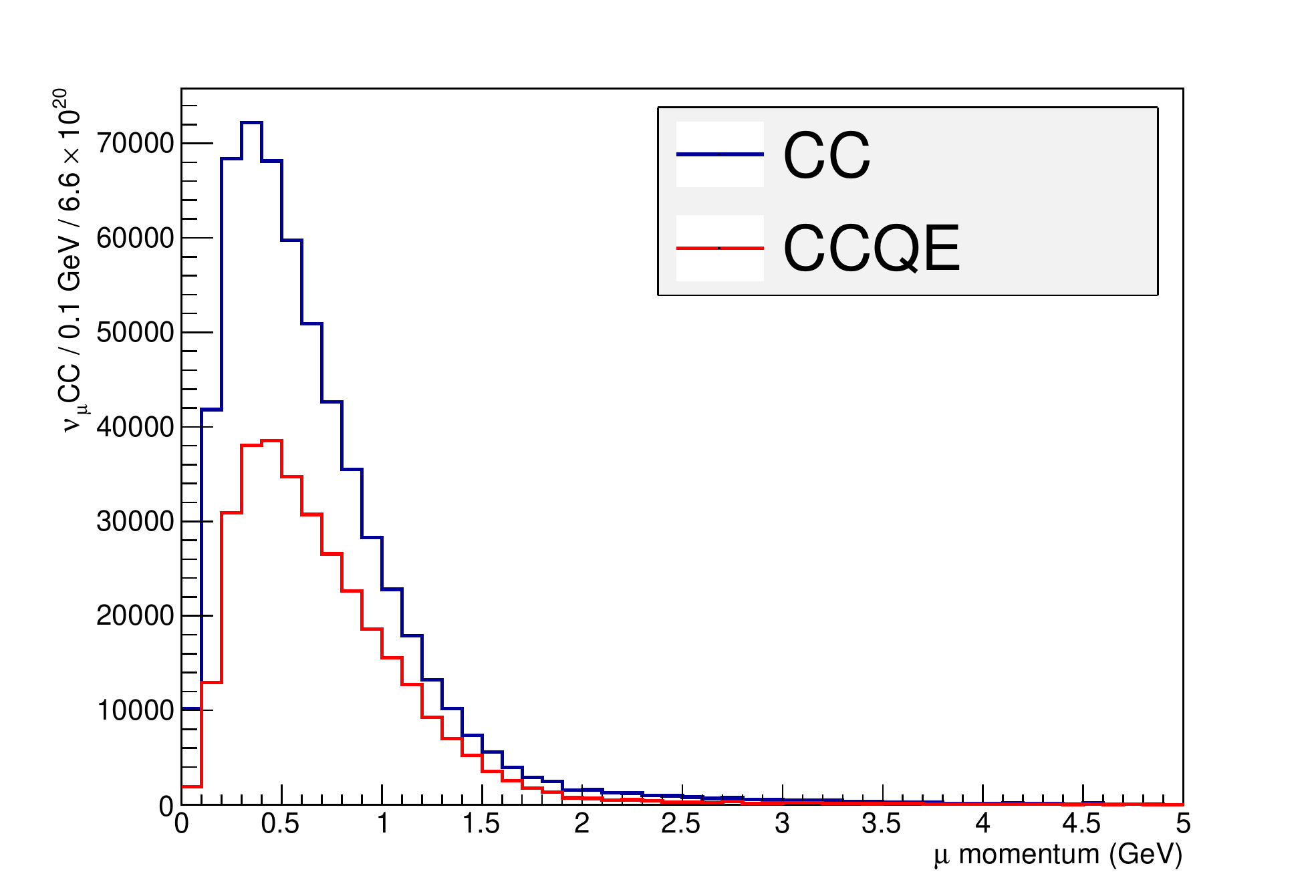}
\caption{The absolute number of \numu CC interactions seen by the Far detector at 710 m, as a function of the
$E_{\nu}$ (top) and the $p_{\mu}$ (bottom). The binning of 100 MeV has been that used further in all the  
statistical analysis.}
\label{fig:norma-interac}
\end{center}
\end{figure}

The final distributions of events, either in the $E_{\nu}$ or the $p_{\mu}$ variables, normalized to the expected
luminosity in 3 years of data taking at FNAL--Booster, or $6.6\times 10^{20}$ p.o.t., are reported in 
Fig.~\ref{fig:norma-interac}.

%\clearpage

\subsection{Standard Sensitivity Analysis}

A first basic estimate of the achievable performance has been obtained as follows:
\begin{enumerate}
\item spectra of $p_\mu$ at the Far and Near detectors ($\vec{N_0}$, $\vec{F_0}$) are generated in the null hypothesis (no disappearance signal)
according to the expected event statistics. We use 50 bins from 0 to 5 GeV/c and only momenta above 0.5 GeV/c are considered. 
\item We define the Far--to--Near ratio in each bin for these
  distributions as $R_{0,i} = F_{0,i}/(kN_{0,i})$ where $k$ is a
  $E_\nu$--independent factor to renormalizes Near and Far
  (i.e. without loss of generality for this study we neglect
  finite--distance effects).
\item For each $\sin^2(2\theta)$ and $\Delta m^2$ sampling point we
  calculate $R_i = N_i / (k F_i)$ obtained reweighing Monte Carlo
  interactions with a 2--flavour oscillation formula using the
  energy of the neutrino at true level.
\item A $\chi^2$ is then defined for each $\sin^2(2\theta)$ and $\Delta m^2$ hypothesis as:
\begin{equation}
	\chi^2  = \sum_{i=1}^{N}\left(\frac{1 - R_i/R_{0,i}}{\sigma_{R_{0,i}}}\right)^2
\end{equation}
where the error in each bin, $\sigma_{R_{0,i}}$, is simply the quadratic sum of the statistical term and a fixed
(and bin--to--bin uncorrelated) systematic error $\sigma_{R_{0,i}} = R_{0,i}\sqrt{\frac{1}{N_{0i}}+ \frac{1}{F_{0,i}}} \otimes \epsilon_{sys}R_{0,i}$.
The sum is intended over the bins of the muon momentum distribution. 

\item We find the point\footnote{Using the gMINUIT ROOT package.} in
  the ($\sin^2(2\theta)$, $\Delta m^2$) plane giving the best
  description of the simulated data i.e. providing the minimum $\chi^2$: $\chi^2_{min}$.

\item We evaluate for each point in the ($\sin^2(2\theta)$, $\Delta
  m^2$) plane the difference $\Delta\chi^2 = \chi^2(\sin^2(2\theta), \Delta m^2) - \chi^2_{min}$
\item The 95\% sensitivity region is then defined by selecting the
  portion of the parameter space for which the
  $\Delta\chi^2(\sin^2(2\theta), \Delta m^2)$ is larger than 5.9915
  (assuming a 2--DOF $\chi^2$ distribution for $\Delta\chi^2$).
\end{enumerate}

\begin{figure}
\centering
\includegraphics[scale=0.45,type=pdf,ext=.pdf,read=.pdf]{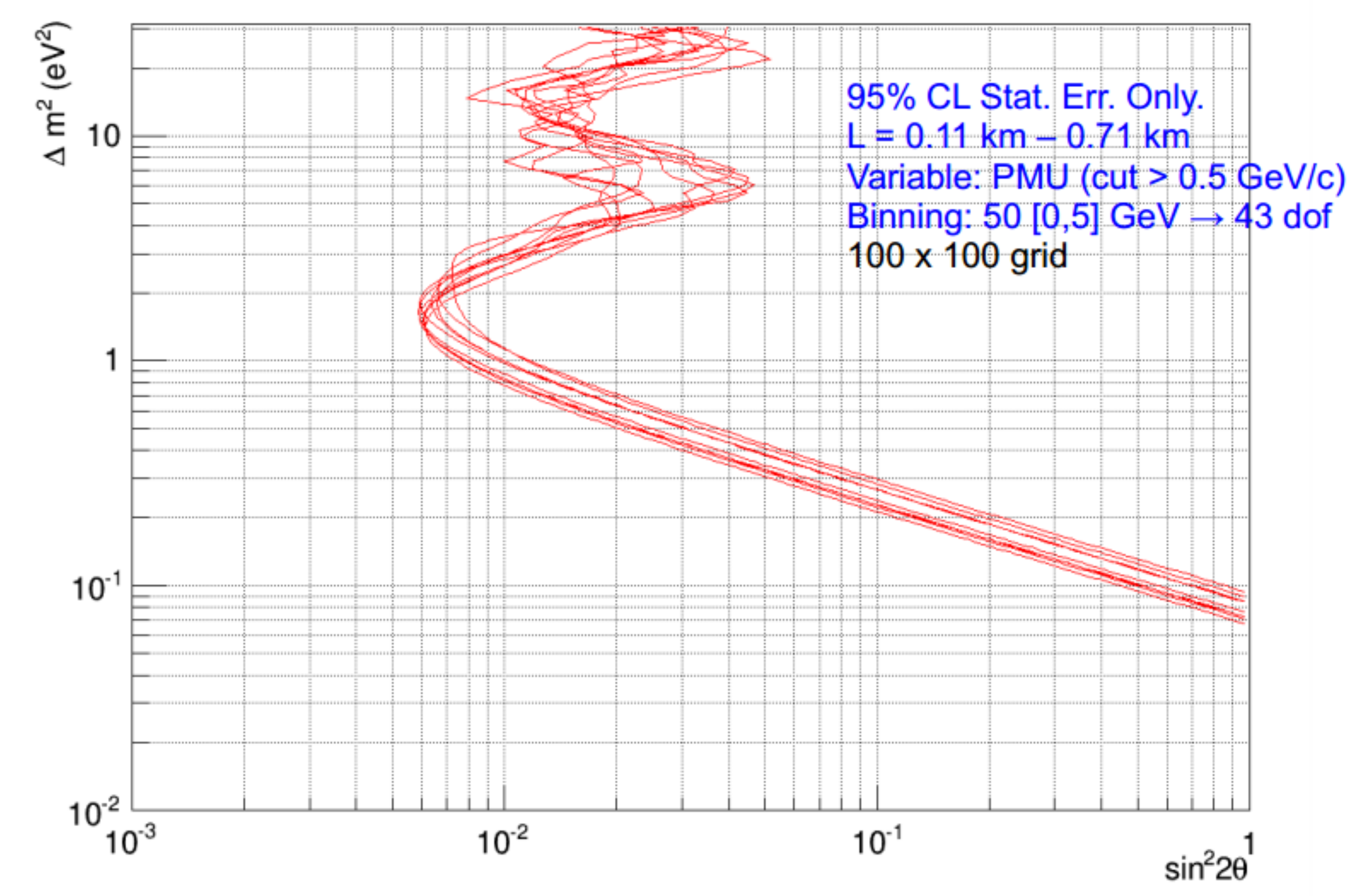}
\includegraphics[scale=0.41,type=pdf,ext=.pdf,read=.pdf]{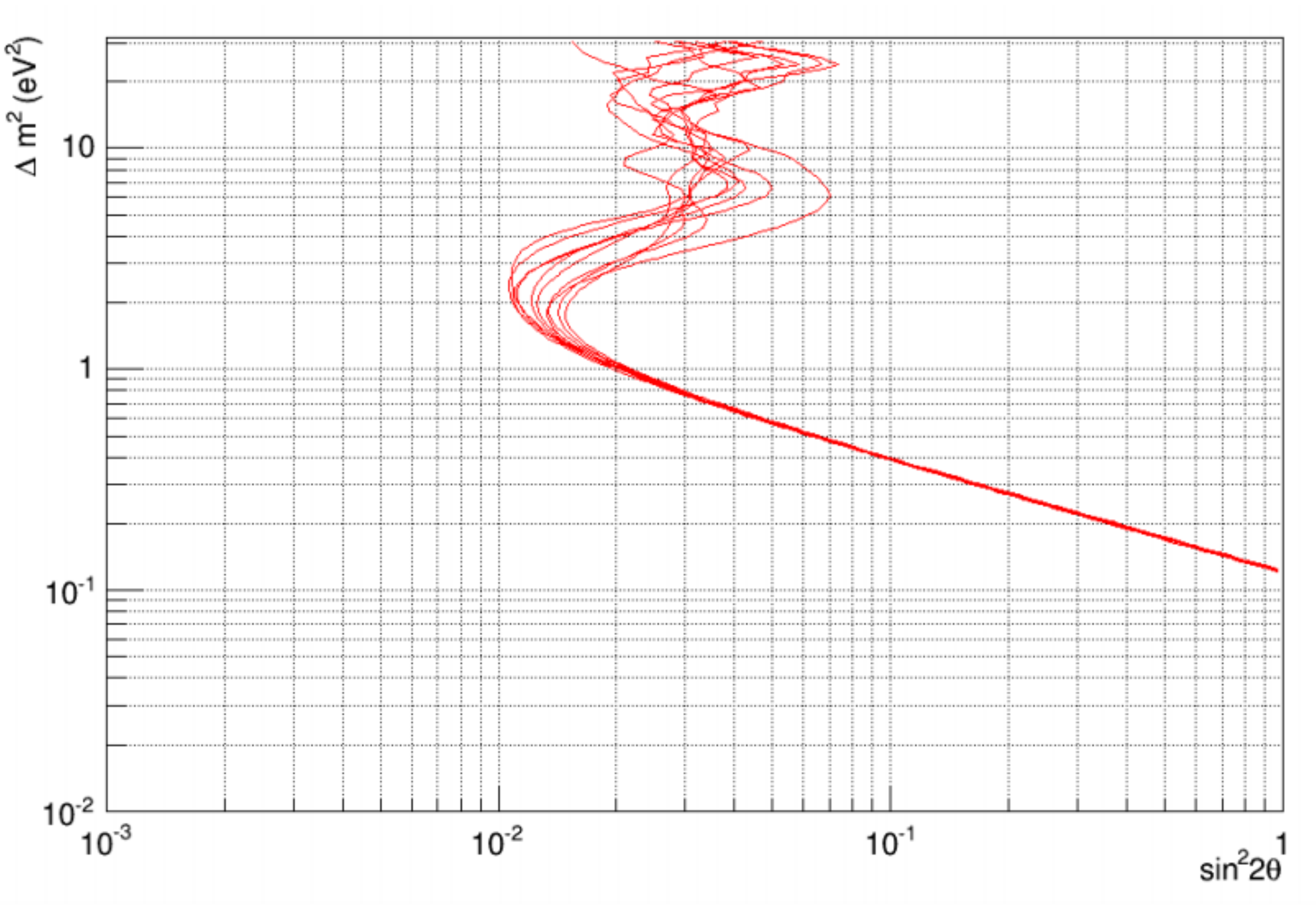}
\caption{Top. Sensitivity curves at 95\% C.L.  with ten simulated toy experiments assuming full statistics 
and no systematic uncertainties.  $p_\mu$ is used as observable above 500~MeV/c in the fit. Baselines are set at 110 and 710 m.
We use total of 10$^4$ sampling points uniformly distributed in log scale. Bottom. As above but using $\epsilon_{sys} = 0.01$.}
\label{fig:sens0}
\end{figure}

Results are shown in Fig.~\ref{fig:sens0} for a set of ten simulated null experiments.
The top plot assumes $\epsilon_{sys}=0$ while the bottom plots assumes $\epsilon_{sys}=0.01$.

In the Feldman\&Cousins approach a cut depending on $\sin^2(2\theta)$
and $\Delta m^2$ is applied in place of a fixed value (5.9915):
$\Delta\chi^2_{cut}(\sin^2(2\theta), \Delta m^2)$.  The critical value
has been determined as follows: for every $\sin^2(2\theta)$, $\Delta m^2$ 
sampling point, oscillated spectra have been generated and
fitted thus defining a $\chi^2_{min}$. The distribution of $\Delta
\chi^2(\sin^2(2\theta), \Delta m^2)$ is taken and $\Delta
\chi^2_{cut}(\sin^2(2\theta), \Delta m^2)$ is the value for which
obtaining a larger $\Delta \chi^2$ has only a 5\% probability.  We
have verified that the limits obtained with a variable critical value
for the $\chi^2$ provide limits which are very close to the ones
obtained using a fixed cut.

%%% Start Laura+Matteo %%%%%%%%%%%%%%%%%%%%%%%%%%%%%%%%%%%%%%%%%%%%%%%%%%%%%%%%%%%%%

\subsection{Full simulation and Matrix--Correlation}\label{sec:full-matrix}

\begin{figure}[htbp]
\centering
\subfloat[][\emph{}]
%{\includegraphics[width=.50\columnwidth]{sensitivity_new_new/smear-matrix-NPLANES_CC}} \quad
%{\includegraphics[width=.50\columnwidth]{sensitivity_new_new/smear-matrix-NPLANES_CCQEL}} \quad
%{\includegraphics[width=.50\columnwidth]{sensitivity_new_new/smear-matrix-NPLANES_NC}} 
%{\includegraphics[width=.50\columnwidth]{simulation/pictures-arc/smear-matrix-NPLANES_CC}} 
{\includegraphics[width=.50\columnwidth]{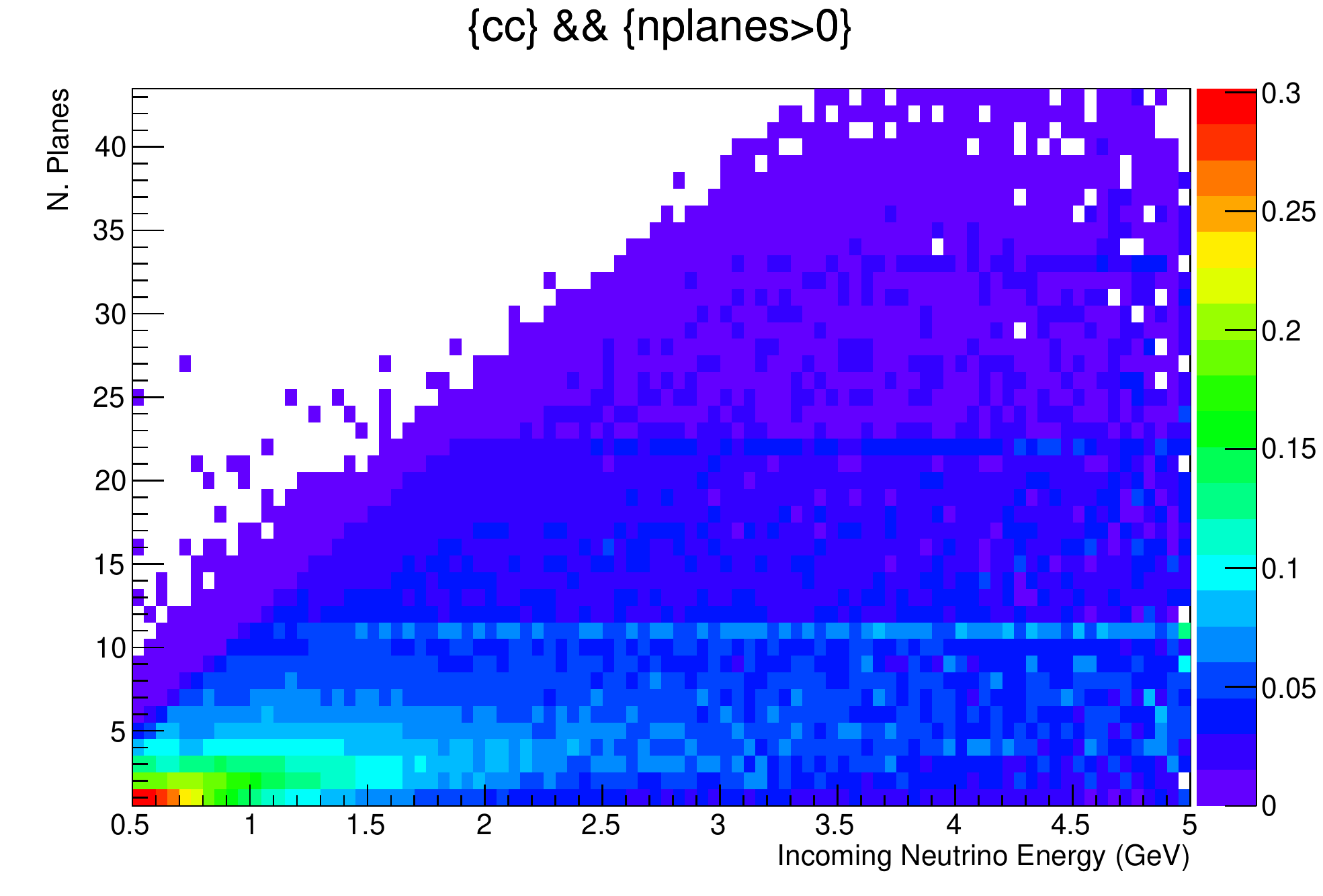}} 
\subfloat[][\emph{}]
%{\includegraphics[width=.50\columnwidth]{simulation/pictures-arc/smear-matrix-NPLANES_CCQEL}} \quad
{\includegraphics[width=.50\columnwidth]{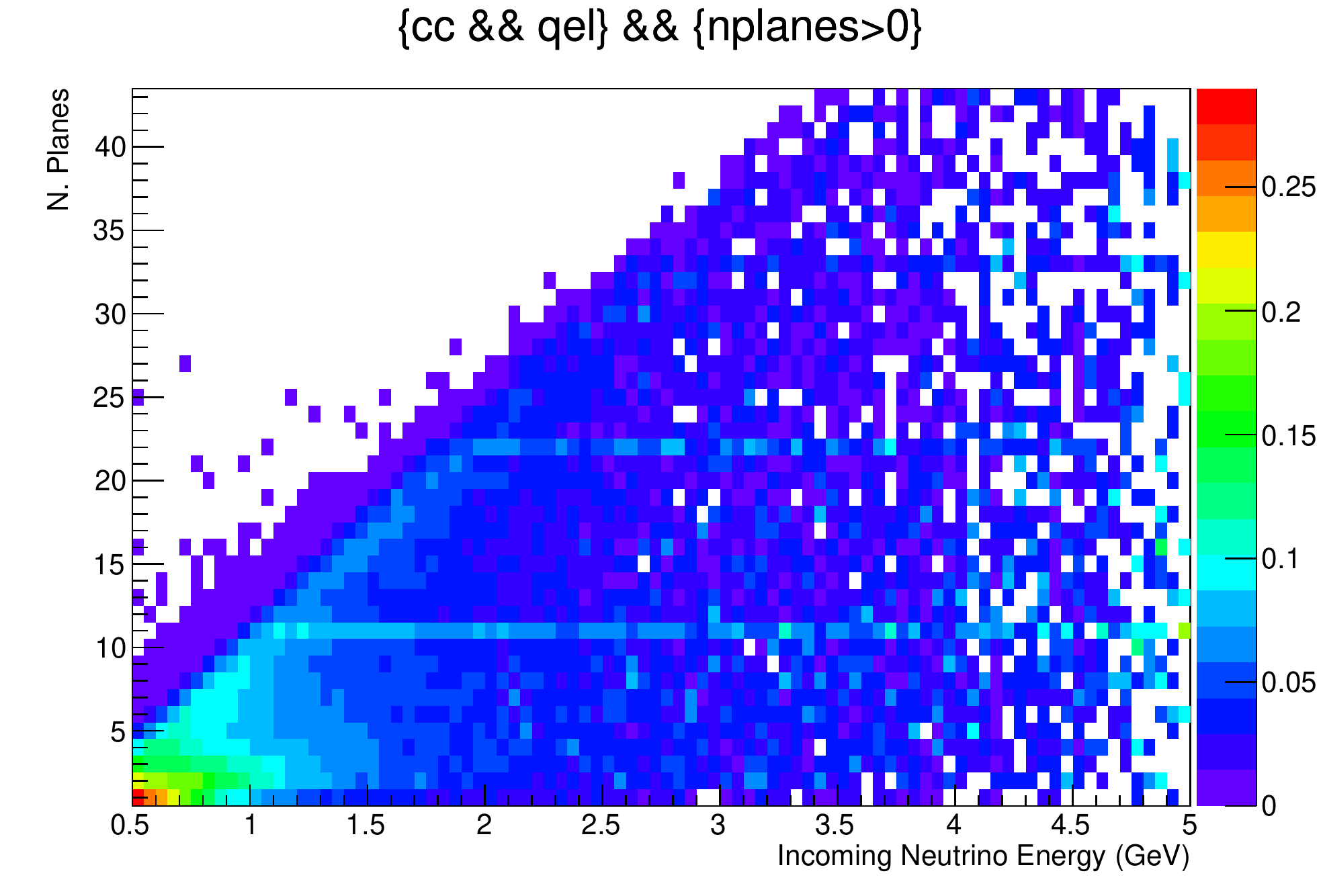}} \quad
\subfloat[][\emph{}]
%{\includegraphics[width=.50\columnwidth]{simulation/pictures-arc/smear-matrix-NPLANES_NC}} 
{\includegraphics[width=.50\columnwidth]{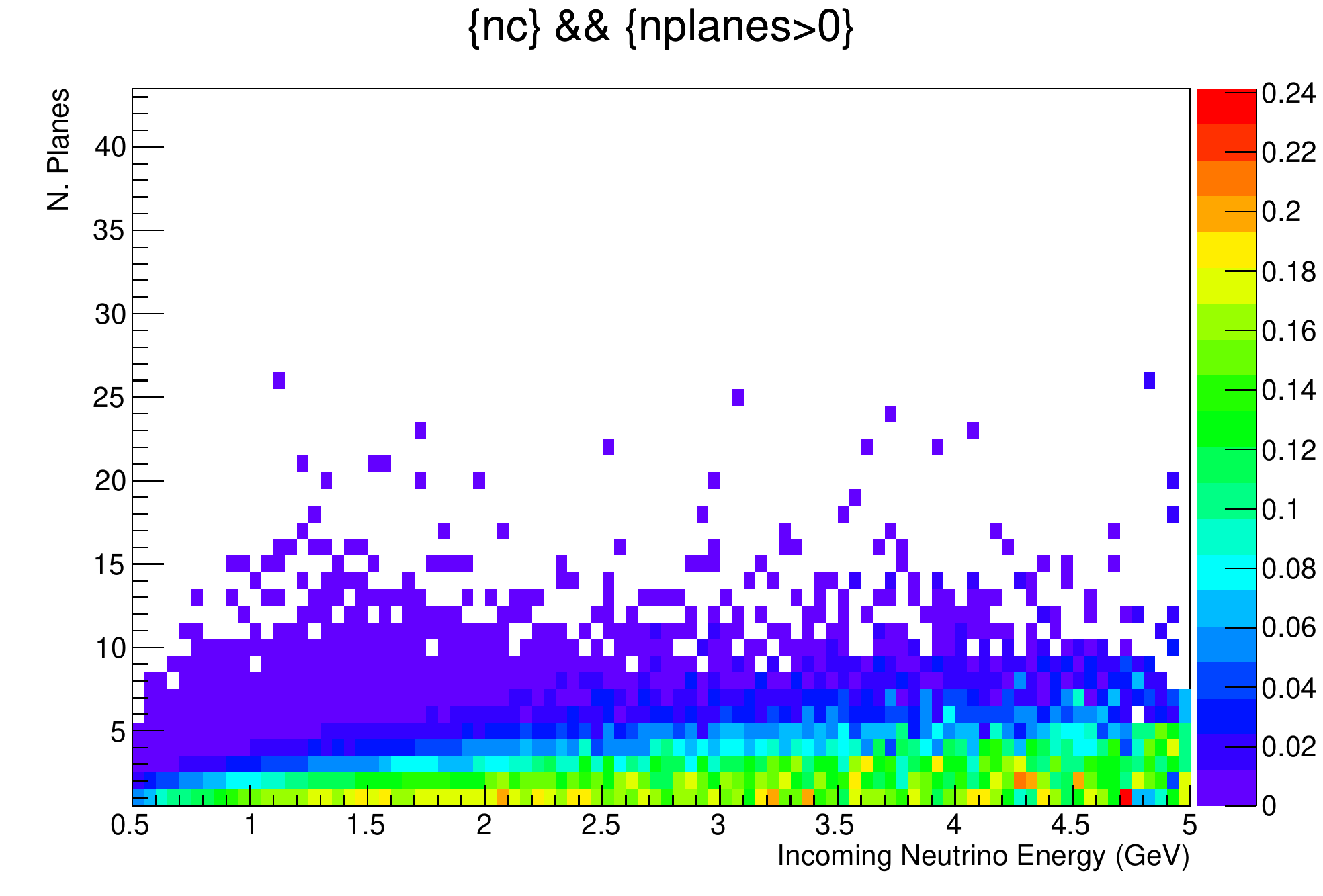}} 
\caption{ Smearing matrix of the reconstructed number of planes for CC (a), NC (b) and CCQE (c) events 
without cut on the number of planes.}
\label{fig:smear_plane}
\end{figure}

\begin{figure}[htbp]
\centering
\subfloat[][\emph{}]
%{\includegraphics[width=.50\columnwidth]{sensitivity_new_new/smear-matrix-RANGE_CC}} \quad
%{\includegraphics[width=.50\columnwidth]{sensitivity_new_new/smear-matrix-RANGE_CCQEL}} \quad
%{\includegraphics[width=.50\columnwidth]{sensitivity_new_new/smear-matrix-RANGE_NC}} 
%{\includegraphics[width=.50\columnwidth]{simulation/pictures-arc/smear-matrix-RANGE_CC}}
{\includegraphics[width=.50\columnwidth]{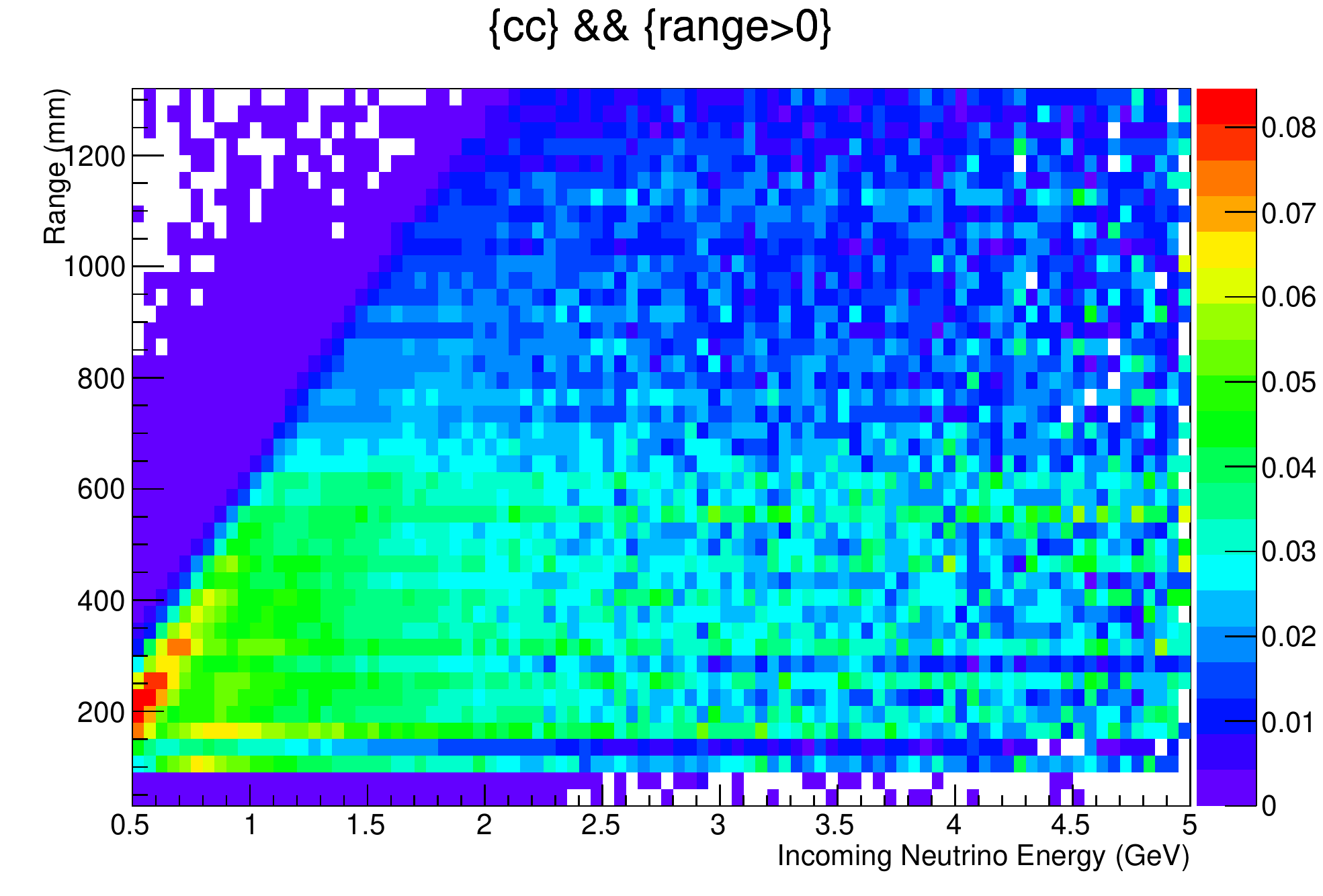}}
\subfloat[][\emph{}]
%{\includegraphics[width=.50\columnwidth]{simulation/pictures-arc/smear-matrix-RANGE_CCQEL}} \quad
{\includegraphics[width=.50\columnwidth]{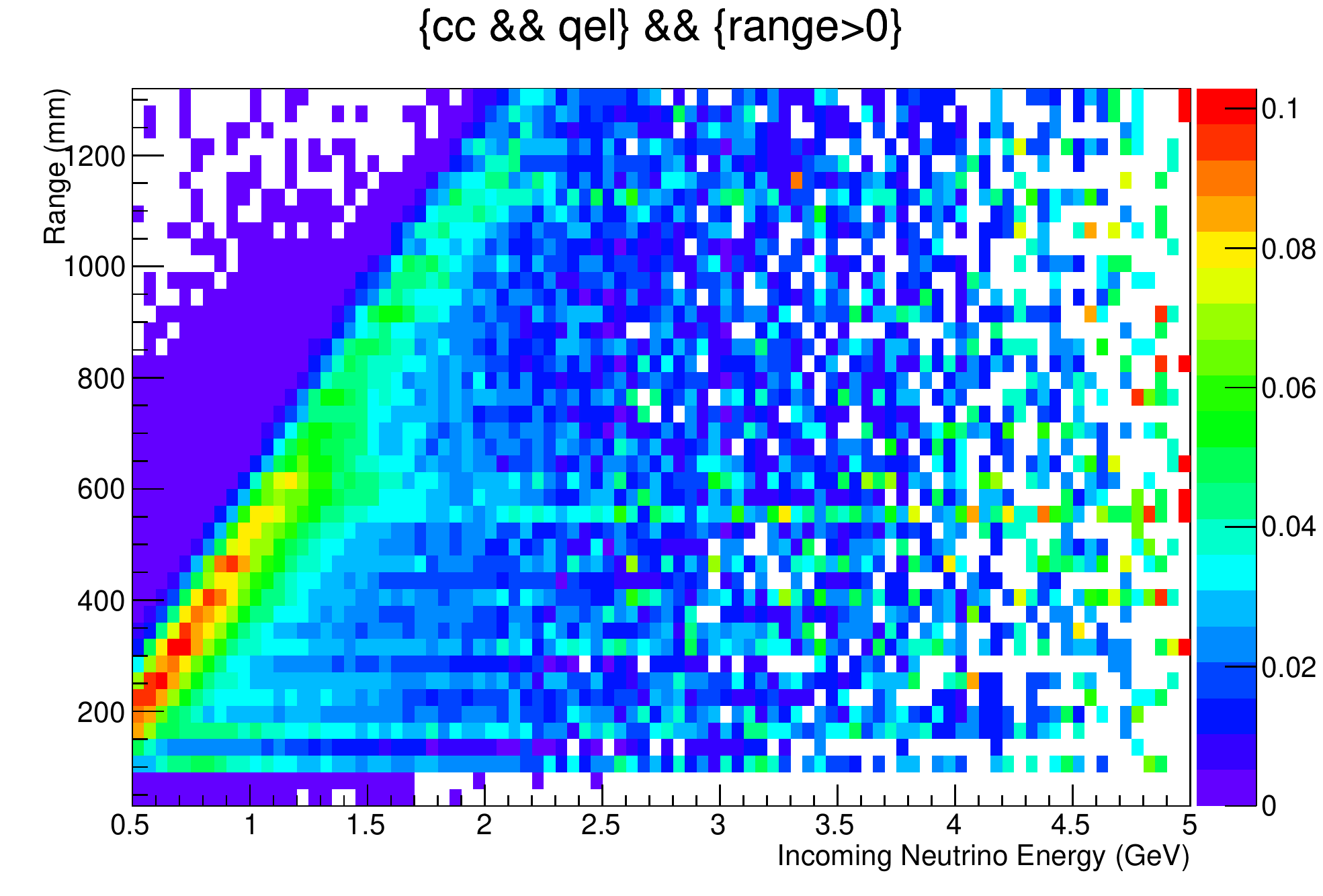}} \quad
\subfloat[][\emph{}]
%{\includegraphics[width=.50\columnwidth]{simulation/pictures-arc/smear-matrix-RANGE_NC}} 
{\includegraphics[width=.50\columnwidth]{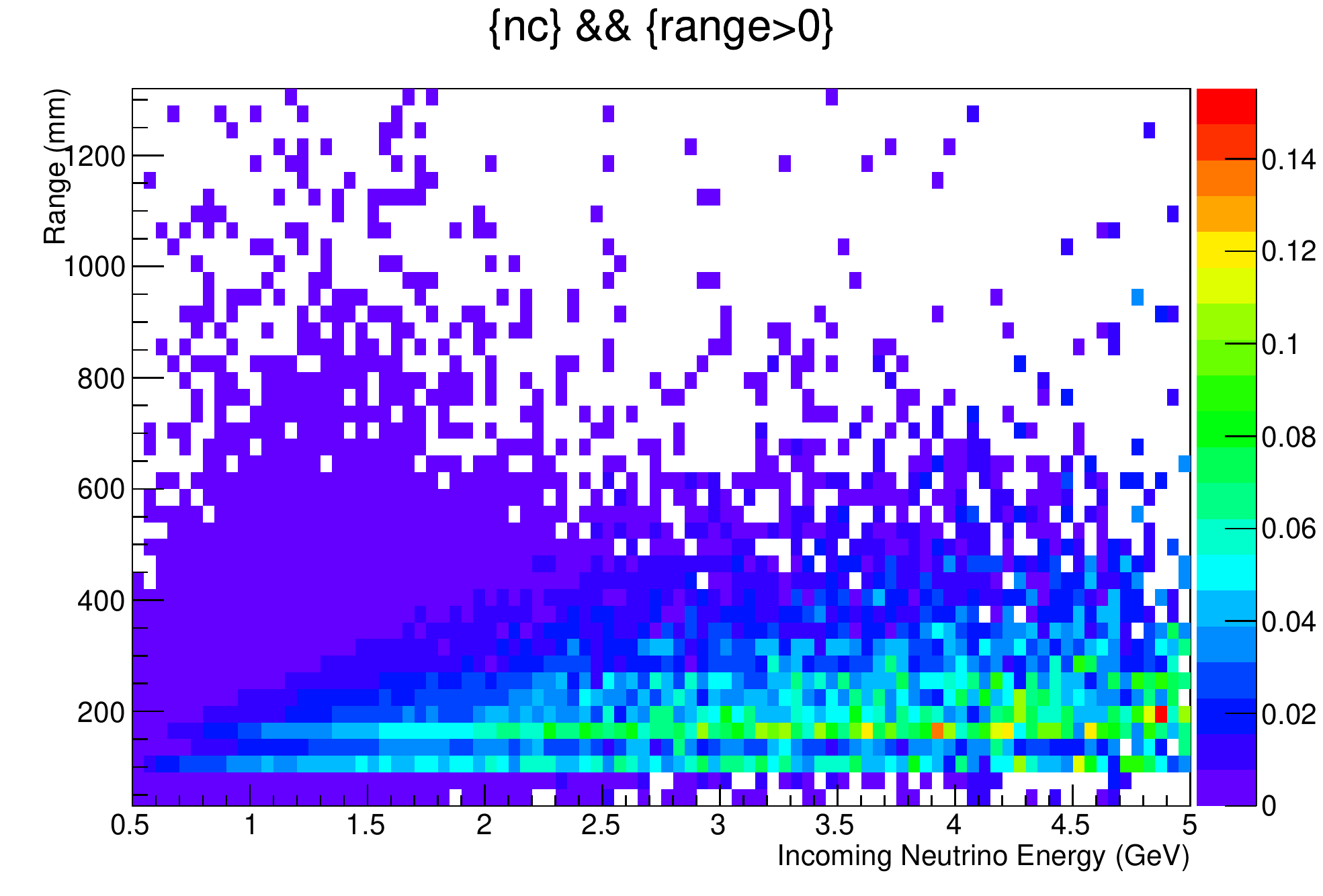}} 
\caption{ Smearing matrix of the reconstructed range for CC (a), NC (b) and CCQE (c) events without cut 
on the range.}
\label{fig:smear_range}
\end{figure}

In this analysis we implemented different smearing matrices for two different observables, the muon {\em range} and the {\em number of crossed planes},
associated with
the true incoming neutrino energy. These matrices were obtained through the Monte Carlo simulation described in Sect.~\ref{sec:MC} and reported here for convenience.
In Fig.~\ref{fig:smear_plane} are shown the smearing matrices for the observable number of planes for CC, CCQE 
and NC events, while in Fig.~\ref{fig:smear_range} are plotted the smearing matrices for the range estimator for CC, CCQE ad NC events. 

\subsubsection{Observables spectrum evaluation}
We studied the sensitivity to the \numu disappearance using two different observables: the range and the number of planes as shown in the previous section. 
We divide range spectrum in 43 uniform bins between 30 and 1320 mm, number of planes spectrum in 43 bins between 1 and 43, while for neutrino energy spectrum 
we choose a 90 equidistant sampling from 0.5 to 5 GeV/c. 

The range and the number of planes spectrum at Near and Far detector were evaluated using GLoBES~\cite{GLoBES}.
%For our analysis we divide muon momentum spectrum in 45 uniform bins between 0.5 GeV/c and 5 GeV/c, while for neutrino energy spectrum we choose a 90 equidistant sampling from 0.5 to 5 GeV/c.\
%The muon momentum spectrum at near and far detector was simulated using GLoBES \cite{GLoBES}. 
If we denote with $E_{i}^{Far}$ and $E_{i}^{Near}$ the number of expected events in $i-th$ bin for the Far and Near detector, respectively, then we can define $N_{i}^{Far}$ and $B_{i}^{Far}$ as:
\begin{equation}
N_{i}^{Far} = E_{i}^{Far}
\end{equation}
\begin{equation}
B_{i}^{Far} = E_{i}^{Near}\left( \frac{m_{Far}}{m_{Near}}\right) \left( \frac{L_{Near}}{L_{Far}} \right)^{2}
\end{equation}
where $m_{Far}$ and $m_{Near}$ denote the mass of Far and Near detector, respectively, and $L_{Near}$ and $L_{Far}$ are the distances for the Near and Far detector from the neutrino source. $B_{i}^{Far}$ represents the number of expected events at Far detector by scaling the number of expected events at the Near detector for the ratio of the masses and the squared ratio of their baseline. $B_{i}^{Far}$ is the number of expected events at Far site without oscillations, while $N_{i}^{Far}$ is the number of expected events at Far detector which depends on the oscillation probability. 

\subsubsection{Results}
\begin{figure}[htbp]
\centering
%\subfloat[][\emph{}]
%{\includegraphics[scale=0.65]{sensitivity_new_new/figure/Sensitivity-range-all}}  \quad
{\includegraphics[scale=0.36]{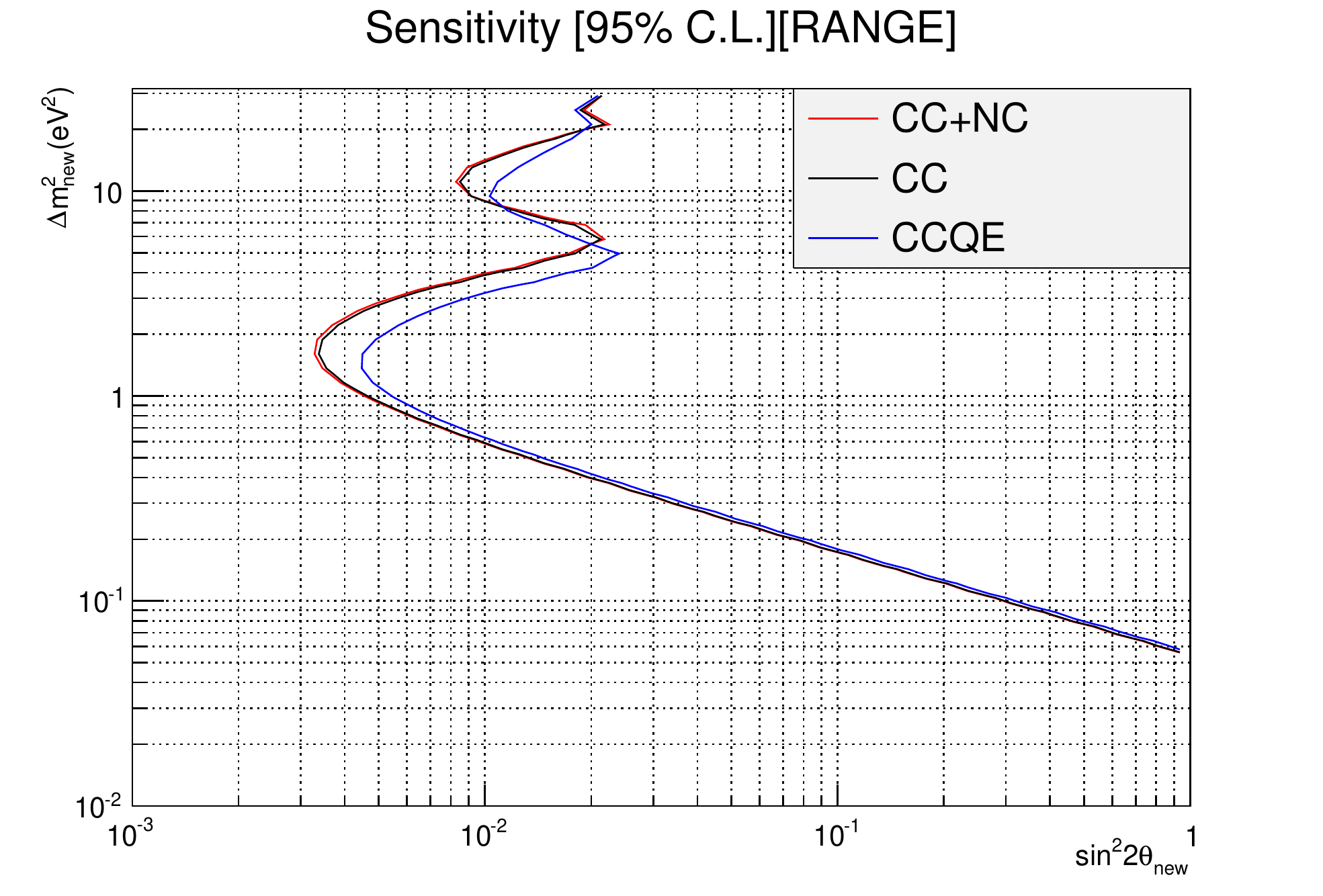}}  \quad
%\subfloat[][\emph{}]
%{\includegraphics[scale=0.65]{sensitivity_new_new/figure/Sensitivity-nplanes-all}} 
{\includegraphics[scale=0.36]{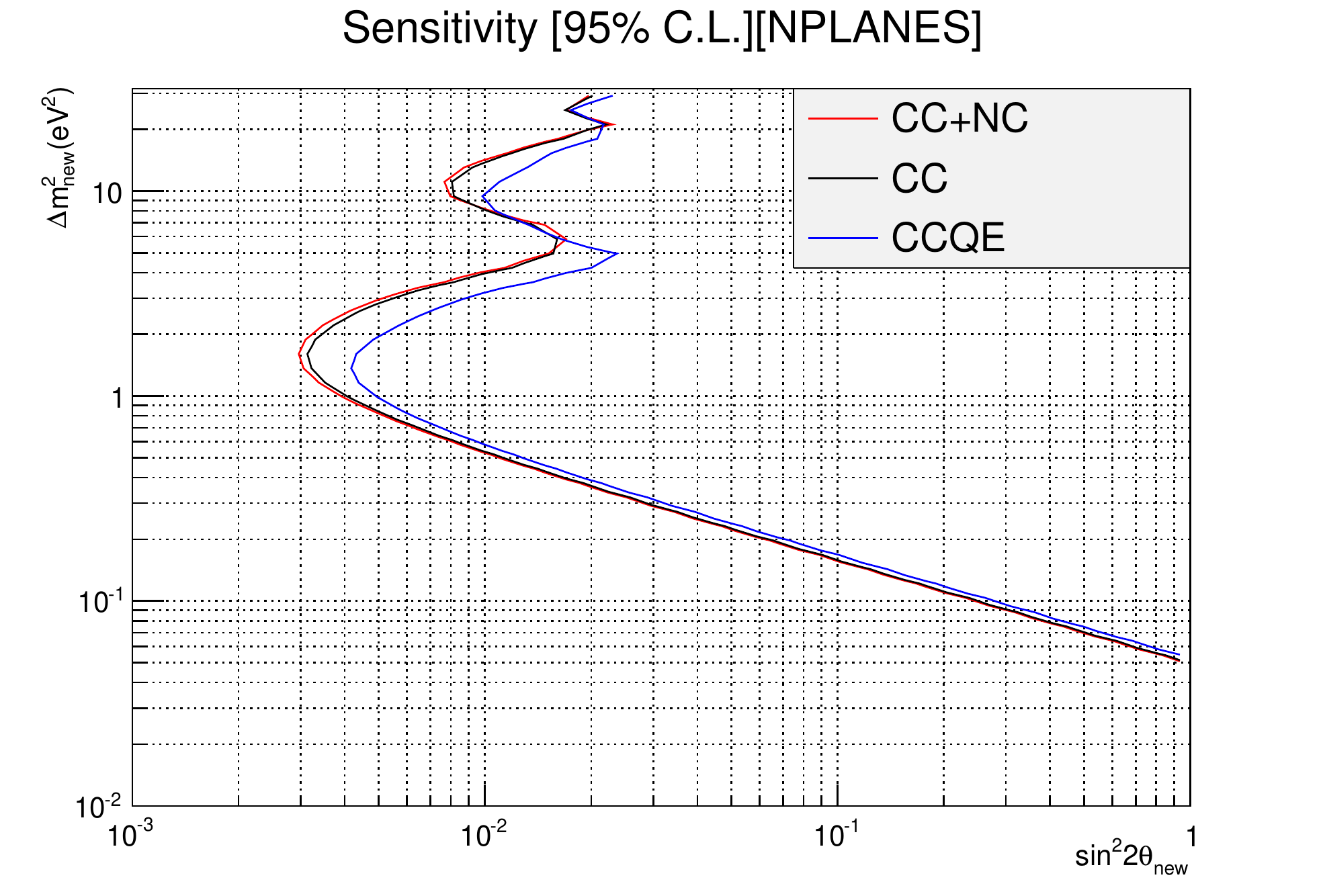}} 
\caption{ 95\% CL sensitivity obtained using range (top) and number of planes (bottom) for all (QE, Res, DIS) CC (black) and CC+NC (red) events and for only CCQE events (blue). In this case we considered no systematic errors and no cuts on the observables. }
\label{fig:sensitivity_plot_all}
\end{figure}

To determine the exclusion region in the oscillation parameter plane $\Delta m^{2}_{new}-\sin^{2}(2\theta_{new})$ we 
evaluate $B_{i}^{Far}$ and $N_{i}^{Far}$ for each value of $\Delta m^{2}_{new}$ and $\sin^{2}(2\theta_{new})$ and we calculate 
the $\chi^{2}$:
\begin{equation}
\chi^{2} = \sum_{i=0}^{N-1}\sum_{j=0}^{N-1}(B_{i}^{Far} - N_{i}^{Far})(M^{-1})_{ij}(B_{j}^{Far} - N_{j}^{Far})
\end{equation}
where $N$ is the number of bins and $M$ is the covariance matrix~\cite{covM} which take into account the uncertainties and their bin--to--bin correlations~\cite{covM2}. The covariance matrix is constructed as:
\begin{equation}\label{eq:smearing}
M = M^{stat}+M^{norm}+M^{shape}
\end{equation}
where $M^{stat}$ is the statistical errors matrix, $M^{norm}$ is the normalization errors matrix and $M^{shape}$ is the shape errors matrix. Statistical errors are added to the diagonals terms of the covariance matrix and are evaluated as:
\begin{equation}
M_{ij}^{stat} = B_{i}^{Far}\delta_{ij}
\end{equation}
It is possible to observe $\nu_{\mu}$ disappearance either from a deficit of events (normalization) or, alternatively, from a distortion of the observable spectrum (shape) which are affected by systematic uncertainties expressed by the normalization errors matrix and the shape errors matrix. The normalization errors matrix is the component of error matrix which is the same for each element. This term is associated with the 
normalization uncertainty applied to each bin as:
\begin{equation}
M_{ij}^{norm}=\epsilon_{norm}^{2}B_{i}^{Far}B_{j}^{Far}
\end{equation}
where $\epsilon_{norm}$ is the normalization error. The shape errors matrix represents a migration of events across the bins. In this case the uncertainties are associated with changes where the total number of event remains unchanged and so a depletion of events in the some region of the spectrum should be compensated by an enhancement in others. In our model we choose a shape error matrix that satisfies the following constrains:
\begin{equation}
\left( N_{i}^{Far}- B_{i}^{Far}\right)\propto i 
\end{equation}
\begin{equation}
\sum_{i=0}^{N-1}\left( N_{i}^{Far}- B_{i}^{Far}\right)=0 
\end{equation}
\begin{equation}
\sum_{i=0}^{N/2}\left( N_{i}^{Far}- B_{i}^{Far}\right) = I\epsilon_{shape} 
\end{equation}
In this case the shape errors matrix elements are:
\begin{equation}
M_{ij}^{shape}=\left[ \left( \frac{N-1}{2}\right) -i\right] \left[ \left( \frac{N-1}{2}\right) -j\right]\left(\frac{8\epsilon_{shape}I}{N^{2}} \right)^{2} 
\end{equation}
where $I = \sum_{i=1}^{N}B_{i}^{Far}$ and $\epsilon_{shape}$ is the shape error.
We use frequentist methods to study the $\chi^{2}$ statistic distribution in order to calculate the sensitivity for oscillation parameters. In Fig.~\ref{fig:sensitivity_plot_all} (top) are shown sensitivity plots obtained using the range as observable without cuts, while in Fig. \ref{fig:sensitivity_plot_all} (bottom) are presented sensitivity plots obtained using the number of planes without cuts on the observable. From these plots it can be seen that sensitivity computed considering CC and NC events is almost the same as the sensitivity obtained with only CC events and therefore NC background events don't affect the sensitivity. 

\begin{figure}[htbp]
\centering
%\subfloat[][\emph{}]
%{\includegraphics[scale=0.65]{sensitivity_new_new/figure/Sensitivity-range-CC-NC-cuts}} \quad
{\includegraphics[scale=0.36]{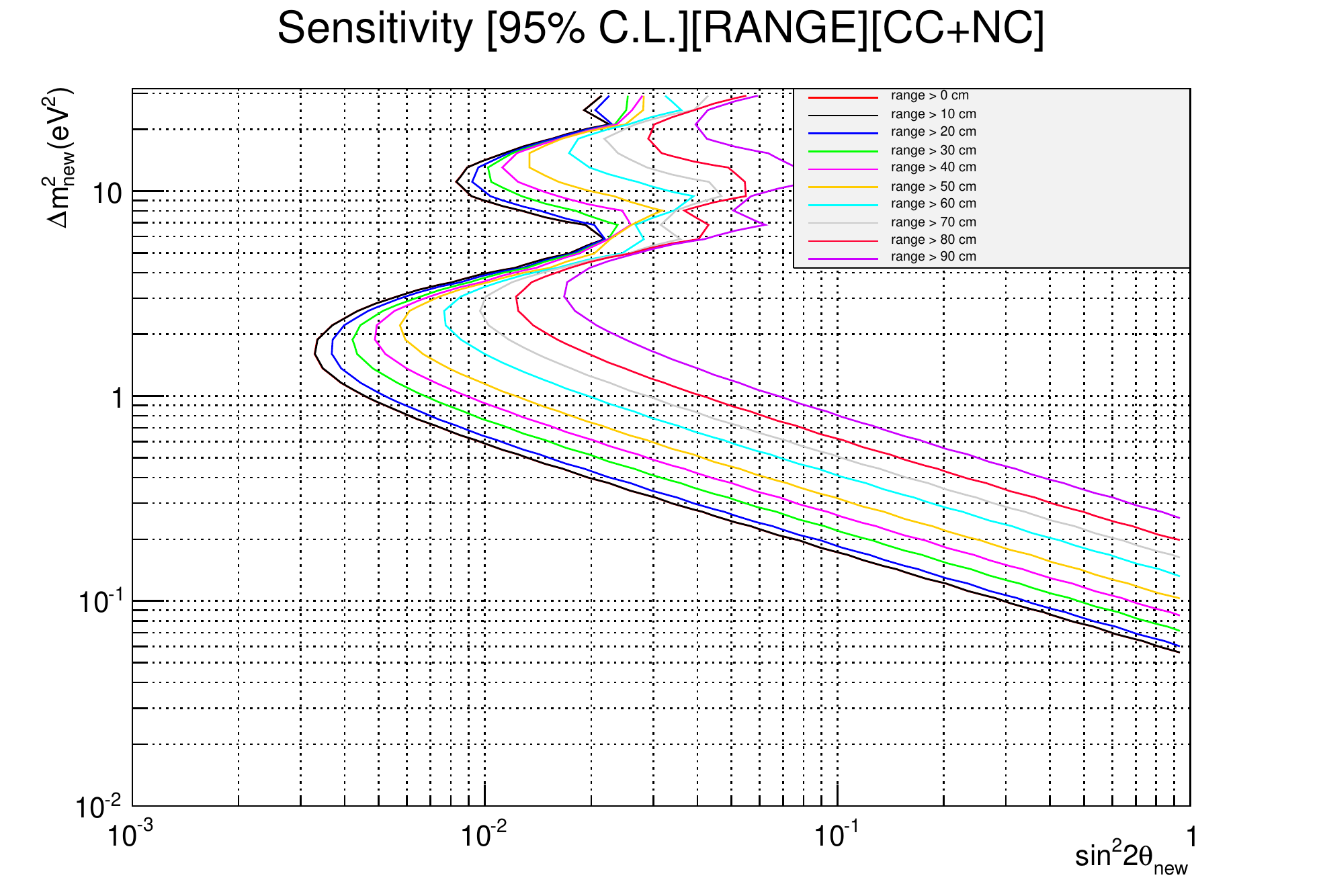}} \quad
%\subfloat[][\emph{}]
%{\includegraphics[scale=0.65]{sensitivity_new_new/figure/Sensitivity-nplanes-CC-NC-cuts}} 
{\includegraphics[scale=0.36]{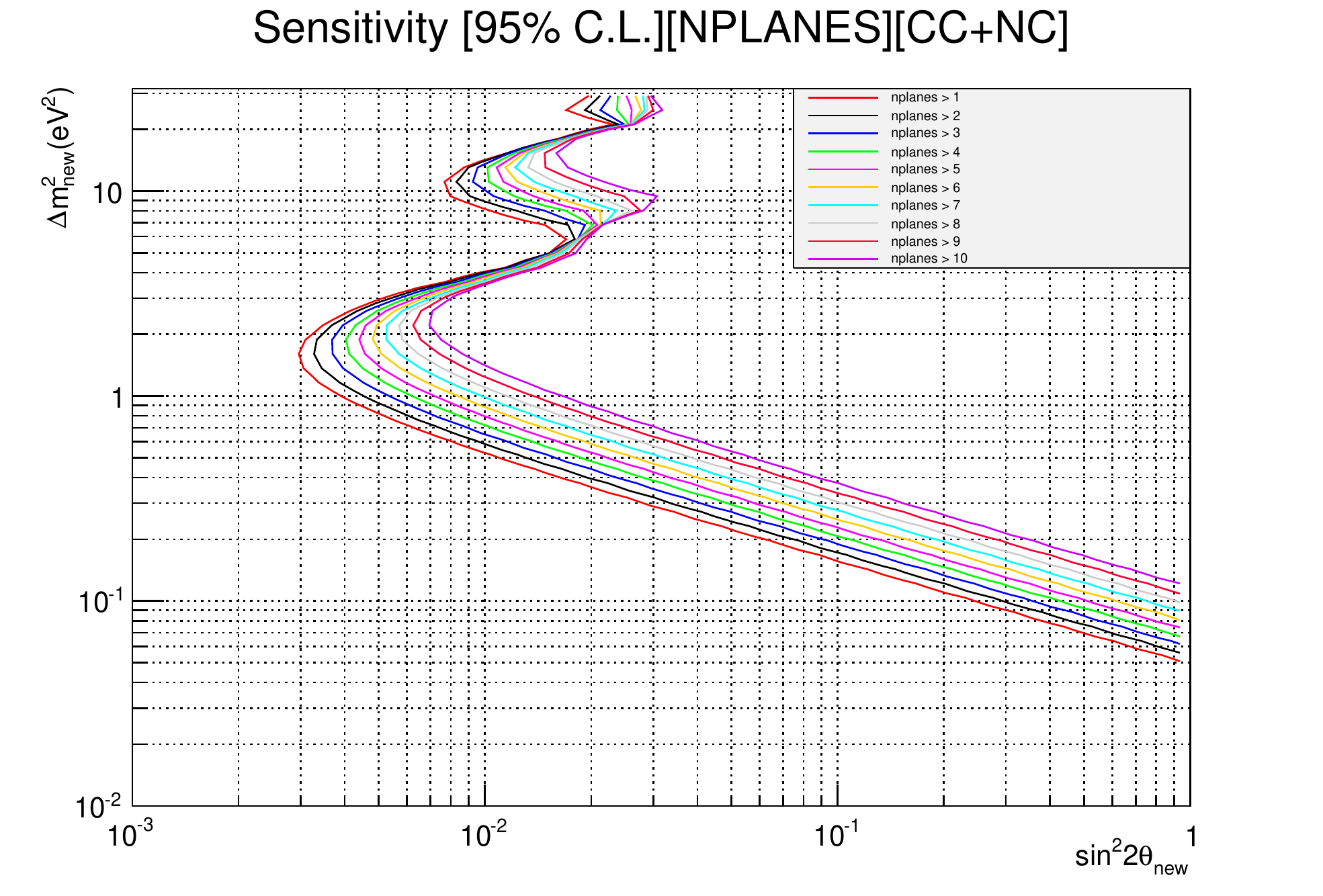}} 
\caption{ 95\% CL sensitivity for different range cuts (top) and for different cuts in the number of planes (bottom) for CC+NC events. The sensitivity was obtained considering no systematic errors.}
\label{fig:sensitivity_plot_cut}
\end{figure}

\begin{figure}[htbp]
\centering
%\subfloat[][\emph{}]
%{\includegraphics[scale=0.65]{sensitivity_new_new/figure/Sensitivity-range-all-norm-1}} \quad
{\includegraphics[scale=0.36]{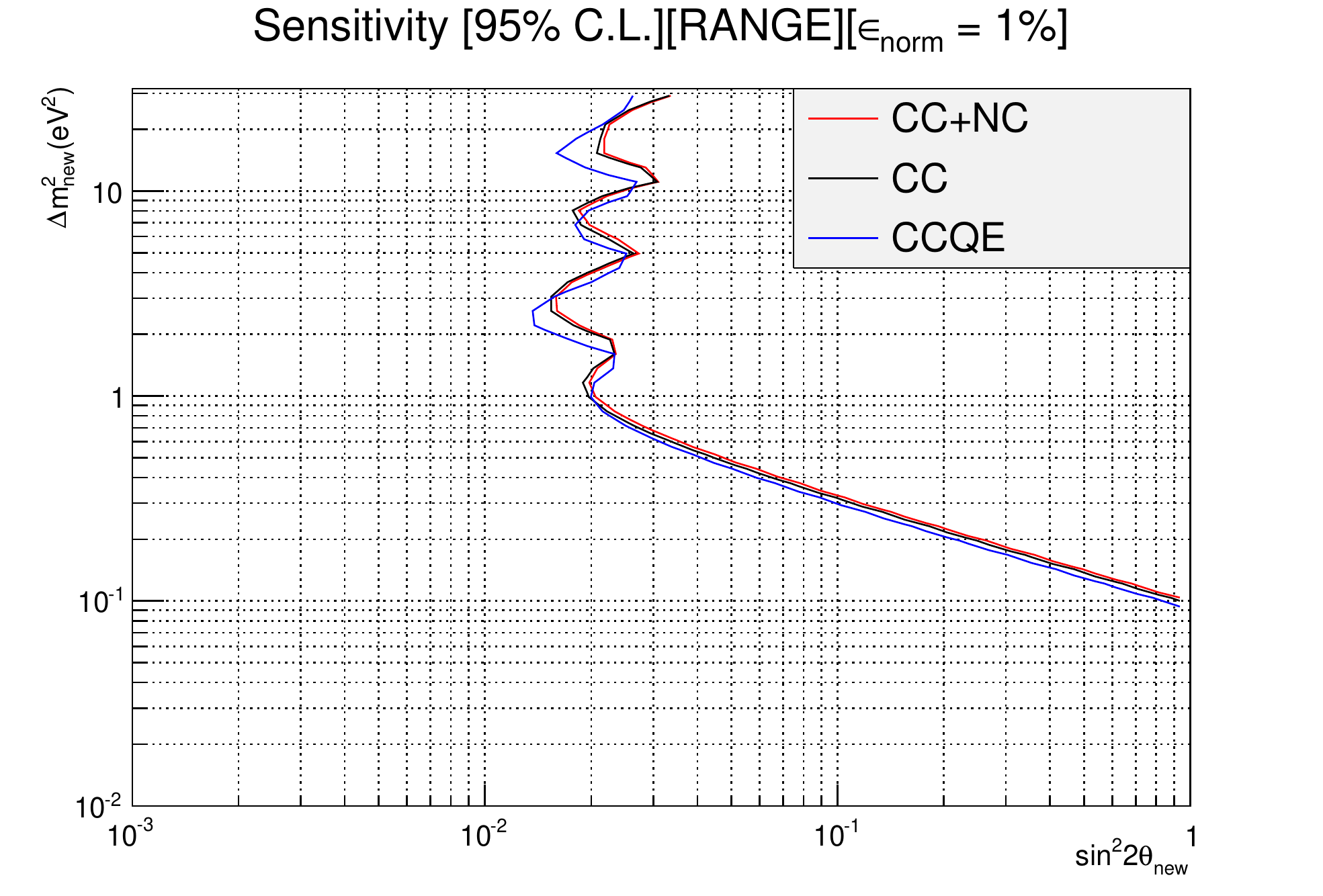}} \quad
%\subfloat[][\emph{}]
%{\includegraphics[scale=0.65]{sensitivity_new_new/figure/Sensitivity-nplanes-all-norm-1}}
{\includegraphics[scale=0.36]{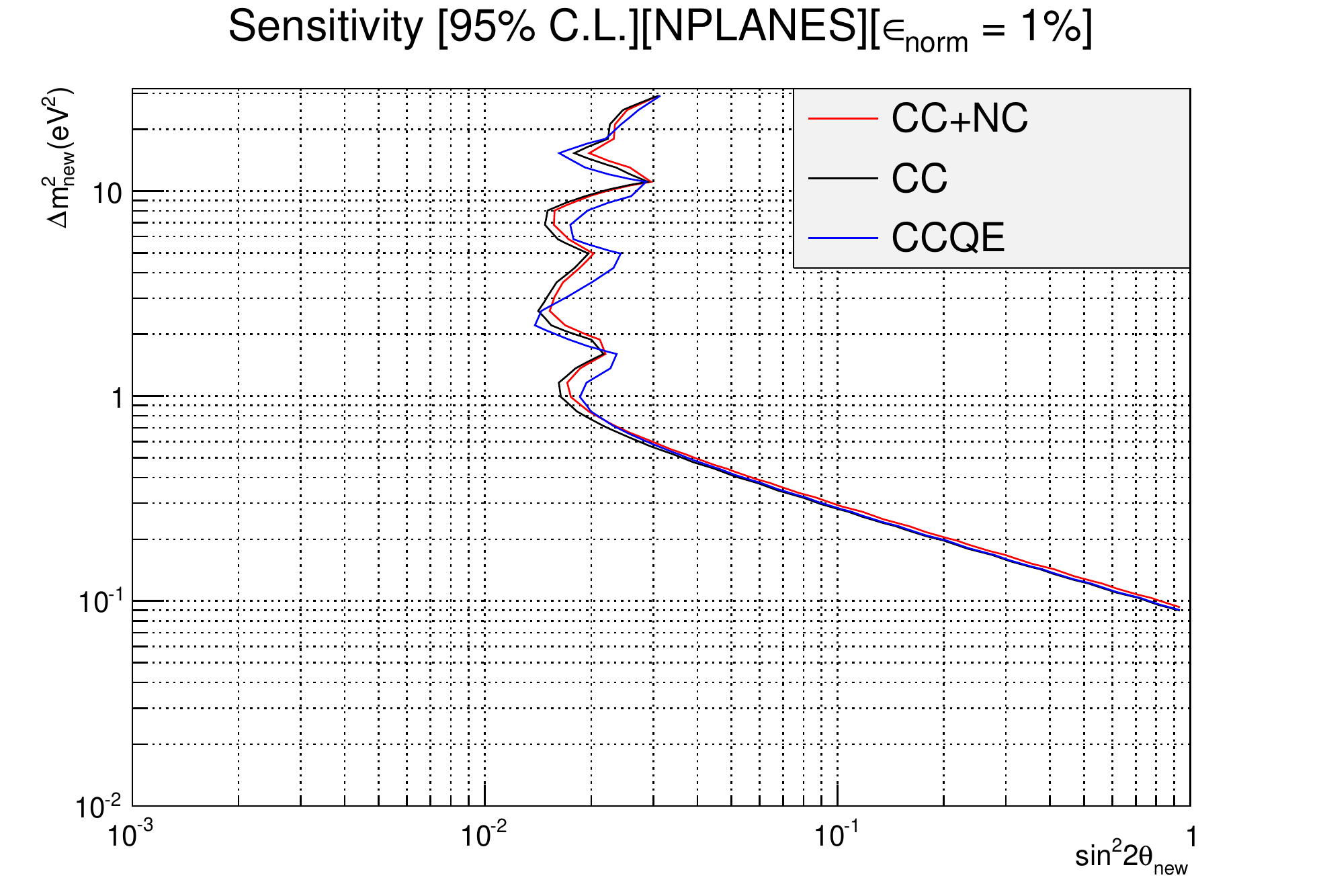}}
\caption{ 95\% CL sensitivity obtained using range (top) and number of planes (bottom) for all  (QE, Res, DIS) CC (black) and CC+NC (red) events and for only CCQE events (blue). In this case we considered 1\% bin-to-bin correlated error in the normalization.}
\label{fig:sensitivity_plot_all_corrNorm}
\end{figure}

\begin{figure}[htbp]
\centering
%\subfloat[][\emph{}]
%{\includegraphics[scale=0.65]{sensitivity_new_new/figure/Sensitivity-range-CC-NC-norm-errors-inset}}
{\includegraphics[scale=0.65]{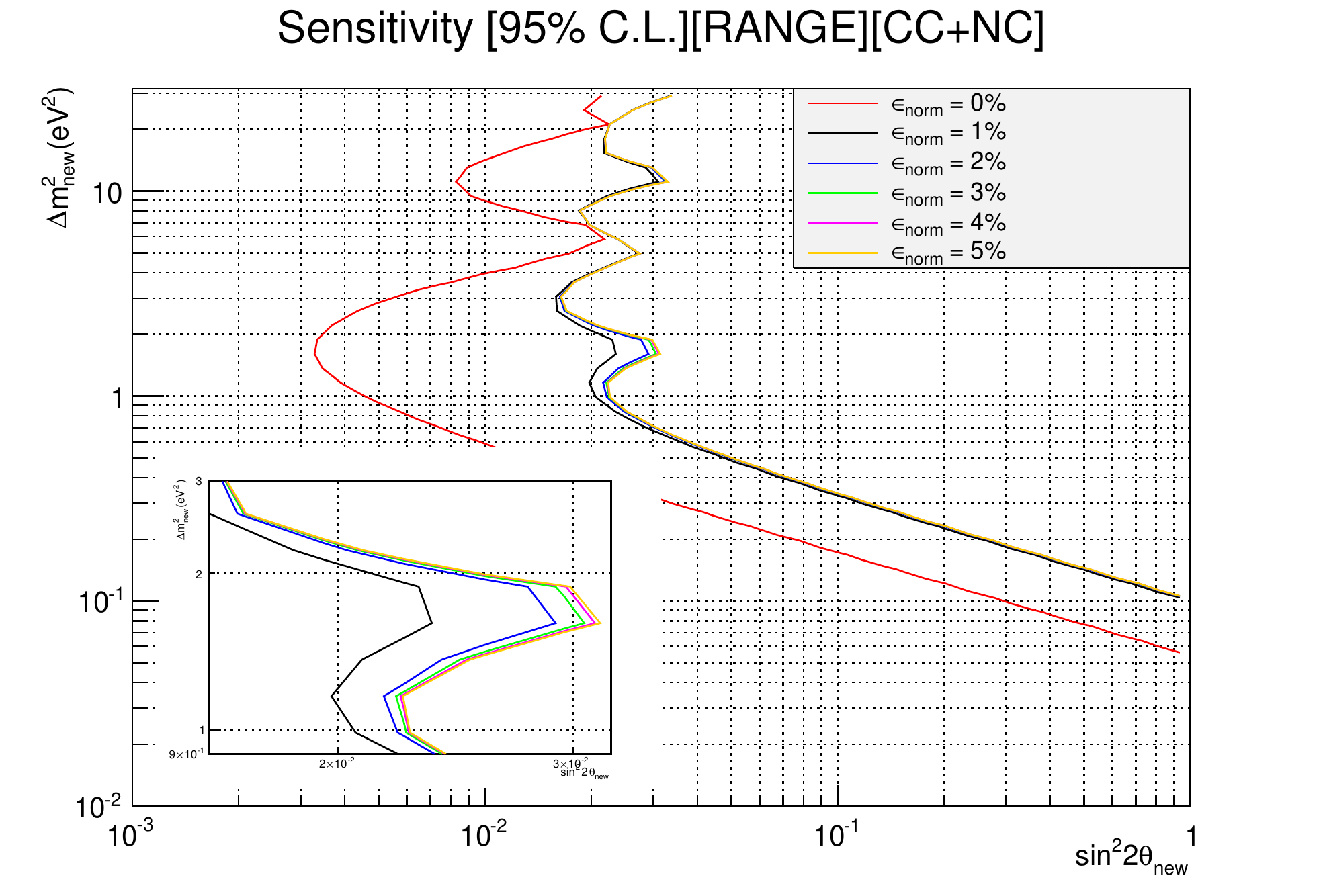}}
%\subfloat[][\emph{}]
%{\includegraphics[scale=0.65]{sensitivity_new_new/figure/Sensitivity-nplanes-CC-NC-norm-errors-inset}}
{\includegraphics[scale=0.65]{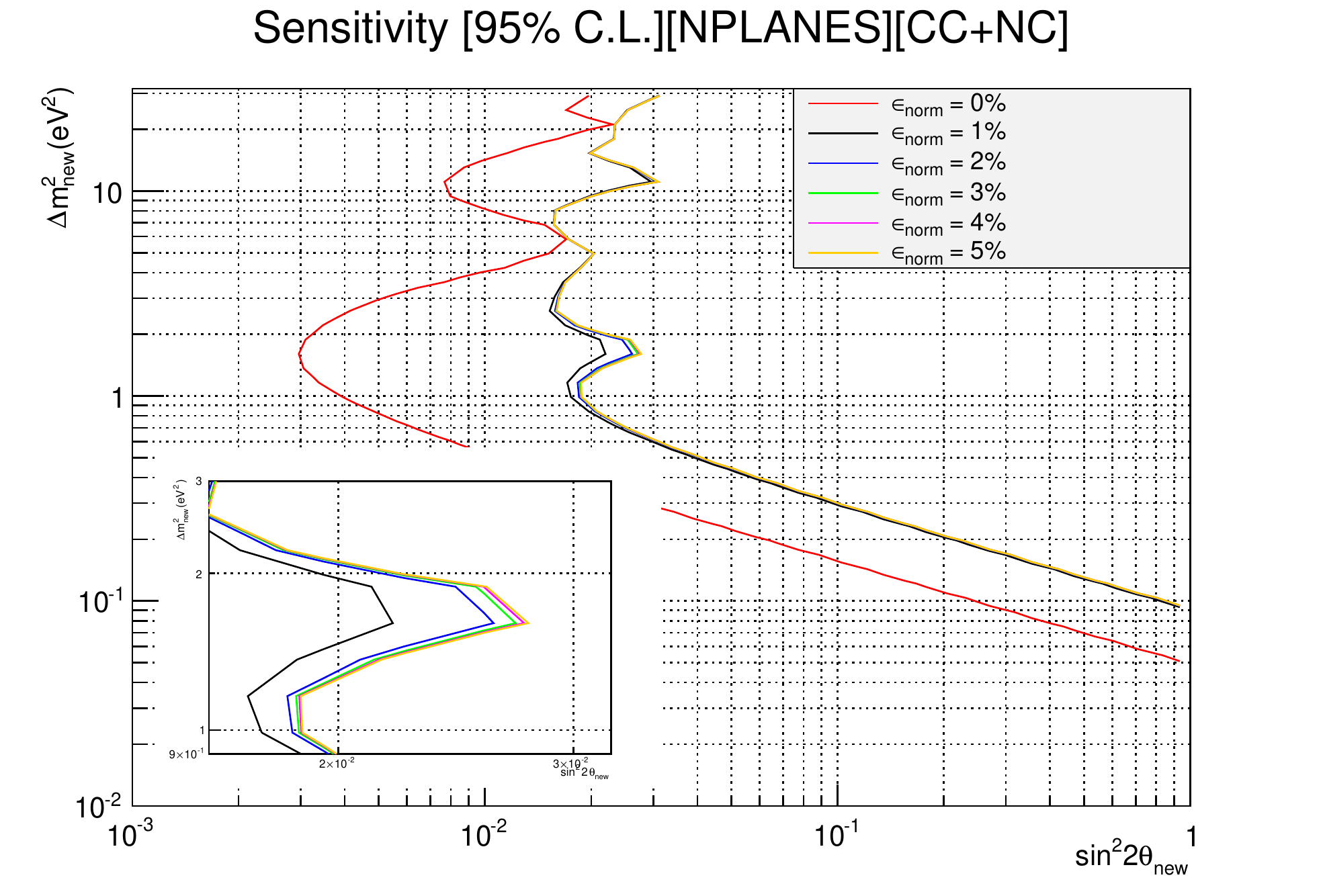}}
\caption{With the same condition of the previous figure (Fig.~\ref{fig:sensitivity_plot_all_corrNorm}:
 95\% CL sensitivity obtained using range (top) and number of planes (bottom) for all  CC+NC events) the
 systematic bin--to--bin correlated error on the normalization has been increased up to 5\%. The 5\% curve 
 actually stays below the 4\% one. In the insets a zoom of the region around 1 eV$^2$ and 0.02 in 
 $\sin^2(2\theta)$is given. }
\label{fig:sensitivity_norma}
\end{figure}

\begin{figure}[htbp]
\centering
%\subfloat[][\emph{}]
%{\includegraphics[scale=0.65]{sensitivity_new_new/figure/Sensitivity-range-all-shape-1}} \quad
{\includegraphics[scale=0.36]{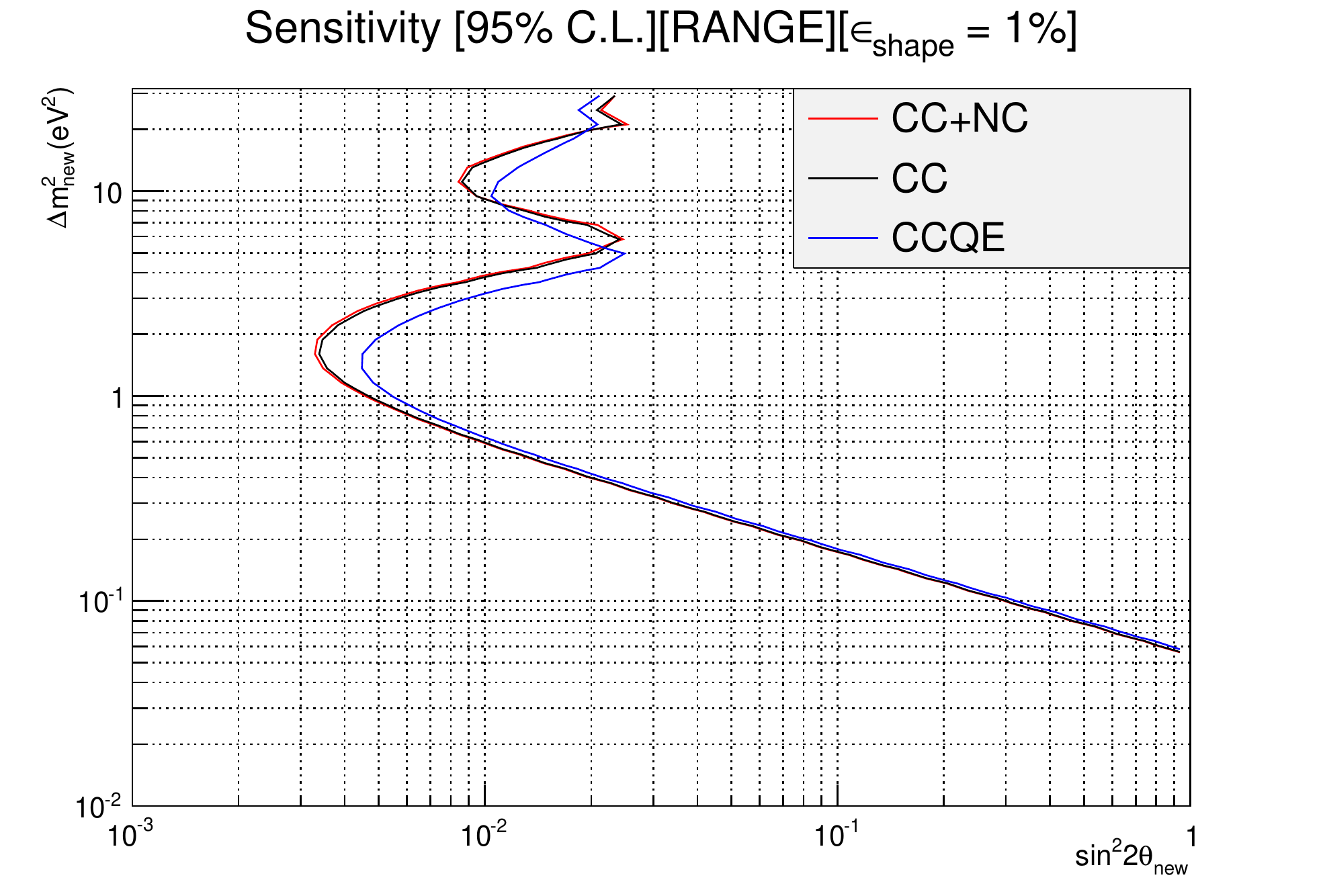}} \quad
%\subfloat[][\emph{}]
%{\includegraphics[scale=0.65]{sensitivity_new_new/figure/Sensitivity-nplanes-all-shape-1}} 
{\includegraphics[scale=0.36]{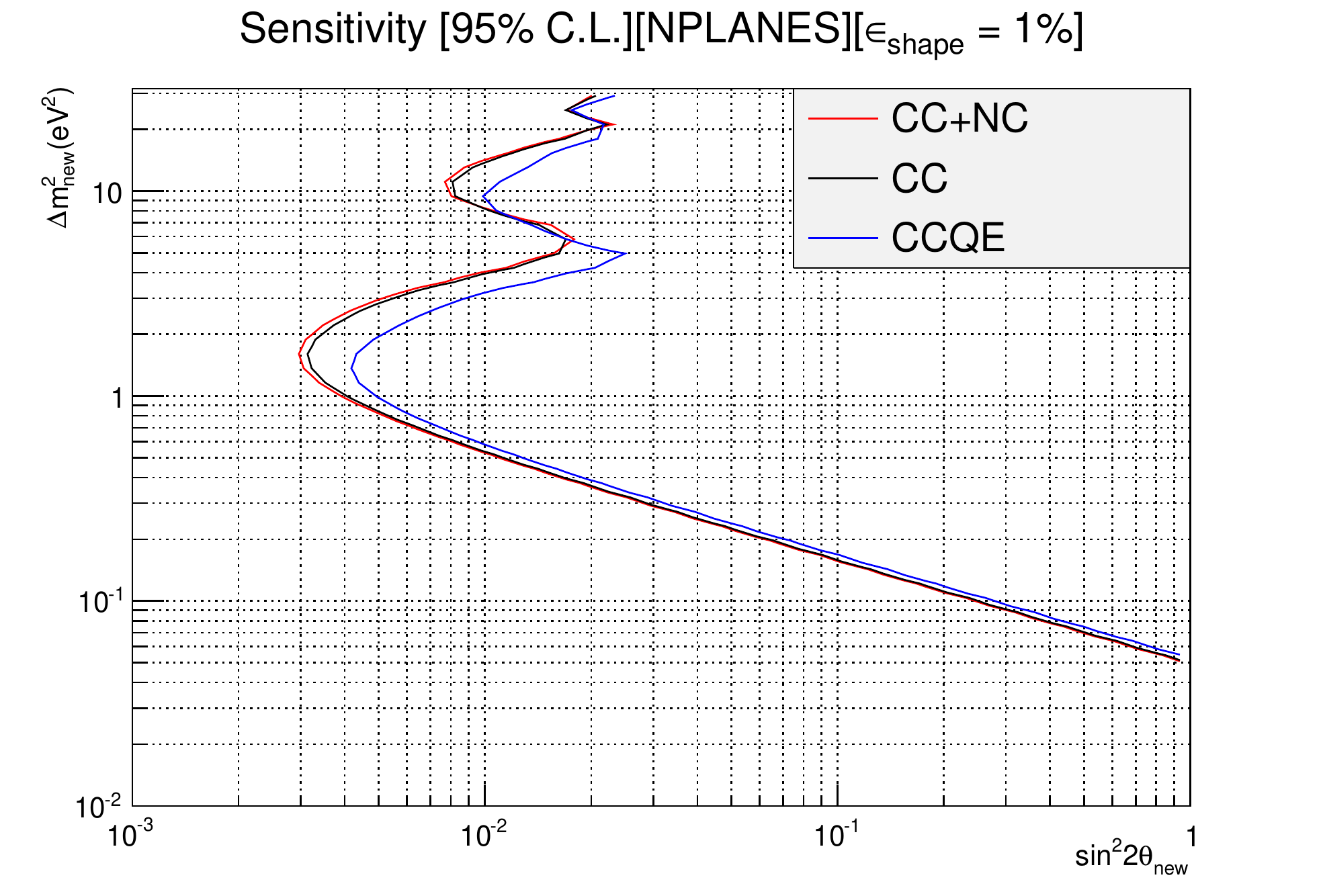}} 
\caption{ 95\% CL sensitivity obtained using range (top) and number of planes (bottom) for all  (QE, Res, DIS) CC (black) and CC+NC (red) events and for only CCQE events (blue). In this case we considered 1\% bin-to-bin correlated error in the shape.}
\label{fig:sensitivity_plot_all_corrShape}
\end{figure}

We studied the sensitivity with different cuts on the range (Fig. \ref{fig:sensitivity_plot_cut} (top)) and on the number of planes (Fig. \ref{fig:sensitivity_plot_cut} (bottom)).
 Then we calculated sensitivity plot introducing bin--to--bin correlated systematic uncertainties as expressed in the covariance matrix in Eq. \ref{eq:smearing}. 
 In Fig.~\ref{fig:sensitivity_plot_all_corrNorm} is shown the sensitivity calculated considering 1\% correlated error in the normalization, while in Fig.~\ref{fig:sensitivity_plot_all_corrShape} is plotted the sensitivity calculated considering 1\% correlated error in the shape. 
It is interesting to outline that the level of the systematic normalization error affects the sensitivity region only at 
the extreme edges at small values of the mixing parameter. This is demonstrated in 
Fig.~\ref{fig:sensitivity_norma}.

\begin{figure}[htbp]
\centering
%{\includegraphics[scale=0.65]{sensitivity_new_new/figure/Sensitivity-range-all-unc_err-1}} \quad
%{\includegraphics[scale=0.65]{sensitivity_new_new/figure/Sensitivity-nplanes-all-unc_err-1}}
{\includegraphics[scale=0.36]{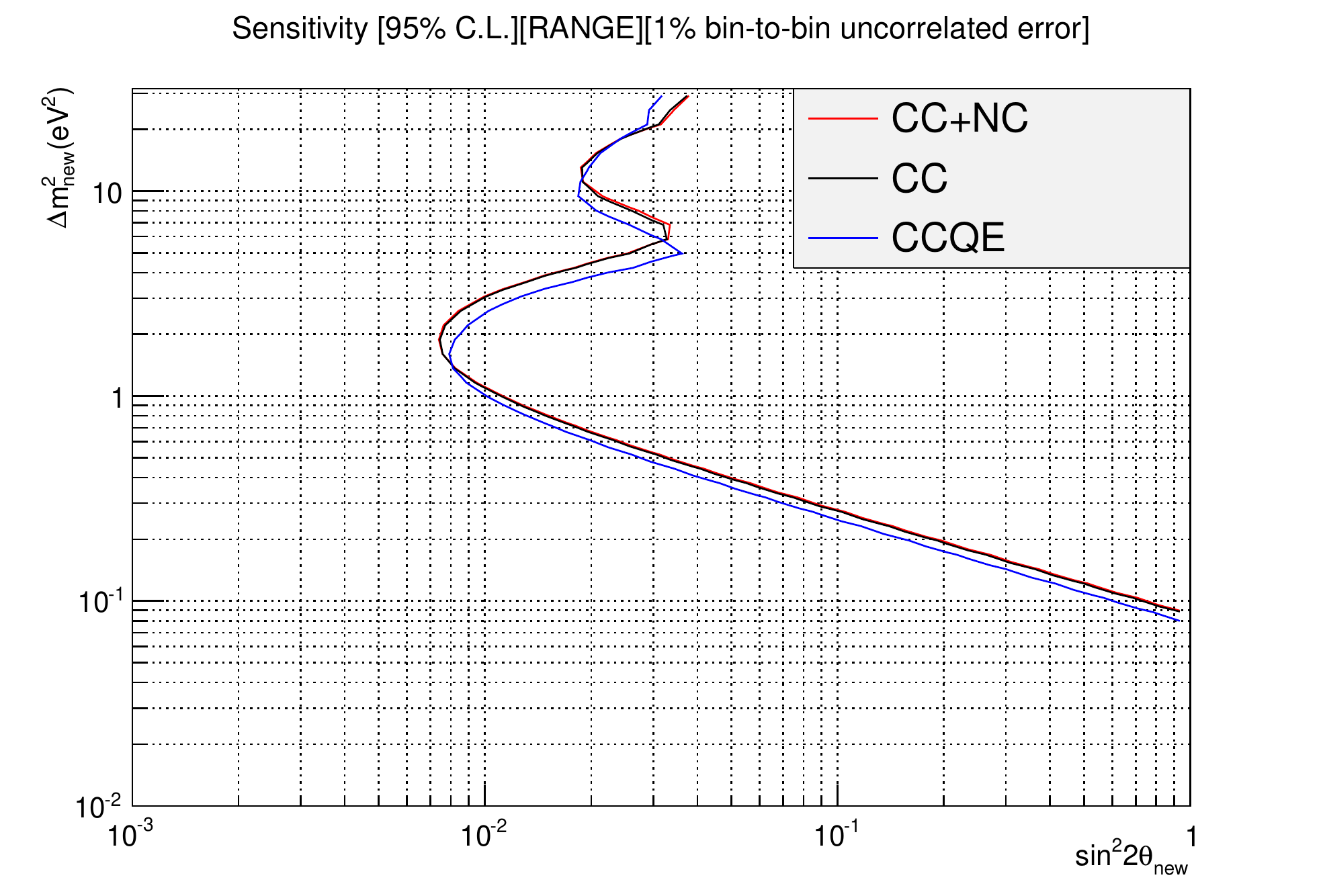}} \quad
{\includegraphics[scale=0.36]{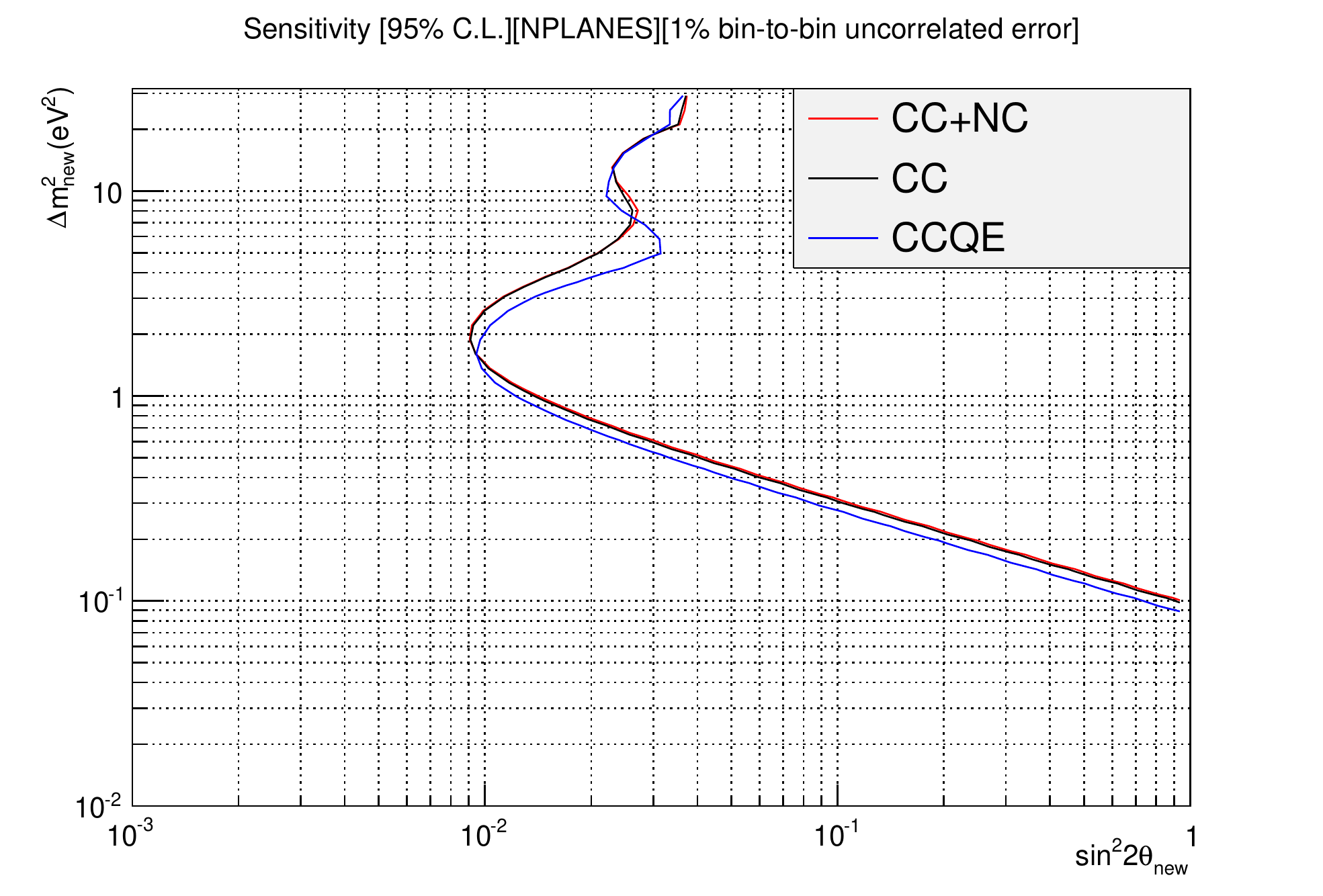}}
\caption{ 95\% CL sensitivity obtained using range (top) and number of planes (bottom) for all  (QE, Res, DIS) CC (black) and CC+NC (red) events and for only CCQE events (blue). In this case we considered only a bin--to--bin uncorrelated error of 1\%.}
\label{fig:sensitivity_plot_all_unc}
\end{figure}

\begin{figure}[htbp]
\centering
%{\includegraphics[scale=0.65]{sensitivity_new_new/figure/Sensitivity-range-all-unc_err-2}} \quad
%{\includegraphics[scale=0.65]{sensitivity_new_new/figure/Sensitivity-nplanes-all-unc_err-2}} 
{\includegraphics[scale=0.36]{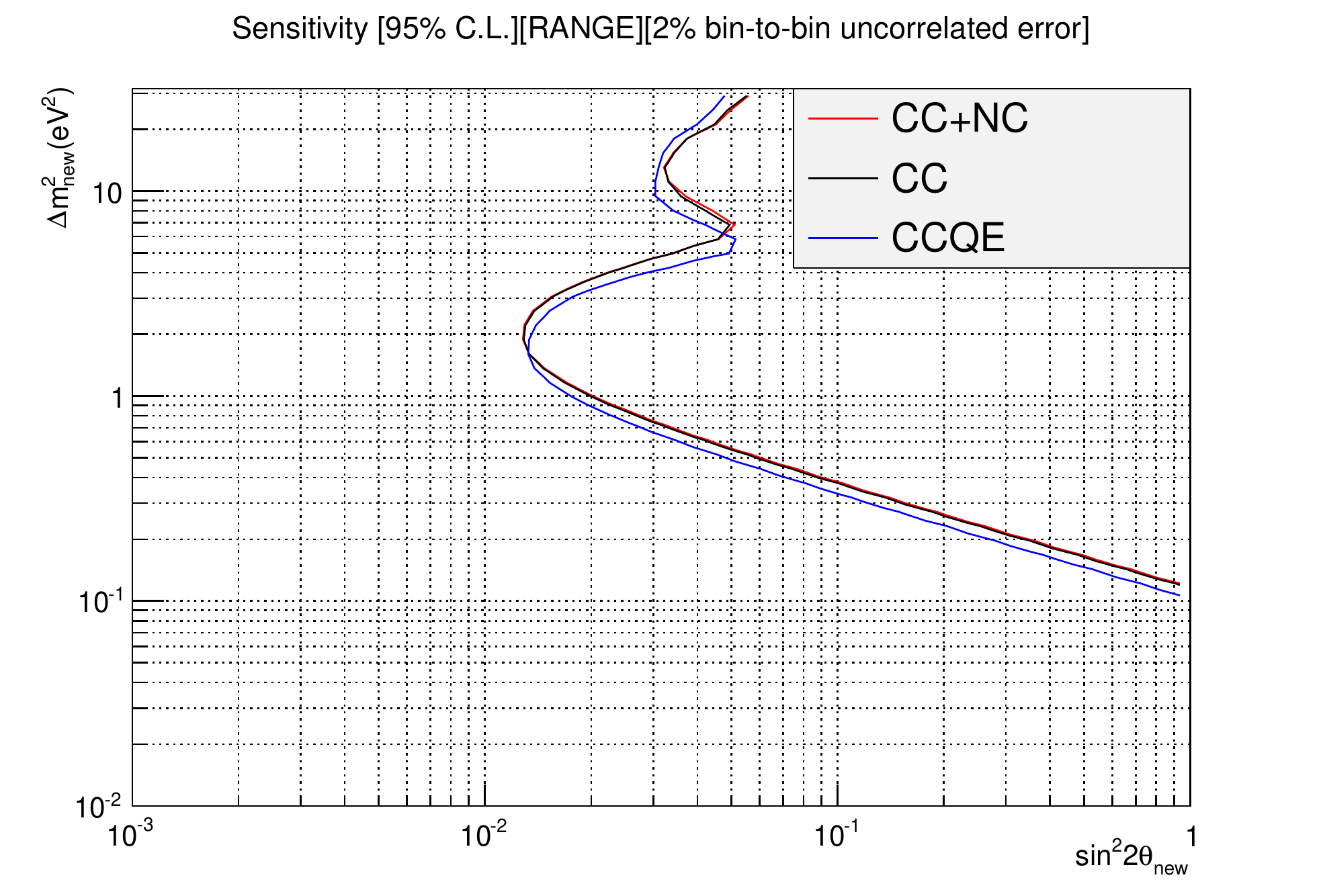}} \quad
{\includegraphics[scale=0.36]{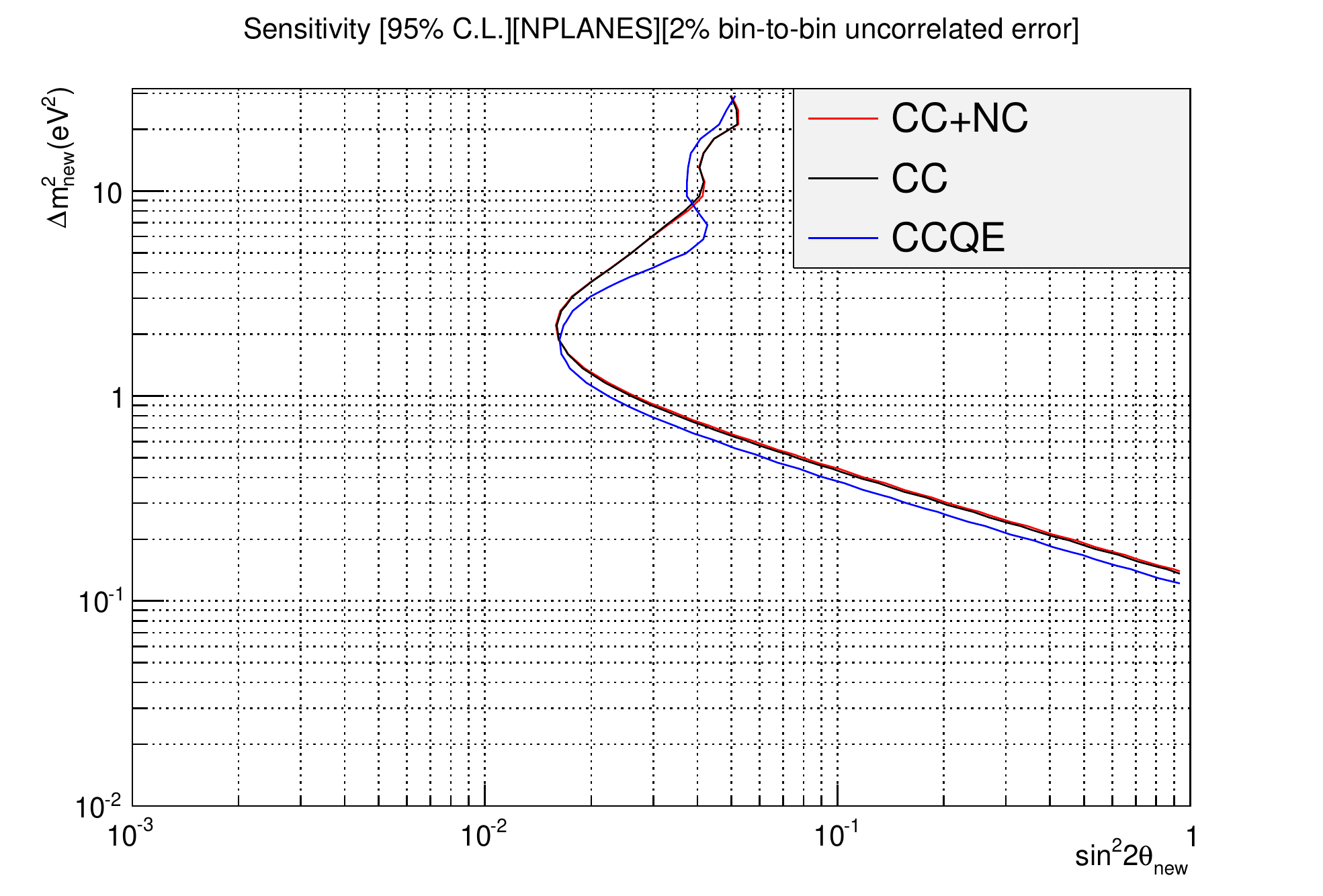}} 
\caption{ 95\% CL sensitivity obtained using range (top) and number of planes (bottom) for all  (QE, Res, DIS) CC (black) and CC+NC (red) events and for only CCQE events (blue). In this case we considered only a bin--to--bin uncorrelated error of 2\%.}
\label{fig:sensitivity_plot_all_unc2}
\end{figure}

  The sensitivity plots calculated without
   correlated errors but with uncorrelated bin--to--bin uncertainties (1\%) are shown in Fig. \ref{fig:sensitivity_plot_all_unc} for both the observables. In 
   Fig.~\ref{fig:sensitivity_plot_all_unc2} we present also sensitivity plot obtained with 2\% uncorrelated error, which in our analysis represents the upper limit for the
    sensitivity.

%%% End Laura+Matteo %%%%%%%%%%%%%%%%%%%%%%%%%%%%%%%%%%%%%%%%%%%%%%%%%%%%%%%%%%%%%

\clearpage

\subsection{Profile CLs}\label{subsec:cls}

\begin{figure}[htbp]
\centering
%{\includegraphics[scale=0.7,type=pdf,ext=.pdf,read=.pdf]{osci-bench}} 
%\includegraphics[scale=0.7]{osci-bench}
\includegraphics[scale=0.5]{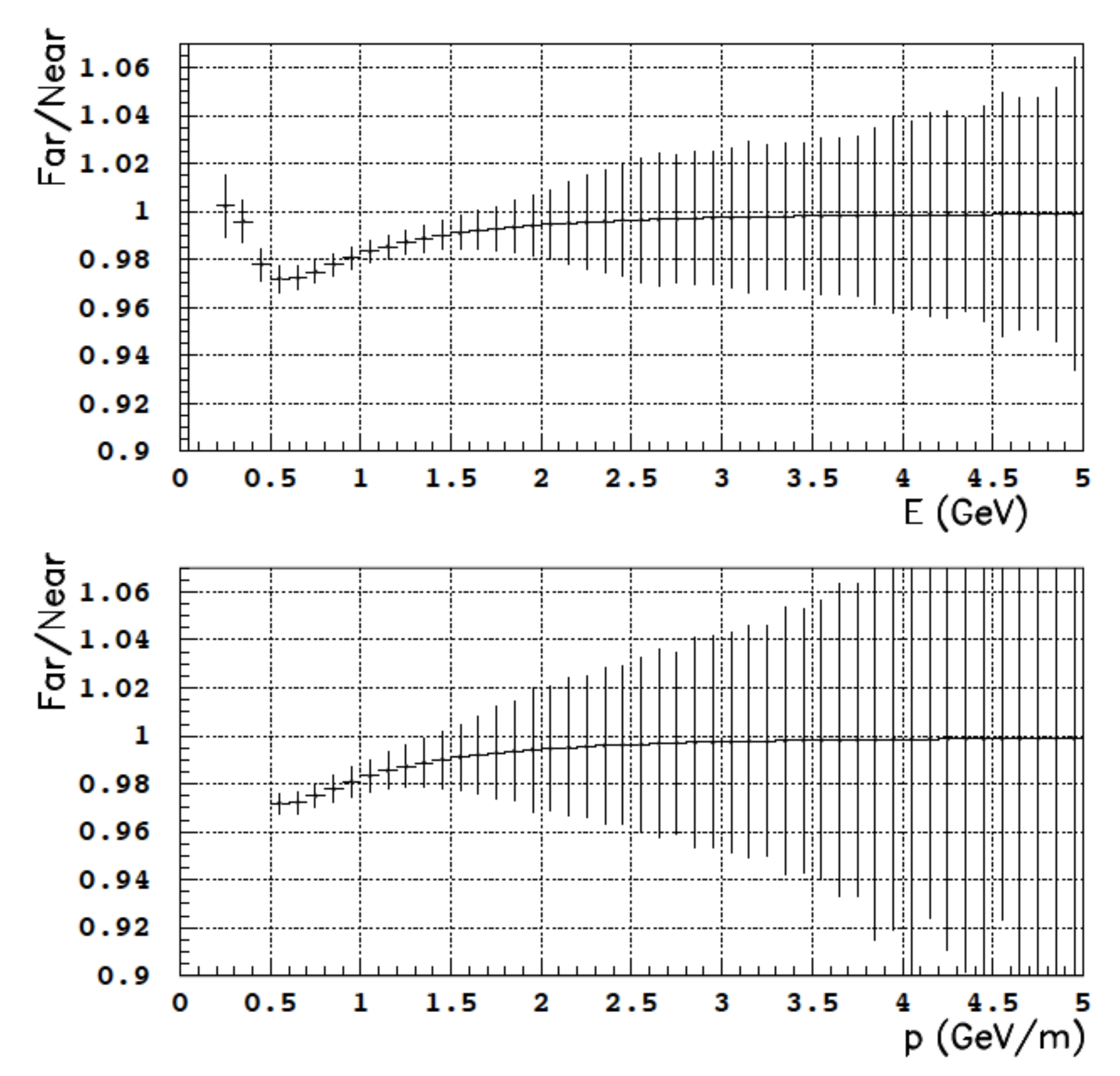}
\caption{The observable oscillation patterns as function of $E_{\nu}$ (top) and $p_{\mu}$ (bottom) for the {\em best} fit point 
of~\cite{Giu-Lav-best}, $\Delta m^{2}=1\ {\rm eV}^2$ and $\sin^{2}(2\theta)=0.03$. The error bars correspond to the full statistical
error given by 3 years of data taking at the FNAL--Booster, or $6.6\times 10^{20}$ p.o.t. A cut of 500 MeV has been 
applied on the muon momentum in the bottom plot.}
\label{fig:osci-bench}
\end{figure}

In this method we introduce a new test--statistics that depends on a {\em signal--strength} variable. We may observe that,
by looking at Eq.~\ref{eq:2flavour}, for a fixed $\Delta m^{2}_{new}$, the first factor, $\sin^{2}(2\theta_{new})$, acts as an amplification 
quantity of the configuration shape of the estimator in using. Then, we may identify the signal--strength $\mu$ with $\sin^{2}(2\theta_{new})$
and construct the estimator function:
\begin{equation}\label{eq:signal-strength}
f = \frac{1-\mu\cdot\sin^{2}(1.27\ \Delta m^{2}_{new}\ L_{Far}/E)}{1-\mu\cdot\sin^{2}(1.27\ \Delta m^{2}_{new}\ L_{Near}/E)}
\end{equation}
Whether, for example, the {\em benchmark} values $\Delta m^{2}_{new}=1\ {\rm eV}^2$ and $\sin^{2}(2\theta_{new})=0.03$ are considered, 
once the Far and Near distributions have been properly normalized, the observed oscillation shapes are depicted in
Fig.~\ref{fig:osci-bench}. 

From the pattern of Fig.~\ref{fig:osci-bench} two different likelihood functions can be constructed, one for the {\em background--only}
hypothesis and one for {\em signal $+$ background}:
\begin{gather*}
L_{bck}(data|0) = \prod_{bins}\ G(data|\alpha_i,\sigma_i) \\
L_{sign+bck}(data|\mu) = \prod_{bins}\ G(data|\alpha_i\otimes\mu,\sigma_i)
\end{gather*}
where $G$ is just the Poisson (or the Gaussian, in our case, provided the large number of events collected) distribution centered in $\alpha_i$
with dispersion $\sigma_i$,
and $\mu$
is the {\em strength}. The usual estimator $q$ can be further elaborated:
\begin{equation*}
q_0=-2\ln \frac{L(data|0)}{L(data|\mu)}
\end{equation*}

If one allows $\mu$ to be as large as possible to be observed at e.g. at a 95\% C.L., then two {\em p--values} can be computed for
the {\em background--only} and the {\em signal $+$ background} hypothesis. The test statistics becomes:
\begin{equation*}
CL_S=\frac{p_{val}(\mu\cdot signal+background)}{1-p_{val}(background)}
\end{equation*}
Finally, $\mu$ has to be adjusted until $CL_S=0.05$ is reached. Note that in our case the possibility to use the Gaussian distribution 
brings to a simple expression for $q_0$:
\begin{flalign*}
q_0 & =  -2\ln L(data|0)+2\ln L(data|\mu_{best}) \\
& = -2\sum\left (\ln\frac{1}{\sqrt{\pi}}+\ln\frac{1}{\sigma}-\frac{x-1}{2\sigma^2}\right ) \\ 
&  \quad + 2\sum\left (\ln\frac{1}{\sqrt{\pi}}+\ln\frac{1}{\sigma}-\frac{x-\mu_{best}}{2\sigma^2}\right ) \\ 
& =  \sum\frac{1}{\sigma^2} \left( (x-1)^2 - (x-\mu_{best})^2 \right) \\
& = \chi^2_0 - \chi^2_{best} 
\end{flalign*}
\noindent where for each bin it holds $\sigma=\left(\frac{N_{Far}}{N_{Near}}\cdot Rate\right)\sqrt{\frac{1}{N_{Far}}+\frac{1}{N_{Near}}}$
with obvious meaning for the symbols,
while $\mu_{best}$ is obtained by fitting over Eq.~\ref{eq:signal-strength}.
Note that this procedure is named {\em raster--scan} by~\cite{Fel-Cou} even if no $CL_S$ estimator is considered there.

\begin{figure}[h]
\centering
%{\includegraphics[scale=0.6,type=pdf,ext=.pdf,read=.pdf]{exclu-enu}} 
%\includegraphics[scale=0.6]{exclu-enu}
\vskip-2cm
\includegraphics[scale=0.6]{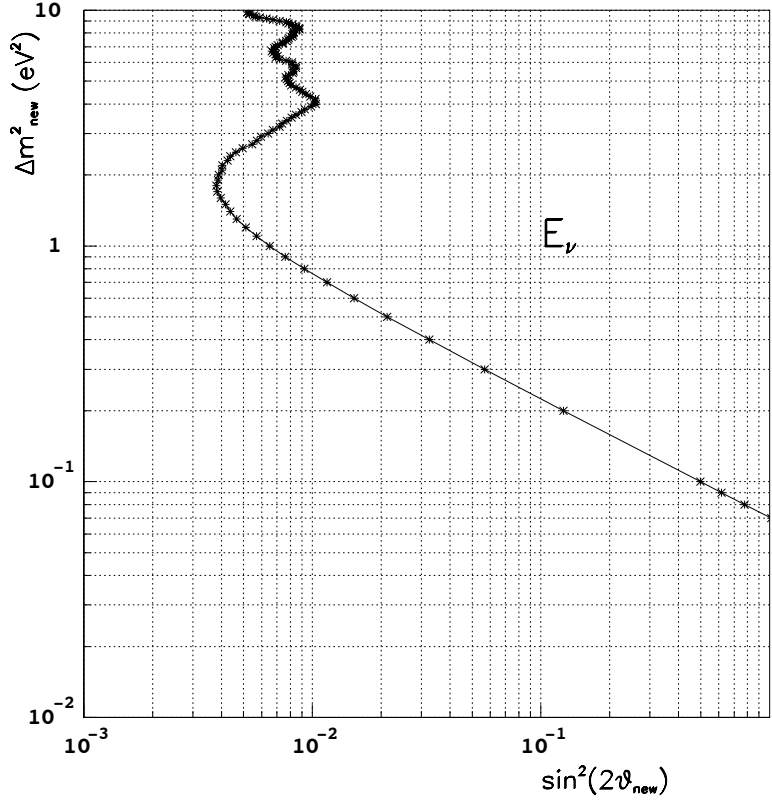}
\vskip-3cm
\caption{The sensitivity plot obtained by computing the modified raster--scan method, in a $CL_S$ framework. The energy of the 
neutrino has been evaluated via Monte Carlo in the hypothesis of CCQE events. A conservative cut of $p_{\mu}\ge\ 500$~MeV has been applied.}
\label{fig:exclu-enu}
\end{figure}

In a simplified way, for each $\Delta m^{2}$ a sensitivity limit can be obtained from the {\em p--value} of the distribution of the estimator
in Eq.~\ref{eq:signal-strength}, in the assumption of {\em background--only} hypothesis. 
The corresponding sensitivity is shown in Fig.~\ref{fig:exclu-enu}.

That procedure does not correspond to
compute the exclusion region of a signal, even if it provides confidence for it. The exclusion plot should be obtained by fully
developing the $CL_S$ procedure as outlined above. However, since we are first interested in
exploiting the sensitivity of our experiment to any oscillation pattern not compatible with the standard 3--neutrino scenario,
the above procedure provides insights to such question, and it is fully compatible with the previous two analysis and the usual neutrino analysis
found in literature.

\begin{figure}[h]
\centering
%{\includegraphics[scale=0.6,type=pdf,ext=.pdf,read=.pdf]{exclu-pmu}} 
%\includegraphics[scale=0.6]{exclu-pmu}
\vskip-3cm
\includegraphics[scale=0.6]{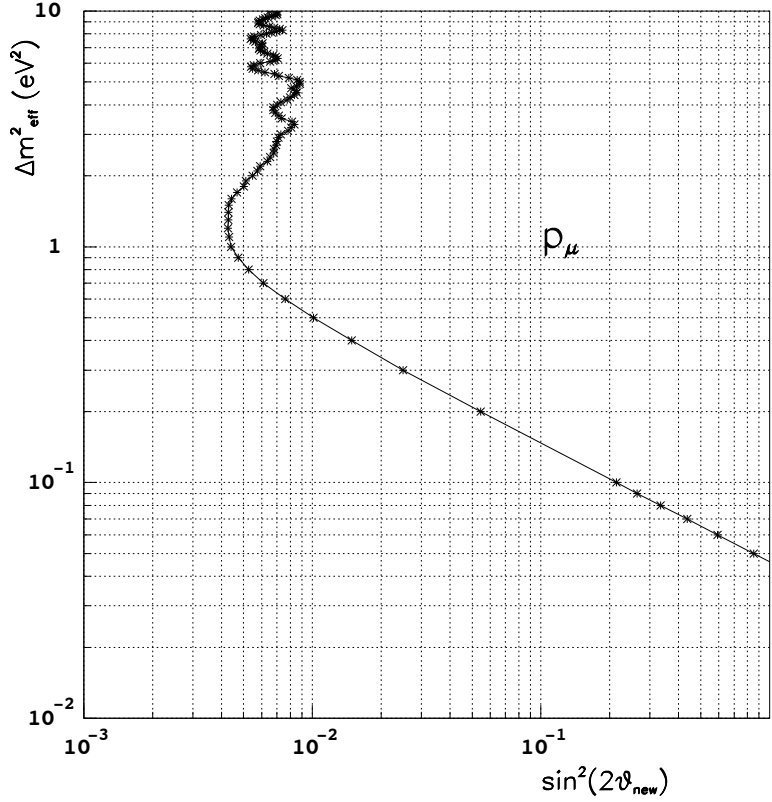}
\vskip-2cm
\caption{The sensitivity plot obtained by computing the modified raster--scan method, in a $CL_S$ framework, and by
using the reconstructed muon momentum as estimator. A conservative cut of $p_{\mu,rec}\ge\ 500$~MeV has been applied.}
\label{fig:exclu-pmu}
\end{figure}

Moreover, following the same attitude, an even more {\em aggressive} procedure can be applied. Since the deconvolution
from $p_{\mu}$ to $E_{\nu}$ introduces obviously a degeneration of the information, we may want to check whether the more
direct and measurable estimator, $p_{\mu}$, is a valuable one. In such a case Eq.~\ref{eq:signal-strength} becomes:
\begin{equation}\label{eq:signal-strength-mu}
f = \frac{1-\mu\cdot\sin^{2}(1.27\ \Delta m^{2}\ L_{Far}/p_{\mu})}{1-\mu\cdot\sin^{2}(1.27\ \Delta m^{2}\ L_{Near}/p_{\mu})}
\end{equation}
A sensitivity plot can be obtained by using $p_{\mu}$ (actually, the reconstructed muon momentum, $p_{\mu,rec}$) instead of $E_{\nu}$. 
The result is shown in Fig.~\ref{fig:exclu-pmu}.

The sensitivity plot in Fig.~\ref{fig:exclu-pmu} actually provides an ``effective'' sensitivity limit in the ``effective'' variables
$\Delta m^{2}$  and $p_{\mu,rec}$. However, comparison between Fig.~\ref{fig:exclu-enu} and Fig.~\ref{fig:exclu-pmu} demonstrates
that the ``effective'' $\Delta m^{2}$ is simply scaled--off towards higher values, not affecting the mixing angle limit.
The latter result is the best sensitivity that our experiment can achieve whether
the systematics can be limited to 1\% level, as we are confident in. A sensitivity to mixing angles in $\sin^2(2\theta_{new})$ 
below $10^{-2}$ can therefore be obtained in a large region of $\Delta m^{2}$, around the $1\ {\rm eV}^2$ scale.

\clearpage

\section{Conclusions}
Existing anomalies in the neutrino sector  may hint to the existence of  one or more additional {\em sterile} neutrino families. 
We performed a detailed study of the physics case in order to set a Short--Baseline experiment at the FNAL--Booster neutrino
beam to exploit the measurement of the charged current events. An independent measurement on \numu, complementary to
the already proposed experiments on \nue, is mandatory
to either prove or reject the existence of sterile neutrinos, even in case of null result for \nue.

The already submitted proposals based on the technology of imaging in ultra--pure cryogenic Liquid Argon (LAr)
may suffer from some experimental limitations, which we deem critical: the measurement of the muon
charge on event--by--event basis extended to the lowest achievable momentum range would be needed.
Moreover, very massive detectors are mandatory to collect a large number of events and therefore improve the disentangling of systematic effects.

The best option in terms of physics reach and funding constraints is provided by two spectrometers based on dipoles iron magnets,
at the Near and Far sites (located at 110 and 710~m from the FNAL--Booster neutrino source, respectively), 
to be eventually placed behind the LAr detectors. 

We plan to perform a full re--use of the OPERA spectrometers that will be started to be dismantled at the end of 2015.
Each site at FNAL will host a part of the two coupled OPERA magnets, base on well know technology. The spectrometers will make use of RPC detectors, already 
available, which have demonstrated their robustness and effectiveness.
The overall cost is two--fold: 1) the arrangement for the Near
and Far sites, including servicing for assembling, is substantial but it may be part of a large SBL project at FNAL; 2) the overall cost of the
experiment stays below 10 M\euro\ and it can be largely covered by in--kind hardware. The remaining expenses can be shared between
a large enough Collaboration.

\vskip 10pt

We believe to have succeeded to develop a substantial proposal that, by keeping the systematic error at the level of $1\div 2$\% 
for the measurements of the \numu interactions, will allow in 3 years of data collection to:
\begin{itemize}
	\item
measure \numu disappearance in the almost entire available momentum range ($p_{\mu}\ge 500$~MeV). This is a key information in 
rejecting/observing the anomalies over the whole expected parameter space of sterile neutrino oscillations;
	\item
measure the neutrino flux at the Near detector, in the relevant muon momentum range, which 
is decisive to keep the systematic errors at the lowest possible values;
\item measure the sign of the muon charge to
separate \numu from \nubarmu to control systematics.
\end{itemize}

A sensitivity to mixing angles in $\sin^2(2\theta_{new})$ 
below $10^{-2}$ can be obtained in a large region of $\Delta m^{2}_{new}$ around the $1\ {\rm eV}^2$ scale.

\newpage

\section*{Acknowledgements}

We are in debit to the FNAL and CERN managements for the encouragements towards the study of this proposal,
in particular to Jim Strait and Sergio Bertolucci. 
%We also thank M. Sioli for pointing out the possible use of the statistical method used in Sect.~\ref{sec:full-matrix}.

\bibliographystyle{my-h-physrev}
%\bibliography{MyBib}

\begin{thebibliography}{100}
%1
\bibitem{theta13-DB} DAYA--BAY Collaboration, F. An et al., ``{\em Observation of electron-antineutrino disappearance at Daya Bay}", 
Phys. Rev.Lett. 108, 171803 (2012), arXiv:1203.1669.
%2
\bibitem{theta13-RE} RENO Collaboration, J. Ahn et al., ``{\em Observation of Reactor Electron Antineutrino Disappearance in the RENO Experiment}", 
Phys. Rev.Lett. 108, 191802 (2012), arXiv:1204.0626.
%3
\bibitem{theta13-DC}  DOUBLE--CHOOZ Collaboration, Y. Abe et al.,``{\em Indication of Reactor $\overline{\nu}_e$ Disappearance in the Double 
Chooz Experiment}", Phys. Rev. Lett. 108, 131801 (2012), arXiv:1207.6632.
%4
\bibitem{theta13-T2} T2K Collaboration, K. Abe et al., ``{\em Indication of Electron Neutrino Appearance from an Accelerator--produced Off--axis 
Muon Neutrino Beam}", Phys. Rev.Lett. 107, 041801 (2011), arXiv:1106.2822.
%5
\bibitem{pontecorvo} B. Pontecorvo, Zh. Eksp. Teor. Fiz. 53, 1717 (1967) [Sov. Phys. JETP 26, 984 (1968)].
%6
\bibitem{whitepaper} K.N. Abazajian et al., ``{\em Light Sterile neutrinos: a White Paper}", (2012), arXiv:1204.5379.
%6bis
\bibitem{bicep2} BICEP2 Collaboration, P. A. R. Ade et al., {\em ``BICEP2 I: Detection Of B--mode Polarization at Degree Angular Scales''}, arXiv:1403.3985 [astro--ph.CO].
%6tris
\bibitem{ster-bicep2} M.~Archidiacono et al., {\em `Light sterile neutrinos after BICEP--2''}, arXiv:1404.1794 [astro--ph.CO].
%7
\bibitem{kopp-tension}
J. Kopp, P. A. N. Machado, M. Maltoni, and T. Schwetz, {\em ``Sterile neutrino oscillations: the global picture''}, 
JHEP 05, 050 (2013), arXiv:1303.3011;\\
T. Schwetz, {\em ``Status of sterile neutrino oscillations''}, Nuclear Physics B, vol. 235-236, pp. 229ﾐ235, 2013;\\
C. Giunti, M. Laveder, Y. F. Li, Q. Y. Liu, and H. W. Long, {\em ``Update of short--baseline electron neutrino and antineutrino disappearance''}, Physical Review D, vol. 86, no. 11, Article ID 113014, 2012.
%8
\bibitem{edms} M. Nessi et al., ``{\em Letter of Intent for the new CERN Neutrino Facility (CENF)}'', 
https://edms.cern.ch/nav/P:CERN-0000077383:V0/P:CERN-0000096728:V0/TAB3.
%9
\bibitem{microboone}
MicroBooNE Experiment, http://www-microboone.fnal.gov/.
%10
\bibitem{LAr1-ND}
LAr1--ND Collaboration, C. Adams et al., ``{\em LAr1-ND: Testing Neutrino Anomalies with Multiple LAr TPC Detectors at Fermilab}'', P--1053, 31 December 2013 
and arXiv:1309.7987v3.
%11
\bibitem{ICARUSFNAL}
ICARUS Collaboration, M. Antonello et al., ``{\em ICARUS at FNAL}'' arXiv:1312.7252.
%12
\bibitem{CDHS}
CDHS Collaboration, F. ~Dydak {\em et al.}, 
{\em ``A Search for $\nu_{\mu}$ Oscillations in the $\Delta m^2$ range 0.3$\div$90 $eV^2$"},
Phys. Lett. {\bf B134}, 281 (1984).
%13
\bibitem{mini-mu} MiniBooNE Collaboration, A. A. Aguilar--Arevalo et al.,  ``{\em A search for muon neutrino and antineutrino disappearance in MiniBooNE}'', 
Phys. Rev. Lett. 103, 061802 (2009), arXiv:0903.2465.
%14
\bibitem{mini-sci-mu1} MiniBooNE and SciBooNE Collaborations,, K.~B.~M.~Mahn et al., ``{\em Dual baseline search for muon neutrino disappearance at $0.5 {\rm eV}^2 <
\Delta m^2 < 40 {\rm eV}^2$}'',  Phys. Rev. D85, 032007 (2012), arXiv:1106.5685.
%15
\bibitem{mini-sci-mu2} MiniBooNE and SciBooNE Collaborations, G. Cheng et al., 
``{\em Dual baseline search for muon antineutrino disappearance at $0.1$ eV$^2$ $< \Delta m^2 < 100$ eV$^2$}'', 
Phys.Rev. D86, 052009 (2012), arXiv:1208.0322.
%16
\bibitem{minos} MINOS Collaboration, P. Adamson et al., ``{\em Search for sterile neutrino mixing in the MINOS long--baseline experiment}'',
Phys.Rev. D81, 052004 (2010), arXiv:1001.0336.
%17
\bibitem{ccfr} CCFR Collaboration, I. E. Stockdale and A. Bodek, F. Borcherding, N. Giokaris, K. Lang, K. {\em et al}, 
{\em ``Limits on Muon Neutrino Oscillations in the Mass Range $55\ eV^2 < \Delta m^2 < 800\ eV^2 $''},
Phys.Rev.Lett. {\bf 52} 1384 (1984).
%18
\bibitem{nessie} NESSiE Collaboration, P. Bernardini et al., ``{\em Prospect for Charge Current Neutrino Interactions Measurements at the CERN--PS}'', 
SPSC--P--343 (2011), arXiv:1111.2242.
%19
\bibitem{larnessie} ICARUS and NESSiE Collaborations, M. Antonello et al., ``{\em Search for anomalies from neutrino and anti--neutrino oscillations at 
$\Delta m^2 \sim 1$ eV$^2$ with muon spectrometers and large LArﾐ-TPC imaging detectors}'', SPSC--P--347 (2012), arXiv:1203.3432.
%20
\bibitem{stancoetal}
L. Stanco et al., ``{\em An Appraisal of Muon Neutrino Disappearance at Short Baseline}, AHEP 2013 (2013) ID 948626, arXiv:1306.3455v2.
%21
\bibitem{bopera}{The OPERA collaboration, R. Acquafredda et al. {\em ``The OPERA experiment in the CERN 
to Gran Sasso neutrino beam''}, JINST {\bf 4} 4018 (2009)}.
%22
\bibitem{G4BNBflux} 
J. R. Sanford and C. L. Wang (1967), BNL Internal Report, BNL--11479;\\
MiniBooNE Collaboration, A.~A.~Aguilar--Arevalo et al., {\em ``The Neutrino Flux prediction at MiniBooNE''}, 
15 May 2009. arXiv:0806.1449v2, arXiv:hep--ex/0601022v1. 13 Jan. 2006.
%22
%\bibitem{Kopp}\emph{Accelerator Neutrino Beams}. S. E. Kopp. arXiv:hep-ex/0609129v1. 14 Sept. 2006.
%23
\bibitem{numikopp} S. E. Kopp., {\em ``Neutrino spectra and uncertainties for MINOS''}, arXiv:hep--ex0712.1280. Dec. 2007.
%23
%\bibitem{SCIB} \emph{Bringing the SciBar Detector to the Booster Neutrino Beam}. A.~A.~Aguilar-Arevalo {\it et al.} (SciBooNE Coll.)
%24
\bibitem{fluka}
G.~Battistoni {\em et al.},
{\em ``The FLUKA code: Description and Benchmarking''},
AIP Conf.\ Proc.\  {\bf 896}, 31 (2007). 
%25
\bibitem{ref_GEN} C.Andreopoulos {\em et al.}, 
{\em ``The GENIE Neutrino Monte Carlo Generator}, Nucl. Instrum. Meth. {\bf A614}, 87 (2010). 
%26
\bibitem{pdg} D.~E.~Groom, N.~V.~Mokhov, and S.~Striganov, \\ http://pdg.lbl.gov/2012/AtomicNuclearProperties/adndt.pdf
%27
\bibitem{deuterium} [S.~J.~Barish et al.,  {\em ``Study of Neutrino Interactions in Hydrogen and Deuterium: Inelastic Charged Current Reactions''}, 
Phys. Rev. D 19, 2521 (1979);\\
N.~J.~Baker et al., {\em `Total cross sections for \numu-n and \numu--p charged--current interactions in the 7--foot bubble chamber''},
Phys. Rev. D 25, 617 (1982).
%28
\bibitem{bib:minerva} MINERvA Collaboration, FERMILAB--DESIGN--2006-01.
%29
\bibitem{Nakajima:2010fp}
SciBooNE Collaboration, Y.~Nakajima et al.,
{\em ``Measurement of inclusive charged current interactions on carbon in a few--GeV neutrino beam''},
  Phys.\ Rev.\ D {\bf 83} (2011) 012005,  arXiv:1011.2131.
%30
\bibitem{Kurimoto:2009wq}
SciBooNE Collaboration, Y.~Kurimoto et al.,
 {\em ``Measurement of Inclusive Neutral Current Neutral $\pi^0$ Production on Carbon in a Few--GeV Neutrino Beam''},
  Phys.\ Rev.\ D {\bf 81} (2010) 033004,  arXiv:0910.5768.
%31
\bibitem{bib:opera-sc}
A. Bergnoli {\em et al.},  {\em `The OPERA Spectrometer Slow Control System"}, IEEE
Trans. Nucl. Sci. {\bf 55} 349 (2008).
%32
\bibitem{bib:origin} R. Santonico, R. Cardarelli, {\em ``Development of Resistive Plate Counters''}, Nucl. Instr. and Meth. {\bf A187} 377 (1981). \\
                     R. Santonico, R. Cardarelli, {\em ``Development of Resistive Plate Counters''}, Nucl. Instr. and Meth. {\bf A263} 20 (1988).
%33
\bibitem{bib:cr} P.K.F. Grieder {\em ``Cosmic Rays at Earth}'', Elsevier (Amsterdam, 2011).
%34
\bibitem{lvds} E. Balsamo, et al., {\em ``The OPERA RPCs front end electronics A novel application of LVDS line receiver as low cost discriminator ''}, JINST 7 (2012) P11007.
%35
\bibitem{Fel-Cou}
G. J. Feldman and R. D. Cousins, {\em ``Unified approach to the classical statistical analysis of small signals''},
 Phys. Rev. D 57, 3873 (1998), http://arxiv.org/abs/physics/9711021v2.
%36
 \bibitem{BNB} http://www-boone.fnal.gov/for\_physicists/data\_release/flux/\\pospolarity\_fluxes.dat.
%37 
 \bibitem{pmns}  B.~Pontecorvo, Sov. Phys. JETP 26, 984 (1968);\\
Z.~Maki, M.~Nakagawa, and S.~Sakata, Prog. Theor. Phys. 28, 870 (1962).
%38
\bibitem{winter} W. Winter,  ``{\em Optimization of a Very Low Energy Neutrino Factory for the Disappearance Into Sterile Neutrinos}'', 
Phys. Rev. D85, 113005 (2012), arXiv:1204.2671.
%39
\bibitem{GLoBES} http://www.mpi-hd.mpg.de/personalhomes/globes/index.html.% P. Huber \textit{et al}., 12 Jan 2012,
%40
\bibitem{covM} K. Mahn, \textit{A search for muon neutrino and antineutrino disappearance with the Booster Neutrino Beam}, Columbia University 2009.
%41
\bibitem{covM2} W. Huelsnitz, \textit{Handling Uncertainties in IceCube with Covariance Matrices}, 26 April 2011.
%42
\bibitem{Giu-Lav-best}
C. Giunti, M. Laveder, Y.F. Li, H.W. Long, {\em ``Pragmatic View of Short-Baseline Neutrino Oscillations''},
Phys.Rev. D88 (2013) 073008, arXiv:1308.5288.

\end{thebibliography}

\end{document}